\documentclass{aa}  

\usepackage{graphicx}
\usepackage{natbib}

\usepackage{array}
\usepackage{gensymb}
\usepackage{xfrac}

\usepackage{txfonts}

\usepackage{hyperref}
\hypersetup{
    colorlinks=true,
    citecolor=blue,
    }
    
\usepackage[capitalise]{cleveref}
\usepackage{appendix}
\usepackage[table,dvipsnames]{xcolor}
\usepackage{xspace}
\usepackage{pdflscape}
\usepackage{rotating}
\usepackage{diagbox}
\usepackage{makecell}
\usepackage{xspace}

\AtBeginDocument{%
    \crefname{section}{Sect.}{Sects.}%
    }

\newcommand\numberthis{\addtocounter{equation}{1}\tag{\theequation}}

\newcommand{\teff}{$T_{\rm eff}$\xspace}
\newcommand{\logg}{$\log g$\xspace}
\newcommand{\yhe}{$x_{\rm He}$\xspace}

\newcommand{\yO}{$n_{\rm O}$\xspace}
\newcommand{\yN}{$n_{\rm N}$\xspace}
\newcommand{\yC}{$n_{\rm C}$\xspace}

\newcommand{\vsini}{$\varv_{\rm eq} \sin i$\xspace}
\newcommand{\vmicro}{$\varv_{\rm micro}$\xspace}
\newcommand{\vmacro}{$\varv_{\rm macro}$\xspace}
\newcommand{\vwindturb}{$\varv_{\rm windturb}$\xspace}
\newcommand{\vclonset}{$\varv_{\rm cl, start}$\xspace}
\newcommand{\fcl}{$f_{\rm cl}$\xspace}
\newcommand{\fic}{$f_{\rm ic}$\xspace}
\newcommand{\fvel}{$f_{\rm vel}$\xspace}
\newcommand{\rv}{$\varv_{\rm rad}$\xspace}
\newcommand{\vinf}{$\varv_{\infty}$\xspace}

\newcommand{\mdot}{$\dot{M}$\xspace}
\newcommand{\angstrom}{\mbox{\normalfont\AA}\xspace}
\newcommand{\logll}{$\log L / {\rm L}_\odot$\xspace}
\newcommand{\loglx}{$L_X/{\rm L_{bol}} \approx 10^{-7}$\xspace}
\newcommand{\vorosity}{vorosity\xspace}
\newcommand{\Msun}{M$_{\odot}$\xspace}
\newcommand{\cmfgen}{{\sc cmfgen}\xspace}

\newcommand{\fw}{F{\sc astwind}\xspace}
\newcommand{\Fw}{F{\sc astwind}\xspace}
\newcommand{\PyGA}{{\sc Kiwi-GA}\xspace}
\newcommand{\pyGA}{{\sc Kiwi-GA}\xspace}
\newcommand{\snr}{{\sc S/N}\xspace}
\newcommand{\chisq}{$\chi^2$\xspace}
\newcommand{\chisred}{$\chi^2_{\rm red}$\xspace}

\defcitealias{2021A&A...648A..36B}{Bj\"ork21}
\defcitealias{2001A&A...369..574V}{Vink01}
\defcitealias{2012A&A...537A..37M}{Muij12}
\defcitealias{2018A&A...612A..20K}{Krti\v{c}18}
\newcommand{\krtic}{\citetalias{2018A&A...612A..20K}\xspace}
\newcommand{\bjork}{\citetalias{2021A&A...648A..36B}\xspace}
\newcommand{\vink}{\citetalias{2001A&A...369..574V}\xspace}

\newcommand{\hei}{He~{\sc i}\xspace}
\newcommand{\heii}{He~{\sc ii}\xspace}
\newcommand{\niii}{N~{\sc iii}\xspace}

\newcommand{\niv}{N~{\sc iv}\xspace}
\newcommand{\nv}{N~{\sc v}\xspace}
\newcommand{\ov}{O~{\sc v}\xspace}

\newcommand{\halpha}{H$\alpha$\xspace}
\newcommand{\hgamma}{H$\gamma$\xspace}
\newcommand{\heiiline}{He{\sc~ii}~$\lambda$4686\xspace}
\newcommand{\cuvline}{C{\sc~iv}~$\lambda$1165-C{\sc iii}~$\lambda$1170\xspace}
\newcommand{\cuvivline}{C{\sc~iv}~$\lambda$1165\xspace}
\newcommand{\cuviiiline}{C{\sc~iii}~$\lambda$1170\xspace}
\newcommand{\nvuvline}{N{\sc~v}~$\lambda$1240\xspace}
\newcommand{\siivline}{Si{\sc~iv}~$\lambda\lambda$1394-1402\xspace}
\newcommand{\oivline}{O{\sc~iv}~$\lambda$1340\xspace}
\newcommand{\ovline}{O{\sc~v}~$\lambda$1371\xspace}
\newcommand{\CIVline}{C{\sc~iv}~$\lambda\lambda$1548-1551\xspace}
\newcommand{\heiiuvline}{He{\sc~ii}~$\lambda$1640\xspace}
\newcommand{\nivuvline}{N{\sc~iv}~$\lambda$1718\xspace}
\newcommand{\nvopt}{N{\sc~v}~$\lambda\lambda$4604-4620\xspace}

\begin{document} 


\title{The R136 star cluster dissected with \textit{Hubble Space Telescope}/STIS. III. The most massive stars and their clumped winds}

\titlerunning{The most massive stars and their clumped winds}

\author{Sarah A. Brands      \inst{1}
           \and 
           Alex de Koter  \inst{1,2}
           \and 
           Joachim M. Bestenlehner \inst{3}
           \and 
           Paul A. Crowther \inst{3}
           \and 
           Jon O. Sundqvist \inst{2}
           \and 
           Joachim Puls \inst{4}
           \and 
           Saida M. Caballero-Nieves \inst{5}
            \and
           Michael Abdul-Masih \inst{2,6}
           \and 
           Florian A. Driessen \inst{2}
           \and 
           Miriam Garc\'{i}a \inst{7}
           \and 
          Sam Geen \inst{1}
          \and 
           G\"{o}tz Gr\"{a}fener \inst{8}
           \and 
           Calum Hawcroft \inst{2}
           \and 
           Lex Kaper  \inst{1}           
           \and 
           Zsolt Keszthelyi \inst{1}           
           \and
           Norbert Langer \inst{8}
           \and 
           Hugues Sana \inst{2}
           \and 
           Fabian R.N. Schneider \inst{9,10}
            \and 
          Tomer Shenar \inst{1}
          \and
           Jorick S. Vink \inst{11}
          }

\institute{ 
        Anton Pannekoek Institute for Astronomy, University of Amsterdam, 1090 GE Amsterdam, The Netherlands
          \email{s.a.brands@uva.nl}
    \and 
          Institute of Astronomy,
          KU Leuven, 
          Celestijnenlaan 200D, 
          3001 Leuven, Belgium
    \and 
          Department of Physics and Astronomy
          University of Sheffield,
          Sheffield, S3~7RH, 
          UK
    \and 
          LMU M\"unchen, Universit\"atssternwarte, Scheinerstr. 1, 81679 M\"unchen, Germany
    \and 
        Department of Aerospace, Physics and Space Sciences, Florida Institute of Technology, 150 W. University Boulevard, Melbourne, FL 32901, USA
    \and 
        European Southern Observatory, Alonso de Córdova 3107, Vitacura, Santiago, Chile
    \and 
          Centro de Astrobiolog\'ia, CSIC-INTA. Crtra. de Torrej\'on a Ajalvir km 4. 28850 Torrej\'on de Ardoz (Madrid), Spain
    \and 
         Argelander-Institut f\"ur Astronomie, Universit\"at Bonn, Auf dem H\"ugel 71, 53121 Bonn, Germany
    \and 
            Heidelberger Institut f\"ur Theoretische Studien, Schloss-Wolfsbrunnenweg 35, 69118 Heidelberg, Germany 
    \and 
            Astronomisches Rechen-Institut, Zentrum f{\"u}r Astronomie der Universit{\"a}t Heidelberg, M{\"o}nchhofstr.\ 12-14, 69120 Heidelberg, Germany    
    \and 
            Armagh Observatory, College Hill, Armagh BT61 9DG, UK
             }

   \date{}


   \abstract
  {The star cluster R136 inside the Large Magellanic Cloud hosts a rich population of massive stars, including the most massive stars known. 
   The strong stellar winds of these very luminous stars impact their evolution and the surrounding environment. 
   We currently lack detailed knowledge of the wind structure that is needed to quantify this impact. 
   }
   {To observationally constrain the stellar and wind properties of the massive stars in R136, in particular the wind-structure parameters related to wind clumping. 
   }
   {We simultaneously analyse optical and ultraviolet spectroscopy of 53 O-type and 3 WNh-stars using the \fw model atmosphere code and a genetic algorithm. 
   The models account for optically thick clumps and effects related to porosity and velocity-porosity, as well as a non-void interclump medium. 
   }
   {We obtain stellar parameters, surface abundances, mass-loss rates, terminal velocities and clumping characteristics and compare these to theoretical predictions and evolutionary models. 
   The clumping properties include the density of the interclump medium and the velocity-porosity of the wind.
   For the first time, these characteristics are systematically measured for a wide range of effective temperatures and luminosities. 
   }
   {We confirm a cluster age of 1.0-2.5~Myr and derive an initial stellar mass of $\geq 250$~\Msun for the most massive star in our sample, R136a1. 
   The winds of our sample stars are highly clumped, with an average clumping factor of $f_{\rm cl} = 29 \pm 15$. 
   We find tentative trends in the wind-structure parameters as a function of mass-loss rate, suggesting that the winds of stars with higher mass-loss rates are less clumped. 
   We compare several theoretical predictions to the observed mass-loss rates and terminal velocities and find that none satisfactorily reproduces both quantities. The prescription of Krti\v{c}ka \& Kub\'at (2018) matches best the observed mass-loss rates. 
   }
   \authorrunning{Brands et al.}

\maketitle


\section{Introduction}

The star cluster R136 inside the Large Magellanic Cloud (LMC) hosts some of the richest populations of high-mass stars in the local universe. 
Nine stars within this cluster have masses around or exceeding $100 \ {\rm M}_{\odot}$ and a few even surpass $\sim 150 \ {\rm M}_{\odot}$ \citep{2010MNRAS.408..731C,2016MNRAS.458..624C,Bestenlehner2020}. 
These massive, very luminous, hot stars play an important role in the universe. 
They strongly influence their surroundings through direct injection of mass, momentum and energy from winds \citep{Weaver1977}, radiation pressure \citep{Mathews1967}, and thermal expansion caused by photoionisation from extreme ultraviolet photons \citep{KahnF.D.1954,SpitzerLyman1978}. This feedback can act to both disperse star-forming clouds \citep[e.g.][]{Dale2014,Kim2018} and cause compressive flows that lead to the formation of new stars \citep[e.g.][]{Inutsuka2015,Rahner2018,Fujii2021}. The mass loss rates and terminal velocities of winds have a direct impact on the ability for stars to regulate their environment.
The deposition rates of stellar wind energy and ionising photons from massive stars are important quantities in driving the multi-phase structure of the interstellar medium and the regulation of future star formation.
Furthermore, massive stars end their lives in supernovae, thereby enriching the interstellar medium with newly formed chemical elements, and leaving behind compact remnants such as black holes \citep[see, e.g., ][for a review]{2009ARA&A..47...63S,2012ARA&A..50..107L}. 
Moreover, with their high masses and at half solar metallicity, the stars in R136 are close observable counterparts to the first stars, that are estimated to have a characteristic mass of tens to hundreds solar masses \citep[see, e.g.,][and references therein]{2014ApJ...781...60H,2015MNRAS.448..568H,2020ApJ...892L..14S,2021MNRAS.508.4175C,2021MNRAS.508.6193P}. 

R136 is residing inside the Tarantula Nebula or 30 Doradus. 
This nearby, unobscured, large and intrinsically very bright star forming region \citep{1984ApJ...287..116K, 2013A&A...558A.134D}, hosting many hundreds of massive stars ($M \gtrsim 8$~\Msun), resembles giant starbursts observed in distant galaxies \citep{2009MNRAS.399.1191C,2017Msngr.170...40C}. 
The massive star content of the Tarantula Nebula was studied in great detail in the VLT Flames Tarantula Survey \citep[VFTS, e.g.,][]{2011A&A...530A.108E,2020Msngr.181...22E,2014A&A...570A..38B,2018Sci...359...69S}, however, due to severe crowding, the central core of the R136 cluster was not part of the observing campaign. 
With the spatial resolution of the Hubble Space Telescope (HST), individual stars in the R136 core can be resolved. This was employed by \citet[][]{2016MNRAS.458..624C}, who used HST to collect optical and UV spectroscopy of the cluster, hereby complementing the VFTS survey and extending the coverage to the most massive stars. Focussing on the UV spectroscopy of this dataset, \citet[][]{2016MNRAS.458..624C} derive a cluster age of $1.5\pm^{0.3}_{0.7}$ Myr, and find that the \heiiuvline emission is completely dominated by stars with initial masses $\gtrsim 100$~\Msun. \citet{Bestenlehner2020} focus on the optical spectra and derive detailed spectral parameters for all sources. Their findings include a top-heavy initial mass function (IMF) for massive stars in the cluster, and a strong helium enrichment for the most luminous stars.  

In order to understand massive star evolution, it is key to know the mass-loss rates of these very massive stars. Moreover, by calibrating theoretical models with observed mass-loss rates and stellar properties, we can improve future large-scale studies of stellar feedback, and hence obtain a more complete picture of how stars shape our universe. 
For very massive stars the effects of mass-loss become especially important, as the rates generally increase with luminosity and thus with mass \citep[e.g.,][]{2008A&ARv..16..209P,2011A&A...531A.132V,2015ASSL..412...77V}. Moreover, their large convective cores ensure that they evolve close to homogeneously, diminishing the relative effect of other processes such as rotational mixing and magnetic fields \citep{2013MNRAS.433.1114Y,2015A&A...573A..71K,2019A&A...625A.104R}. Unfortunately, due to a lack of empirical constraints and proximity to the Eddington limit, the mass-loss rates in this regime are uncertain  \citep[e.g.,][]{2012ARA&A..50..107L}. 
Furthermore, obtaining accurate empirical mass-loss rates is hampered by the presence of small-scale inhomogeneities in the wind, also called `clumps'  \citep[see e.g.,][for a review]{2008A&ARv..16..209P}. 
The origin of this wind structure, or so-called `wind clumping', is theoretically attributed to the line-deshadowing instability (LDI), an inherent property of the line-force that drives the winds of these stars \citep[e.g.,][and references therein]{1984ApJ...284..337O,1985ApJ...299..265O}. 

Since the wind clumping determines how our diagnostics respond to mass-loss rates, it is imperative to take it into account properly when studying massive star winds. 
The simplest approach to account for wind clumping in diagnostic models is to assume that the out-flowing gas is concentrated in clumps that are small and rarefied enough so that they stay optically thin, and that the interclump medium is void \citep[e.g.,][]{1998A&A...335.1003H,1999ApJ...519..354H,2006A&A...454..625P}. 
For O-stars, this `optically thin clumping' or `micro-clumping' approach leads to a downward revision of empirical mass-loss rates compared to the assumption of a smooth wind for processes that are dependent on the square of the wind density, such as the formation of the \halpha line, but can also affect lines indirectly due to changes in the ionisation/excitation equilibrium \citep[see e.g.,][and references therein]{2008A&ARv..16..209P}.
If clumps become optically thick (for the considered process), the clumping affects diagnostics in a more complicated way. In this case, light can be blocked by clumps, but can also leak through porous channels in the wind and in this way escape without interacting \citep{1998ApJ...494L.193S,2000ApJ...532L.137S,2004ApJ...616..525O}. 
These velocity-porosity or `vorosity' effects impact mostly resonance lines; neglecting these phenomena can lead to an underestimation of mass-loss rates \citep[e.g.,][]{2006ApJ...637.1025F,2007A&A...476.1331O,2011A&A...528A..64S,2013A&A...559A.130S}. In order to obtain reliable mass-loss rate measurements it is thus essential to consider all the aforementioned effects. 
To date, only one sample of O4-O7.5 supergiants was studied using an optically thick clumping description in a model atmosphere code \citep{Hawcroft21}. 

In this paper, we reanalyse the R136 sample of \citet[][]{2016MNRAS.458..624C} and \citet[][]{Bestenlehner2020}, but now combine the optical and UV spectroscopy, allowing us to study in detail 
the mass-loss rates and wind structure. For the wind structure, we assume the two-component formalism of \citet{2014A&A...568A..59S} implemented in \fw \citep{2018A&A...619A..59S}, allowing for optically thick clumps and thus including the effects of porosity, velocity-porosity and a non-void interclump medium. 
This will yield the most accurate mass-loss rate measurements possible with current model atmosphere codes, and furthermore will allow us for the first time to investigate wind structure for a wide range of stellar properties. 

The remainder of this paper is structured as follows. We start by presenting the R136 sample and our dataset in \Cref{sec:sample_data}. 
In \cref{sec:method} we lay out our methodology. Here, we introduce our fitting algorithm \pyGA and describe the model atmosphere code \Fw. In particular, we emphasise the parameterisation of the wind structure parameters (\cref{sec:method_structure}). The results of our analysis are presented in \cref{sec:results}. This section is concluded with several tests of robustness (\cref{sec:robust}). We discuss our results in the context of theoretical predictions and evolutionary models in \cref{sec:discussion};  \Cref{sec:mdot_vs_theory} is dedicated to mass-loss rates and wind momentum; \Cref{sec:windstructuretrends} to the potential trends that we observe in the wind structure parameters, and in \cref{dis:evolution} we consider our findings in the context of stellar evolution. 
Two methods for measuring terminal velocities are compared in \cref{dis:terminal}. We conclude with a summary and outlook (\cref{sec:conclusion_outlook}).

\begin{figure}
    \includegraphics[width=0.50\textwidth]{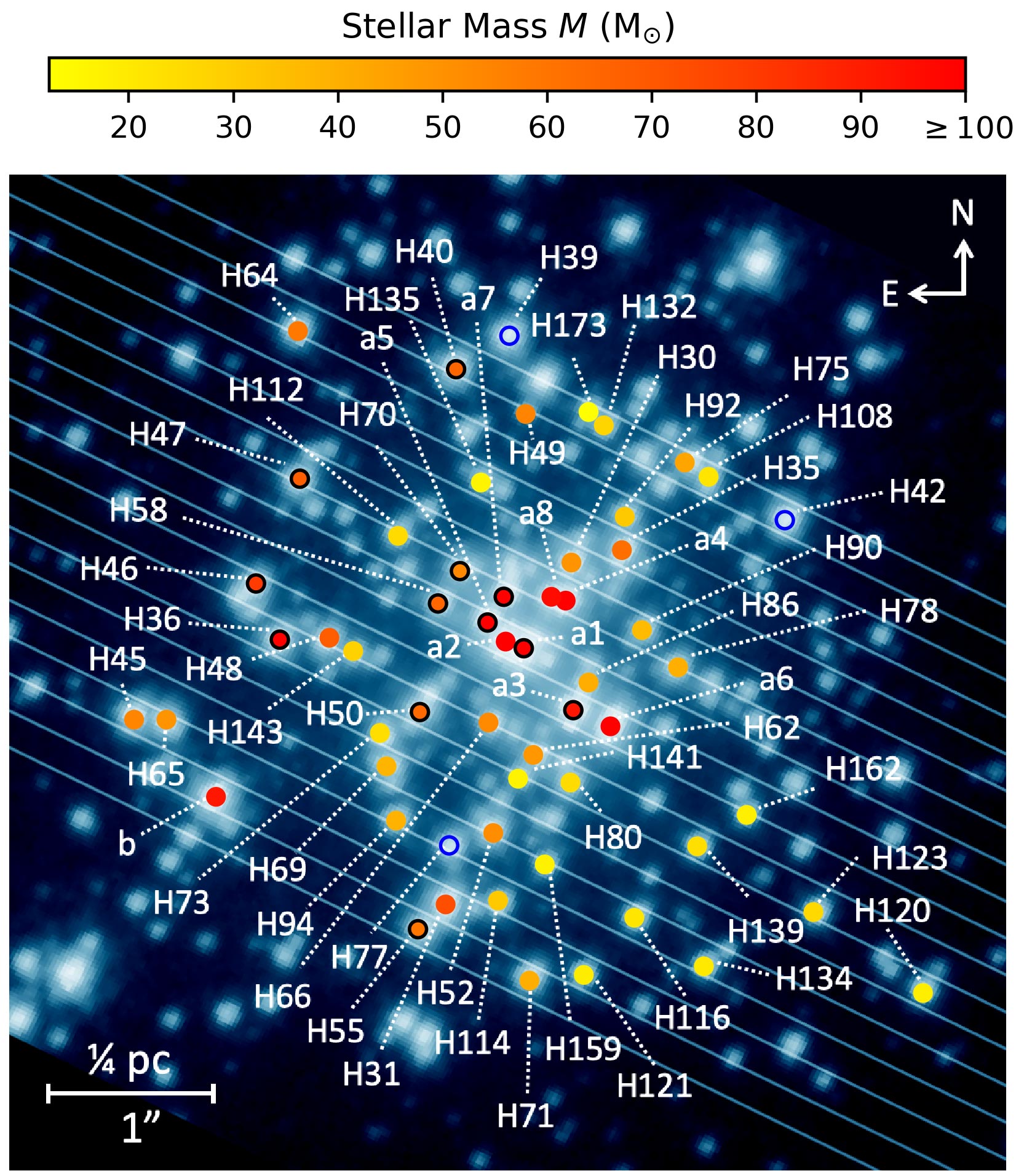}
    \caption{HST/WFC3 V-band (F555W) photometry of the core of R136 \citep{2010AAS...21522205O}. Positions of the stars in our sample are indicated (yellow to red closed circles) with respect to the slits of the HST/STIS observations (light blue lines). The colour of the circles indicates the current (evolutionary) mass of each source as derived in this paper. Identifications starting with `H' from \citet{1995ApJ...448..179H}; those starting with `a' and `b' from \citet{1985A&A...150L..18W}. H68 and H129 (located $4.61"$ and $\sim4.10"$ from a1, respectively) are part of our sample but fall outside the region shown here. Three stars identified here (blue open circles) are not in our sample: H42 and H77 (SB2s), and H39 \citep[analysed in][outside our slit coverage]{1998ApJ...509..879D}. Twelve stars of our sample overlap with that of \citet[black open circles]{1997ApJ...477..792D,1998ApJ...509..879D}. Not indicated in the image are H17, north of a1, and the two components of a6, H19 (north) and H26 (south); see \cref{sec:stis_data_r136} for more details. }
    \label{fig:positions}
\end{figure}

\section{Sample and data\label{sec:sample_data}}

Our sample consists of 56 stars residing in the core of the R136 cluster. Spectral types range from late to early O-type, plus three hydrogen-rich Wolf-Rayet (WNh) stars of subtype WN5h. Of the O-type stars, four are supergiants, five are giants, and the rest are dwarfs \citep{Bestenlehner2020, Saidainprep}. \Cref{fig:positions} shows their positions with their \citet{1995ApJ...448..179H} or \citet{1985A&A...150L..18W} identification, projected onto an HST/WFC3 image \citep{2010AAS...21522205O}. 
The figure shows that the area is very crowded; the high spatial resolution of HST is thus a necessity to resolve individual stars in the core of the cluster. Recent advances in adaptive optics have made it possible to obtain such high-resolution observations of the core of R136 also with ground based instruments, as is done by \citet[][imaging with SPHERE]{2017A&A...602A..56K,2021MNRAS.503..292K} and \citet[][optical spectroscopy with MUSE-NFM]{2021A&A...648A..65C}. 

Previous spectroscopic analyses of several sample stars have been carried out. A sub sample consisting of bright members of the cluster core has been studied by \citet{1997ApJ...477..792D,1998ApJ...509..879D}, who measured stellar parameters as well as mass loss rates of 14 sources using HST/GHRS/FOS optical and UV spectroscopy (overlap with our sample is indicated in \cref{fig:positions}). \citet{1998ApJ...493..180M} analyse optical HST/GHRS/FOS spectra, focussing on stars in the outskirts and surroundings of the cluster. 
\citet{2009MNRAS.397.2049S} obtained time-resolved NIR spectra of 5 stars in the core, searching for binarity and reporting a dearth of short period binaries in their sample. 
We assume in this study that the sources we observe are either single or that the light is dominated by the brightest component; however, the multiplicity properties of the sample remain an open question and require further investigation \citep{2019hst..prop15942S}. 
Combining aforementioned UV, optical and NIR spectroscopy, \citet{2010MNRAS.408..731C} re-derived physical properties of the WNh stars, finding a present day mass of 265 \Msun for the most massive star, R136a1. 

A comprehensive view of the cluster core was first given by \citet{2016MNRAS.458..624C}, who secured optical and UV HST/STIS spectroscopy of the central cluster and obtained temperatures, wind velocities and spectral types from the UV spectra. The 55 optical spectra of this dataset\footnote{For one star, R136a8, there are no optical spectra available except the \halpha line. This star was not included in the sample of \citet{Bestenlehner2020}, but is included in our optical~+~UV analysis.}  were analysed by \citet{Bestenlehner2020}, who determined detailed stellar parameters for all stars and found that at least seven stars have current masses of 100 \Msun or more, reporting 215 \Msun for R136a1. 

\subsection{HST/STIS data \label{sec:stis_data_r136}}

\begin{table}
\caption{Observational setup of the HST/STIS long-slit spectroscopy. \label{table:datagratings}}
\begin{tabular}{l l l l}
\hline \hline
\textrm{Grating} &  Grating positions &\textrm{Wavelength} &  $\lambda/\Delta \lambda$ \\  \hline
G140L & 1425 & 1150 -- 1717 \AA & 1250 \\
G430M & 3936, 4194, 4451, 4706 & 3795 -- 4743 \AA & 7700 \\
G750M & 6581 & 6297 -- 6866 \AA & 5850 \\ \hline
\end{tabular}
\end{table}

For our analysis we use blue-optical, \halpha, and far-UV HST-STIS spectroscopy (PI: Crowther,  \citealt{2016MNRAS.458..624C}). For this, 17 HST-STIS long-slit (52''x0.2'') contiguous pointings were done for six different gratings; technical details are summarised in \Cref{table:datagratings}. The setup is depicted in \cref{fig:positions}. 
The orientation of the slits (at position angle 64\degree/244\degree) was chosen to align with R136a1 and R136a2, that lie only 0.1'' apart. The image shows that crowding can occur elsewhere, which can cause contamination of spectra of stars that lie close together. While the spectral extraction process was designed to avoid this, several spectra might still be affected (see \Cref{tab:app:more_uv_restuls}).
The contamination is severe only in the case of R136a6. This source can be resolved into two sources, H19 and H26, having a flux ratio of 0.78 in the V-band and a separation of only 70 mas\footnote{These sources are clearly visible in the extreme adaptive optics VLT/SPHERE K-band images of \citet[][their Fig. 2]{2021MNRAS.503..292K}, but were already identified by \citet{1995ApJ...448..179H} with HST/WFC2.} \citep{1995ApJ...448..179H,2017A&A...602A..56K}. The brighter component was likely located partially out of the slit, and we therefore expect that H19 and H26 have an approximately equal contribution to the flux of what we call `R136a6'. We have no way of separating these components and therefore analyse R136a6 as if it were single; however, we exclude it from the analysis of the sample as a whole regarding mass-loss and clumping properties. 

The spectra were extracted with {\sc multispec}, a package tailored to extracting spectra from crowded regions \citep{2005stisreport,2007multispec}. Exceptions are the UV spectra of H70 and H141, where the extraction was done with {\sc calstis} \citep{2011stis.book.....B}. Since more than two sources are in each pointing the sources are not necessarily centred in the slit, which causes an uncertainty in the absolute wavelength scale that can be as large as $\pm$2 pixels (see also \cref{sec:opticaldataprepR136}). \halpha suffers from strong nebular emission, for which we correct by interpolating the \halpha emission of off-source spectra and subtracting this from the source spectra. The signal-to-noise-ratio (\snr) of the resulting spectra is in the range 7-70, with average values of 23, 19, and 19 for UV, blue-optical and \halpha. \Cref{fig:app:snr_dist} shows a distribution of the \snr of the sample stars per wavelength range. A more comprehensive description of the UV data reduction can be found in \citet{2016MNRAS.458..624C}, and the optical reduction will be described in more detail in \citet{Saidainprep}.

\subsection{GHRS data \label{sec:ghrs_data_r136}}

\begin{figure*}
    \centering
    \includegraphics[width=1.0\textwidth]{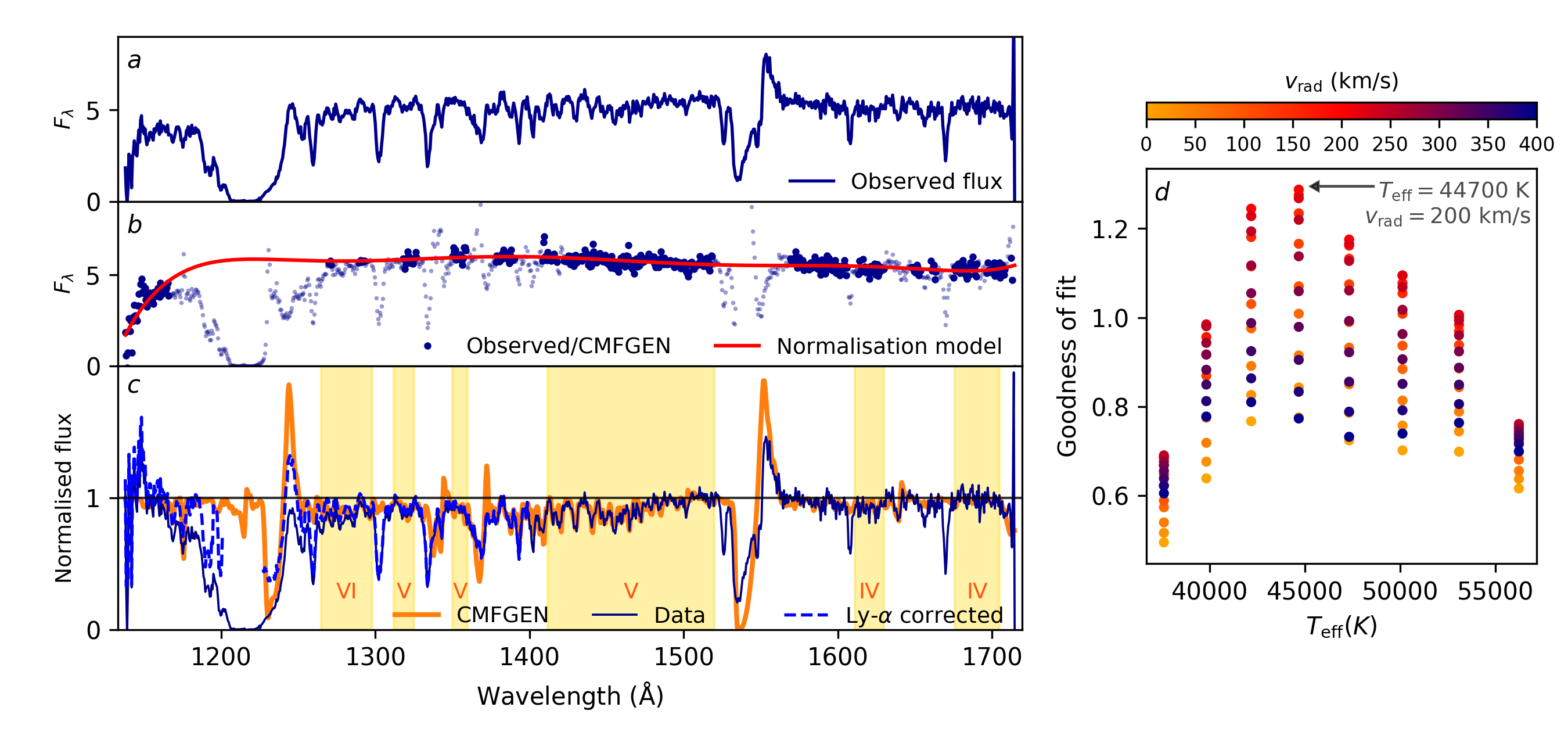}
    \caption{UV normalisation and \teff fitting procedure for H35. The left panels show \emph{a)} the observed flux in units of $10^{-14}~{\rm ergs \ cm}^{-2} {\rm s}^{-1} $ \AA$^{-1}$, \emph{b)} the observed flux divided by the normalised \cmfgen model (dark blue points, masked points are shown semi-transparent) and a fit through the non-masked points (red solid line), and \emph{c)} the normalised spectrum (dark blue solid line), the normalised, Ly-$\alpha$ corrected spectrum (blue dashed line) and the \cmfgen model used for obtaining the normalisation model (solid orange line). The yellow shaded areas indicate the regions used for determining the goodness of fit of the iron pseudo-continuum of the model; roman numerals indicate the dominant iron ion in each region. The right panel shows for each model \teff and \rv the dimensionless goodness of fit to the normalised iron pseudo-continuum of the data (defined as the inverse of the squared sum of the difference between the normalised fluxes of the model and the data in the iron continuum). At the peak lie the \teff and \rv that give the best fit: these are the values that are used for the final normalisation shown on the left.
    \label{fig:uvnormalisation}}
\end{figure*}

For 10 sources we complement the HST/STIS spectra with archival HST/GHRS UV spectra \citep[PIs: Heap \& Ebbetts, ][]{1997ApJ...477..792D,1998ApJ...509..879D}. The wavelength range of these spectra spans from 1150 to 1750 \angstrom, which means that they include the full \nivuvline line, contrary to our UV HST/STIS spectra where this line is positioned right on the edge of the grating. We do not use the spectra of R136a1 and a3 because they are possibly contaminated in view of the size of the aperture (0.22'')\footnote{For the same reason, \citet{1997ApJ...477..792D,1998ApJ...509..879D} do not analyse the spectrum of R136a2.}. Furthermore, the sample of \citet{1998ApJ...509..879D} includes H39 which is not covered by our slits (see \cref{fig:positions}) and H42, which is an SB2. We include the \nivuvline in our fitting for the 10 remaining spectra of their sample, while using HST/STIS data for all other lines. 
Moreover, the resolving power ($\lambda / \Delta \lambda$) of these spectra is approximately $5000$, which allows us to resolve the interstellar \CIVline lines and use this for the correction of the HST/STIS data (see \cref{sec:ismcorrection}). 
For more details of the HST/GHRS observations and data reduction we refer to \citet{1997ApJ...477..792D,1998ApJ...509..879D}. 

\subsection{Optical data preparation \label{sec:opticaldataprepR136}}

Our spectral fitting code needs a set of normalised spectral lines. To this end we have normalised the spectra of the optical and H$\alpha$ gratings locally around the diagnostic lines, assuming that the continuum can be approximated as a straight line. 
For each line we obtain the \snr from the continuum selected for the normalisation, and define the uncertainties on the normalised flux as the inverse of the \snr. In some cases we exclude data points, this includes isolated points that deviate a factor 2 or more from the value of the surrounding points, or data points in the centre of H$\alpha$ where nebular subtraction may have been imperfect \footnote{The removal of points in the core of \halpha may increase uncertainties on the mass-loss rate determination. Given the UV wind lines that we take into account in our fits, we expect this effect to be of minor importance.}.

The radial velocity shift is determined by fitting to the spectral lines in each grating a set of Gaussian functions with centres corresponding to the rest wavelengths of the lines considered. 
For lines that are affected strongly by the stellar wind (\halpha and \heiiline) we use the synthetic lines of a small grid of \fw models to determine the radial velocity shift. 

As described in \cref{sec:stis_data_r136}, the absolute wavelength scale of all observations can deviate up to 2 pixels. We correct for this by assuming that the wavelength deviation behaves similar to a Doppler shift. In practice, this means that we correct the spectra for a radial velocity without considering the aforementioned wavelength deviation, that is, we measure the `radial velocity', but this value includes both the true radial velocity, as well as an adjustment for the absolute wavelength deviation. The latter adjustment, typically on the order of $0-100$~km~s$^{-1}$ (or 2 pixels), is not physical, but we simply do not have a better model to describe the offset. This approach is thus a pragmatic one, merely to correct the wavelength scale for several effects in order to bring the diagnostic lines in the rest frame of the synthetic spectra. An overview of the derived velocities used for the correction can be found in \cref{sec:app:dataquality_rv}.

\subsection{UV data preparation \label{sec:uvnormalisationR136}}

The continuum of hot star UV spectra is hidden by a forest of lines, most notably lines of highly ionised iron group elements such as Fe~{\sc iv}-{\sc v}. Locating the continuum is thus not trivial, especially since the depth of the so-called pseudo-continuum, formed by the iron lines, depends on stellar properties, in particular on the effective temperature \teff. 
To assist the normalisation, we use a grid of \cmfgen models in which the iron lines are modelled \citep[][]{1998ApJ...496..407H,2014A&A...570A..38B}. 
With the normalised synthetic spectra that these models provide we can recover the shape of the true continuum. 
Our approach is as follows: 
\begin{itemize}
    \item Select a \cmfgen model from the grid that matches best the stellar and wind parameters from \citet{Bestenlehner2020}. \emph{But see also below}. 
    \item Mask the wind lines and interstellar lines, so that only the pseudo-continuum is left.
    \item Divide the observed UV flux by the normalised flux of the model and fit this quantity with a polynomial, getting a so-called normalisation model.
    \item Divide the observed UV flux by the normalisation model in order to obtain the normalised flux.
\end{itemize}
Getting a reliable normalisation hinges on the first step. It is especially important that the \teff of the model matches that of the observed spectrum. In order to assure this, we treat \teff as a free parameter. We repeat the above steps for all \teff in the \cmfgen grid \citep[ranging from 35-56~kK in 9 steps, LMC metallicity; see][]{2014A&A...570A..38B} and assess for which temperature the iron pseudo-continuum has the best fit. We also vary the radial velocity \rv of each model (in steps of 25 km~s$^{-1}$, which is about a tenth of the resolution element) and assess which value fits best. Just as for the optical, this \rv value includes a possible correction for the wavelength calibration (see also \cref{sec:opticaldataprepR136}). For the micro-turbulent velocity \vmicro we assume 10 km~s$^{-1}$ for all models, as this is the only value included in the grid of \citet{2014A&A...570A..38B} and because the exact value of \vmicro has little influence on this specific exercise. An example of the UV normalisation process is shown in \cref{fig:uvnormalisation}. 

The final output is a spectrum normalised with the best fitting \teff, and corrected for \rv. Before accepting a fit we check it visually; extra care is taken for sources where the fit value of \teff falls outside the uncertainty margins of the \teff derived by \citet{Bestenlehner2020}. For six sources the model with \teff as derived from fitting the iron pseudo-continuum did not result in a good fit and in these cases we adopted the value of \citet{Bestenlehner2020} for \teff (see \Cref{tab:app:more_uv_restuls}). The GHRS data are normalised in a similar way, but instead of fitting \teff we assume the value found from the HST-STIS iron forest fit.

A by-product of the normalisation process is a measurement of \teff from the iron lines alone; this measure is independent from the H, He, C, N, O diagnostic lines used for the rest of this work and the analysis of \citet{Bestenlehner2020}. These values can be found in \Cref{tab:app:more_uv_restuls} and are compared to the H, He, C, N, O temperature measurements in \cref{app:teff_discussion}.

We obtain the \snr of the HST-STIS UV spectra by using the HST-STIS exposure time calculator\footnote{\url{https://etc.stsci.edu/etc/input/stis/spectroscopic/}}, assuming the F555W magnitude from \citet{2011ApJ...739...27D}, and exposure times and $A_{\rm F555W}$ from \citet{2016MNRAS.458..624C}. Using the \snr we get from the calculator, we estimate the uncertainty on the flux points we use for the fit. For the GHRS data we use the provided error spectra. 

\subsection{Corrections for interstellar absorption lines in the UV \label{sec:ismcorrection}}

Three interstellar absorption lines blend with important diagnostics: H~\textsc{i} at 1215.67~\angstrom (Ly-$\alpha$), the Si~\textsc{iv} doublet at 1393.76-1402.77~\angstrom, and C~\textsc{iv} doublet at 1548.20-1550.77~\angstrom. 
We correct for Ly-$\alpha$ and \CIVline by recovering the H~\textsc{i} and  C~\textsc{iv} column density in the line of sight ($N_{\rm H}$ and $N_{\rm C \sc{iv}}$, respectively), by fitting the interstellar profiles with a Voigt-Hjerting function\footnote{Often simply called Voigt function (e.g., \citealt{1978stat.book.....M}, \citealt{2014bookHubeny})} \citep{2006MNRAS.369.2025T, 2007MNRAS.382.1375T}. For the damping factors of the Lorentzian component of the profiles we use the radiative damping constants of the transitions. We fit the interstellar components of multiple spectra and correct the spectra with averaged values rather than the values of the individual fits, since we expect the uncertainties in this fitting process to dominate over the difference in column density from star to star.

\begin{figure}
    \centering
    \includegraphics[width=0.46\textwidth]{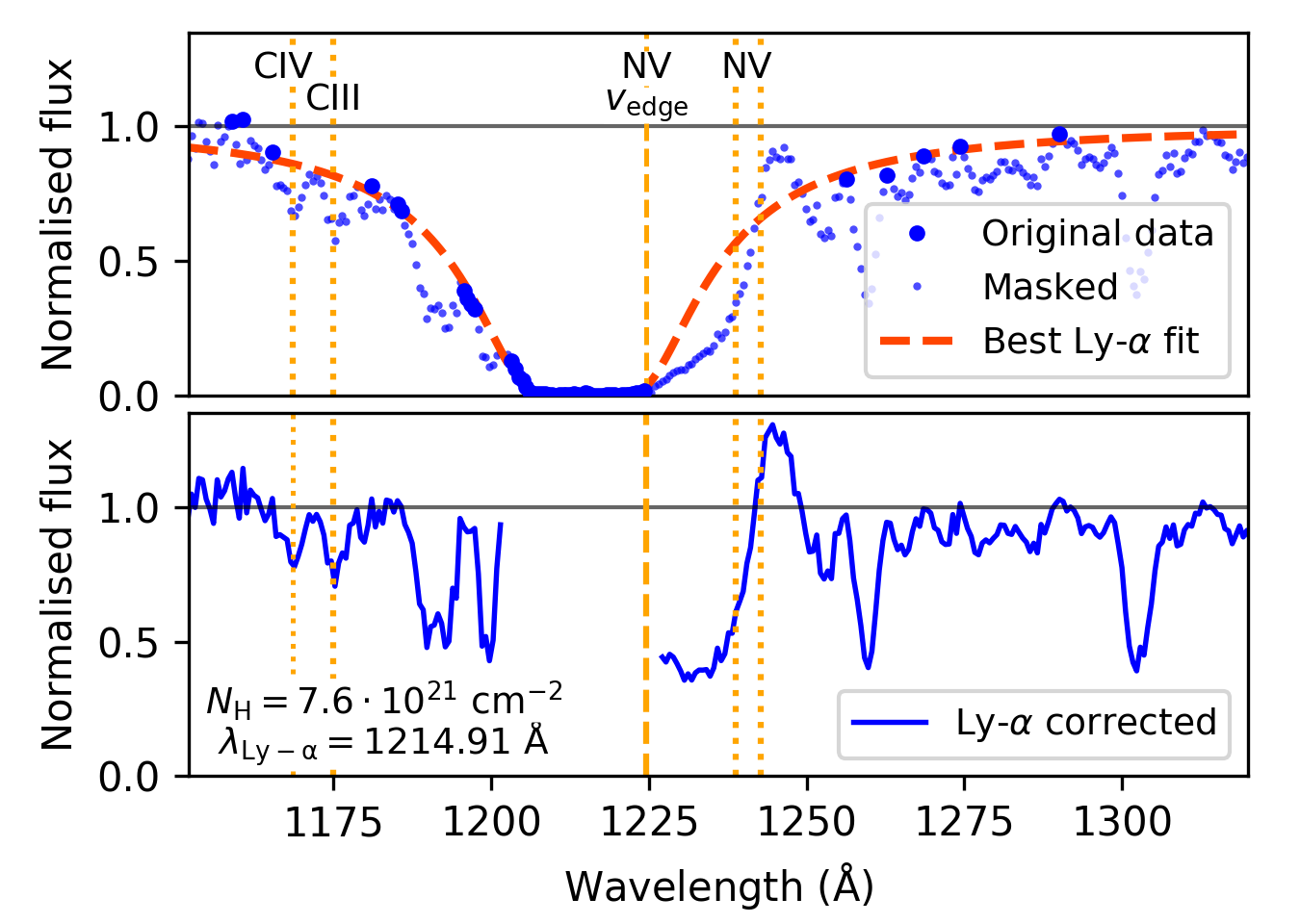}
    \caption{The Ly-$\alpha$ fitting procedure for H35. The upper panel shows the original data (blue: flux points indicated with a large dot are considered in the fit, small ones are not) and the best fit Ly-$\alpha$ profile (orange dashed line). The lower panel shows the Ly-$\alpha$ corrected flux and the values of $N_\textrm{H}$ and $\lambda_{Ly-\alpha}$ used for the correction. Indicated with vertical lines are the rest wavelengths of transitions of diagnostic lines (yellow dotted) and the position of the edge of the \nvuvline line (yellow dashed). }
    \label{fig:lya}
\end{figure}

In the case of Ly-$\alpha$, the Voigt-Hjerting profile is fitted to a subset of the points of the normalised UV flux. We do not fit those parts of the spectrum where the Ly-$\alpha$ profile might be blended with the N~\textsc{v}~1240 doublet. To estimate where this line ends, we use for each source the edge velocity $v_{\rm edge}$ from \citet{2016MNRAS.458..624C}. Furthermore, for fitting the wings, we select points that trace the stellar continuum: parts of the spectrum that seem free of absorption lines. \Cref{fig:lya} shows an example of a fit of the Ly-$\alpha$ profile. For finding the average of $N_{\rm H}$ we fit the Ly-$\alpha$ profiles of the 29 stars brighter than $M_V = -5.50$ (values from \citealt{Bestenlehner2020}) and obtain a value of $\log (N_{\rm H} \ [{\rm cm}^{-2}]) = 21.88 \pm 0.07 $, in good agreement with other $N_{\rm H}$ measurements towards 30 Doradus (e.g., \citealt{1980ApJ...236..769D}, who find $\log (N_{\rm H} \ [{\rm cm}^{-2}]) = 21.85\pm^{0.10}_{0.15}$).

\begin{figure}
    \centering
    \includegraphics[width=0.48\textwidth]{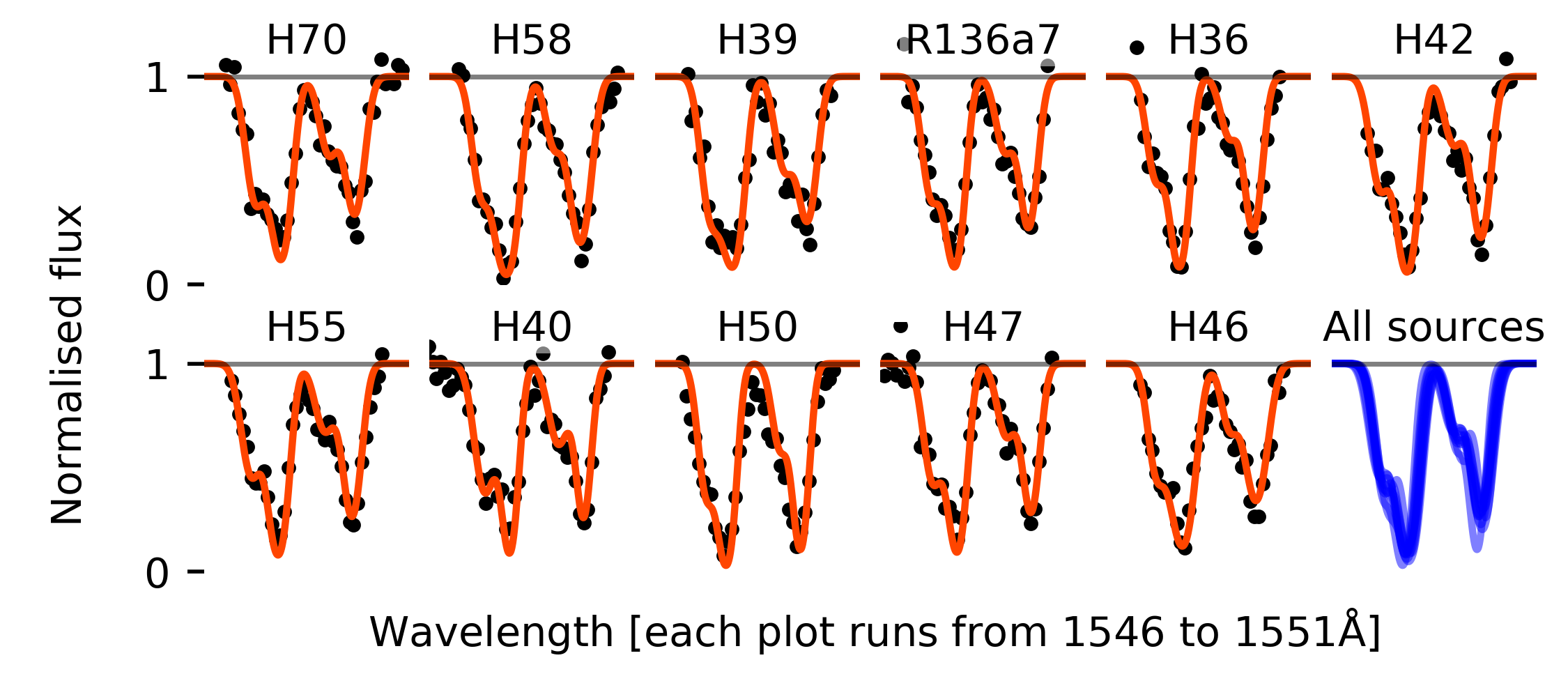}
    \caption{Best fits of the interstellar \CIVline lines for all sources of \citet{1998ApJ...509..879D}. For each source we show the normalised flux of the interstellar lines (black dots) and the best fit (orange lines). The last panel contains the best fit of all sources (blue lines). }
    \label{fig:CIVism}
\end{figure}

For \CIVline, we use the higher resolution GHRS spectra of \citet{1998ApJ...509..879D}. Before fitting the interstellar profile we fit a polynomial through the stellar P-Cygni profile of each star and subtract it from the spectrum, so that the interstellar component remains. We resolve two interstellar components with a different velocity in each part of the doublet, so we fit two double Voight-Hjerting profiles, where the ratio of the strength of each of the doublet components and the distance between them is set by the oscillator strengths and the rest wavelength difference, respectively. We fit all spectra of \citet{1998ApJ...509..879D}, except for R136b, where we had problems correcting for the stellar line. From this we find column densities of $\log (N_{\rm C \sc{iv}} [{\rm cm}^{-2}]) = 14.72 \pm 0.10 $ and $14.21 \pm 0.07 $.  The individual and mean fits are shown in \cref{fig:CIVism}.

The derived average profiles are used for computing the interstellar line optical depth (as a function of wavelength), which we then subtract from the observed optical depth of each star, obtaining the corrected optical depth, which we convert back to normalised flux. In the case of \CIVline, the average profile is convolved with an instrumental profile corresponding to $R=1250$ before it is used for the interstellar corrections of the HST/STIS data. The uncertainty margins on the column densities are used to estimate an uncertainty on the corrected flux, in addition to photon noise. 

For all sample stars but one, the \siivline stellar lines are in absorption and can, even in the higher resolution data of \citet{1998ApJ...509..879D} not be distinguished from the interstellar components. Only in R136b the interstellar component is resolved, however here we are not able to accurately correct for the stellar profile. We therefore cannot correct for the interstellar \siivline and do not use this line. The exception is R136b where the line is strongly in emission, and we can clip the interstellar part.

\subsection{VLT/SPHERE $K_s$ photometry \label{sec:dataphot}}

In our fitting procedure we calibrate the luminosity with an observed stellar flux (see \cref{sec:lumanchor}). 
For this we use the absolute, dereddened $K_s$-band magnitudes as presented in \citet{Bestenlehner2020}. They use VLT/SPHERE $K_s$ magnitudes from \citet{2017A&A...602A..56K} and in addition $B$ or $U$ and $V$ magnitudes for the extinction correction \citep{1995ApJ...448..179H,2011ApJ...739...27D} and an LMC distance modulus of 18.48 mag \citep{2019Natur.567..200P}. 
The $K_s$-band is the optimal choice for a luminosity anchor because at these wavelengths ($2.2~\mu\textrm{m}$) the extinction is low, while thermal radiation of dust is not yet an issue. 

\section{Methods\label{sec:method}}

For the analysis we use the model atmosphere code \fw (version: V10.3.1) to compute synthetic spectra and the genetic algorithm \pyGA for the fitting. In this section we introduce both tools and also describe our fitting setup and related assumptions. 

\subsection{\Fw \label{sec:fastwind}}

\Fw is a model atmosphere code tailored to hot stars with winds \citep{1997A&A...323..488S,2005A&A...435..669P,2012A&A...537A..79R,2016A&A...590A..88C, 2018A&A...619A..59S}. 
It solves the NLTE number-density rate equations and takes into account the effects of line-blocking and line-blanketing\footnote{See, e.g., \citealt{2001A&A...375..161P}, for an explanation of these concepts.}. The atmosphere consists of a spherically extended photosphere in (pseudo-)hydrostatic equilibrium that is connected to an expanding stellar wind at a velocity transition point $\varv_{\rm 0}$ near the sonic point. The stellar wind is parameterised by a mass loss rate $\dot{M}$, a terminal velocity $\varv_{\infty}$, and a wind acceleration parameter $\beta$. The wind velocity $\varv_{\rm r}$ as a function of radius $r$ is expressed by the classic $\beta$-velocity law:
\begin{equation}\label{eq:betalaw}
    \varv_{\rm r}(r)=\varv_{\infty} (1-b/r)^{\beta},
\end{equation} 
where $b$ is a radius close to the stellar radius\footnote{As defined in \citet{1997A&A...323..488S}, their Eq. (10).} $R_{\star}$, the exact value of $b$ depending on $\varv_{\rm 0}$ \citep[see][]{1997A&A...323..488S}. Under these assumptions the structure and ionisation/excitation state of the atmosphere and the wind are computed, resulting in a so-called atmosphere model. Using this model, the individual spectral lines are synthesised. 

\begin{figure*}
    \centering
    \includegraphics[width=1.0\textwidth]{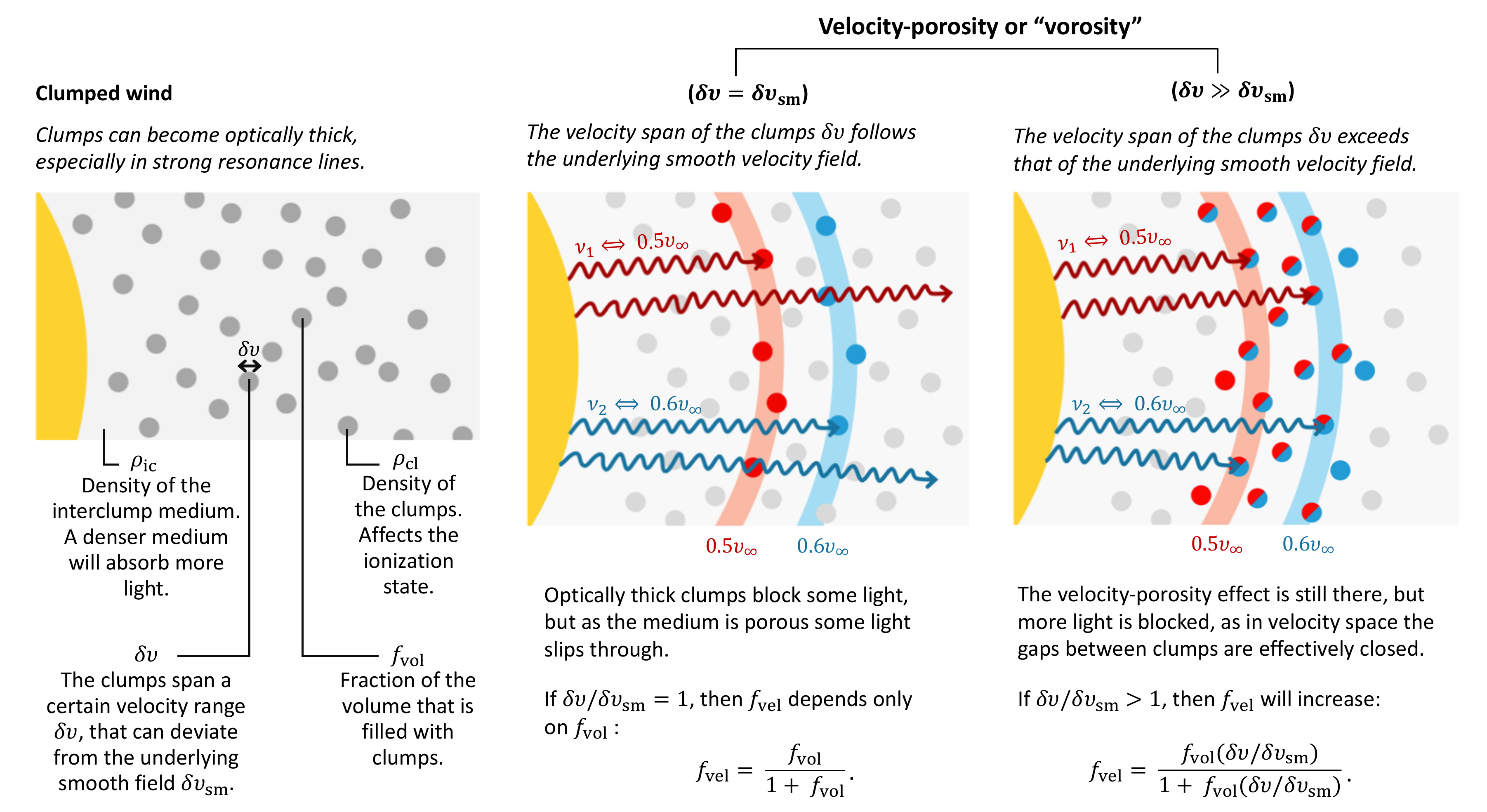}
    \caption{Illustration of the effects of clumping and porosity in velocity space. Each sketch shows the star (yellow) and its clumpy wind (different shades of grey). In the left figure we indicate different parameters that describe the wind structure. In the middle and right sketches we illustrate how the effective transparency of spectral lines in the wind depends on volume filling fraction as well as on the velocity span of the clumps. Depicted are photons of different frequencies ($\nu_1$ in red and $\nu_2$ in blue, with $\nu_1 < \nu_2$) that are intercepted by a strong absorption line if they are Doppler shifted with the right velocity. 
    The corresponding resonance zones lie at 0.5$v_\infty$ and 0.6$v_\infty$ (shaded red and blue, respectively). 
    }
    \label{fig:vorosity_explained}
\end{figure*}

\Fw stands out in terms of speed, as one model is computed in approximately 15-45 minutes on a single modern CPU. 
Such a short computation time allows one to compute many models, and thus explore the parameter space thoroughly (see \cref{sec:pyEA}).
Precision and speed are achieved simultaneously by splitting up the atomic elements in \emph{explicit} and \emph{background} elements. 
The explicit elements are computed in the co-moving frame using detailed atomic models for the spectral lines, while the background elements are only computed in an approximate way. 
Individual transitions of the latter are not synthesised, but their radiation field is taken into account, which is essential for the treatment of effects of line-blocking/blanketing. To speed up the computation, \fw calculates a representative mean radiation background instead of a detailed field \citep[for details, see][]{2005A&A...435..669P}. 

For this work we use \fw version V10.3.1 \citep{2018A&A...619A..59S}, including explicit elements H, He, C, N, O, Si, P.  This version is suitable for the analysis of stars with winds that are moderately optically thin in the optical continuum. This condition is met for all sample stars, including the three WNh stars. While the WNh stars do have the densest winds of our sample stars, they are not as dense as those of classical Wolf-Rayet stars. Indeed, when we run a \fw model with WNh parameters, we find that the electron scattering optical depth at the sonic point is well below unity at $\tau_{\rm e} = 0.28$ (for a typical O-star, we find $\tau_{\rm e} \approx 0.02$).

\subsubsection{Wind clumping, porosity, and vorosity\label{sec:method_structure}}

Clumping is implemented in \fw V10.3.1 as detailed in \citet{2018A&A...619A..59S}, employing the two-component formalism introduced in \citet{2014A&A...568A..59S}. In this prescription, the clumped wind consists of overdense clumps with a density $\rho_{\rm cl}$ and an interclump medium with a lower density $\rho_{\rm ic}$. The clumps occupy a certain fraction of the total wind volume $f_{\rm vol}$, referred to as the volume filling factor. The clumping factor $f_{\rm cl}$ of the medium can be written as:
\begin{equation}
    f_{\rm cl} \equiv \frac{\langle \rho^2 \rangle}{\langle \rho \rangle^2} = \frac{f_{\rm vol}\rho_{\rm cl}^2 + (1-f_{\rm vol})\rho_{\rm ic}^2}{(f_{\rm vol}\rho_{\rm cl} + (1-f_{\rm vol})\rho_{\rm ic})^2}.
\end{equation}
We note that for the conventional assumption (not adopted here) of a void interclump medium ($\rho_{\rm ic} = 0$) this would lead to $f_{\rm cl} = 1/ f_{\rm vol}$. 
The parameters describing the clumped medium are illustrated in the left panel of \cref{fig:vorosity_explained}. 

The formalism of \citet{2014A&A...568A..59S} allows for the possibility of the clumps becoming optically thick. 
When clumps become optically thick porosity effects come into play, both in physical and velocity space. Spatial porosity allows photons that would have normally interacted with a slab of gas to escape freely. Some photons will be absorbed by gas that is compressed into dense clumps, but the separation between the clumps allows others to escape without interaction. 
The velocity field of the wind plays a crucial role in this; the fact that the outflow accelerates results in increasing Doppler shifts throughout the wind and causes the spectral line resonance zones in which the clumps are optically thick for a certain frequency to be very narrow in the radial direction (at least as long as the velocity is not
close to \vinf). 
If one then assumes that all clumps follow the underlying average velocity field, the amount of leakage for line photons can be directly linked to the spatial volume filling factor $f_{\rm vol}$. This effect is illustrated in the middle panel of \cref{fig:vorosity_explained}. 

However, the clumps do not necessarily follow the average velocity field\footnote{The velocity field can, for example, be severely altered by shocks (see \cref{sec:meth:shocks}).}. For example, the velocity span of the clumps, $\delta\varv$, can be larger than that of the underlying smooth field $\delta\varv_{\rm sm}$ (Eq. \ref{eq:betalaw}). 
In this case the Doppler shifted gas in the clumps spans a wider range of velocities than does the smooth field, which means that effectively the resonance zone in the wind for a given transition becomes larger. In other words, if the clumps are at least somewhat optically thick more gas has the right Doppler shift to absorb a photon of a given frequency, and thus more light is absorbed. 
 This effect of porosity in velocity space, is also called velocity-porosity or `vorosity' (as coined by \citealt{2008cihw.conf..121O}). 
 The non-normalised velocity filling factor, $f_{\rm vor}$, depends on both  $f_{\rm vol}$ and the relative velocity span of the clumps $\delta\varv/\delta\varv_{\rm sm}$:
\begin{equation}\label{eq:vor}
     f_{\rm vor} = f_{\rm vol}\left|\frac{\delta\varv}{\delta\varv_{\rm sm}}\right|. 
\end{equation}
In \fw a normalised version of this factor is implemented:
\begin{equation}\label{eq:vel}
     f_{\rm vel} = \frac{f_{\rm vor}}{1+f_{\rm vor}} = \frac{f_{\rm vol}(\delta\varv/\delta\varv_{\rm sm})}{1 + f_{\rm vol}(\delta\varv/\delta\varv_{\rm sm})},
\end{equation}
taking values from 0 to 1. The parameter \fvel is called the velocity filling factor. Note that this equation reduces to the purely geometrical effect, depending only on $f_{\rm vol}$, when $\delta\varv$ follows the underlying smooth field:
\begin{equation}
     f_{\rm vel} = \frac{f_{\rm vol}}{1 + f_{\rm vol}}, \hspace{1cm} [\delta\varv/\delta\varv_{\rm sm} \rightarrow 1].
\end{equation}
The effect of \vorosity is illustrated in the middle and right panels of \cref{fig:vorosity_explained}. 

These clumping and \vorosity effects are implemented in \fw by means of an `effective opacity formalism'. In this formalism, various properties of the clumps and the interclump medium (such as temperature) are assumed to be similar, and the rate equations, etc., are evaluated for a fiducial clump density $\rho = \langle \rho \rangle f_{\rm cl}$.  This allows one to approximate the clumpy wind as a one component medium with a certain average ionisation state and a single effective opacity. Essentially, the expensive computation of the NLTE occupation numbers is done only once, obtaining an average opacity $\langle \chi \rangle$ for a mean clump density, and then re-scaled in order to infer the effective opacity of the two-component clumped wind. 
The effective opacity $\chi_{\rm eff}$ can be expressed as: 
\begin{equation} \label{eq:chieff}
    \chi_{\rm eff} = \langle \chi \rangle \frac{1 + \tau_{\rm cl} f_{\rm ic}}{1 + \tau_{\rm cl}},
\end{equation}
with $\tau_{\rm cl}$ the clump optical depth \citep{2014A&A...568A..59S}. The interclump density contrast, $f_{\rm ic}$, is defined as:
\begin{equation}
    f_{\rm ic} \equiv \rho_{\rm ic}/\langle \rho \rangle.
\end{equation}
The formalism accounts for the \vorosity effects by adjusting the clump optical depth. For line opacity the clump optical depth in the rapidly accelerating winds is then computed in the Sobolev approximation:  
\begin{equation}\label{eq:clumptau}
    \tau_{\rm cl} = \frac{\tau_{\rm S}}{f_{\rm vor}} (1 - (1- f_{\rm vol})f_{\rm ic}),
\end{equation}
with $ \tau_{\rm S}$ the Sobolev optical depth for the mean wind. 
In the case of continuum opacity, on the other hand, the clump optical depth will depend on the porosity length $h$ ($\equiv l_{\rm cl} / f_{\rm vol}$, with $l_{\rm cl}$ the characteristic length scale of clumps). This parameter, describing the spatial porosity, can impact optically thick continua, where it can affect, for example, the ionisation rates. By default in \fw a radial variation of this parameter in the form of a so-called `velocity-stretch' law is assumed: 
\begin{equation}
h(\varv_{\rm r}) = h_\infty \varv_{\rm r}/\varv_\infty,
\end{equation} 
with $h_\infty$ the porosity length at the terminal wind velocity, given as input by the user. In this work we adopt $h_\infty = R_{\star}$, following \citet{2018A&A...619A..59S}. 
We refer the reader to \citet{2014A&A...568A..59S} and \citet{2018A&A...619A..59S} for a more detailed and quantitative explanation of the effective opacity formalism and its implementation in \fw. 

We conclude our description of the wind structure implementation by noting that we assume a stratified clumping factor, that is, we assume the clumping factor to vary throughout the wind. Several stratifications are implemented in \fw. 
In this work we adopt the implementation used by \citet{2018A&A...619A..59S} and \citet{Hawcroft21}, where the clumping is described by three parameters: 
the onset velocity of clumping $\varv_{\rm cl, start}$, the maximum clumping factor $f_{\rm cl}$, and the velocity at which this maximum clumping factor is reached, $\varv_{\rm cl, max}$. At the base of the wind the medium is assumed to be unclumped, its structure being only affected by micro-turbulence. Then, from $\varv_{\rm r}=\varv_{\rm cl, start}$ until $\varv_{\rm r}=\varv_{\rm cl, max}$ the clumping factor increases linearly with wind velocity from 1 to $f_{\rm cl}$, staying constant at $f_{\rm cl}$ for $\varv_{\rm r}>\varv_{\rm cl, max}$. 
This assumption for the clumping stratification is conform empirical findings in at least the lower and intermediate wind (e.g., \citealt{2006A&A...454..625P,2021arXiv210811734R}). 
At  $\varv_{\rm r}>\varv_{\rm cl, max}$ the clumping stays constant at the maximum value, $f_{\rm cl}$. The values for $\varv_{\rm cl, start}$, $\varv_{\rm cl, max}$ and $f_{\rm cl}$ are specified by the user (see \cref{sec:fitstrategy}). A summary of the wind structure parameters is given in \Cref{tab:fw_clumping}.

\subsubsection{Wind turbulence}

The wind structure parameters described in \cref{sec:method_structure} are all used in the process of computing the ionisation/excitation structure of the model atmosphere. An additional parameter, the wind turbulence velocity \vwindturb, is used only during the synthesis of spectral lines. This parameter introduces a depth-dependence of the micro-turbulent velocity throughout the wind. During the computation of the ionisation/excitation structure the micro-turbulent velocity is assumed to be constant, but when the lines are synthesised, the micro-turbulent velocity increases linearly with wind velocity from \vmicro at the base of the wind, to \vwindturb, at the point where the wind reaches its terminal velocity \citep{1995A&A...295..136H}. 
The wind turbulence velocity is typically on the order of $0.1\varv_\infty$ \citep[e.g.,][]{1989A&A...221...78G}, and is used here to mimic the effects of a large wind velocity dispersion upon the spectral line formation. Evidence for such a velocity dispersion is found in both LDI simulations \citep[e.g.,][]{1980A&A....84..342H,1993A&A...279..457P,2019A&A...631A.172D} as well as in observations  \citep[e.g.,][]{1982ApJ...255..278L,1989A&A...221...78G,1990ApJ...361..607P}. 

\begin{table}
    \centering
        \caption{Parameters that describe the wind structure in \fw. All these parameters can directly affect the shape of a spectral line, except for $h$, which is only impacting continuum opacity. Note that $f_{\rm cl}$ is in fact the maximum value of clumping that is reached at $\varv_{\rm cl, max}$, and \vwindturb the maximum value of the wind turbulence reached at the terminal velocity. Furthermore, note that not all parameters listed here are treated as free parameters in our analysis (see \cref{sec:fitstrategy}). 
        }
    \begin{tabular}{p{1cm}p{4.8cm}}
        \hline \hline
        \multicolumn{2}{l}{Wind structure parameters in \fw}\\\hline
         $f_{\rm cl}$ & Clumping factor \\ 
         $f_{\rm ic}$ & Interclump density contrast \\
         $f_{\rm vel}$ & Velocity filling factor \\
         $\varv_{\rm cl, start}$ & Onset of clumping\\
         $\varv_{\rm cl, max}$ & Clumping reaches maximum ($f_{\rm cl}$)\\
         $h$ & Porosity length \\ 
         $\varv_{\rm windturb}$ & Wind turbulence \\ \hline 
    \end{tabular}
    \label{tab:fw_clumping}
\end{table}

\subsubsection{Wind-embedded shocks \& X-rays \label{sec:meth:shocks}}

Instabilities in the winds of massive stars can cause shocks to form in the wind \citep[e.g.,][]{1988ApJ...335..914O,1997A&A...322..878F}. These wind-embedded shocks give rise to X-ray emission, which, both by direct and Auger ionisation, can alter the ionisation balance of the wind, as well as the velocity fields of the interclump medium and the clumps. 
Wind-embedded shocks and associated X-ray emission are implemented into \fw, and their characteristics can be tweaked. 
In our analysis we include X-ray emission by assuming canonical values for each star. Details about the implementation of X-rays in \fw and our assumptions regarding the canonical values are detailed in \cref{app:sec:X-rays}.

\subsection{A new genetic algorithm: \pyGA \label{sec:pyEA}}

\begin{figure}
    \centering
    \includegraphics[width=0.48\textwidth]{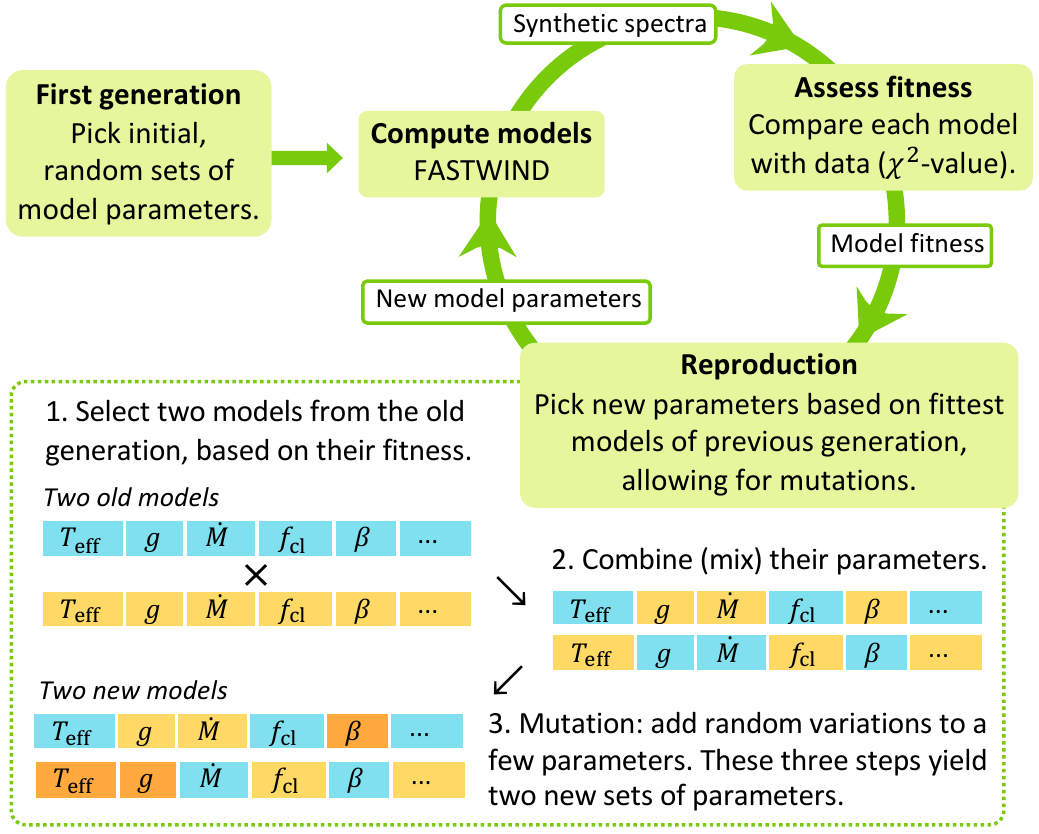}
    \caption{Flowchart portraying the workings of \pyGA. In each generation, depicted by the green circle with arrows, a large amount of models is computed (typically $50-250$). The reproduction step is a means of natural selection towards fitter models, which eventually leads to convergence of the algorithm towards the best fitting solution (typically after $30-100$ generations).}
    \label{fig:KiwiFlowchart}
\end{figure}

In order to find the best fitting \fw models we developed a Genetic Algorithm (GA), which we call \pyGA. GAs employ the concept of natural selection or survival of the fittest \citep[][]{Darwin1859}. The goal is to find, for a given dataset (stellar spectrum), the best fitting model. For this, the algorithm starts by computing a group (generation) of random models (individuals). 
After computing the first generation of models (parents), the fitness of each model is assessed by comparing each model to the data. In our case, we use the \chisq value as a fitness measure: 
\begin{equation}
    \chi^{2} = \sum_{i=0}^N \left( \frac{\mathcal{F}_{{\rm obs},i} - \mathcal{F}_{{\rm mod},i}}{ \mathcal{E}_{{\rm obs},i}} \right)^2, 
\end{equation}
where $N$ is the number of data points of the spectrum that is considered in the fit, $\mathcal{F}_{{\rm obs},i}$ the observed normalised flux,  $\mathcal{F}_{{\rm mod},i}$ the normalised flux of the model, and $\mathcal{E}_{{\rm obs},i}$ the uncertainty on the observed flux. Generally, models that have parameters that resemble the properties of the observed spectrum will be fitter (have a lower \chisq) than models with parameters that are far off. By picking new (offspring) models by combining the parameters of the fittest models of the previous (parent) generation, the offspring models will generally fit the observed spectrum better than the parent generation. For example, a model with a \teff that is similar to that of the observed star will generally have a better fit than a model with a \teff that is far off, and this value of \teff will thus have an increased chance of being selected for models of the new generation. We note that in the process of combining the parameters of two parent models, a fraction of the parameters is altered randomly (we call this mutation), in order to maintain and introduce diversity in the model parameters. By iterating this procedure (the offspring generation becomes the new parent generation), we eliminate parameter values that differ greatly from value that matches the data, while parameter values that match the data well will be kept. This way, the algorithm converges towards models with a better fit to the data.
\Cref{fig:KiwiFlowchart} illustrates the workings of the algorithm. Especially for large parameter spaces, this is a very efficient search method. In the past GAs have been successfully used for the analysis of massive star spectra \citep[e.g.,][]{2005A&A...441..711M,2006A&A...456.1131M,2014A&A...572A..36T,2017A&A...600A..81R,2021A&A...651A..96A,Hawcroft21}. 

\PyGA is written in {\sc python} and uses elements of the algorithm of \citet{2005A&A...441..711M}, who in turn use the {\sc pikaia} algorithm of \citet{1995ApJS..101..309C}. The new aspects of the algorithm are introduced after careful assessment of considerations laid out by \citet{GAbasics}, who presents an overview of possible structures and operators that can be part of a GA. For \PyGA we selected structures and operators that seemed beneficial for solving our specific optimisation problem; for details, we refer the reader to \cref{sec:app:pyga}.
For the parallel processing we use the {\sc schwimmbad} package, following \citet{2021A&A...651A..96A}. Within \pyGA a python command initiates the execution of \fw, a Fortran executable. \PyGA is publicly available and has a comprehensive documentation in order to be accessible to new users\footnote{\url{https://github.com/sarahbrands/Kiwi-GA}}. 

\subsection{Stellar radius \& luminosity\label{sec:lumanchor}}

In our \pyGA runs the stellar radius is estimated for each model individually using an observed, extinction corrected magnitude (\cref{sec:dataphot}), following the procedure described in \citet{2005A&A...441..711M}. Based on the temperature of each model, a Planck curve with temperature $T = 0.9T_{\rm eff}$ is computed and compared to the observed magnitude, after which a radius is chosen such that the Planck curve matches the observed anchor magnitude (\citealt{2005A&A...441..711M}, who follow \citealt{2004A&A...413..693M}). For this, we use a transmission curve of the adopted filter, in our case the VLT/SPHERE $K_s$ filter\footnote{The filter is specified by the user and any filter of which a transmission curve is available can be chosen. Currently the following filters are implemented: Johnson V, HST F555W, and VLT/SPHERE $K_s$. The transmission curves are taken from the SVO Filter Profile Service: \url{http://svo2.cab.inta-csic.es/theory/fps/}}. The Planck curve thus serves as an `SED estimate' during the run and the radius is an output of the run, as is luminosity\footnote{\PyGA also has the option to set the radius to a fixed value for all models.}. When a run is finalised we compute the real SED of the best fitting model and use this to correct the radius that was estimated during the run. Furthermore we scale the obtained mass-loss rates using the optical depth invariant wind-strength parameter $Q=\dot{M}/(R_{\star} \varv_\infty)^{3/2}$ \citep[][\citealt{2018A&A...613A..65H}, their Appendix B]{1996A&A...305..171P}. Note that there is no need to recompute all models, as the radius has very little impact on the normalised spectrum. The radius corrections for our stars range from 0 to +8\% with an average of 3.6\% for the O-stars and from $-20$ to $-9$\% for the WNh stars. We note that for future runs, where the $K_s$-band is used as an anchor magnitude, a Planck curve with $T = 0.83T_{\rm eff}$ would be a better guess 
-- the previous estimate of $T = 0.9T_{\rm eff}$ was tailored to $V$-band magnitude anchors. 

\subsection{Best fit parameters and error bars}

From the output of each \pyGA run we derive best fit parameters and uncertainties thereof (error bars) with the method of \citet{2014A&A...572A..36T}. 
For this, we identify the best fitting model (that with the lowest \chisq) and use this model to implicitly adjust the uncertainty of each flux point that is fitted, such that the \chisred value of the best fitting model equals unity. These adjusted flux uncertainties are then used for recomputing the \chisq of each model in the run. This procedure is equivalent to dividing all \chisq values of the run by the (original) \chisred of the best model. 
After the flux uncertainty adjustment we find the models that should be considered statistically indistinguishable from the best model, which we call the family of best fitting models. We do this by computing for each model the probability $P$ that the difference between the two models is caused by random fluctuations: 
\begin{equation} \label{eq:probability}
    P\left(\chi^2, \nu\right) = 1 - \Gamma\left(\chi^2/2, \nu/2\right)
\end{equation}
with $\Gamma(\chi^2/2, \nu/2)$ the incomplete gamma function, representing the cumulative distribution function of the \chisq distribution, evaluated at \chisq, for $\nu = n_{\rm data} - n_{\rm free}$ the degrees of freedom, where  $n_{\rm data}$ is the number of flux points that is taken into account during the fit and $n_{\rm free}$ the number of free parameters. 
The best fitting models are all models where $P > 0.32$~(i.e., the 68\% confidence interval, we will call this $1\sigma$) or $P > 0.05$~(i.e., the 95\% confidence interval). From this group we derive error bars by identifying for each parameter what is the lowest and the highest value that is present. In other words, the parameter space spanned by the family of best fitting models determines the size of the error bars. In case the distribution from which we derive the confidence intervals is symmetric and Gaussian, the 68\% and 95\% confidence intervals translate directly into standard deviations of $1\sigma$ and $2\sigma$, respectively. For convenience, we will refer to the 68\% and 95\% confidence intervals as $1\sigma$ and $2\sigma$ uncertainties, even though the confidence intervals we derive are not necessarily symmetric and Gaussian. 
In all tables we present 1$\sigma$ uncertainties, unless explicitly stated otherwise. For practical purposes, we generally mark the parameter of the best fitting model, however, we stress that all models in the family of best fitting models should be considered as statistically equivalent. 

The normalisation of \chisq values that is part of this method relies on the assumption that the best fitting model has a good fit to the data. This condition is satisfied for all stars in our sample except for the three WNh stars, where clear deviations between the best fitting model and data can be seen. In this case the method described above underestimates the error bars and therefore we assume increased error bars for these three stars, such that the error region covers the width of the peak in the \chisq distributions. This way, the error bars of the WNh stars are more in line with those of the O-stars of the sample (\cref{sec:results}).  

For luminosity, radius and mass-loss rate we increase the error bars given the uncertainty on the magnitudes (as presented in \citealt{2017A&A...602A..56K}). The uncertainty in radius and luminosity is directly related to the uncertainty in the observed flux at the K-band. For the mass-loss rate, we increase the errors propagating the uncertainty on the stellar radius, assuming a scaling of $\dot{M} \propto R_{\star}^{3/2}$ \citep[see e.g.,][]{1996A&A...305..171P}. 

Lastly, we stress that the uncertainties that we derive from the \pyGA runs, as described in this section, are only statistical uncertainties. Systematic uncertainties, that could arise, for example, due to assumptions regarding extinction, normalisation, or the modelling, are not included in these values. 

\subsection{Fitting strategy \label{sec:fitstrategy}}

We fit the full sample two times with \pyGA. The first time we consider only the optical parts of the data, the second time we fit the optical and UV data simultaneously. Ultimately, we are interested in the values of the optical~+~UV analysis, but the optical fitting still serves a threefold purpose. First, we use the derived values for rotational broadening and helium abundance as fixed values for the optical~+~UV fitting, reducing the amount of free parameters of those runs. Second, it provides mass-loss rates as derived from recombination lines only, assuming a smooth wind. Third, it allows us to compare our analysis method, fitting with \pyGA, to the spectroscopic analysis with IACOB-GBAT \citep{2011JPhCS.328a2021S} of the same data by \citet[][]{Bestenlehner2020}. The second and third point are addressed in \cref{app:comparisons}. 
The details of each fitting setup are summarised in \cref{tab:fitsetup_summary} and explained in detail below. 

\begin{table}
    \centering
    \small 
    \caption{Free parameters in the optical-only and optical + UV fits. Parameters in brackets are free only in a subset of the runs, see text. The last column contains for each fit a reference to the table where the best fit parameters are presented. Parameter names are defined in the text. \label{tab:fitsetup_summary} }
    \renewcommand{\arraystretch}{1.3}
    \begin{tabular}{p{1.7cm}>{\raggedright\arraybackslash}p{4.4cm}>{\raggedright\arraybackslash}p{1.4cm}}
    \hline \hline 
    Fit & Free parameters & Results\\ \hline 
    Optical-only & \teff, $g$, \vsini, \mdot, \yhe, ($n_{\rm N}$, $\beta$) & \cref{tab:app:bestfit:optical}\\ 
    Optical~+~UV (high \snr) & \teff, $g$, \mdot, $\beta$, \vinf, \fcl, \vclonset, \vwindturb, \fic, \fvel, $n_{\rm N}$, $n_{\rm C}$, ($n_{\rm O}$, \yhe) & \cref{tab:bestfit:12free}, \ref{tab:WNHoxy}\\ 
    Optical~+~UV (low \snr) & \teff, $g$, \mdot, $\beta$, \vinf, \fcl &  \cref{tab:bestfit:6free} \\ \hline 
    \end{tabular}
\end{table}

\subsubsection{Optical-only setup \label{sec:setupoptical}} 

The optical-only runs have 5 to 7 free parameters $n_{\rm free}$, as specified in \Cref{tab:fitsetup_summary}. Here, $g$ is the gravitational acceleration, \vsini the projected rotational broadening, \mdot the mass-loss rate, and $x_{\rm He}$ the helium surface abundance, where $x_{\rm He} = n_{\rm He}/n_{\rm H}$, with $n_{\rm He}$ and $n_{\rm H}$ the helium and hydrogen number density. If any line is (partially) in emission, we also fit the wind acceleration parameter $\beta$, and when we see a nitrogen line above the noise we fit the nitrogen abundance \yN (with $n_{\rm N}$ the number density). The other parameters are held fixed at the values discussed below. 

We assume a smooth wind ($f_{\rm cl} = 1$) for all stars except for the WNh stars, for which we assume $f_{\rm cl} = 10$. Furthermore, we assume \vmicro~$= 10$~km~s$^{-1}$, and in case $\beta$ is fixed we assume $\beta =0.9$. 
Because the resolution and \snr of the data do not allow us to distinguish between broadening due to rotation versus broadening due to macro-turbulence, we only fit \vsini, assuming \vmacro~$= 0$~km~s$^{-1}$. In practice this means that all broadening is captured in a single parameter \vsini and, since for our stars likely \vmacro~$> 0$~km~s$^{-1}$ \citep[see  e.g.,][]{2017A&A...597A..22S}, the projected rotational velocities that we find are upper limits of the actual \vsini. 
The derived \vsini is thus an upper limit. We adopted surface abundances of the CNO-elements using the evolutionary grids of \citet{2011A&A...530A.115B} and \citet{2015A&A...573A..71K}, based on stellar parameters of \citet{Bestenlehner2020}, and other abundances are fixed to $Z = 0.5~{\rm Z}_\odot$. We assume the values of \citet{2016MNRAS.458..624C} for the terminal velocities of the winds \vinf. For 12 stars \vinf was not available and in these cases we estimate the velocities by inter- and extrapolating the dependence of \vinf on luminosity $L$, that we empirically find using the values of \citet{2016MNRAS.458..624C} and \citet{Bestenlehner2020}, for $\log L/L_\odot < 5.6$. 

The optical-only \pyGA runs have a population of 71 to 95 individuals, with the exception of the WNh stars, where we have 191 individuals. The runs of most stars converge in approximately 20 generations. To ensure that all runs are fully converged, we iterate for 30 generations. The runs of sources with strong emission lines converge later and we run them for 40-60 generations. 

\subsubsection{Optical~+~UV setup\label{sec:optUVsetup}} 

The optical~+~UV runs have 6 to 14 free parameters. For 39 stars with relatively high \snr we fit 12 free parameters as listed in \Cref{tab:fitsetup_summary}, in which \yC refers to the carbon abundance (by number). For the WNh-stars we fit two additional free parameters (see also below): oxygen abundance \yO (by number) and \yhe. 
The other 17 stars have too low \snr and too weak wind lines to distinguish between 12 free parameters and we therefore only consider 6 free parameters for these stars (\Cref{tab:fitsetup_summary}). In this case, the CNO-abundances are fixed to LMC baseline values, for which we assumed $\log ({n}_{\rm C}/{n}_{\rm H}) + 12 = 7.75$ for carbon, $\log ({n}_{\rm N}/{n}_{\rm H}) + 12 = 6.9$ for nitrogen and $\log ({n}_{\rm O}/{n}_{\rm H}) + 12 = 8.35$ for oxygen \citep[][as in \citealt{2011A&A...530A.115B, 2015A&A...573A..71K}]{1998RMxAC...7..202K}. The wind structure parameters are fixed based on typical values we find from the 12-free-parameter runs with lower mass-loss rates: \fic$ = 0.05$, \fvel$=0.15$, \vclonset/$\varv_\infty=0.05$, \vwindturb/$\varv_\infty=0.15$ (\cref{sec:windstructuretrends}). For all optical~+~UV runs the velocity at which the maximum clumping factor is reached is given by $\varv_{\rm cl, max}/\varv_\infty = {\rm max}(0.3, \varv_{\rm cl, start}/\varv_\infty)$. 

Oxygen abundance is only a free parameter for the WNh stars. Test runs with free oxygen abundance for the other stars resulted in extremely high values of $\log ({n}_{\rm O}/{n}_{\rm H}) + 12=9-10$. We suspect that this is related to the fact that we have only two oxygen lines in our spectra, both in UV, where there is overlap with various iron lines. 
We therefore fix it based on the evolutionary grids of \citet{2011A&A...530A.115B} and \citet{2015A&A...573A..71K}, based on stellar mass, rotation and age as derived by \citet{Bestenlehner2020}. 

The optical~+~UV \pyGA runs have a population of 95 or 191 individuals (6 or 12 free parameters, respectively). For the WNh stars we have 239 individuals (14 free parameters). The runs of most stars converge in approximately 20 or 40 generations (6 or 12-14 free parameters), so to be on the safe side, we iterate for 30 or 60 generations (6 or 12-14 free parameters). The limits within which each parameter is allowed to vary can be read off from fitness plots shown in the run overview of each star, which can be found in \Cref{sec:app:GA_summaries}. We discuss the robustness of this setup in \cref{sec:robust}.

\begin{figure*}
    \centering
    \includegraphics[width=0.98\textwidth]{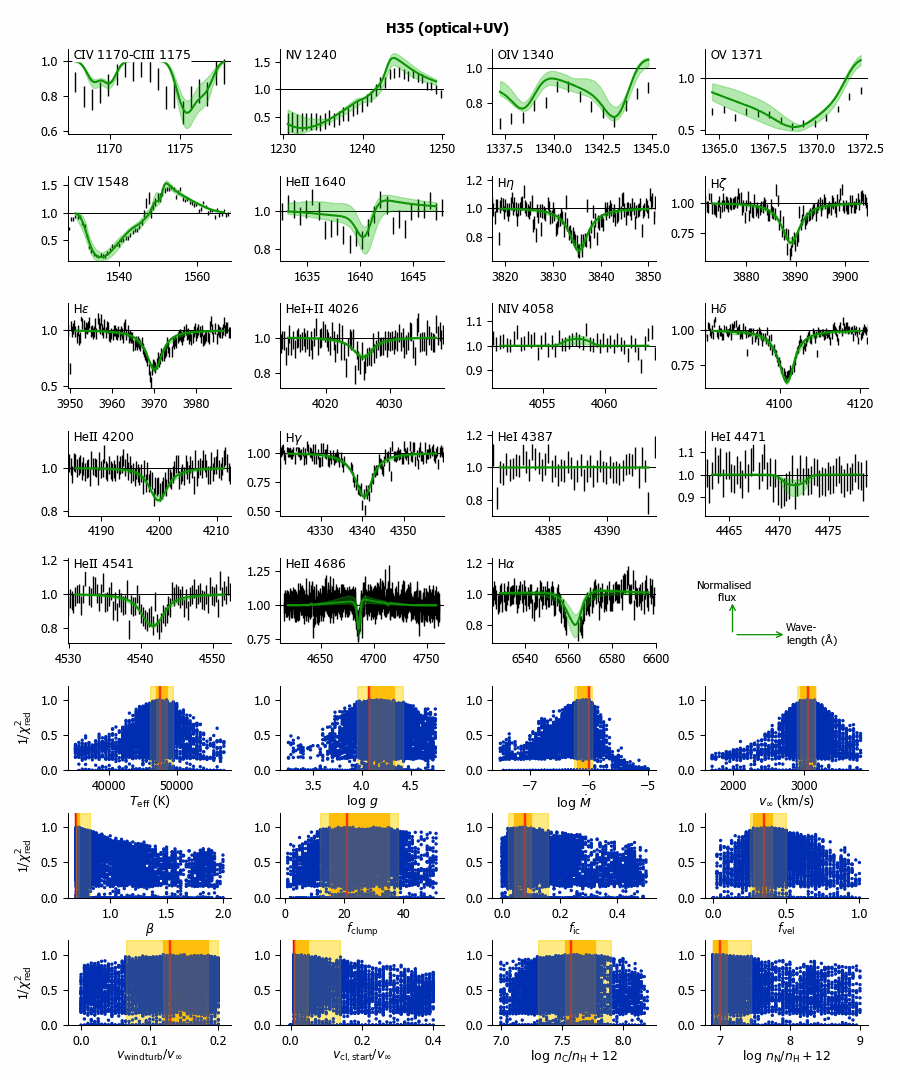}
    \caption{\PyGA output summary for the optical~+~UV run of H35 (60 generations). The top of the figure shows the line profiles that were considered in the fit. For each spectral line we show the observed spectrum (black bars), the best fit model (green solid line), and the family of best fitting models, that is, the region spanned by all models in the $2 \sigma$ confidence interval (light green shaded area). The bottom of the figure shows for each free parameter the goodness-of-fit (expressed as $1/$\chisred) of each model of the run represented by a dark blue dot). The position of the best model, as well as the 1$\sigma$ and 2$\sigma$ error regions (dark and light shaded yellow, respectively) are indicated. Output summaries for the other runs can be found in \Cref{sec:app:GA_summaries}.}
    \label{fig:fitspec_example_H35}
\end{figure*}

\begin{sidewaysfigure*}
  \includegraphics[width=25cm]{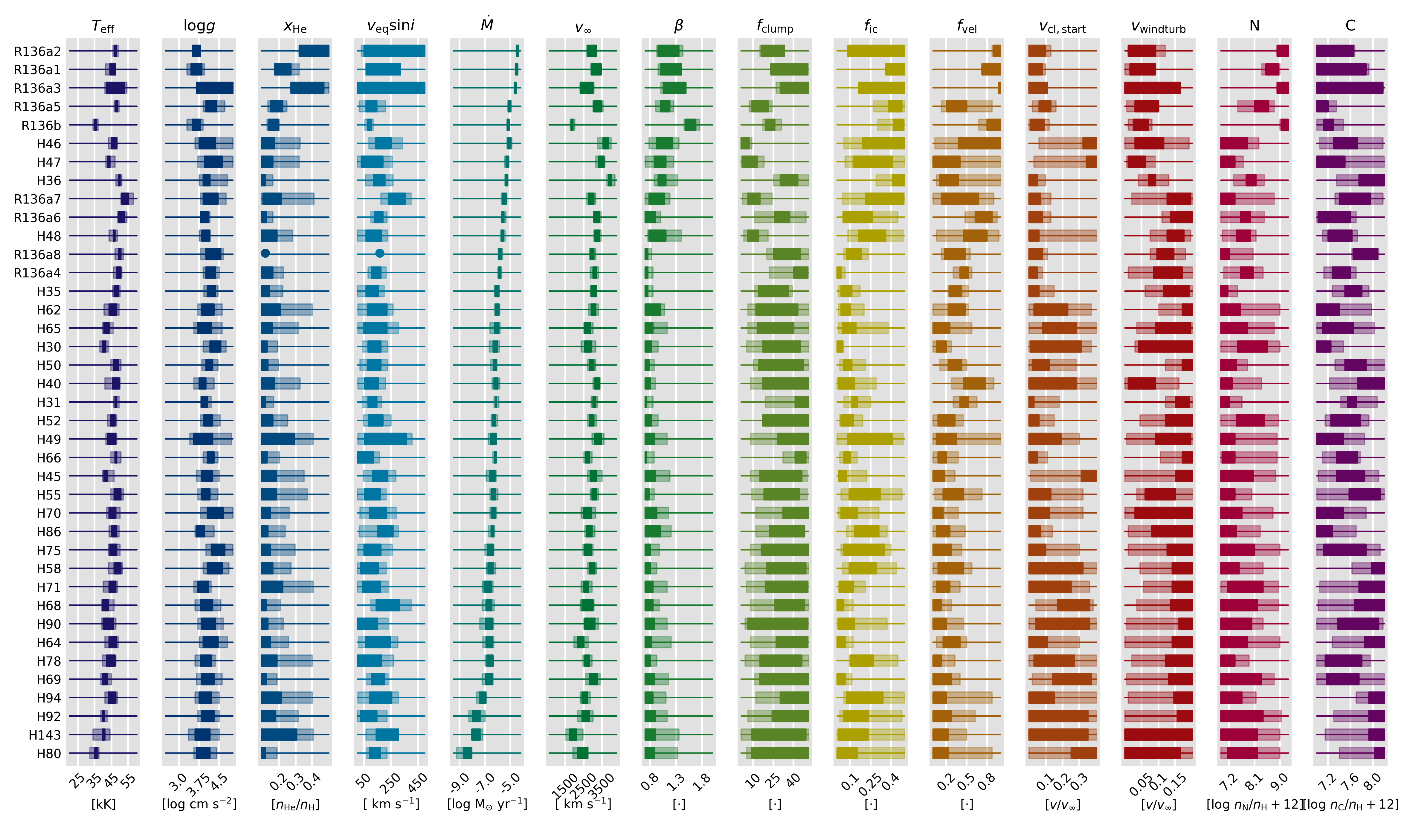}
      \caption{Best fit parameter ranges for the 39 stars where a 12-free-parameter optical~+~UV fit was done, ordered by decreasing $\dot{M}$. First four columns (blue colours) show basic stellar parameters \teff and \logg as well as \vsini and \yhe derived from the optical-only fit. The next eight columns (green to red colours) show the parameters that describe the wind and wind structure. The last two columns (pink and purple) concern the C and N abundances. The darker and lighter shaded regions correspond to $1 \sigma$ and $2 \sigma$ uncertainties, respectively. Note that R136a8 has no optical data and thus no optical-only fit: in this case \vsini and \yhe are indicated with a $\bullet$.}
         \label{fig:param_overview_allfree}
\end{sidewaysfigure*}

\begin{figure*}
\centering
\includegraphics[width=0.9\textwidth]{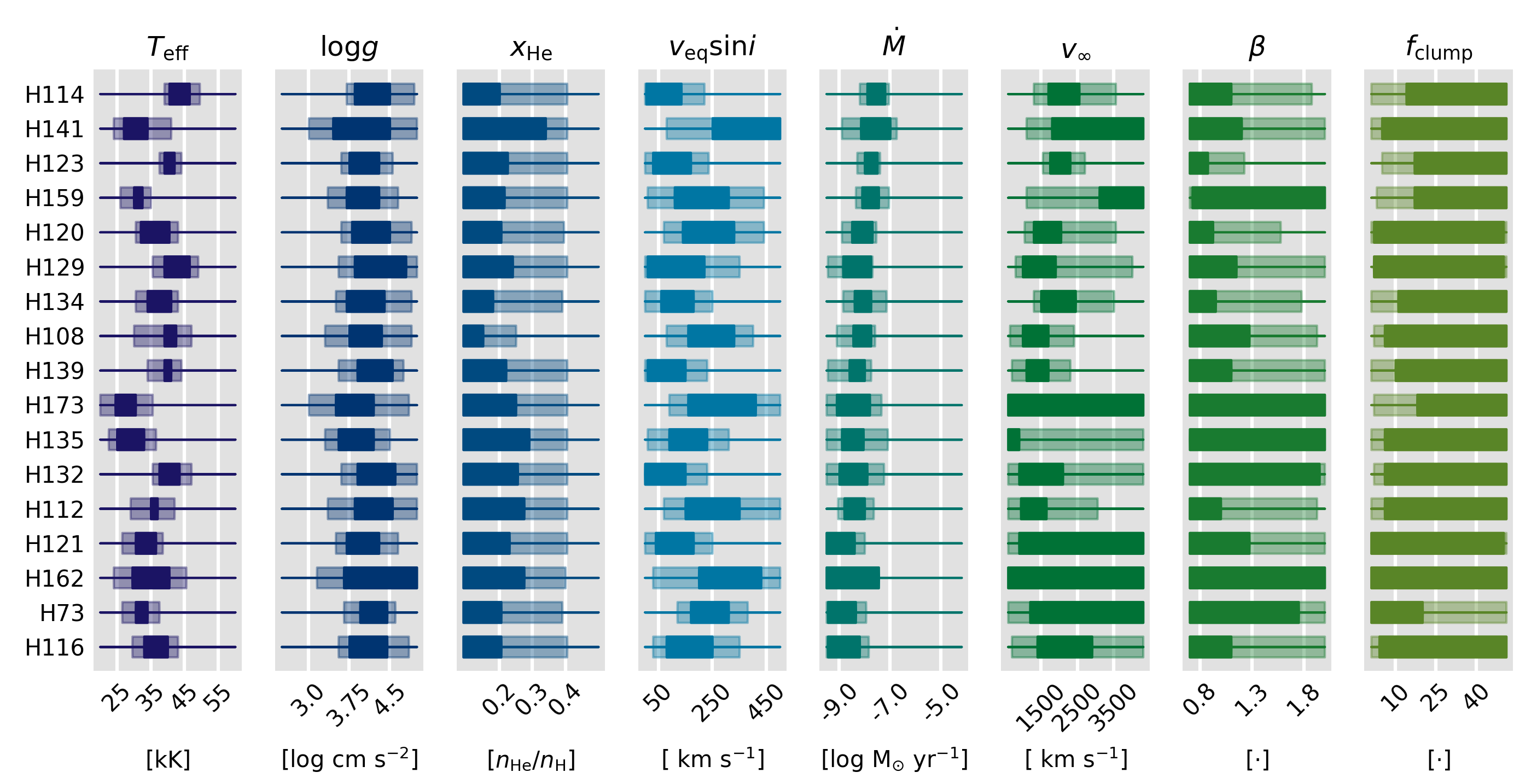}
      \caption{As \cref{fig:param_overview_allfree}, but for the best fit parameter ranges for the 17 stars where a 6-free-parameter optical~+~UV fit was done. The columns correspond to the first eight columns of \cref{fig:param_overview_allfree}. }
         \label{fig:param_overview_6free}
\end{figure*}

\subsubsection{Diagnostic line selection}

We use all strong spectral lines that are present in our spectra that can be synthesised with \fw V10.3.1 and where interstellar absorption or emission did not pose a problem. For the optical, these are lines of H, \hei, \heii, \niii, \niv and/or \nv. No optical C or O lines were available due to the limited optical wavelength range and moderate \snr and resolution. In some cases data quality was too poor to include a certain line and the line was omitted from fitting; in particular this concerned \halpha for 11 sources. We included the following UV lines in the fits of most stars: \cuvivline, \cuviiiline, \nvuvline, \oivline, \ovline, \CIVline, \heiiuvline. Note that:
\begin{itemize}
    \item For \nvuvline we only fit the part that could be recovered after the Ly-$\alpha$ correction (see also \cref{sec:ismcorrection}).
    \item In one case we fitted \siivline (R136b, see also \cref{sec:ismcorrection}). 
    \item Part of \oivline is clipped because of the presence of the strong interstellar C{\sc ii}~1336 line. 
    \item For stars where \ovline was very weak (cooler stars), we omit it from the analysis completely, as, in those cases, the iron pseudo-continuum dominates the absorption. 
    \item \nivuvline was included for about half of the stars. In cases where we had GHRS data we included the full line. In other cases, we included the blue part of the line from the HST data, but only if this was clearly visible in absorption. 
\end{itemize}  
An overview of the spectral lines used for the analysis of each individual star is presented in \Cref{tab:app:linelist}. 

We note that the UV spectroscopy includes diagnostics for \mdot, \vinf as well as for the wind structure parameters \fcl, \fic, \fvel, so that we can break the degeneracy between these parameters. 
For example, the strength of \halpha depends on the density squared, whereas resonance lines typically depend linearly on density, and so respond to clumping differently \citep[e.g.,][]{2006A&A...454..625P}. Clumps often become optically thick in strong resonance lines, while recombination lines are generally less affected; nonetheless, vorosity effects can sometimes also result in extra light leakage in recombination lines, resulting in weaker profiles \citep[][]{2010A&A...510A..11S,2011A&A...528A..64S,2007A&A...476.1331O,2012A&A...541A..37S}.  
\citealt{2005A&A...438..301B} find that \ovline and \nivuvline are also indirectly (due to a modified ionisation structure) sensitive to optically thin clumping, where typically the absorption part of the lines get weaker for higher clumping factors.  
A non-void interclump density further affects line saturation, for example in the case of \nvuvline \citep[][]{2008ApJ...685L.149Z,2012A&A...541A..37S,2013A&A...559A.130S,2018A&A...619A..59S}. In particular, both the absorption and emission parts of the line profiles get stronger with a larger interclump density, where \citet{2012A&A...541A..37S} find that the effect is most pronounced for the strong lines.  
Lastly, the onset of clumping affects the line shape close to the line centre \citep{2003ApJ...595.1182B,2012A&A...541A..37S}. 

\section{Results \label{sec:results}}

For 39 stars, we obtain 14 stellar and wind parameters per star, for the remaining 17 stars, 8 parameters. For the WNh stars we additionally obtain oxygen surface abundance, as their oxygen lines are very strong and dominate the iron pseudo-continuum. A representative example of an output summary is presented in \cref{fig:fitspec_example_H35}. The top half of the figure shows that the agreement between models and data is good: the best fitting model and the family of best solutions ($2 \sigma$ confidence region) cover the error bars on the data both for the optical as well as the UV data. 
The bottom half of the figure contains the goodness of fit for all computed models. This is illustrative for the way we derive uncertainties on all parameters: if the fitness distribution of a certain parameter is strongly peaked, the uncertainties on that parameter are small; if it is wide, the uncertainties are large. Output summaries for the other stars can be found in \Cref{sec:app:GA_summaries}. 

The best fit parameters of the optical~+~UV runs for all stars are presented in \cref{fig:param_overview_allfree,fig:param_overview_6free} and \Cref{tab:all_params_UVopt,tab:6free_params_UVopt}. Notes on peculiarities of individual sources can be found in \cref{app:notespersource}. In \cref{sec:app:fitvalues} we present additional values derived from the optical~+~UV runs such as stellar masses, ages and ionising fluxes, as well as best fit values of the optical-only runs.
Note that we derive several parameters from both the optical-only as well as the optical~+~UV analysis. 
In the remainder of the paper, unless specified otherwise, we show and interpret only one set of values: \yhe and \vsini were taken from the optical-only analysis\footnote{We do not fit this in the optical~+~UV analysis.} while the remaining parameters were taken from the optical~+~UV analysis. The WNh stars are an exception: here we do measure \yhe in the optical~+~UV fit, so in this case we use this value instead of the optical-only value. Lastly, our values generally agree well with the stellar properties derived by \citet{Bestenlehner2020} based on optical spectroscopy only. A detailed discussion of the comparison of different methods can be found in \cref{app:comparisons}. 
In the rest of this section we highlight several results that deserve special attention, and conclude with a robustness analysis. 
\subsection{Hertzsprung-Russell diagram\label{res:teffs}}

\begin{figure}
    \centering
    \includegraphics[width=0.47\textwidth]{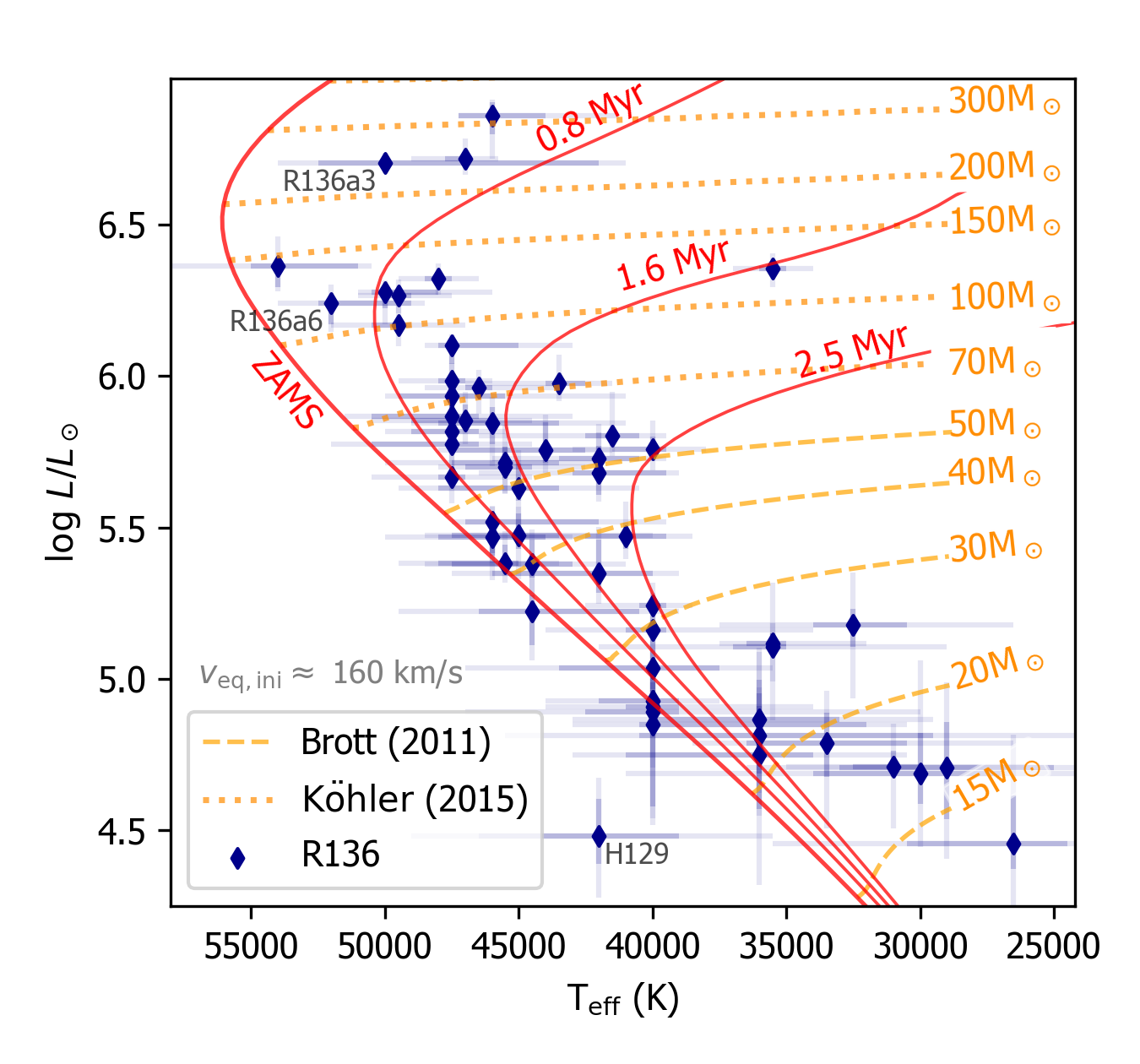}
    \caption{Positions of the R136 sources (optical~+~UV analysis, dark blue points) in the Hertzsprung-Russell diagram. Yellow dashed and dotted yellow lines are LMC evolutionary tracks of \citet{2011A&A...530A.115B} and \citet{2015A&A...573A..71K}, respectively. Red solid lines are isochrones. All tracks have an initial rotation of $\approx$ 160 km~s$^{-1}$, representative for the O-stars in the sample. Note that the tracks shown in \cref{fig:a1a2a3mass} have a higher initial rotation. }
    \label{fig:hrd_uvopt_all}
\end{figure}

\Cref{fig:hrd_uvopt_all} shows the derived temperatures and luminosities in a Hertzsprung-Russell diagram (HRD). We review the HRD positions of the stars and inspect the fit of all sources by eye to check for irregularities. From this we conclude that our temperature and luminosity measurements are reliable for all stars except H129 and R136a3. For R136a3 we find a low temperature (42000~K), but see in the spectrum strong lines of higher ionised ions such as \nv and \ov. These lines are not matched by the best fit model, where they are weak. A higher temperature for this star thus seems more likely and we test this with an additional run where we fix \teff to 50000~K, the value found by \citet{Bestenlehner2020}. Although this decreases the fitness of several other lines -- and in such a way that we obtain a worse overall fitness -- a higher temperature does improve the fit to the  \nv and \ov lines, and places the star closer to the other WNh stars in the HRD. 
Lastly, from the fit of the iron lines in the UV we found a best fitting temperature of 47000~K, where the fitness of the 50000~K model is almost as good, but the fit to the 45000~K model significantly less, even worse for the 42000~K model (\cref{fig:app:a3_iron}). 
Taking all this into account, we consider the higher \teff for this star more likely and accept the parameters of the fixed-\teff run (50000~K) as the parameters which we use for further analysis (for more details and the spectral fits, see \cref{sec:app:a3_analysis}). 
The best fitting models of H129 seem to fit the data well, however, the \snr of this source is very low and its position in HRD is far left of the main sequence where we do not expect any O-type stars. \citet{Bestenlehner2020} reported for this star a total-to-selective extinction $R_V$ that was $5\sigma$ below the average for R136. This could point to NIR-excess, although this would imply an even lower luminosity for this star, keeping it in the improbable HRD region. 

For the subsequent analysis we only use temperatures from our optical+UV analysis.  In \cref{app:teff_discussion} we present a detailed comparison of the different temperature measurements (our three different measurements, plus those of \citealt{Bestenlehner2020}). Generally, the temperatures agree within their uncertainties. 

\begin{sidewaystable*}
\small 
\caption{Best fit parameters and $1\sigma$ error bars$^{a)}$ for the optical~+~UV fits of 39 stars where we fitted 12 free parameters. \label{tab:bestfit:12free}}

\begin{tabular}{l  r@{$\pm$}l r@{$\pm$}l r@{$\pm$}l r@{$\pm$}l r@{$\pm$}l r@{$\pm$}l r@{$\pm$}l r@{$\pm$}l r@{$\pm$}l r@{$\pm$}l r@{$\pm$}l r@{$\pm$}l r@{$\pm$}l r@{$\pm$}l}
\hline \hline
Source  & \multicolumn{2}{c}{$\log L/{\rm L}_{\odot}$} & \multicolumn{2}{c}{$T_{\rm eff}$(K)} & \multicolumn{2}{c}{$\log g$} & \multicolumn{2}{c}{$R_{\star}/{\rm R}_{\odot}$} & \multicolumn{2}{c}{$\log\dot{M}$} & \multicolumn{2}{c}{$v_{\infty}$ (km/s)} & \multicolumn{2}{c}{$v_{\rm windturb}/v_{\infty}$} & \multicolumn{2}{c}{$f_{\rm cl}$} & \multicolumn{2}{c}{$f_{\rm ic}$} & \multicolumn{2}{c}{$f_{\rm vel}$} & \multicolumn{2}{c}{$\beta$} & \multicolumn{2}{c}{$v_{\rm cl, start}/v_{\infty}$} & \multicolumn{2}{c}{$\log(\frac{n_{\rm C}}{n_{\rm H}})$+12} & \multicolumn{2}{c}{$\log(\frac{n_{\rm N}}{n_{\rm H}})$+12} \\ \hline 

R136a1$^{b)}$ & 6.86& $^{0.04}_{0.07}$ & 46000& $^{1250}_{2375}$ & 3.65& $^{0.20}_{0.20}$ & 42.7& $^{1.6}_{0.9}$ & -4.57& $^{0.13}_{0.03}$ & 3150& $^{300}_{250}$ & 0.04& $^{0.05}_{0.03}$ & 43& $^{7}_{20}$ & 0.48& $^{0.03}_{0.12}$ & 0.97& $^{0.05}_{0.25}$ & 1.18& $^{0.23}_{0.17}$ & 0.03& $^{0.05}_{0.03}$ & 7.60& $^{0.28}_{0.60}$ & 8.75& $^{0.20}_{0.25}$ \\
R136a2$^{b)}$ & 6.71& $^{0.03}_{0.03}$ & 47000& $^{1000}_{625}$ & 3.55& $^{0.25}_{0.05}$ & 34.7& $^{0.7}_{0.8}$ & -4.48& $^{0.12}_{0.06}$ & 2900& $^{300}_{200}$ & 0.07& $^{0.03}_{0.06}$ & 29& $^{4}_{13}$ & 0.50& $^{0.01}_{0.42}$ & 1.00& $^{0.03}_{0.10}$ & 1.18& $^{0.17}_{0.23}$ & 0.03& $^{0.07}_{0.03}$ & 7.20& $^{0.45}_{0.23}$ & 9.05& $^{0.25}_{0.15}$ \\
R136a3$^{b,c)}$ & 6.70& $^{0.02}_{0.02}$ & 50000& $^{2500}_{8000}$ & 4.05& $^{1.00}_{0.40}$ & 30.2& $^{0.6}_{0.6}$ & -4.64& $^{0.08}_{0.05}$ & 2700& $^{300}_{400}$ & 0.07& $^{0.10}_{0.07}$ & 46& $^{5}_{16}$ & 0.41& $^{0.10}_{0.25}$ & 1.00& $^{0.03}_{0.03}$ & 1.30& $^{0.17}_{0.25}$ & 0.02& $^{0.09}_{0.02}$ & 7.75& $^{0.42}_{0.78}$ & 9.15& $^{0.25}_{0.25}$ \\
R136a4 & 6.28& $^{0.02}_{0.05}$ & 50000& $^{500}_{2000}$ & 4.15& $^{0.20}_{0.15}$ & 18.5& $^{0.5}_{0.3}$ & -5.84& $^{0.08}_{0.10}$ & 3150& $^{75}_{150}$ & 0.14& $^{0.03}_{0.07}$ & 49& $^{1}_{14}$ & 0.00& $^{0.03}_{0.01}$ & 0.45& $^{0.07}_{0.07}$ & 0.72& $^{0.06}_{0.05}$ & 0.03& $^{0.03}_{0.03}$ & 7.38& $^{0.23}_{0.12}$ & 7.90& $^{0.25}_{0.45}$ \\
R136a5 & 6.32& $^{0.03}_{0.03}$ & 48000& $^{750}_{750}$ & 4.35& $^{0.17}_{0.35}$ & 21.1& $^{0.4}_{0.4}$ & -5.09& $^{0.10}_{0.10}$ & 3250& $^{200}_{200}$ & 0.06& $^{0.04}_{0.03}$ & 16& $^{5}_{7}$ & 0.44& $^{0.04}_{0.09}$ & 0.30& $^{0.29}_{0.10}$ & 1.10& $^{0.07}_{0.10}$ & 0.11& $^{0.03}_{0.05}$ & 7.03& $^{0.17}_{0.05}$ & 8.30& $^{0.30}_{0.40}$ \\
R136a6 & 6.24& $^{0.03}_{0.07}$ & 52000& $^{1000}_{3000}$ & 4.00& $^{0.10}_{0.20}$ & 16.4& $^{0.5}_{0.3}$ & -5.60& $^{0.11}_{0.08}$ & 3200& $^{150}_{150}$ & 0.18& $^{0.01}_{0.05}$ & 34& $^{8}_{12}$ & 0.19& $^{0.13}_{0.14}$ & 0.70& $^{0.17}_{0.11}$ & 0.78& $^{0.12}_{0.10}$ & 0.05& $^{0.04}_{0.05}$ & 7.53& $^{0.09}_{0.50}$ & 7.80& $^{0.33}_{0.45}$ \\
R136a7 & 6.36& $^{0.05}_{0.07}$ & 54000& $^{2000}_{3000}$ & 4.30& $^{0.23}_{0.40}$ & 17.5& $^{0.5}_{0.4}$ & -5.52& $^{0.18}_{0.10}$ & 2900& $^{150}_{150}$ & 0.15& $^{0.04}_{0.07}$ & 12& $^{6}_{6}$ & 0.39& $^{0.10}_{0.18}$ & 0.25& $^{0.42}_{0.12}$ & 0.93& $^{0.15}_{0.15}$ & 0.02& $^{0.06}_{0.02}$ & 7.70& $^{0.25}_{0.30}$ & 7.10& $^{0.85}_{0.25}$ \\
R136a8 & 6.17& $^{0.03}_{0.03}$ & 49500& $^{1250}_{1250}$ & 4.25& $^{0.30}_{0.25}$ & 16.6& $^{0.4}_{0.3}$ & -5.82& $^{0.10}_{0.10}$ & 3000& $^{75}_{150}$ & 0.11& $^{0.04}_{0.02}$ & 37& $^{8}_{12}$ & 0.13& $^{0.05}_{0.06}$ & 0.42& $^{0.06}_{0.25}$ & 0.70& $^{0.06}_{0.03}$ & 0.02& $^{0.06}_{0.02}$ & 7.72& $^{0.35}_{0.07}$ & 6.95& $^{0.55}_{0.10}$ \\
R136b & 6.35& $^{0.03}_{0.03}$ & 35500& $^{750}_{750}$ & 3.55& $^{0.25}_{0.12}$ & 40.0& $^{0.8}_{0.8}$ & -5.15& $^{0.05}_{0.10}$ & 1850& $^{75}_{50}$ & 0.06& $^{0.01}_{0.03}$ & 20& $^{6}_{2}$ & 0.45& $^{0.04}_{0.07}$ & 0.95& $^{0.05}_{0.15}$ & 1.50& $^{0.17}_{0.03}$ & 0.03& $^{0.06}_{0.02}$ & 7.28& $^{0.10}_{0.15}$ & 9.20& $^{0.10}_{0.15}$ \\
H30 & 5.76& $^{0.05}_{0.03}$ & 40000& $^{1500}_{1000}$ & 4.20& $^{0.35}_{0.15}$ & 15.9& $^{0.3}_{0.4}$ & -6.09& $^{0.11}_{0.29}$ & 2650& $^{250}_{150}$ & 0.13& $^{0.07}_{0.09}$ & 31& $^{14}_{14}$ & 0.01& $^{0.03}_{0.01}$ & 0.10& $^{0.10}_{0.05}$ & 0.70& $^{0.12}_{0.03}$ & 0.28& $^{0.04}_{0.27}$ & 7.15& $^{0.16}_{0.17}$ & 8.50& $^{0.25}_{1.00}$ \\
H31 & 5.98& $^{0.03}_{0.03}$ & 47500& $^{1000}_{750}$ & 4.00& $^{0.10}_{0.15}$ & 14.6& $^{0.3}_{0.3}$ & -6.15& $^{0.11}_{0.08}$ & 3050& $^{100}_{75}$ & 0.15& $^{0.04}_{0.02}$ & 42& $^{9}_{12}$ & 0.14& $^{0.06}_{0.04}$ & 0.47& $^{0.09}_{0.09}$ & 0.70& $^{0.07}_{0.03}$ & 0.02& $^{0.08}_{0.02}$ & 7.67& $^{0.20}_{0.12}$ & 6.90& $^{0.38}_{0.05}$ \\
H35 & 5.82& $^{0.04}_{0.03}$ & 47500& $^{1500}_{1000}$ & 4.08& $^{0.28}_{0.07}$ & 12.1& $^{0.2}_{0.3}$ & -5.97& $^{0.05}_{0.25}$ & 3050& $^{150}_{150}$ & 0.13& $^{0.06}_{0.04}$ & 21& $^{15}_{7}$ & 0.08& $^{0.04}_{0.05}$ & 0.35& $^{0.09}_{0.10}$ & 0.70& $^{0.07}_{0.03}$ & 0.01& $^{0.07}_{0.01}$ & 7.58& $^{0.23}_{0.15}$ & 7.00& $^{0.25}_{0.15}$ \\
H36 & 6.27& $^{0.03}_{0.03}$ & 49500& $^{750}_{1000}$ & 4.10& $^{0.35}_{0.20}$ & 18.6& $^{0.4}_{0.4}$ & -5.29& $^{0.05}_{0.10}$ & 3900& $^{150}_{75}$ & 0.08& $^{0.03}_{0.02}$ & 31& $^{12}_{3}$ & 0.49& $^{0.02}_{0.10}$ & 0.25& $^{0.38}_{0.15}$ & 1.05& $^{0.14}_{0.10}$ & 0.03& $^{0.04}_{0.03}$ & 8.07& $^{0.12}_{0.36}$ & 8.05& $^{0.20}_{0.35}$ \\
H40 & 5.93& $^{0.05}_{0.09}$ & 47500& $^{2000}_{3250}$ & 3.90& $^{0.20}_{0.17}$ & 13.8& $^{0.6}_{0.3}$ & -6.12& $^{0.14}_{0.18}$ & 3250& $^{100}_{150}$ & 0.04& $^{0.06}_{0.03}$ & 44& $^{7}_{27}$ & 0.02& $^{0.14}_{0.01}$ & 0.60& $^{0.17}_{0.16}$ & 0.70& $^{0.10}_{0.03}$ & 0.27& $^{0.07}_{0.27}$ & 8.18& $^{0.03}_{0.47}$ & 6.95& $^{0.70}_{0.10}$ \\
H45 & 5.80& $^{0.07}_{0.03}$ & 41500& $^{2500}_{1000}$ & 4.15& $^{0.17}_{0.30}$ & 15.5& $^{0.3}_{0.5}$ & -6.27& $^{0.08}_{0.35}$ & 3100& $^{200}_{250}$ & 0.17& $^{0.03}_{0.09}$ & 28& $^{18}_{13}$ & 0.03& $^{0.10}_{0.02}$ & 0.05& $^{0.14}_{0.05}$ & 0.70& $^{0.24}_{0.03}$ & 0.39& $^{0.01}_{0.19}$ & 7.47& $^{0.38}_{0.23}$ & 7.20& $^{0.85}_{0.30}$ \\
H46 & 6.10& $^{0.02}_{0.06}$ & 47500& $^{500}_{2500}$ & 3.90& $^{0.55}_{0.15}$ & 16.7& $^{0.5}_{0.3}$ & -5.15& $^{0.18}_{0.06}$ & 3650& $^{200}_{200}$ & 0.09& $^{0.05}_{0.06}$ & 4& $^{3}_{4}$ & 0.39& $^{0.11}_{0.20}$ & 0.82& $^{0.17}_{0.45}$ & 1.05& $^{0.17}_{0.14}$ & 0.36& $^{0.05}_{0.15}$ & 7.40& $^{0.39}_{0.17}$ & 7.80& $^{0.23}_{0.95}$ \\
H47 & 5.98& $^{0.05}_{0.03}$ & 43500& $^{1750}_{1000}$ & 4.45& $^{0.30}_{0.50}$ & 17.3& $^{0.4}_{0.4}$ & -5.25& $^{0.08}_{0.11}$ & 3450& $^{150}_{150}$ & 0.03& $^{0.04}_{0.01}$ & 6& $^{7}_{4}$ & 0.30& $^{0.11}_{0.18}$ & 0.05& $^{0.49}_{0.05}$ & 0.95& $^{0.15}_{0.12}$ & 0.40& $^{0.01}_{0.18}$ & 7.33& $^{0.45}_{0.33}$ & 7.05& $^{0.35}_{0.15}$ \\
H48 & 5.96& $^{0.03}_{0.04}$ & 46500& $^{1000}_{1500}$ & 3.90& $^{0.23}_{0.07}$ & 14.9& $^{0.4}_{0.3}$ & -5.60& $^{0.09}_{0.10}$ & 3200& $^{150}_{100}$ & 0.17& $^{0.01}_{0.05}$ & 9& $^{6}_{3}$ & 0.18& $^{0.18}_{0.05}$ & 0.60& $^{0.20}_{0.28}$ & 0.93& $^{0.24}_{0.15}$ & 0.02& $^{0.20}_{0.02}$ & 7.25& $^{0.38}_{0.07}$ & 7.60& $^{0.35}_{0.33}$ \\
H49 & 5.76& $^{0.09}_{0.04}$ & 44000& $^{3500}_{1500}$ & 3.85& $^{0.55}_{0.30}$ & 13.1& $^{0.3}_{0.5}$ & -6.22& $^{0.10}_{0.33}$ & 3300& $^{150}_{200}$ & 0.12& $^{0.07}_{0.06}$ & 38& $^{13}_{15}$ & 0.18& $^{0.23}_{0.10}$ & 0.20& $^{0.40}_{0.15}$ & 0.72& $^{0.20}_{0.05}$ & 0.02& $^{0.17}_{0.02}$ & 7.12& $^{0.36}_{0.15}$ & 7.15& $^{0.88}_{0.30}$ \\
H50 & 5.85& $^{0.05}_{0.04}$ & 47000& $^{2000}_{1250}$ & 4.15& $^{0.15}_{0.15}$ & 12.8& $^{0.3}_{0.3}$ & -6.12& $^{0.03}_{0.23}$ & 2850& $^{200}_{75}$ & 0.20& $^{0.01}_{0.04}$ & 15& $^{32}_{2}$ & 0.07& $^{0.07}_{0.03}$ & 0.25& $^{0.17}_{0.07}$ & 0.70& $^{0.09}_{0.03}$ & 0.06& $^{0.11}_{0.04}$ & 7.83& $^{0.20}_{0.33}$ & 7.05& $^{0.40}_{0.15}$ \\
H52 & 5.70& $^{0.04}_{0.04}$ & 45500& $^{1500}_{1500}$ & 4.05& $^{0.24}_{0.12}$ & 11.5& $^{0.3}_{0.2}$ & -6.22& $^{0.10}_{0.20}$ & 2900& $^{150}_{100}$ & 0.20& $^{0.01}_{0.07}$ & 36& $^{15}_{19}$ & 0.08& $^{0.06}_{0.05}$ & 0.17& $^{0.15}_{0.10}$ & 0.75& $^{0.10}_{0.07}$ & 0.02& $^{0.07}_{0.02}$ & 7.45& $^{0.33}_{0.20}$ & 7.75& $^{0.70}_{0.45}$ \\
H55 & 5.77& $^{0.08}_{0.04}$ & 47500& $^{3000}_{1500}$ & 3.95& $^{0.25}_{0.12}$ & 11.5& $^{0.3}_{0.4}$ & -6.27& $^{0.14}_{0.20}$ & 3150& $^{75}_{200}$ & 0.10& $^{0.05}_{0.04}$ & 28& $^{16}_{10}$ & 0.22& $^{0.13}_{0.13}$ & 0.30& $^{0.21}_{0.15}$ & 0.72& $^{0.07}_{0.05}$ & 0.02& $^{0.15}_{0.02}$ & 7.75& $^{0.38}_{0.39}$ & 7.05& $^{0.47}_{0.20}$ \\
H58 & 5.87& $^{0.08}_{0.06}$ & 47500& $^{3000}_{2250}$ & 4.40& $^{0.23}_{0.35}$ & 12.8& $^{0.4}_{0.4}$ & -6.52& $^{0.18}_{0.25}$ & 3000& $^{100}_{200}$ & 0.15& $^{0.06}_{0.05}$ & 32& $^{17}_{14}$ & 0.11& $^{0.18}_{0.04}$ & 0.30& $^{0.15}_{0.23}$ & 0.70& $^{0.11}_{0.03}$ & 0.22& $^{0.10}_{0.22}$ & 8.18& $^{0.05}_{0.28}$ & 7.00& $^{0.70}_{0.15}$ \\
H62 & 5.63& $^{0.08}_{0.07}$ & 45000& $^{3000}_{2250}$ & 4.05& $^{0.28}_{0.20}$ & 10.8& $^{0.3}_{0.4}$ & -6.02& $^{0.10}_{0.15}$ & 3050& $^{125}_{150}$ & 0.20& $^{0.01}_{0.05}$ & 15& $^{28}_{4}$ & 0.05& $^{0.07}_{0.02}$ & 0.38& $^{0.10}_{0.19}$ & 0.72& $^{0.28}_{0.05}$ & 0.09& $^{0.14}_{0.06}$ & 7.15& $^{0.41}_{0.17}$ & 7.15& $^{0.93}_{0.30}$ \\
H64 & 5.85& $^{0.05}_{0.07}$ & 46000& $^{2000}_{2500}$ & 4.30& $^{0.25}_{0.40}$ & 13.3& $^{0.4}_{0.3}$ & -6.67& $^{0.25}_{0.25}$ & 2250& $^{250}_{150}$ & 0.20& $^{0.01}_{0.10}$ & 37& $^{13}_{14}$ & 0.02& $^{0.05}_{0.02}$ & 0.30& $^{0.10}_{0.15}$ & 0.70& $^{0.25}_{0.03}$ & 0.01& $^{0.14}_{0.01}$ & 8.05& $^{0.17}_{0.39}$ & 7.25& $^{0.88}_{0.40}$ \\
H65 & 5.73& $^{0.06}_{0.04}$ & 42000& $^{2000}_{1500}$ & 3.85& $^{0.38}_{0.15}$ & 13.9& $^{0.3}_{0.4}$ & -6.07& $^{0.15}_{0.28}$ & 2700& $^{150}_{150}$ & 0.14& $^{0.06}_{0.05}$ & 25& $^{15}_{12}$ & 0.10& $^{0.14}_{0.06}$ & 0.03& $^{0.28}_{0.03}$ & 0.72& $^{0.20}_{0.05}$ & 0.14& $^{0.14}_{0.07}$ & 7.30& $^{0.36}_{0.20}$ & 7.45& $^{0.68}_{0.55}$ \\
H66 & 5.66& $^{0.04}_{0.04}$ & 47500& $^{1500}_{1500}$ & 4.10& $^{0.20}_{0.10}$ & 10.1& $^{0.2}_{0.2}$ & -6.22& $^{0.10}_{0.09}$ & 2700& $^{125}_{100}$ & 0.20& $^{0.01}_{0.05}$ & 42& $^{7}_{5}$ & 0.07& $^{0.04}_{0.03}$ & 0.10& $^{0.14}_{0.05}$ & 0.70& $^{0.09}_{0.03}$ & 0.01& $^{0.05}_{0.01}$ & 7.70& $^{0.04}_{0.35}$ & 7.00& $^{0.95}_{0.07}$ \\
H68 & 5.68& $^{0.07}_{0.07}$ & 42000& $^{2250}_{2500}$ & 3.95& $^{0.30}_{0.17}$ & 13.2& $^{0.4}_{0.4}$ & -6.62& $^{0.18}_{0.30}$ & 2650& $^{350}_{250}$ & 0.18& $^{0.02}_{0.07}$ & 30& $^{18}_{11}$ & 0.01& $^{0.06}_{0.01}$ & 0.05& $^{0.11}_{0.05}$ & 0.75& $^{0.11}_{0.07}$ & 0.31& $^{0.05}_{0.14}$ & 8.15& $^{0.05}_{0.56}$ & 8.15& $^{0.40}_{1.25}$ \\
H69 & 5.47& $^{0.06}_{0.04}$ & 41000& $^{2000}_{1500}$ & 4.15& $^{0.23}_{0.30}$ & 10.9& $^{0.3}_{0.3}$ & -6.83& $^{0.40}_{0.20}$ & 3050& $^{200}_{250}$ & 0.20& $^{0.01}_{0.10}$ & 50& $^{1}_{35}$ & 0.04& $^{0.04}_{0.04}$ & 0.05& $^{0.23}_{0.05}$ & 0.78& $^{0.19}_{0.10}$ & 0.35& $^{0.03}_{0.21}$ & 7.58& $^{0.31}_{0.40}$ & 7.00& $^{1.35}_{0.15}$ \\
H70 & 5.71& $^{0.06}_{0.05}$ & 45500& $^{2250}_{2000}$ & 4.15& $^{0.50}_{0.17}$ & 11.7& $^{0.3}_{0.3}$ & -6.32& $^{0.13}_{0.15}$ & 2600& $^{300}_{125}$ & 0.16& $^{0.04}_{0.13}$ & 49& $^{2}_{18}$ & 0.09& $^{0.11}_{0.06}$ & 0.03& $^{0.17}_{0.03}$ & 0.70& $^{0.23}_{0.03}$ & 0.02& $^{0.15}_{0.02}$ & 7.15& $^{0.36}_{0.17}$ & 7.05& $^{0.85}_{0.20}$ \\
H71 & 5.47& $^{0.07}_{0.07}$ & 45000& $^{2500}_{2500}$ & 3.90& $^{0.20}_{0.20}$ & 9.1& $^{0.3}_{0.3}$ & -6.57& $^{0.10}_{0.50}$ & 2650& $^{150}_{150}$ & 0.15& $^{0.04}_{0.05}$ & 21& $^{29}_{8}$ & 0.05& $^{0.08}_{0.03}$ & 0.23& $^{0.09}_{0.17}$ & 0.70& $^{0.21}_{0.03}$ & 0.01& $^{0.24}_{0.01}$ & 7.83& $^{0.40}_{0.39}$ & 7.75& $^{0.65}_{0.60}$ \\
H75 & 5.47& $^{0.05}_{0.05}$ & 46000& $^{2000}_{2000}$ & 4.45& $^{0.28}_{0.25}$ & 8.6& $^{0.2}_{0.2}$ & -6.47& $^{0.10}_{0.35}$ & 2700& $^{200}_{150}$ & 0.15& $^{0.05}_{0.07}$ & 26& $^{25}_{10}$ & 0.11& $^{0.24}_{0.06}$ & 0.05& $^{0.23}_{0.05}$ & 0.72& $^{0.12}_{0.05}$ & 0.04& $^{0.13}_{0.04}$ & 7.53& $^{0.35}_{0.40}$ & 7.35& $^{0.82}_{0.45}$ \\
H78 & 5.52& $^{0.03}_{0.11}$ & 46000& $^{1000}_{4000}$ & 4.00& $^{0.20}_{0.20}$ & 9.1& $^{0.4}_{0.2}$ & -6.67& $^{0.30}_{0.18}$ & 2700& $^{125}_{150}$ & 0.15& $^{0.05}_{0.07}$ & 44& $^{4}_{29}$ & 0.16& $^{0.14}_{0.06}$ & 0.07& $^{0.12}_{0.07}$ & 0.70& $^{0.11}_{0.03}$ & 0.06& $^{0.21}_{0.03}$ & 7.58& $^{0.23}_{0.42}$ & 7.25& $^{0.33}_{0.40}$ \\
H80 & 5.12& $^{0.03}_{0.06}$ & 35500& $^{1000}_{1750}$ & 3.90& $^{0.25}_{0.25}$ & 9.6& $^{0.3}_{0.2}$ & -8.29& $^{0.20}_{0.48}$ & 2400& $^{300}_{250}$ & 0.07& $^{0.09}_{0.08}$ & 32& $^{19}_{23}$ & 0.08& $^{0.21}_{0.08}$ & 0.03& $^{0.42}_{0.03}$ & 0.70& $^{0.31}_{0.03}$ & 0.34& $^{0.07}_{0.17}$ & 8.18& $^{0.05}_{0.39}$ & 7.60& $^{0.70}_{0.45}$ \\
H86 & 5.38& $^{0.06}_{0.03}$ & 45500& $^{2500}_{1000}$ & 3.85& $^{0.23}_{0.20}$ & 8.0& $^{0.2}_{0.3}$ & -6.35& $^{0.05}_{0.19}$ & 2750& $^{200}_{100}$ & 0.15& $^{0.05}_{0.07}$ & 26& $^{22}_{7}$ & 0.15& $^{0.16}_{0.04}$ & 0.15& $^{0.16}_{0.10}$ & 0.72& $^{0.28}_{0.03}$ & 0.03& $^{0.06}_{0.03}$ & 7.00& $^{0.35}_{0.03}$ & 7.15& $^{0.57}_{0.30}$ \\
H90 & 5.35& $^{0.11}_{0.06}$ & 42000& $^{4000}_{2000}$ & 3.95& $^{0.40}_{0.12}$ & 9.0& $^{0.3}_{0.4}$ & -6.62& $^{0.20}_{0.40}$ & 2800& $^{300}_{250}$ & 0.20& $^{0.01}_{0.10}$ & 22& $^{28}_{16}$ & 0.04& $^{0.17}_{0.03}$ & 0.07& $^{0.19}_{0.07}$ & 0.70& $^{0.21}_{0.03}$ & 0.21& $^{0.15}_{0.17}$ & 7.62& $^{0.47}_{0.33}$ & 7.40& $^{0.60}_{0.55}$ \\
H92 & 5.24& $^{0.04}_{0.03}$ & 40000& $^{1250}_{750}$ & 4.15& $^{0.17}_{0.30}$ & 8.8& $^{0.2}_{0.2}$ & -7.69& $^{0.35}_{0.30}$ & 2700& $^{150}_{300}$ & 0.20& $^{0.01}_{0.10}$ & 50& $^{1}_{26}$ & 0.09& $^{0.20}_{0.04}$ & 0.03& $^{0.30}_{0.03}$ & 0.78& $^{0.17}_{0.10}$ & 0.19& $^{0.16}_{0.19}$ & 8.18& $^{0.05}_{0.44}$ & 7.75& $^{0.65}_{0.85}$ \\
H94 & 5.38& $^{0.08}_{0.05}$ & 44500& $^{3000}_{1750}$ & 3.95& $^{0.28}_{0.20}$ & 8.3& $^{0.2}_{0.3}$ & -7.19& $^{0.25}_{0.25}$ & 2550& $^{150}_{150}$ & 0.18& $^{0.03}_{0.09}$ & 43& $^{8}_{16}$ & 0.12& $^{0.22}_{0.05}$ & 0.05& $^{0.41}_{0.05}$ & 0.75& $^{0.17}_{0.07}$ & 0.03& $^{0.18}_{0.03}$ & 8.20& $^{0.03}_{0.28}$ & 6.95& $^{0.70}_{0.10}$ \\
H143 & 5.16& $^{0.06}_{0.10}$ & 40000& $^{2000}_{3000}$ & 3.75& $^{0.40}_{0.20}$ & 8.0& $^{0.4}_{0.2}$ & -7.79& $^{0.40}_{0.25}$ & 1950& $^{250}_{250}$ & 0.20& $^{0.01}_{0.20}$ & 18& $^{31}_{9}$ & 0.10& $^{0.20}_{0.09}$ & 0.07& $^{0.21}_{0.07}$ & 0.72& $^{0.31}_{0.05}$ & 0.01& $^{0.34}_{0.01}$ & 8.20& $^{0.03}_{0.53}$ & 6.95& $^{1.25}_{0.10}$ \\ \hline 
\multicolumn{29}{l}{} \\ 
\multicolumn{29}{l}{\footnotesize$^{a)}$ Note that when the error bar reaches the edge of the parameter space, the best fit value is in fact an upper or lower limit. \Cref{fig:param_overview_allfree} shows when this is the case.} \\
\multicolumn{29}{l}{\footnotesize$^{b)}$ Run with 14 free parameters. Values for the helium and oxygen surface abundance can be found in \cref{tab:WNHoxy}. } \\
\multicolumn{29}{l}{\footnotesize$^{c)}$ Formal uncertainties from the \pyGA run with fixed, estimated \teff of 50~kK (see text). In reality uncertainties on all parameters for this source are larger due to the uncertain \teff.} \\

    \end{tabular}
    \label{tab:all_params_UVopt}

\end{sidewaystable*}

\begin{table*}
\centering
\small 
\caption{Best fit parameters and $1\sigma$ error bars$^{a)}$ for the optical~+~UV fits of the 17 stars where we fitted 6 free parameters.
\label{tab:bestfit:6free}}

\begin{tabular}{l  r@{$\pm$}l r@{$\pm$}l r@{$\pm$}l r@{$\pm$}l r@{$\pm$}l r@{$\pm$}l r@{$\pm$}l r@{$\pm$}l}
\hline \hline
Source  & \multicolumn{2}{c}{$\log L/{\rm L}_{\odot}$} & \multicolumn{2}{c}{$T_{\rm eff}$(K)} & \multicolumn{2}{c}{$\log g$} & \multicolumn{2}{c}{$R_{\star}/{\rm R}_{\odot}$} & \multicolumn{2}{c}{$\log\dot{M}$} & \multicolumn{2}{c}{$v_{\infty}$ (km/s)} & \multicolumn{2}{c}{$f_{\rm cl}$} & \multicolumn{2}{c}{$\beta$} \\ \hline
H73 & 5.18& $^{0.09}_{0.12}$ & 32500& $^{2500}_{3000}$ & 4.10& $^{0.35}_{0.23}$ & 12.4& $^{0.7}_{0.5}$ & -8.48& $^{0.28}_{0.95}$ & 4000& $^{600}_{2800}$ & 1& $^{25}_{1}$ & 0.90& $^{0.85}_{0.20}$ \\
H108 & 4.89& $^{0.10}_{0.18}$ & 40000& $^{3500}_{5000}$ & 4.05& $^{0.42}_{0.38}$ & 5.9& $^{0.5}_{0.2}$ & -8.09& $^{0.35}_{0.50}$ & 1350& $^{525}_{350}$ & 31& $^{20}_{25}$ & 0.70& $^{0.61}_{0.03}$ \\
H112 & 5.11& $^{0.11}_{0.12}$ & 35500& $^{3250}_{3250}$ & 4.10& $^{0.47}_{0.38}$ & 9.5& $^{0.5}_{0.4}$ & -8.40& $^{0.40}_{0.40}$ & 1300& $^{875}_{425}$ & 37& $^{14}_{31}$ & 0.70& $^{0.61}_{0.03}$ \\
H114 & 5.22& $^{0.07}_{0.11}$ & 44500& $^{2500}_{4000}$ & 4.20& $^{0.38}_{0.35}$ & 6.9& $^{0.3}_{0.2}$ & -7.48& $^{0.30}_{0.40}$ & 2100& $^{725}_{400}$ & 46& $^{5}_{32}$ & 0.70& $^{0.59}_{0.03}$ \\
H116 & 4.87& $^{0.13}_{0.12}$ & 36000& $^{4000}_{3250}$ & 4.05& $^{0.40}_{0.30}$ & 7.0& $^{0.4}_{0.4}$ & -8.68& $^{0.50}_{0.75}$ & 1900& $^{1350}_{600}$ & 45& $^{6}_{41}$ & 0.70& $^{0.66}_{0.03}$ \\
H120 & 4.85& $^{0.05}_{0.27}$ & 40000& $^{1500}_{8000}$ & 4.45& $^{0.23}_{0.65}$ & 5.6& $^{0.7}_{0.1}$ & -7.99& $^{0.30}_{0.50}$ & 1600& $^{975}_{300}$ & 37& $^{13}_{35}$ & 0.72& $^{0.42}_{0.05}$ \\
H121 & 4.79& $^{0.10}_{0.14}$ & 33500& $^{3000}_{3500}$ & 4.05& $^{0.30}_{0.35}$ & 7.4& $^{0.5}_{0.3}$ & -8.42& $^{0.23}_{1.10}$ & 4300& $^{300}_{3400}$ & 1& $^{49}_{1}$ & 0.75& $^{0.64}_{0.07}$ \\
H123 & 4.93& $^{0.06}_{0.04}$ & 40000& $^{2000}_{1250}$ & 4.00& $^{0.30}_{0.25}$ & 6.1& $^{0.1}_{0.2}$ & -7.69& $^{0.20}_{0.30}$ & 2000& $^{350}_{250}$ & 40& $^{11}_{23}$ & 0.70& $^{0.26}_{0.03}$ \\
H129 & 4.48& $^{0.12}_{0.10}$ & 42000& $^{4500}_{3250}$ & 4.25& $^{0.55}_{0.40}$ & 3.3& $^{0.2}_{0.2}$ & -8.09& $^{0.35}_{0.75}$ & 1300& $^{1350}_{300}$ & 9& $^{41}_{7}$ & 0.70& $^{0.66}_{0.03}$ \\
H132 & 5.04& $^{0.10}_{0.07}$ & 40000& $^{3500}_{2500}$ & 4.20& $^{0.40}_{0.30}$ & 6.9& $^{0.2}_{0.3}$ & -8.34& $^{0.55}_{0.65}$ & 1300& $^{1650}_{500}$ & 43& $^{8}_{37}$ & 0.70& $^{1.25}_{0.03}$ \\
H134 & 4.75& $^{0.16}_{0.10}$ & 36000& $^{5000}_{2750}$ & 4.05& $^{0.42}_{0.35}$ & 6.1& $^{0.3}_{0.4}$ & -8.09& $^{0.48}_{0.38}$ & 2000& $^{750}_{500}$ & 36& $^{15}_{25}$ & 0.70& $^{0.54}_{0.03}$ \\
H135 & 4.71& $^{0.15}_{0.17}$ & 29000& $^{4000}_{4000}$ & 3.90& $^{0.30}_{0.35}$ & 9.0& $^{0.7}_{0.5}$ & -8.25& $^{0.60}_{0.63}$ & 400& $^{1950}_{300}$ & 41& $^{10}_{35}$ & 1.30& $^{0.72}_{0.60}$ \\
H139 & 4.91& $^{0.06}_{0.10}$ & 40000& $^{2000}_{3000}$ & 4.15& $^{0.40}_{0.30}$ & 6.0& $^{0.3}_{0.2}$ & -8.20& $^{0.23}_{0.63}$ & 1300& $^{500}_{300}$ & 42& $^{9}_{32}$ & 0.70& $^{0.66}_{0.03}$ \\
H141 & 4.69& $^{0.19}_{0.12}$ & 30000& $^{5500}_{3000}$ & 3.80& $^{0.70}_{0.40}$ & 8.3& $^{0.5}_{0.6}$ & -7.56& $^{0.60}_{0.65}$ & 4300& $^{300}_{2500}$ & 47& $^{4}_{42}$ & 0.70& $^{0.66}_{0.03}$ \\
H159 & 4.71& $^{0.07}_{0.10}$ & 31000& $^{2000}_{2500}$ & 3.95& $^{0.35}_{0.30}$ & 7.9& $^{0.4}_{0.3}$ & -7.71& $^{0.35}_{0.35}$ & 4100& $^{500}_{1500}$ & 42& $^{9}_{25}$ & 1.30& $^{0.72}_{0.57}$ \\
H162 & 4.81& $^{0.15}_{0.25}$ & 36000& $^{4750}_{6500}$ & 4.30& $^{0.75}_{0.65}$ & 6.6& $^{0.8}_{0.4}$ & -8.47& $^{1.05}_{1.05}$ & 100& $^{4400}_{100}$ & 46& $^{5}_{46}$ & 1.02& $^{1.00}_{0.35}$ \\
H173 & 4.46& $^{0.18}_{0.23}$ & 26500& $^{4500}_{4250}$ & 3.85& $^{0.50}_{0.42}$ & 8.1& $^{0.9}_{0.6}$ & -8.15& $^{0.43}_{0.90}$ & 3700& $^{900}_{3400}$ & 49& $^{2}_{31}$ & 0.70& $^{1.32}_{0.03}$ \\ \hline
\multicolumn{17}{p{13cm}}{} \\
\multicolumn{17}{p{13cm}}{\footnotesize$^{a)}$ Note that when the error bar reaches the edge of the parameter space, the best fit value is in fact an upper or lower limit. \Cref{fig:param_overview_6free} shows when this is the case.} \\
    \end{tabular}
    \label{tab:6free_params_UVopt}

\end{table*}

\begin{table}
    \centering
\caption{Best fit parameters and $1\sigma$ error bars for the oxygen and helium abundances of the optical~+~UV runs of the WNh stars. Only for these stars oxygen and helium abundance were fitted in the optical~+~UV runs.\label{tab:WNHoxy}}
\begin{tabular}{l r@{$\pm$}l r@{$\pm$}l }
    \hline \hline
    Source & \multicolumn{2}{l}{\yhe} & \multicolumn{2}{c}{$\log n_{\rm O}/n_{\rm H}$+12}\\ \hline
    R136a1 & 0.22&$0.05$ & 8.30&$^{0.05}_{0.65}$\\
    R136a2 & 0.39&$^{0.11}_{0.07}$ & 7.80&$^{0.50}_{0.20}$ \\
    R136a3 & 0.37&$0.10$ & 7.45&$^{0.70}_{0.75}$ \\ \hline
\end{tabular}
\end{table}

\subsection{Stellar mass \& age \label{sec:bonnsai}}

In order to derive the evolutionary mass $M_{\rm evol}$, the initial mass $M_{\rm ini}$ and the age $\tau$ we use {\sc Bonnsai}\footnote{The BONNSAI web-service is available at \url{https://www.astro.uni-bonn.de/stars/bonnsai/}.} \citep{2014A&A...570A..66S} combined with the grids of \citet{2011A&A...530A.115B} and \citet{2015A&A...573A..71K}. {\sc Bonnsai} is a Bayesian tool that allows us to compare observed stellar parameters to stellar evolution models in order to infer full posterior distributions of key model parameters such as initial mass and stellar age. Our input parameters are observed luminosity, temperature, helium surface abundance and surface gravity. We use standard settings except for the prior of the initial rotational velocity, for which we assume the distribution of \citet{2013A&A...560A..29R} instead of a flat distribution. 
We find a robust output for all stars except three. For those the observed values match poorly with the posterior distribution of the {\sc Bonnsai} run. In the case of H129 the value for \teff cannot be reproduced given the observed \logll and \logg. For this star, we deemed our derived luminosity measurement unreliable (\cref{res:teffs}), and we exclude this source from further analysis. In two other cases, R136b and H30, our observed \logg value lies in the $P< 0.05$ tail of the posterior distribution. Therefore both spectroscopic and evolutionary parameters should be treated with care, although the spectroscopic fits of these stars look good. We do include both sources in further analyses that need $M_{\rm evol}$ as an input, but check whether the results change drastically upon inclusion/exclusion of R136b and H30, which is not the case. 
The derived ages and masses can be found in \Cref{tab:app:more_uv_restuls}. We cross-check our {\sc Bonnsai} output with that of \citet{Bestenlehner2020} and find generally good agreement, see \Cref{sec:app:bonnsai} for details.
In the remainder of the paper we will use the {\sc Bonnsai} evolutionary masses when we need stellar masses for our analysis. 

\subsection{Terminal velocity\label{sec:res:terminal}}

For all stars we have set the terminal wind velocity \vinf as a free parameter in the optical~+~UV fit. For 46 sources we were able to accurately constrain \vinf, albeit with large uncertainties for the stars with lower mass-loss rates (see \cref{fig:param_overview_allfree,fig:param_overview_6free}). For 3 of the remaining sources (R136a2, R136b, H36) we do find a tightly constrained value, but see that the fit to the blue wing of \CIVline is not good: the saturated absorption edge of the best model for these stars extends about 400 km~s$^{-1}$ more to the blue as the absorption edge we see in the data. 
For the remaining 7 sources (H73, H121, H135, H141, H159, H162, H173) the  wind lines crucial for determining \vinf, especially \CIVline, turn out to be too weak to get a constraint: the \chisq distribution for \vinf was flat. In these cases, while we do find some best fit value \vinf, this value is not meaningful and we regard \vinf as unconstrained.  

\subsection{Wind acceleration parameter $\beta$}

The wind acceleration parameter $\beta$ is fitted for all stars. For stars with $\log \ \dot{M} > -5.7$, we find values up to $\beta=1.50$ with an average of $1.08\pm 0.20$, whereas for stars with lower mass-loss rates we find that for all but two sources $\beta$ is consistent with 0.7 within $1\sigma$ errors, with an average of $0.72\pm 0.06$ (see \cref{fig:param_overview_allfree,fig:param_overview_6free}). We note that $\beta=0.7$ was the lowest allowed value during the fitting, this is discussed in \cref{sec:robust}. 

\begin{figure*}
    \centering
    \includegraphics[width=1.0\textwidth]{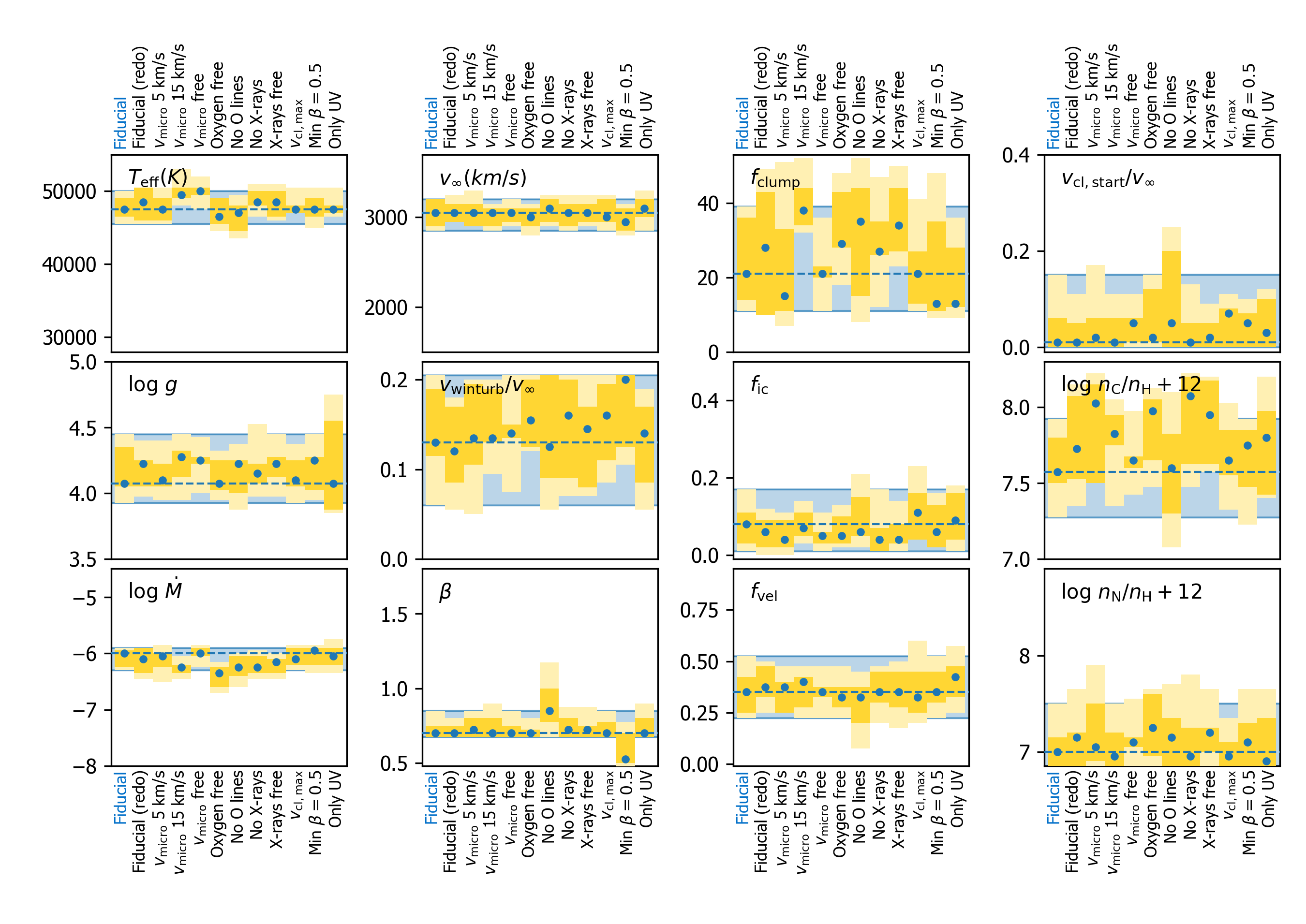}
    \caption{Comparison of parameters from the different test runs, fitting in each instance the spectrum of H35. For each run and each parameter we indicate the best fit value (blue dots) and 1$\sigma$ and 2$\sigma$ error bars (dark and light yellow). In the first column we show the parameters of our `fiducial run': this is the setup as used throughout the paper. The other columns show parameters of runs were we changed the setup, one aspect at the time: `Fiducial (redo)' -- different initial random population of models, `$\varv_{\rm micro} 5$ km~s$^{-1}$' and `$\varv_{\rm micro} 15$ km~s$^{-1}$' -- assumed value for $\varv_{\rm micro}$ to 5 and 15 km~s$^{-1}$, respectively, instead of 10 km~s$^{-1}$, `$\varv_{\rm micro}$ free'  --  $\varv_{\rm micro}$ a free parameter, `Oxygen free' -- oxygen abundance a free parameter,  `No O lines' -- exclude both \oivline and \ovline from the fitting, `No X-rays ' -- do not include any X-rays, `X-rays free' -- $f_{\rm X}$ a free parameter, `$\varv_{\rm cl,max}$' -- assume $\varv_{\rm cl,max} = 2\varv_{\rm cl,start}$ instead of $\varv_{\rm cl, max}/\varv_\infty = {\rm max}(0.3, \varv_{\rm cl, start}/\varv_\infty)$, `Min $\beta = 0.5$' - set lower limit of $\beta$ to 0.5 instead of 0.7 and `Only UV' -- only fit the UV spectra. 
    For reference, the best fit value and $2\sigma$ error region of the fiducial run are shown in blue throughout all columns.}
    \label{fig:test_H35}
\end{figure*}

\subsection{Onset of clumping}

We derive the onset of clumping for 39 sources and find a value of \vclonset~$ = 0.07\pm^{0.10}_{0.07}$ \vinf, translating into $R_{\rm cl,start} = 1.02\pm^{0.07}_{0.02} \ R_{\star}$ on average. There is not much variation across the sample: two thirds of the stars have a value of \vclonset~$<0.1$~\vinf or $R_{\rm cl,start} < 1.08~R_{\star}$, the higher values that we derive have large error bars -- within $2 \sigma$ all sources are consistent with \vclonset~$\leq0.05$~\vinf. This is visible in \cref{fig:param_overview_allfree}. 

\subsection{Ionising flux}

We derive H, \hei and \heii ionising fluxes $Q_0$, $Q_1$, $Q_2$ for each star\footnote{Here, by convention, $Q_x = q_x 4 \pi R_{\star}^2$, with $q_x$ the ionising radiation (number of photons) per unit surface area per second and $x \in \{1,2,3\}$.} based on the best fitting model (\Cref{tab:app:more_uv_restuls}). We estimate the errors on these values by computing, for one star (H35, spectral type O3~V), the ionising flux for each model in a full \pyGA run; afterwards we apply an error analysis based on the \chisq value of each model, as we do for all other parameters\footnote{Uncertainties on ionising fluxes are now a standard output of \pyGA, but this functionality was implemented only after we had done all fits. Since carrying out a fit is computationally expensive, we decided to do a rerun with the new functionality only for one star.}. From this we find $1\sigma$ uncertainties on the derived ionising fluxes to be, approximately, $0.07$ dex for $\log \ Q_0$,  $0.1$ dex for $\log \ Q_1$, $0.4$ dex for $\log \ Q_2$. We assume that the uncertainties of the ionising fluxes of the other sources scale with their relative uncertainties (compared to those of H35) on effective temperature and luminosity. Lastly, in \Cref{tab:app:more_uv_restuls} we provide also the H-{\sc i} ionising luminosity, that is, the energy of each star emitted by photons capable of ionising hydrogen, a quantity relevant for large scale simulations involving radiative feedback of massive stars. 

\subsection{Robustness and systematic errors\label{sec:robust}}

In order to check whether our results are robust under small changes in our setup we carry out several test runs. We picked the O3~V star H35, a typical source\footnote{We consider H35 a `typical source' because O3~V is the most common spectral type in our sample and the data quality of H35 is representative: there are sources with higher \snr, but the H35 data is good enough so that we can constrain each of our 12 free parameters. 
}, and fit its spectrum many times, each time changing one aspect of the setup. As a reference point, we compare all runs with the `fiducial run' for H35, that is, the run with the setup as used for all optical~+~UV runs in this paper. In \cref{fig:test_H35} this comparison is presented. 

First, we show the robustness of the genetic algorithm itself by redoing our fiducial run. We must be sure that the initial pool of individuals contains enough variation. If the variation is large enough to cover the full parameter space, then with exactly the same setup but different random initial parameters, we should get the same or very similar results. Indeed, when we do this test we do see small differences, but generally the agreement between the two runs is very good: for each parameter the $1\sigma$ and $2\sigma$ regions are similar and the best fitting parameters of each runs lie in the $1\sigma$ error regions of the other run. 

Having done this initial test we then vary the setup, changing one aspect at the time. We see that within uncertainties the different setups show consistent results and our setup is robust to most changes. 
The choice for micro-turbulence \vmicro seems to have the largest effect on the derived parameters, especially \teff.  Changing the value from our fiducial fixed value of 10~km~s$^{-1}$ to 5~km~s$^{-1}$ does not have much effect, but changing it to either a fixed value of 15~km~s$^{-1}$ or leaving it as a free variable results in a best fit value of \teff that is $\sim 2000$~K higher than that of our fiducial run, just on the edge of the $2\sigma$ error bars. From the data that we have we cannot determine what is the actual value of \vmicro and thus of \teff: the run with \vmicro as a free parameter resulted in a velocity exceeding the typical sound speed in the atmospheres of these stars (\vmicro = 30~km~s$^{-1}$) and thus seemed not reliable (though for a recent finding, see \citealt{2022ApJ...924L..11S}). Apart from changes in \teff we note that also abundances change when we assume a different value for \vmicro, the largest change being seen in the carbon abundance when \vmicro is lowered to 5 km~s$^{-1}$. This is expected as \vmicro impacts the equivalent width of lines. Considering the above, we must thus keep in mind that the lack of atmospheric lines from which we can accurately determine micro-turbulence leads to systematic uncertainties in \teff and abundances, which we estimate to be about 2000~K and 0.5 dex, respectively.  

The mass-loss rates and high clumping factors that we derive are robust within the optically thick clumping framework. We consistently find \fcl~$>10$, but distinguishing between the higher clumping factors proves difficult. This could be due to the fact that the clumping sensitivity saturates towards higher clumping factors for several of the clumping diagnostics. Since leaving oxygen abundance as a free parameter consistently leads to very high oxygen abundances, we decided to keep the oxygen abundance fixed during our fits. Possible causes for the high obtained oxygen abundances could be blends with iron lines, which are not present in our synthesised model spectrum, or specific shortcomings in the \fw oxygen model atom, so that ionisation structure of these ions is not well reproduced by \fw. Regardless of the cause we checked the robustness of our results given this uncertainty by doing a run with oxygen abundance left free, and one where we left out both oxygen lines. In both cases, the resulting fit parameters change slightly but are within errors consistent with our fiducial run. This also holds for the wind structure parameters \fic and \fvel, which show to be unaffected by either leaving the abundance free nor leaving out the lines (see \cref{fig:test_H35}). 

For stars with comparatively low mass-loss rates of $\log \ \dot{M} \lesssim -5.8$ we find an average wind acceleration parameter of $\beta=0.72 \pm 0.06$, with many of the derived values at 0.7. Since this is the lowest value allowed in our fits, we test the effect of extending our parameter space: for four stars we do another run with the same setup except that now we extend the allowed range of $\beta$ values up to as low as 0.5. We find in all cases that the distribution is nearly flat between 0.5 and 0.7. In these runs $\beta = 0.7$ remains in the 2$\sigma$ error range. 
\cref{fig:test_H35} shows that for H35 with the lower best fit value of $\beta$ the other parameters do not change significantly given the uncertainties. 

Apart from the aspects discussed so far, we do also change the prescription for $\varv_{\rm cl, max}$ and the X-ray setup. Lastly, we do a run with only UV data. The results seem robust to all these changes. The run with UV data only shows the diagnostic power of these relatively few spectral lines. The optical data adds most to the accuracy of the gravity, though one has to keep in mind that in this `UV-only run' rotation and helium abundance are fixed to values derived from optical data. 
In conclusion, our fitting setup seems generally robust to the assumptions we made. However, one should be aware of possible systematic errors, especially with regard to uncertainty due to the micro-turbulence that seems to affect the derived \teff and abundances.

\section{Discussion\label{sec:discussion}}

We discuss our results in the context of theoretical predictions and evolutionary models. In \cref{sec:mdot_vs_theory} we compare the mass-loss rates that we obtained to the predictions of \citet[][]{2000A&A...362..295V,2001A&A...369..574V,2018A&A...612A..20K} and \citet{2021A&A...648A..36B}. Here, we also compare the observed and predicted terminal velocities and the modified wind momenta. We conclude this section with a comparison to the CAK-type mass-loss theory of \citet{2020MNRAS.493.3938B}, and provide an equation for mass-loss rate as a function of the Eddington parameter for electron scattering. The mass-loss rates used in this section rely on the simultaneous fit of the wind structure (clumping) parameters. We observe weak trends in these parameters as a function of mass-loss rate, which is discussed in \cref{sec:windstructuretrends}.

The stellar evolution, mass and age of the sources based on the optical data is already discussed in detail by \citet{Bestenlehner2020}. After briefly reviewing consistency with their results (\cref{dis:WNh_stars}), we add to their discussion based on additional clues we can get from the abundances based on UV spectroscopy (\cref{dis:abundances_evolution}). Furthermore, \cref{dis:WNh_stars} contains a discussion of the surface gravities and mass estimates that we find for the three WNh stars. 
We conclude our discussion with a more technical topic, namely the comparison of the terminal velocity measurements from by-eye fitting \citep{2016MNRAS.458..624C} to our optical~+~UV spectral analysis (\cref{dis:terminal}). 

\subsection{Mass-loss rates, wind momentum and terminal velocities \label{sec:mdot_vs_theory}}

\begin{figure}
    \includegraphics[width=0.46\textwidth]{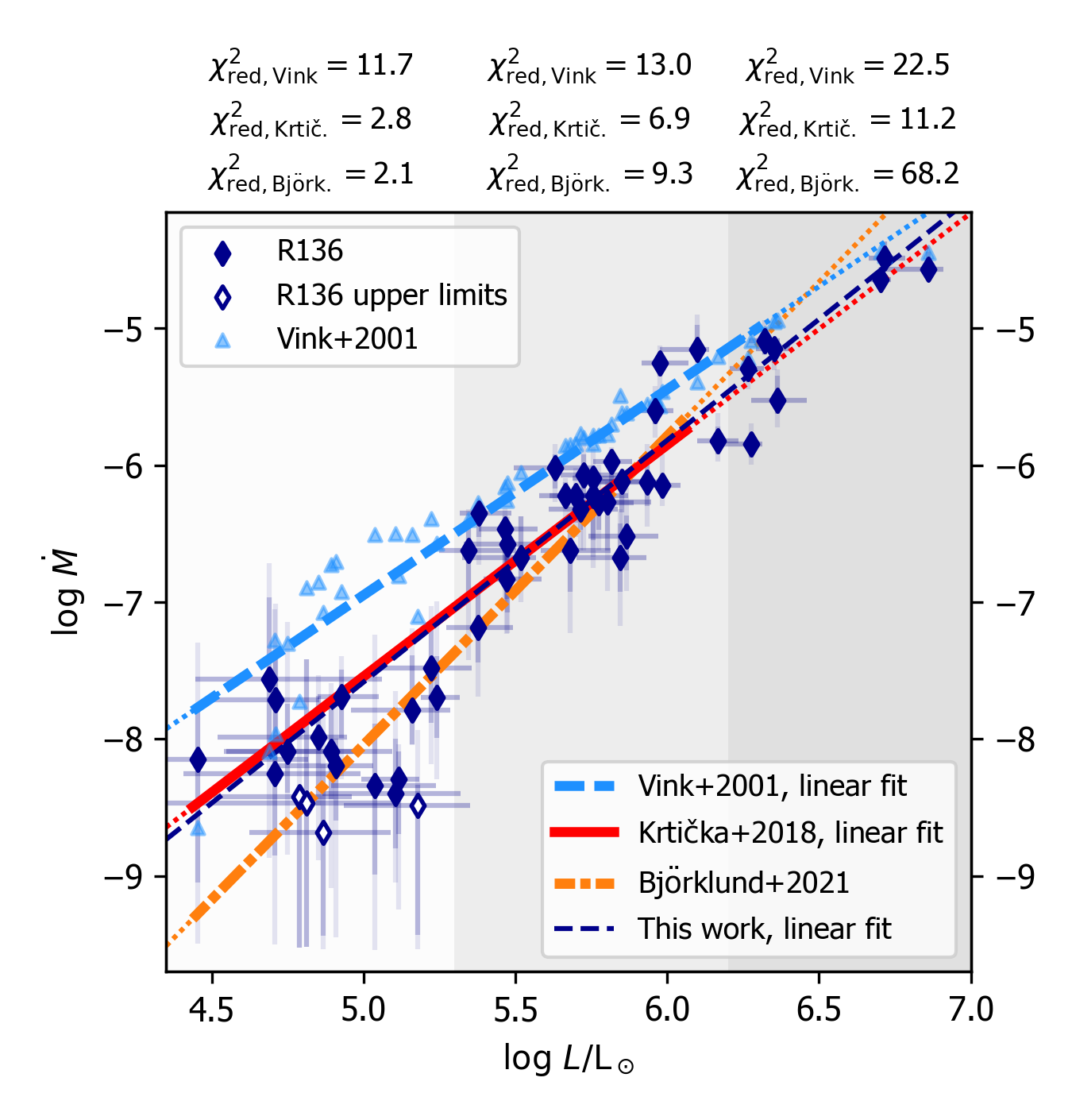}
    \caption{Mass-loss rates from the optical~+~UV fits (dark blue solid diamonds) compared to the mass-loss predictions of \citet[][light blue triangles]{2001A&A...369..574V}, \citet[][red solid line]{2018A&A...612A..20K} and \citet[][orange dot-dashed line]{2021A&A...648A..36B}. The prescription of \citet{2001A&A...369..574V} depends on more parameters than luminosity; a linear fit through the individual points (light blue dashed line) is plotted to guide the eye. Thin dotted lines in corresponding colours show the extrapolation of each prescription beyond the coverage of their respective model grids. For reference, a linear fit through the data points is shown (thin dashed darkblue line). For all mass-loss prescriptions we assess the goodness of fit (\chisq-values, top) for three ranges of luminosity.}
    \label{fig:mdot_recipes}
\end{figure}

\begin{figure}
    \includegraphics[width=0.46\textwidth]{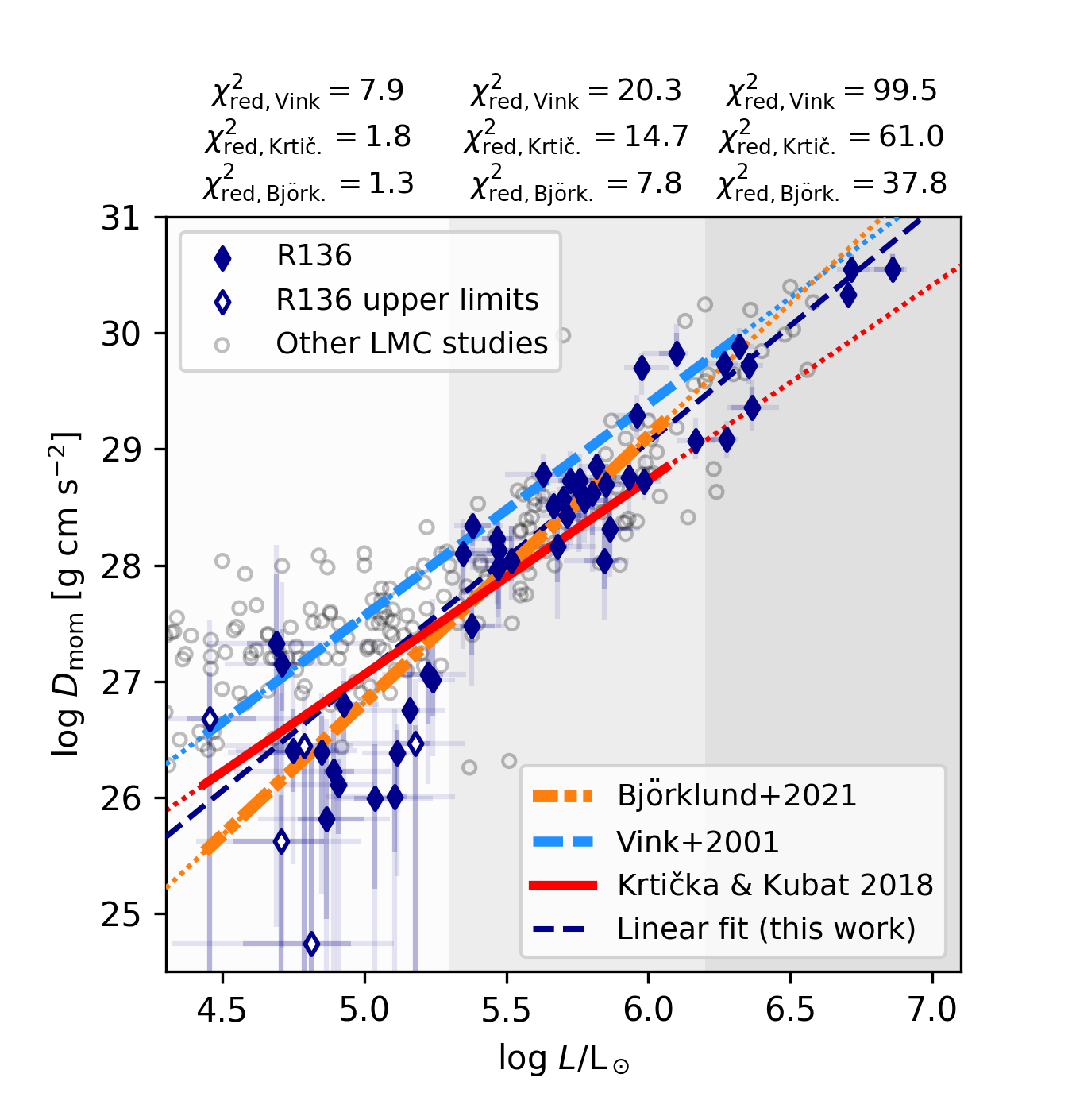}
    \caption{As \cref{fig:mdot_recipes} but now for the modified wind momentum. 
    In grey circles we show the values or upper limits of \citet[][]{2007A&A...465.1003M,2014A&A...570A..38B,2017A&A...600A..81R,2017A&A...601A..79S}, which we all lowered by 0.7 dex, assuming $f_{\rm cl} = 25$.}
    \label{fig:dmom_recipes}
\end{figure}

In \cref{fig:mdot_recipes} we compare our observed mass-loss rates to theoretical predictions of  \citet[][hereafter \vink]{2001A&A...369..574V}, \citet[][hereafter \krtic]{2018A&A...612A..20K} and  \citet[][hereafter \bjork]{2021A&A...648A..36B}. 
All three predict mass-loss rates of hot luminous stars by computing the line force based on the density and ionisation structure of a model atmosphere: \vink use {\sc isa-wind} \citep{1993A&A...277..561D}, \krtic use {\sc Metuje} \citep{2010A&A...519A..50K,2017A&A...606A..31K} and \bjork use \fw \citep[version 11:][]{2020A&A...642A.172P}. Where \vink uses a Monte-Carlo method in combination with the Sobolev approximation for the line computation, \krtic and \bjork perform radiative transfer in the co-moving frame, though the former effectively places their critical point upstream from that of \bjork (see \citet{2019A&A...632A.126S}). Furthermore the codes differ in their approach regarding the wind dynamics; \krtic and \bjork solve numerically the equations of motion and \vink use a prescribed velocity structure assuming conservation of total radiative and kinetic energy.
 We note that for this comparison we use the \vink, \krtic and \bjork prescriptions as presented in their respective papers, even though they assume different values for the solar abundances: \vink on the one hand, and \krtic and \bjork on the other hand assume solar metal mass fractions of $Z_{\odot} = 0.019$ and $Z_{\odot} = 0.013$, respectively. 
 For a mass-loss prediction at $Z=0.5~Z_{\odot}$ in the \vink prescription one thus implicitly assumes a metal content of a factor 1.46 higher than one would under the same assumption ($Z=0.5~Z_{\odot}$) in the \bjork or \krtic prescriptions. Correcting for this would bring the \vink relation approximately 0.13 dex closer to the \bjork prescription, on average. Other differences in assumptions between the approaches may also result in systematic differences, for example, related to micro-turbulence, line lists and temperature structure. 
 
We assess the goodness of fit of each prescription to our observations for three luminosity regimes: low (\logll$< 5.3$), intermediate ($ 5.3 \leq $~\logll$\leq 6.2$) and high (\logll$> 6.2$). Results are shown in \cref{fig:mdot_recipes}. Note that the highest luminosities that we observed lie outside the grids of \vink, \krtic and \bjork, which extend only to \logll~$=6.00$, \logll~$=6.07$ and \logll~$=6.25$, respectively. 
Note furthermore, that for four sources (H73, H116, H121, H162), for which we derive mass-loss rates in the range log~\mdot~$= -8.68$ to $-8.42$, the $1 \sigma$ error bars range (nearly) to the edge of our parameter space, log~\mdot~$= -9.50$. In our fitting and with the derivation of the \chisq values, we treat the derived values and uncertainties the same as those of the other points, even though these are technically upper limits. Removing these 4 points from the fit does not alter the results significantly. The same holds for fitting and \chisq values of the modified wind momentum (see below); here we have one additional source with only an upper limit, H135, for which we derive an upper limit only for \vinf. 

\begin{figure*}
    \includegraphics[width=0.995\textwidth]{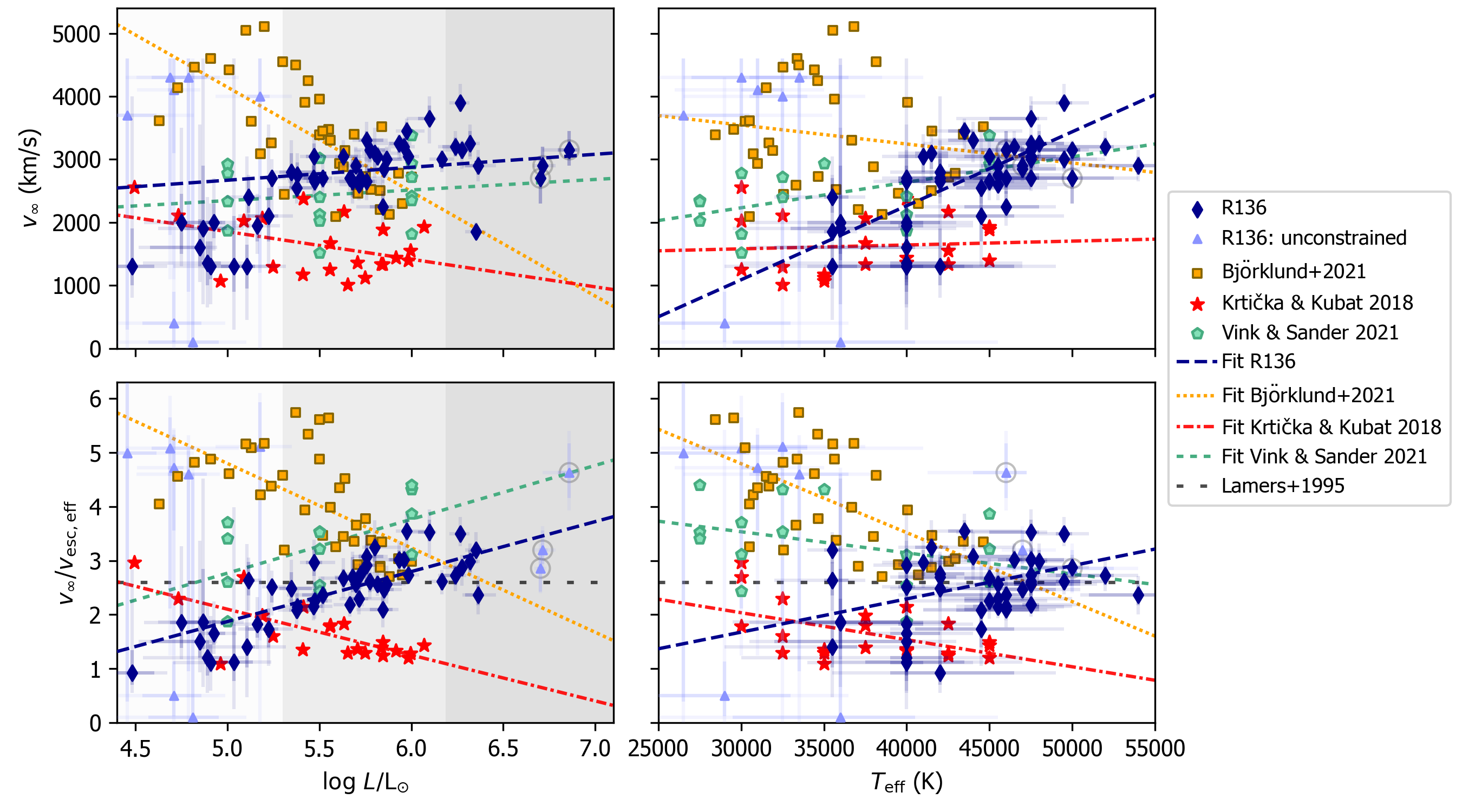}
    \caption{Terminal velocity $\varv_\infty$ (top) and the ratio $\varv_\infty/\varv_{\rm esc, eff}$ (bottom) against \logll (left) and \teff (right) for the R136 sample (solid dark blue diamonds) compared to predicted values (yellow squares, red stars, and green pentagons \citealt{2021A&A...648A..36B}, \citealt{2018A&A...612A..20K}, and \citealt{2021MNRAS.504.2051V}, respectively). Light purple triangles denote R136 sources for which we could not derive $\varv_\infty$ values (see \cref{sec:res:terminal}), and in the bottom plot also includes the WNh stars, where we cannot be too confident about $\varv_{\rm esc, eff}$. Grey circles around the points of the WNh stars allows to distinguish them from the O-type stars. Dark and light coloured error bars denote $1\sigma$ and $2\sigma$ uncertainties, respectively. The lines, plotted in different styles (see legend) show linear fits to both the observed (dark blue) and predicted values (red, orange and green). In the bottom panel, the black dashed line shows the empirically derived  $\varv_\infty/\varv_{\rm esc, eff}$ = 2.6 \citep{1995ApJ...455..269L}. The grey shaded regions in the luminosity plots correspond to those in \cref{fig:mdot_recipes}. Note that we do not show any points of models with ${T}_{\rm eff}=35000$~K for the \citealt{2021MNRAS.504.2051V} predictions; in this regime the predicted terminal velocities largely exceed the velocity scale of this plot. The linear fit through the \citealt{2021MNRAS.504.2051V} predictions also excludes these points. }
    \label{fig:vinf_vesc}
\end{figure*}

Comparing observed and theoretical mass-loss rates (\cref{fig:mdot_recipes}), we see that, overall, the predictions of \krtic fit best to our observations, outperforming the other two in the regime where mass-loss impacts evolution the most (\logll~$> 5.3$), and matching almost perfectly with the linear fit through our observations. 
In the low-luminosity regime the \bjork rates match better with our observations: the prediction is in good agreement with the group of low-luminosity observations, reflected in the low \chisred. The \vink rates there are on average an order of magnitude too high in this regime, though several individual points lie well below that average, close to the observed points. Observing weak winds in this luminosity regime is not unusual \citep[e.g.,][]{1996A&A...305..171P,2008A&ARv..16..209P,2005A&A...441..735M,2019A&A...628A..36D}.
In the intermediate regime, the performance of the three predictions is similar. The \krtic rates perform best; their predictions are, on average, only a factor 1.3 lower than the observed rates. The \bjork rates are a close second, being too low with, on average, a factor 1.5. 
The \vink rates in the intermediate regime are, on average, a factor 2.0 too high, consistent with previous findings \citep[e.g.,][]{2012A&A...544A..67B,2013A&A...559A.130S,2014MNRAS.439..908C}. 
However, note that the overestimation of the \vink rates in this regime decreases to only a factor 1.5 if one would apply an average downward shift to correct for the different assumptions of \vink versus \bjork and \krtic regarding LMC metallicity. 
Inspecting the high-luminosity end we see that the \krtic rates best match the observations. Both \bjork and \vink overestimate the mass-loss rates of the most luminous stars; of those two, the prediction of \vink lies closest to the observed rates. 

We conclude the mass-loss rate comparison with noting that recently \citet{2021MNRAS.504.2051V} updated the \vink Monte Carlo mass-loss recipe with dynamically consistent computations of the terminal wind velocity. The mass-loss rates they predict are similar to those of \vink, but typically lie a bit higher; for our sample, the \citet{2021MNRAS.504.2051V} rates lie on average 0.11 and 0.17 dex above the \vink rates, with the average absolute difference being 0.19 and 0.17 dex, using the predicted and observed terminal velocities, respectively. 
Given that the \vink rates generally overpredict the observed mass-loss rates, the updated recipe does so even more and therefore the \vink recipe outperforms the \citet{2021MNRAS.504.2051V} recipe for our sample.

Another way of comparing the predictions to our observations is by their modified wind momentum, $D_{\rm mom} = \dot{M} \varv_\infty \sqrt{R_{\star}/{\rm R}_\odot}$ (with \mdot and \vinf in cgs-units), shown in \cref{fig:dmom_recipes}. \bjork provide an explicit equation for the wind momentum as a function of luminosity and metallicity. \krtic do not provide this, instead we compute $D_{\rm mom}$ for all the models in their LMC grid, and obtain a linear fit through these points as a function of luminosity. 
Since \vink do not predict terminal velocities, their relation for wind momentum \citep[][their Eq. 15]{2000A&A...362..295V} is semi-empirical; they use observed values for \vinf and $R_{\star}$. The relation plotted in \cref{fig:dmom_recipes} is corrected for lowered \mdot and \vinf as a result of the lower metallicity in the LMC compared to the Milky Way (\vink; \citealt{1992ApJ...401..596L}).  
For the modified wind momentum the \bjork predictions match best our observations in all luminosity regimes. The \vink prediction is too high over the full luminosity range, which is to be expected given their overprediction of the mass-loss rates. The \krtic predictions lie close to that of \bjork, but their prediction is less steep, translating in underpredicting $D_{\rm mom}$ for the higher luminosities, and overpredicting it for the less bright stars. 

\Cref{fig:vinf_vesc} shows observed and predicted terminal velocities as a function of both \logll and \teff. Neither the predictions of \bjork nor those of \krtic fit the observed values well. Both predict decreasing terminal velocities as a function of \teff and \logll, whereas observations show the opposite. Also in absolute sense the predicted velocities deviate from the observed values. Looking at the left side of the figure, we see that the difference between the observed velocities and those predicted by \bjork is especially large in the lower luminosity regime, where the \bjork and \vink mass-loss rates diverge the most. The predicted terminal velocities of \krtic in this regime match reasonably well our observations, but their predictions are too low in the intermediate regime. 

Of all terminal velocity predictions considered here, those of \citet{2021MNRAS.504.2051V} are most in line with observations\footnote{In \cref{fig:vinf_vesc} we plot the \vinf of the $Z=1/3\ {\rm Z}_\odot$ models, multiplied by a factor $(0.5/(1/3))^{0.19} = 1.08$ so that they correspond to our assumed LMC metallicity of $Z = 0.5\ {\rm Z}_\odot$.}; both the absolute values as well as the trend as a function of luminosity are in line with observations (see the green dashed lines in the left panels of \cref{fig:vinf_vesc}), except around 35000~K, where they predict extremely high velocities. In \Cref{fig:vinf_vesc} we do not show their predictions for \teff~$=35000$~K, and do not include them in the fit. 

It is clear that none of the theoretical models considered here can reproduce both the observed mass-loss rates and the terminal velocities that we observe. It has to be noted, however, that while the luminosity range spanned by the hydrodynamic models covers our observations reasonably well, this is not the case for the temperatures of the models: the maximum model \teff in the grids of \bjork, \krtic and \citealt{2021MNRAS.504.2051V} is 45000~K, while about 50\% of our sample has a \teff higher than that. The \vink grid extends further to 50000~K, however they do not predict terminal velocities. 

We will now look more quantitatively at the modified wind momentum, and fit a powerlaw to our observations as a function of luminosity (dark blue dashed line):
\begin{equation}\label{eq:Dmomfit}
    \log \ D_{\rm mom} = \log \ D_{\rm 0} + x \log \ L / {\rm L}_\odot. 
\end{equation}
For this and other fits where we have to take into account uncertainties in both coordinates we use the orthogonal distance regression (ODR) routine of {\sc scipy}. 
Our best fit yields $x=2.00\pm0.11$ and $\log \ D_{\rm 0} = 17.05\pm0.65$. 
A comparison to other LMC studies shows that this is relatively high: observations of \citet[][grey circles in \cref{fig:dmom_recipes}]{2007A&A...465.1003M,2014A&A...570A..38B,2017A&A...600A..81R,2017A&A...601A..79S} give slopes in the range 1.45-1.87. Note however that not all these analyses take into account uncertainties in $\ D_{\rm mom}$ and/or $L$ in the same way, which might affect the derived slopes (see \citealt{2004A&A...413..693M}). Furthermore, beware that the relatively high wind momenta for low-luminosity stars found by other LMC studies (grey circles in the leftmost part of the plot) are likely only upper limits, as these studies lack UV coverage and are therefore rely only on \halpha and \heiiline for their mass-loss determinations. 

As shown by \citet[][]{1996A&A...305..171P}, the slope $x$ of the modified wind momentum-luminosity relation can be interpreted as a measure for the distribution of line strengths of the spectral lines contributing to the wind driving: it is the inverse of the force multiplier $\alpha_{\rm eff}$ in (modified) CAK-theory (\citealt{1975ApJ...195..157C}, with subsequent modifications by \citealt{1982ApJ...259..282A,1986A&A...164...86P}): $x = 1/\alpha_{\rm eff}$. Here, $\alpha_{\rm eff}$ captures both $\alpha$, the slope of the line strength distribution, as well as the force multiplier $\delta$, that accounts for the ionisation state of the wind in an approximate way \citep{1982ApJ...259..282A}, that is, $x^{-1} = \alpha_{\rm eff} = \alpha - \delta$ in \cref{eq:Dmomfit}. Our slope of $x = 2.00\pm0.11$ then translates into a mean value of $\alpha_{\rm eff} = 0.50\pm0.03$ or $\alpha = 0.60\pm0.03$ if we assume the typical value of 0.1 for $\delta$ \citep{1982ApJ...259..282A,2008A&ARv..16..209P}. 
This is in line with typical values expected for O-stars, $\alpha \approx 0.5-0.6$ \citep{1982ApJ...259..282A,1994A&A...283..525P,2000A&AS..141...23P,2008A&ARv..16..209P}.  

\begin{figure}
    \centering
    \includegraphics[width=0.45\textwidth]{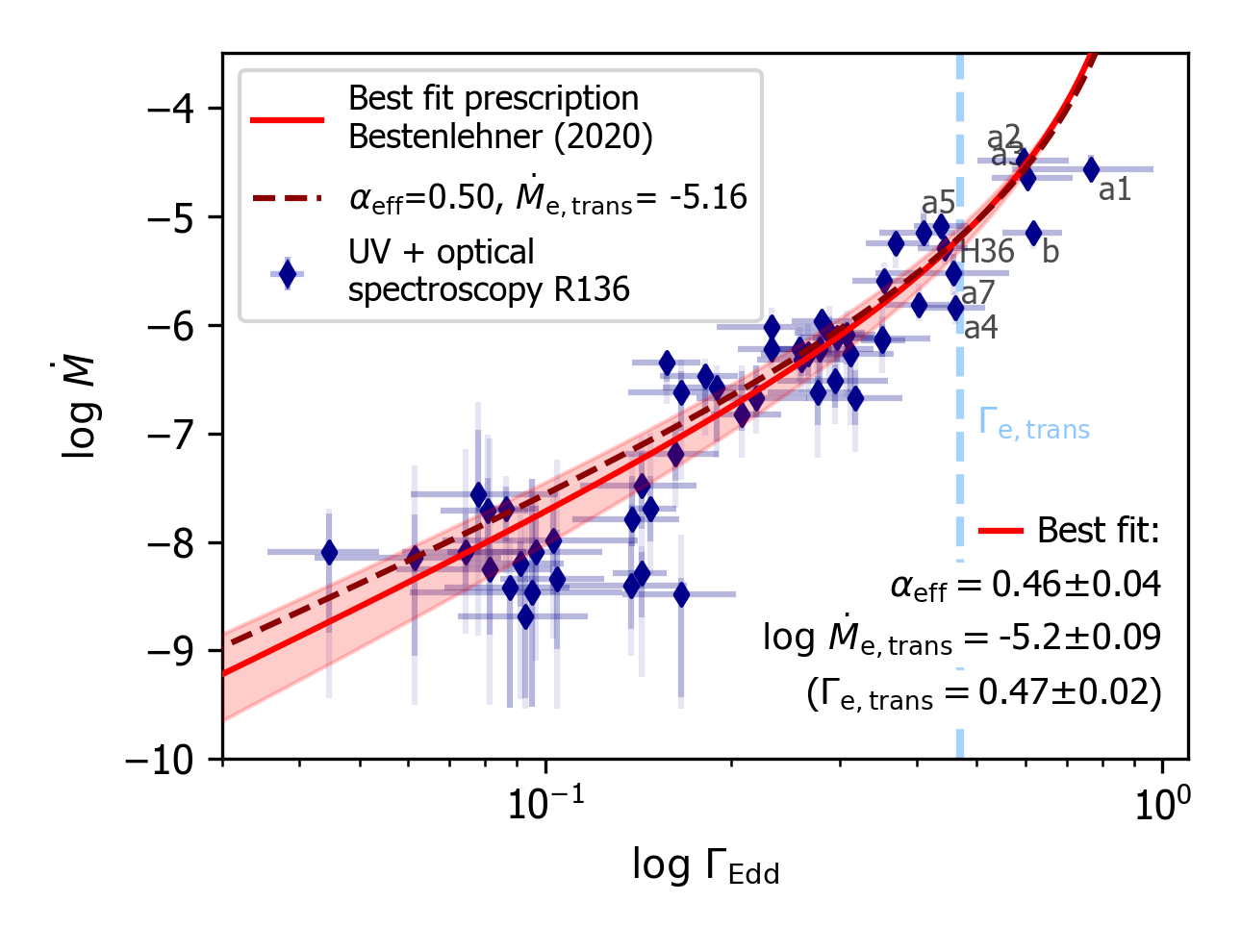}
    \caption{Best fit and $1\sigma$ error region that we obtain by fitting the CAK-type mass-loss prescription as described in \citet[][red solid line and shaded area]{2020MNRAS.493.3938B} to our observed mass-loss rates (blue circles, dark and light error bars denote $1\sigma$ and $2\sigma$ uncertainties). 
    The best fit values we derive are shown in the bottom right corner. For comparison we show the CAK-type prescription also for the case of $\alpha_{\rm eff} = 0.50$, as obtained by fitting the slope of the modified wind momentum (dark red dashed line). Eight sources that lie close or above $\Gamma_{\rm e,trans}$ are labelled with their abbreviated identifications (e.g., a1 is R136a1).}
    \label{fig:mdot_cak} 
\end{figure}

We can obtain $\alpha_{\rm eff}$ from our data in a different manner by inspecting the dependence of mass-loss rate on the Eddington parameter for electron scattering. 
\citet{2020MNRAS.493.3938B} extended the CAK mass-loss prescription \citep[][]{1975ApJ...195..157C} from the regime of optically thin to that of optically thick winds, by expressing the stellar mass in terms of the electron scattering Eddington parameter $\Gamma_{\rm e}$. The mass-loss rate can then be expressed as a function of $\Gamma_{\rm e}$, the transition mass-loss rate $\dot{M}_{\rm e,trans}$ and the force multiplier $\alpha_{\rm eff}$:
\begin{equation} \label{eq:bj2020_eq12}
\begin{split}
\log \dot{M} = & \log \dot{M}_{\rm e,trans} \hspace{0.1cm}+ \\ & \underbrace{\left(\frac{1}{\alpha_{\rm eff}} + 0.5 \right) \ \log(\Gamma_{\rm e})}_{\textrm{I. Dominates when }\Gamma_{\rm e} \ll 1} -  \underbrace{\left(\frac{1-\alpha_{\rm eff}}{\alpha_{\rm eff}} + 2\right) \ 
\log(1 - \Gamma_{\rm e})}_{\textrm{II. Dominates when }\Gamma_{\rm e} \rightarrow 1}. 
\end{split}
\end{equation}
Our $\alpha_{\rm eff}$ is equal to what \citet{2020MNRAS.493.3938B} calls $\alpha$, and what \citet{1994A&A...283..525P} and \citet{1996A&A...305..171P,2008A&ARv..16..209P} call $\alpha^\prime$. The transition mass-loss rate corresponds to the transition Eddington parameter $\Gamma_{\rm e,trans}$. At $\Gamma_{\rm e} = \Gamma_{\rm e,trans}$ the first (I) and the second (II) terms in \cref{eq:bj2020_eq12} are equal. At this point the mass-loss dependency changes from being dominated by the first term to being dominated by the second term. 

\begin{figure*}
    \centering
    \includegraphics[width=\textwidth]{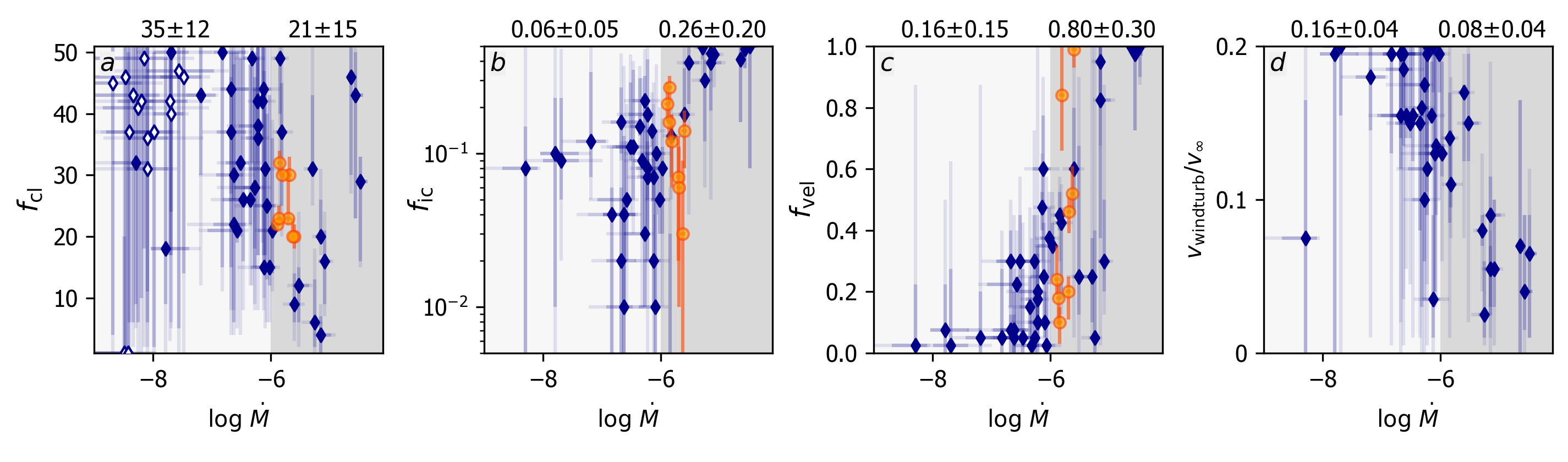}
    \caption{Wind structure parameters plotted against mass-loss rate of stars in R136 (blue diamonds, dark and light shaded error bars denote $1\sigma$ and $2\sigma$ uncertainties) and 8 Galactic stars of the sample of \citet[][orange/yellow cirlces, $2\sigma$ uncertainties]{Hawcroft21}. The limits of the $y$-axis of each plot coincide with the range of values that was allowed during the fit. The panels $a-d$ show, from left to right, the clumping factor, the interclump density contrast, the \vorosity and the wind turbulence. At the top of each panel the average value of the parameter ($\pm 1 \sigma$ uncertainty) is quoted for two \mdot regimes: low ($\log \dot{M}< -6$, light grey shaded) and high ($\log \dot{M}> -6$, dark grey shaded). In the leftmost panel, the diamonds with a white interior denote sources that are not present in the other three panels, as the \fic, \fvel and \vwindturb values were not fitted in their optical~+~UV runs. No values of \citet{Hawcroft21} are shown in the rightmost panel, as they do not fit \vwindturb.}
    \label{fig:windstructure1}
\end{figure*}

We compute $\Gamma_{\rm e}$ for all our sources\footnote{$\Gamma_{\rm e} = L \kappa_{\rm e} / 4 \pi {\rm c G} M$ with $L$ the luminosity, $M$ the stellar mass, $\kappa_{\rm e}$ the electron scattering opacity, $c$ the speed of light and $G$ the gravitational constant.} using the {\sc Bonnsai} evolutionary masses and fit \cref{eq:bj2020_eq12} to our observations, in order to derive a mean value for $\alpha_{\rm eff}$ for the sample, as well as the transition mass-loss rate. We find $\alpha_{\rm eff} = 0.46\pm 0.04$ ($\alpha \approx 0.56$) and $\log \ \dot{M}_{\rm e, trans} = -5.19\pm 0.10$ (see \cref{fig:mdot_cak}, red solid line). Our value for $\dot{M}_{\rm e, trans}$ matches well with the the mass-loss rate of $\log \dot{M} = -5.2 \pm 0.2$ that \citet{2012ApJ...751L..34V} find for transition objects in the Arches cluster. This agreement was not strictly expected as the definitions for the transition mass-loss rates of \citet{2020MNRAS.493.3938B} and  \citet{2012ApJ...751L..34V} differ, with the transition of \citet{2020MNRAS.493.3938B} relating to different terms in \cref{eq:bj2020_eq12}, and   \citet{2012ApJ...751L..34V} deriving their rate based on the inference that stars with a spectral type O4-6If+ correspond to the transition from optically thin to optically thick winds. 
The $\alpha_{\rm eff}$ we find from the fit with the \citet{2020MNRAS.493.3938B} prescription is lower than we found before using the wind momentum relation, however, taking into account errors on both $\alpha_{\rm eff}$ and $\dot{M}_{\rm e, trans}$ we see that $\alpha_{\rm eff} = 0.50$ found from the wind momentum is just within the $1\sigma$ uncertainty range (\cref{fig:mdot_cak}, shaded area and dashed line). In the high mass-loss regime the relations using the different values for $\alpha_{\rm eff}$ barely differ, for the low mass-loss regime one sees that the rates match better the low value of $\alpha_{\rm eff}=0.46$. 

Filling in the best fit values for \cref{eq:bj2020_eq12}, we obtain the following empirical mass-loss rate dependence on the Eddington parameter for electron scattering\footnote{For this we use the non-rounded value from our fit: $\alpha_{\rm eff} = 0.456$.}:
\begin{equation}
    \log \ \dot{M} = -5.19 + 2.69 \log(\Gamma_{\rm e}) - 3.19\log(1-\Gamma_{\rm e}).
\end{equation}
This equation could be used as a mass-loss rate prescription for stellar evolutionary computations for massive stars in the LMC. While our relation is derived based on stars of $M \geq 15~{\rm M}_\odot$, the scatter up to $M = 40~{\rm M}_\odot$ ($\Gamma_{\rm e,trans} = 0.2$) is large and the best results will be obtained for stars with $M \geq 40~{\rm M}_\odot$. 

\subsection{Wind structure \label{sec:windstructuretrends}}

\begin{figure*}
    \centering
    \includegraphics[width=1.0\textwidth]{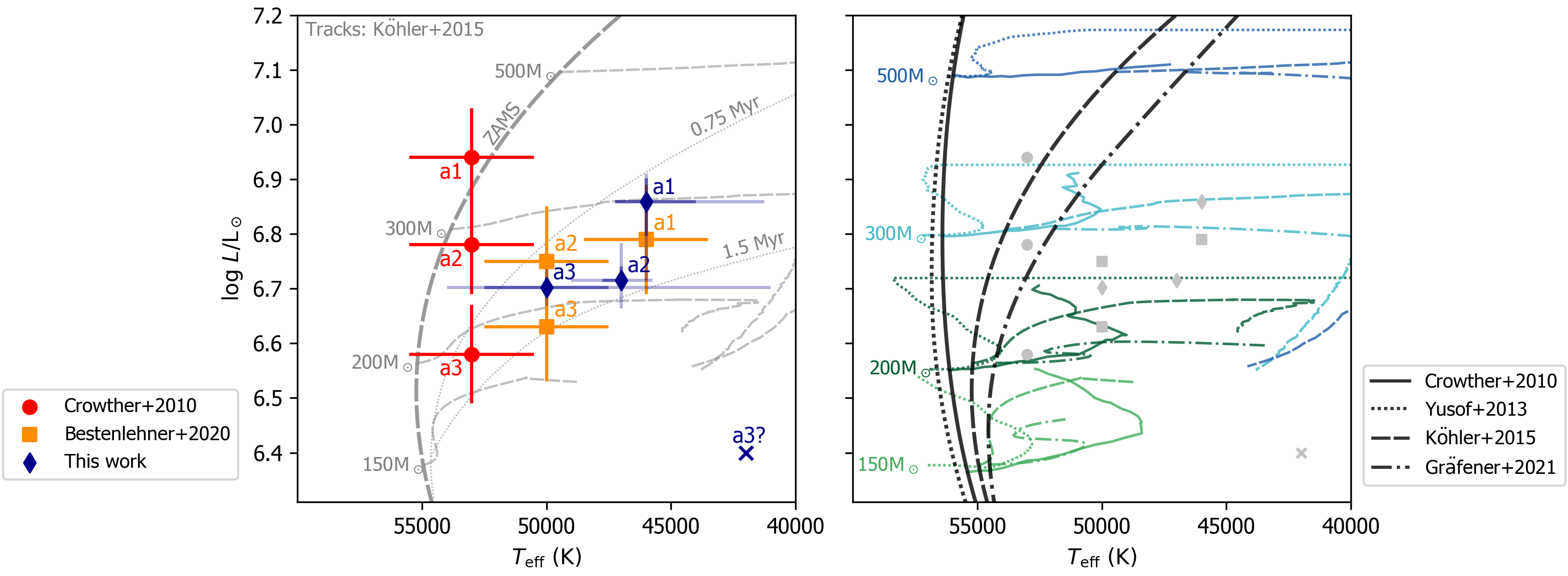}
    \caption{Positions of the WNh stars and evolutionary tracks in the Hertzsprung-Russell diagram. \emph{Left:} Temperature and luminosity of the WNh stars as found from this analysis (dark blue diamonds), as derived by \citet[red circles]{2010MNRAS.408..731C} and as derived by \citet[orange squares]{Bestenlehner2020}. The cross indicates an alternative (but unlikely, see text) position for R136a3.  Shown in the background is a subset of the evolutionary tracks of \citet[][thin grey dashed lines]{2015A&A...573A..71K}, on which the evolutionary masses and ages derived in this paper are based. Also shown is the corresponding Zero Age Main Sequence (ZAMS, thick grey dashed line) and the 0.75 Myr and 1.5 Myr isochrones (grey dotted) of the \citet{2015A&A...573A..71K} models. 
   \emph{Right:} Comparison of the stellar evolution models of \citet[][solid lines]{2010MNRAS.408..731C}, \citet[][dotted lines]{2013MNRAS.433.1114Y}, \citet[][dashed lines]{2015A&A...573A..71K} and \citet[][dashed-dotted lines]{2021A&A...647A..13G}. For each grid we show tracks of models with an initial mass of 150, 200, 300 and 500 \Msun in light green, dark green, light blue and dark blue, respectively. Black thick lines denote the ZAMS positions of each grid. The notable difference in ZAMS positions of the tracks of \citet{2010MNRAS.408..731C}, \citet{2013MNRAS.433.1114Y} on the one hand, and \citet{2015A&A...573A..71K} and \citet{2021A&A...647A..13G} on the other hand is related to their treatment of convection in the inflated envelopes of the most massive stars. For reference, the observed positions as in the left panel are also shown in the right panel (grey and without error bars). 
   All tracks have an initial rotation $\varv_{\rm eq,ini}$ of $0.4~\varv_{\rm eq,crit}$, with $\varv_{\rm eq,crit}$ the critical velocity, except for those of \citet{2015A&A...573A..71K}, for which we show the models with $\varv_{\rm eq,ini} = 350$~km~s$^{-1}$ for 150-300~\Msun models and $300$~km~s$^{-1}$ for the 500~\Msun model, corresponding to $\varv_{\rm eq,ini}/\varv_{\rm eq,crit} = 0.38 \pm 0.01$.} 
    \label{fig:a1a2a3mass}
\end{figure*}

Upon investigating possible trends in wind structure parameters, we plot the obtained values against mass-loss rate. The results are shown in \cref{fig:windstructure1}. For computing the averages $\mathcal{M}$ and the errors on the averages  $\mathcal{U}$ quoted in this section we weigh the best fit value for each individual star, $x_i$, with the inverse of its $2\sigma$ uncertainty ($w_i$): 
\begin{equation}
    \mathcal{M} = \frac{\sum^{N_x}_{i=0} w_i x_i}{\sum^{N_x}_{i=0} w_i} \hspace{0.7cm} \mathrm{and} \hspace{0.7cm}  \mathcal{U} = \frac{\sum^{N_x}_{i=0} w_i (x_i -  \mathcal{M})^2}{\sum^{N_x}_{i=0} w_i},
\end{equation}
where $N_x$ is the number of measurements of the quantity under consideration. 

The left panel ($a$) of \cref{fig:windstructure1} shows the derived clumping factors. We find that all but five stars have a best fitting clumping factor of $f_{\rm cl}>10$, with an average for all stars of $\langle f_{\rm cl}\rangle=29\pm15$.  When dividing the sample in two groups based on their mass-loss rates being lower or higher than $\log \dot{M} = -6$, we find typically lower values for \fcl for the stars in the higher mass-loss rate group, although this difference is barely significant; average values and uncertainties are displayed at the top of \cref{fig:windstructure1}. \citet{Hawcroft21}, who analyse a sample of 8 Galactic O-supergiants with mass-loss rates $\log \ \dot{M} > -6$ in a similar fashion, find $\langle f_{\rm cl}\rangle=25\pm~4$ for their sample. This is consistent with our findings. 

The middle panels ($b$ and $c$) show our values for \fic and \fvel. We find averages of  $\langle f_{\rm fic}\rangle=0.13\pm^{0.15}_{0.13}$ and  $\langle f_{\rm vel}\rangle=0.46\pm0.39$. 
 As before, we also divide the points in two groups based on $\log \ $\mdot and compute the averages of each group (displayed at the top of \cref{fig:windstructure1}). For the interclump density contrast we find a significant difference between the two groups. The stars with a lower mass-loss rate typically have a lower \fic than stars with a higher mass-loss rate (values and errors at the top of \cref{fig:windstructure1}). 
 In other words for stars with lower mass-loss rate we find a stronger density contrast between the clumps and the interclump medium, or, equivalently, a relatively lower interclump medium density. 
We also find a trend in the velocity filling factor, where for the groups with a lower and higher mass-loss rate we find lower and higher values of \fvel respectively. 
We compare our values of \fic and \fvel to those of \citet{Hawcroft21} and find that these are generally in agreement with our findings. However, given the small range of mass-loss rates of the stars of the sample that \citet{Hawcroft21} consider, this comparison cannot give any confirmation of the differences we find between the low- and high mass-loss rate groups. 

The last parameter that is related to the wind structure is the wind turbulence \vwindturb, shown in the rightmost panel ($d$).
Also here, we observe a weak trend where the turbulence seems to be less strong in the stars with higher mass-loss rates. \citet{Hawcroft21} do not measure this parameter. 

For all wind structure parameters, the observations show tentative trends as a function of mass-loss. 
Overall, it appears that the stars with higher mass-loss rates typically have smoother winds than the stars with lower mass-loss rates. All wind structure diagnostics indicate this: the stars with higher-mass loss rates have on average lower clumping factors, a lower contrast between the density in the interclump and clump medium, less wind turbulence, and higher velocity filling factors. The latter may sound like evidence for stronger clumping effects, however a high velocity filling factor too can indicate smooth wind. Namely, as \fvel~$\rightarrow 1$, this `erases' the density contrast, so that the wind is fully consisting of absorbing material and there are no gaps in velocity space through which the light can escape. In other words, the velocity-porosity effects are no longer present, as it would in either a smooth wind or a wind with only optically thin clumps. This follows explicitly from the equations for the effective opacity \citep{2018A&A...619A..59S}, summarised in our \cref{sec:method_structure}: if \fvel~$\rightarrow 1$, then $f_{\rm vor} \rightarrow \infty$ (\cref{eq:vel}), leading to $\tau_{\rm cl} \rightarrow 0$ (\cref{eq:clumptau}), which then means that $\chi_{\rm eff} \rightarrow \langle \chi \rangle$ (\cref{eq:chieff}), such that the limit for either optically thin clumping (if \fcl~$>1$) or a smooth wind (if \fcl~$=1$) is recovered. We do stress that, although the wind structure parameters of the higher mass-loss rate stars imply that their winds are on average smoother than those of the low mass-loss rate stars, their clumping factors are still significant ($\geq 4$ with an average of $21 \pm 15$). The high measured velocity filling factors are thus pointing to optically thin clumps, rather than a smooth wind.  

We note that while we discuss these trends only as a function of mass-loss rate, we obtain very similar results when we plot the wind structure parameters against other stellar or wind properties, such as luminosity. This is due to strong correlations between mass-loss rate and stellar properties. The interested reader can find the wind structure parameters plotted against luminosity, temperature, Eddington parameter for electron scattering or wind acceleration in \cref{fig:app:windstructure_teff} to \ref{fig:app:windstructure_gamma}. 

\subsection{Evolution\label{dis:evolution}}

Since the stellar parameters we derive are generally consistent with those of \citet[][]{Bestenlehner2020}, we do expect similar results for the age and initial mass distributions of the cluster. Indeed this is what we find when analysing the HRD positions of our sources (see \cref{fig:hrd_uvopt_all} and \cref{sec:bonnsai}): we derive a cluster age of $1-2.5$~Myr (median: 1.46~Myr) and an initial mass function with a power law slope of $\gamma = 1.99\pm0.11$ in the range $30-200~{\rm M}_{\odot}$. Detailed comparisons between this work and \citet[][]{Bestenlehner2020} can be found in \cref{app:comparisons,sec:app:bonnsai}. The rest of this section will focus on mass determination of the WNh stars and the evolution of the stars in the context of surface abundances. 

\subsubsection{Mass and age of the WNh stars\label{dis:WNh_stars}}

\Cref{fig:a1a2a3mass} shows the HRD positions of these stars as we find from our analysis, and compares these to the positions derived by \citet[][UV/optical/near-IR spectroscopy]{2010MNRAS.408..731C} and \citet[][optical spectrocopy]{Bestenlehner2020}. 
The position of R136a3 as found from our analysis is indicated twice as its effective temperature is hard to constrain; the diamond marks the higher temperature HRD-position that we adopt in this discussion, the cross the less likely alternative (see \cref{res:teffs} and \cref{sec:app:a3_analysis} for details). 

\begin{table}[]
    \centering
    \caption{Ages and masses of the WNh stars with 1$\sigma$ errors.}
    \begin{tabular}{l l l l }
    \hline \hline 
                & R136a1 & R136a2 & R136a3 \\ \hline 
         ${M}_{\rm spec}$ (\Msun) & $303 \pm^{123}_{79}$ & $159 \pm^{93}_{5}$ & $\geq 179$ \\
         ${M}_{\rm evol} $ (\Msun)& $222 \pm^{28}_{29}$  & $186 \pm^{17}_{15}$ & $179 \pm^{16}_{11}$  \\ 
         ${M}_{\rm ini}$   (\Msun)& $273 \pm^{25}_{36}$  & $221 \pm^{16}_{12}$ & $213 \pm^{12}_{11}$ \\ 
        Age (Myr) & $1.14 \pm^{0.17}_{1.34}$ & $1.34 \pm^{0.13}_{0.18}$ & $1.28 \pm^{0.17}_{0.21}$ \\ \hline
    \end{tabular}
    \label{tab:masses}
\end{table}
\renewcommand{\arraystretch}{1.}

Looking at the left hand side of \cref{fig:a1a2a3mass} we see that there is a considerable spread in observed temperatures and luminosities for the WNhs stars. However, while the temperature we derive for R136a1 is 7000~K lower than that of \citet{2010MNRAS.408..731C}, the initial masses from our and their analyses agree rather well: 273$\pm^{25}_{36}$~\Msun and $320 \pm^{100}_{40}$~\Msun, respectively. This is can be explained by differences in the evolutionary tracks used to derive these masses. The right hand side of \cref{fig:a1a2a3mass} shows how different assumptions for the evolutionary computations can lead to divergent theoretical predictions. 
Nonetheless, it is clear that regardless uncertainty in both observations and theory, the previously accepted initial mass-limit \citep{2005Natur.434..192F} challenged by \citet{2010MNRAS.408..731C} is indeed well exceeded by a1 and a2, and likely by a3 too; all different tracks in the left of \cref{fig:a1a2a3mass} point to an initial mass of $\geq250$~\Msun for R136a1, the most massive star in our sample, and, conservatively, $\geq150$~\Msun for R136a2 and R136a3. 
Of course, these results might not hold if the sources turn out to be in multiples, something that is currently being investigated with radial velocity measurements using HST-observations \citep{2019hst..prop15942S}. 
Furthermore, we note that our masses not only rely on the adequate determination of effective temperatures, and the used evolutionary tracks, but also on the flux calibration and reddening of the anchor magnitude used for our analysis. In this context, we note that \citet{2017IAUS..329..131R} focus in particular on the infrared (K-band) flux calibration of R136a1, R136a2 and R136a3, and find considerably lower initial masses for these stars compared to our analysis and that of \citet{2010MNRAS.408..731C}, their highest derived initial mass being an upper limit of 194~\Msun for R136a1. For their analysis, \citet{2017IAUS..329..131R} use VLT/SINFONI K-band spectrophotometry, effective temperatures of \citet{2010MNRAS.408..731C} and evolutionary tracks of \citet{2015A&A...573A..71K}.  

\begin{figure}
    \centering
    \includegraphics[width=0.40\textwidth]{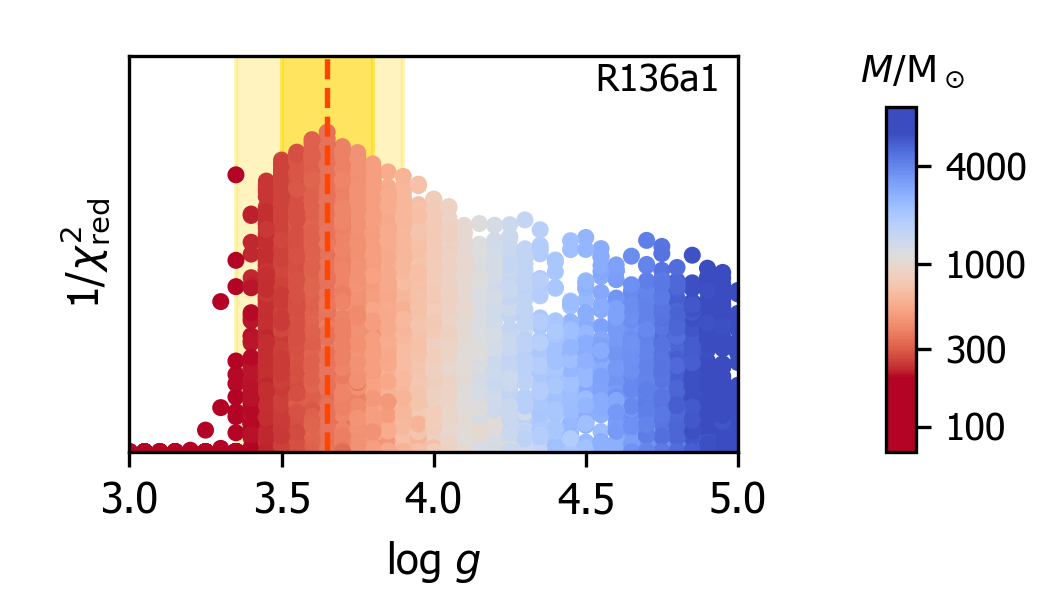}
    \caption{Measurements of the surface gravity \logg of the WNh star R136a1. Plotted is the fitness (1/\chisred) as a function of \logg, where each dot is a model in the \pyGA run. The colour corresponds to the (spectroscopic) stellar mass matching each model. The yellow shaded regions correspond to $1\sigma$ and $2\sigma$ uncertainties, and the orange dashed line the position of the best fit. 
    }
    \label{fig:logg_WNh}
\end{figure}

We measure the current masses of the WNh stars in a second manner. By measuring the surface gravity, we can obtain the spectroscopic mass $M_{\rm spec}$ ($= g_e  R_{\star}^2/G$, with $g_e = g + (v_{\rm eq}\sin i)^2/R_*$ surface gravity corrected for centrifugal accelerations, and $G$ the gravitational constant). Contrary to the analysis of \citet{Bestenlehner2020}, in our fits \logg is a free parameter also for the WNh stars. The winds of the WNh stars are so dense that we do not expect to see a very strong signature of the surface gravity in the spectrum. Still, from our \pyGA fits we do constrain \logg of both R136a1 and R136a2, for which we find a $2\sigma$ range of $\log \ g = 3.35-3.9$ and $\log \ g = 3.55-3.75$, respectively. For R136a3 we only find a lower limit, $\log \ g > 3.4$. \Cref{fig:logg_WNh} shows that the fitness distribution of the gravity of R136a1 clearly favors a value lower than 4.0. 
When using the measured \logg values in combination with the derived radii to derive spectroscopic masses, these compare well with the evolutionary masses (see \Cref{tab:masses}), further supporting very high masses for these stars.  

\subsubsection{CNO abundances\label{dis:abundances_evolution}}

\begin{figure}
    \centering
    \includegraphics[width=0.46\textwidth]{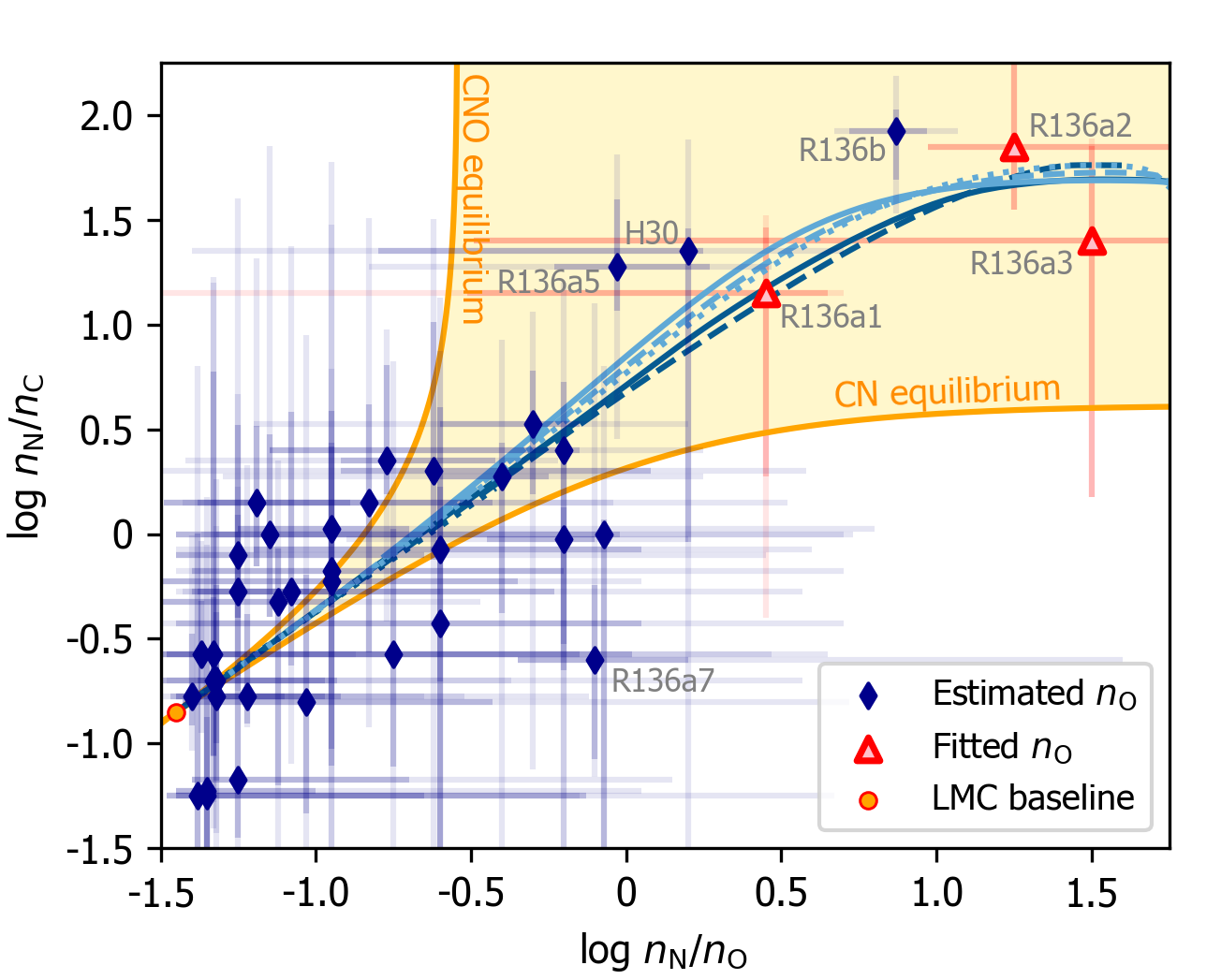}
    \caption{Comparison of our observed CNO-abundances to the theory of CNO processing. Only for three sources all abundances were measured (red triangles), for the rest the oxygen abundance is fixed to the LMC baseline value of \yO~$ =8.35$ (see text). 
    The yellow shaded region marks the regime between the analytical limiting solutions (CNO- and CN-equilibrium) of \citet{2014A&A...565A..39M}, their Eq. (14) and (17), respectively. Dotted, dashed and solid blue lines show evolutionary tracks of 30, 60 and 150~\Msun, respectively \citep[][]{2011A&A...530A.115B, 2015A&A...573A..71K}, where light and dark blues indicate models with low ($\sim100$~km~s$^{-1}$) and high ($\sim500$~km~s$^{-1}$) initial rotation velocities. }
    \label{fig:dis:CNOcycle}
\end{figure}

\begin{figure}
    \centering
    \includegraphics[width=0.49\textwidth]{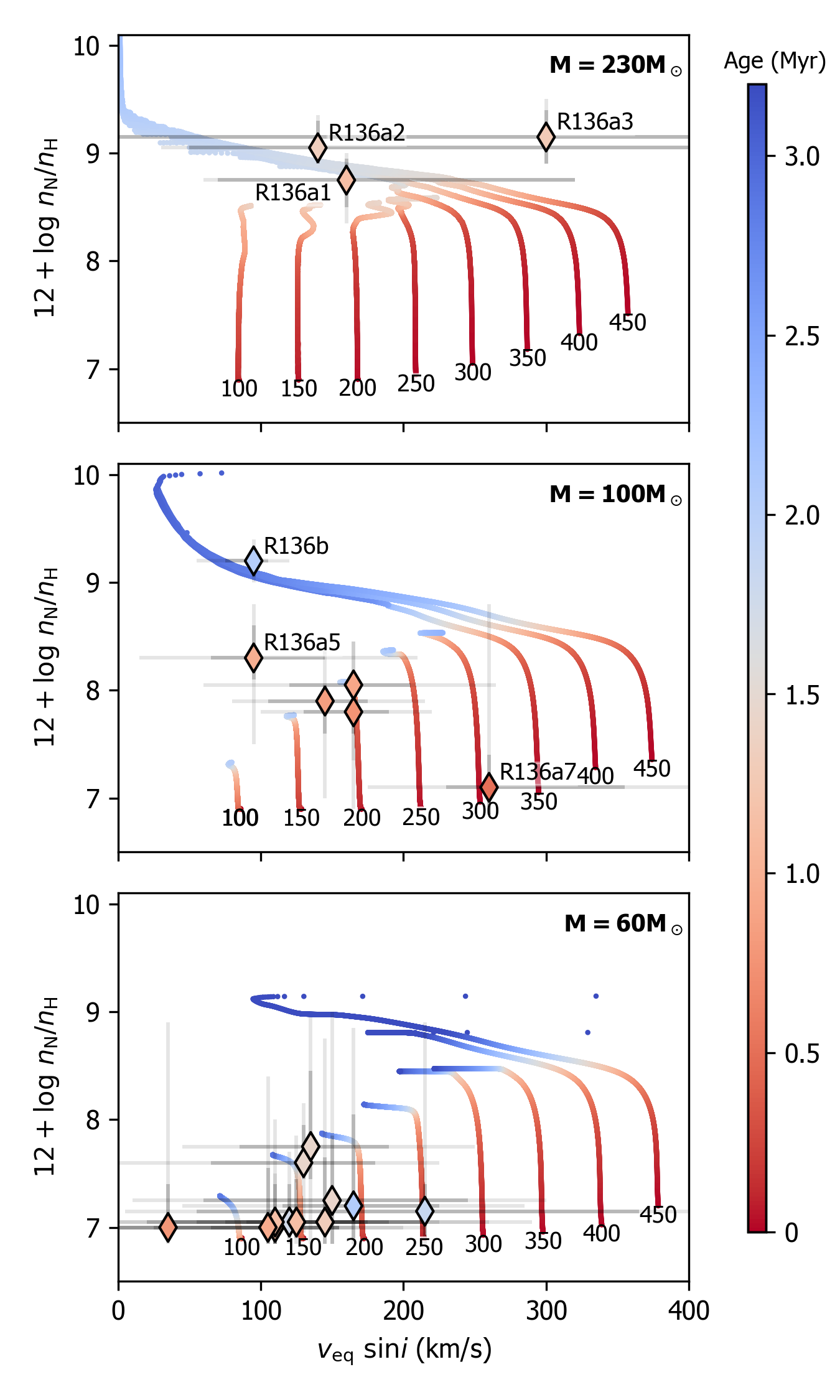}
    \caption{Hunter diagrams, showing stellar age (red to blue colour bar) as a function of rotation and nitrogen surface abundance. Evolutionary tracks are taken from \citet{2015A&A...573A..71K} and depict the evolution of stars with an initial mass of 230~\Msun (top), 100~\Msun (center), and 60~\Msun (bottom) with a range of initial rotation rates. Note that the theoretical rotation rates are scaled by $\pi/4$ to correct for the average projection angle, and that the tracks are cut off at \teff$< 29000$~K. The diamond markers indicate the observed positions of sources in the respective mass-regimes, where their colour maps to age via the same coding as the tracks. For a discussion on the error bars on the observed rotation rates of the WNh stars, see text. }
    \label{fig:dis:hunter_100_200}
\end{figure}

As CNO surface abundances are expected to change over the course of stellar evolution \citep[e.g.,][]{2015A&A...573A..71K,2021A&A...652A.137E}, they could be used to set apart the more evolved stars from the rest. Before we assess this for our sample, we check whether the derived abundances are consistent with the theory of CNO-processing. This is an especially important check because without exception the C and N abundances are derived from lines that are (mostly) formed in the stellar wind, and thus their strength and shape not only depends on abundance and temperature, but also on a handful of wind properties. Since all those properties can vary independently from each other, consistency with CNO-processing theory is not guaranteed intrinsically and needs to be checked. The diagram in \cref{fig:dis:CNOcycle} allows for such a consistency check, by comparing the ratios $n_{\rm N}/n_{\rm C}$ to $n_{\rm N}/n_{\rm O}$. \citet{2014A&A...565A..39M} derive analytically the limits of these ratios given CNO and CN equilibrium: we expect all observations to fall somewhere between the region bordered by those. Furthermore, evolutionary tracks of different masses and initial rotation \citep{2011A&A...530A.115B,2015A&A...573A..71K} predict quite a narrow range in which we expect points to lie. This method has also been applied by \citet{2015A&A...575A..34M} and \citet{2019A&A...623A...3C}. 

The WNh stars -- the only sources for which we measure oxygen abundance -- lie in the area predicted by \citet{2014A&A...565A..39M}, and moreover match well the evolutionary predictions depicted in \cref{fig:dis:CNOcycle}. 
For the other sources, we do not measure oxygen abundance and we therefore assumed for all stars the LMC baseline value, $\log \ n_{\rm O}/n_{\rm H} + 12 = 8.35$ \citep[][as in \citealt{2011A&A...530A.115B, 2015A&A...573A..71K}]{1998RMxAC...7..202K}, so no enrichment or depletion. With this assumption and considering the uncertainties on the measurements, our observations are generally consistent with CNO-processing, except for R136a7. For this source evolutionary models predict oxygen depletion to \yO~$= 7.20$, assuming an initial mass of 100~\Msun and an initial rotation of 300 km~s$^{-1}$. This would move this source into the region consistent with CNO-processing. 
If we also assume oxygen depletion for other sources, as was done in the optical+UV fitting\footnote{For each star we estimate the amount by combining the mass, projected rotation and age of the stars according to \citet{Bestenlehner2020}, as described in \cref{sec:optUVsetup}.}, the shift along the $\log n_{\rm N}/n_{\rm O}$ axis is small ($0.1-0.2$ dex to the right). We note that there is an uncertainty in the observed LMC baseline abundances of CNO \citep[see, e.g.,][]{2019AJ....157...50D} that could affect \cref{fig:dis:CNOcycle}. However, we find that changes in the diagram that could result from this are smaller than the typical error bars on our observations. 
Overall, we conclude that no violation of CNO-processing is observed and that abundances and wind parameters can be disentangled from a set of 9-11 (metal) wind lines, albeit with large uncertainties.  

Six sources stand out in \cref{fig:dis:CNOcycle}. These are the WNh stars, R136a5, R136b, and H30.  \citet{Bestenlehner2020} show, based on the helium abundance, that the enrichment of these stars is mainly driven by mass-loss (see also \citealt{2014A&A...570A..38B}). 
Here, we further investigate the nitrogen enrichment by placing our sources in a Hunter diagram. This diagram, introduced by \citealt{2008ApJ...676L..29H}, shows the nitrogen surface abundance of stars versus their projected rotational velocity. 
By comparing these quantities, one can gain insight into mixing processes that occur within the star; rotational mixing being one of the mechanisms that can bring processed elements to the surface. 
In the Hunter diagrams in \cref{fig:dis:hunter_100_200} we compare our observed values with evolutionary tracks of different rotation rates \citep{2015A&A...573A..71K}. In order to account for the different masses of the sources, we compare three subgroups of sources with tracks of three different masses. 

All WNh stars ($M_{\rm ini} = 213-273$~\Msun, top panel of \cref{fig:dis:hunter_100_200}) show strong nitrogen enrichment; we find a similar nitrogen surface abundance and age for all three. Given the large uncertainty on the rotation rate, the nitrogen abundance and age of R136a1 match the tracks quite well. For R136a2 and R136a3 the obtained nitrogen abundances are, given the obtained ages, slightly too high for any of the tracks. However, the difference is not large and within uncertainties the single star models and observations match. We emphasise that, while the best fit values for the \vsini are prominently marked (based on the best fitting model of the optical only run for each star), one should not overlook the large error bars. In the most extreme case of R136a3 these span all displayed \vsini values, implying that we cannot put any constraints on the current \vsini of this source. 
Adding to the uncertainty (not captured in the statistical error bars shown here) is the fact that we derive all rotation rates by convolution of the line profiles, hereby implicitly assuming that the emergent radiation is emitted from one single rotating layer. This assumption likely breaks down for all available spectral lines in the WNh spectra, casting further doubt on the validity of the WNh rotation rates that we derive. A more sophisticated approach would be to include the effects of rotation on the velocity field of the wind into the formal integral \citep{2014A&A...562A.118S}, however this is not within the current capabilities of \fw. 
Regardless of the observed \vsini values, a high initial rotation is suggested for all WNh stars if one compares the tracks to the observed ages and nitrogen abundances. This was also found from the {\sc Bonnsai} runs for these sources based on a comparison of the observed luminosity, temperature, surface gravity and helium surface abundance to the \citet{2015A&A...573A..71K} tracks. 

The middle panel of \cref{fig:dis:hunter_100_200} shows stars in the mass range $M_{\rm ini} = 92-127$~\Msun. The age and position of the sources generally show good agreement with the evolutionary tracks. The three points that lie close together (R136a4, R136a6 and H36), all seem to have started out with a moderate initial rotation rate. R136b is highly nitrogen enriched, which suggests that the star must have had an initial rotation in the range of $300-450$ km~s$^{-1}$ and is older than the other stars. Indeed, the age of the star, derived from the observed luminosity, temperature, surface gravity and helium abundance, is 2~Myr. Looking at the physical position of this star in the cluster (\cref{fig:positions}) we see that it is located a bit on the outskirts of the cluster, further away from the centre than  most other very massive stars (\cref{fig:positions}). Possibly this could be related to its somewhat higher age, however, we do not have any evidence for this; \citet{Bestenlehner2020} considers the ages and positions of all sources and does not find a correlation of age and position (we confirm this finding, see \cref{sec:app:bonnsai}). 
The outlier in this panel is R136a5, that, within its $1\sigma$ errors, does not seem to fall on any of those tracks. We note that this star has a slightly higher initial mass than that of the tracks shown here ($116$~\Msun). Tracks with higher mass would have more enrichment (see top panel), and thus would bring the tracks and the observations of R136a5 closer together. R136a7 could also be considered an outlier as it is the only source in this mass-range not showing significant enrichment. Yet, it is consistent with the tracks, as, with an age of 0.5$\pm^{0.37}_{0.46}$~Myr, R136a7 is one of the youngest, if not the youngest, star in the sample\footnote{See \Cref{tab:app:more_uv_restuls}. For several other stars we find from {\sc Bonnsai} a formal age of almost zero, however in all these cases the errors bars are extremely large (1-2.5 Myr), so we consider the age of these sources unconstrained.}. 

The bottom panel of \cref{fig:dis:hunter_100_200} shows stars in the mass range $M_{\rm ini} = 52-68$~\Msun. Most of them are not yet nitrogen enriched. For a few we do observe a slight enrichment, though with very large uncertainties. Within errors, the observations are consistent with the $60$~\Msun tracks of \citet{2015A&A...573A..71K}. 

It is interesting that while for the WNh stars we require high initial surface rotation rates ($\varv_{\rm eq,ini} > 300$~km~s$^{-1}$) in order to match the tracks, this is not the case for most O-stars. The O-stars in \cref{fig:dis:hunter_100_200}, except R136a7 and R136b, suggest initial surface rotation rates $\varv_{\rm eq,ini} < 300$~km~s$^{-1}$. The inferred initial rotation rates are, for all stars, consistent with the values inferred using {\sc Bonnsai} based on different observables (luminosity, temperature, surface gravity and helium abundance); with this we find 100~km~s$^{-1}$ for all with the exception of the WNh stars, R136a7 and R136b, for which we find $\varv_{\rm eq,ini} > 300$~km~s$^{-1}$. The difference between the WNh stars and R136b and R136a7 on the one hand, and the other O-stars on the other hand might indicate that the two groups of stars are formed through a different channel. We note that only for sources for which we observed helium enrichment ($x_{\rm He} \geq 0.14 $), a high initial rotation rate was derived using {\sc Bonnsai}. 

In summary, the collection of Hunter diagrams shows the enrichment of young (very) massive stars. At an age of $\sim 1.5$~Myr the $\sim 60$~\Msun stars are barely nitrogen enriched, while the $\sim 100$~\Msun and $\sim 200$~\Msun stars show enrichment of about one and two orders of magnitude, respectively. 
This is roughly consistent with the single star models of \citet{2015A&A...573A..71K}. We note that \citet{2008ApJ...676L..29H,2011A&A...530A.116B,2017A&A...600A..82G} report a population of slowly spinning nitrogen enriched stars, which we do not identify in our study. This may potentially point to binary interaction as a source of such stars; given the young age of our population it could be expected that such interactions have not yet occurred frequently.

\begin{figure}
    \centering
    \includegraphics[width=0.40\textwidth]{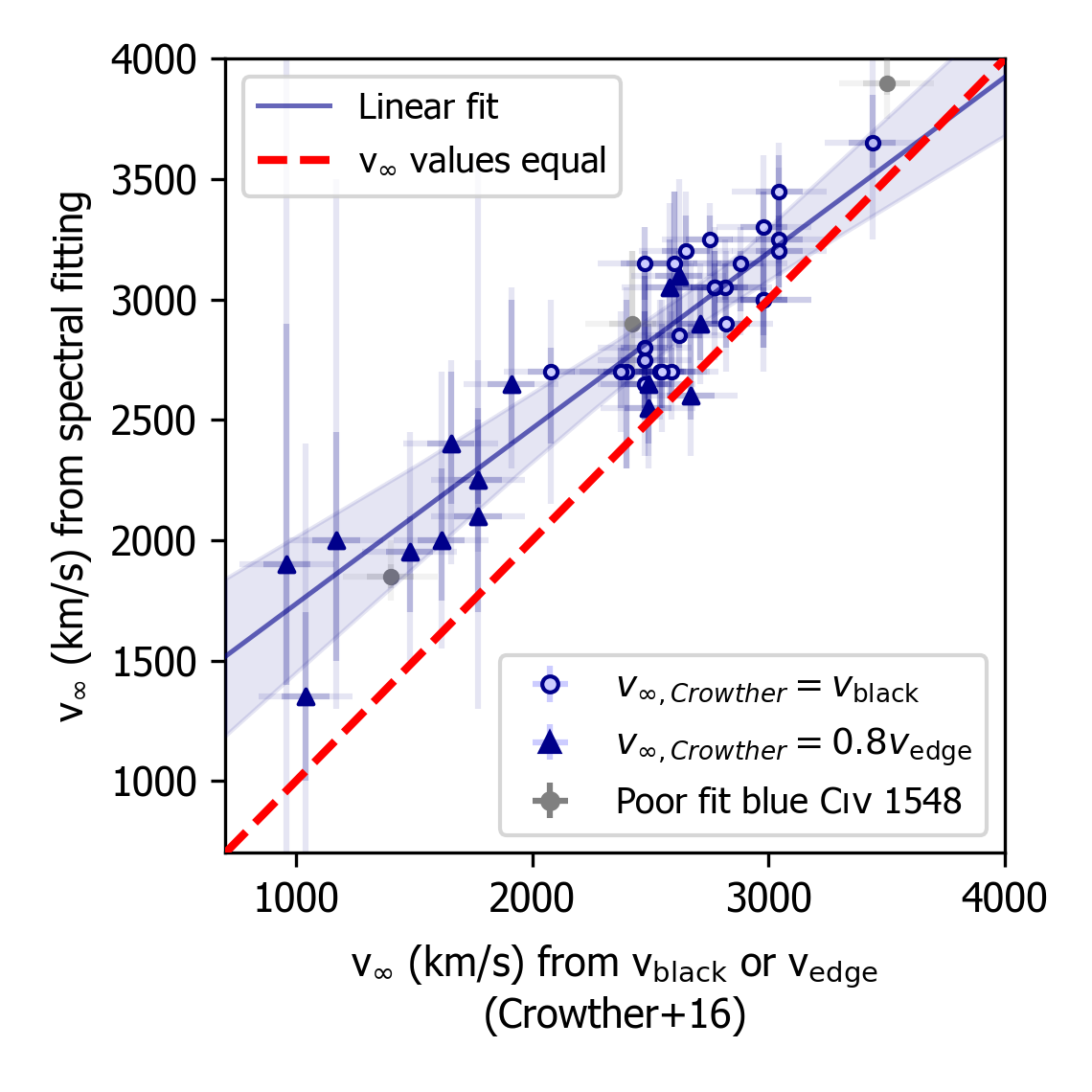}
    \caption{Comparison of terminal velocities $\varv_\infty$ determined by spectral fitting (this work) to those determined by locating $\varv_{\rm black}$ \citep[][]{2016MNRAS.458..624C}.  Circles indicate sources for which \citet{2016MNRAS.458..624C} estimated \vinf based on $\varv_{\rm black}$, solid triangles indicate sources for which they used $0.8\varv_{\rm edge}$. Grey symbols indicate sources for which we could not obtained a good fit to the \CIVline profile. The thick red dashed line shows where the two methods agree, while the blue solid line shows a linear fit through the data. See also \cref{fig:vinf_thiswork_crow}.
    }
    \label{fig:vinf_vinf}
\end{figure}

\begin{figure}
    \includegraphics[width=0.42\textwidth]{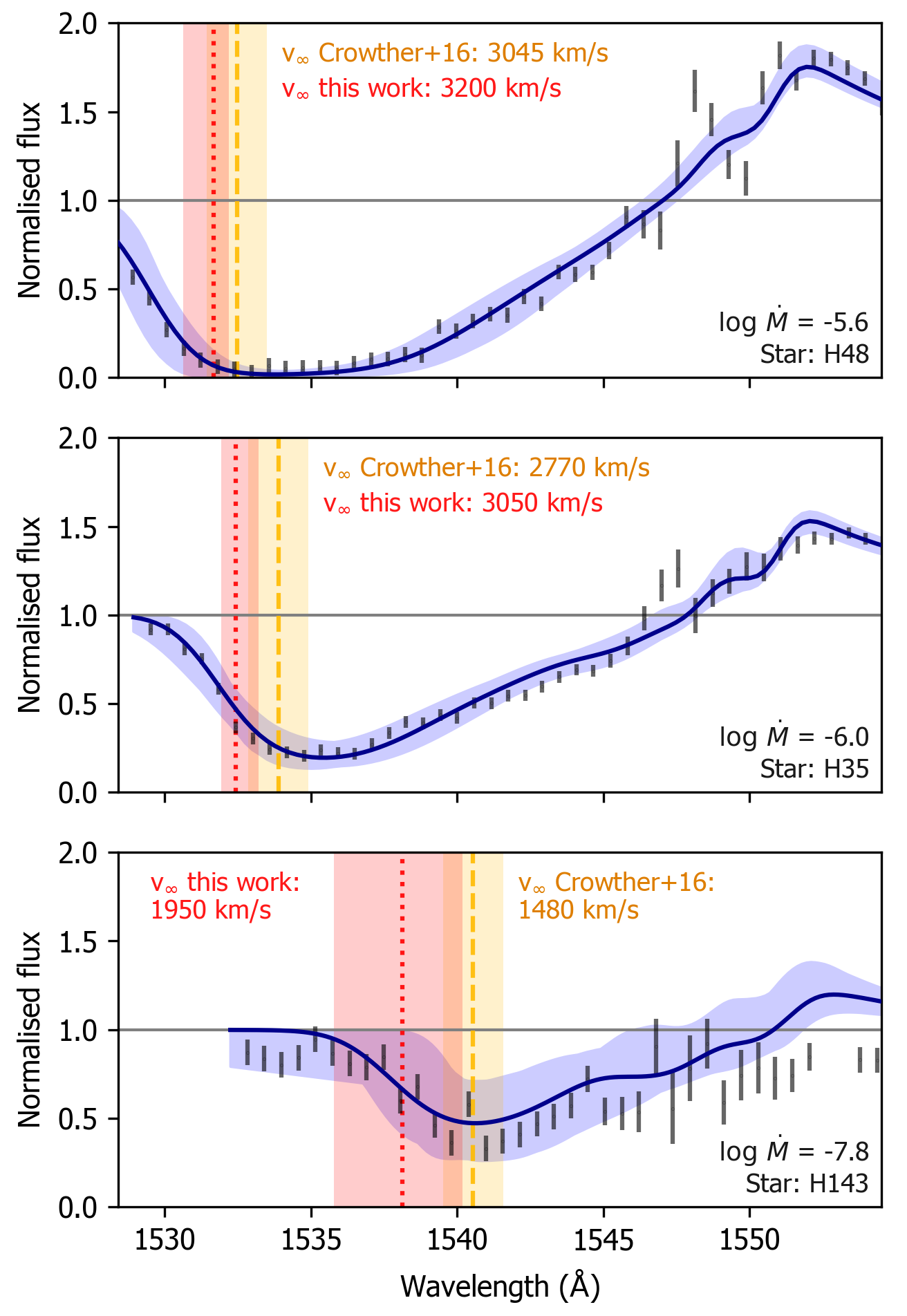}
    \caption{Line profiles of \CIVline and the best fit models ($2\sigma$) compared to the values of $v_\infty$ determined by spectral fitting (this work, red dotted line) and those of \citet[][yellow dashed line]{2016MNRAS.458..624C}. The shaded regions indicate the $2 \sigma$ errors on the derived $v_\infty$. We used the blue transition of the doublet ($\lambda = 1548.19$\angstrom) for indicating velocities. 
    We show profiles with different \CIVline appearances: strong and saturated (top), strong but not saturated (middle) and weak (bottom). In all cases $v_\infty$ from spectral modelling lies bluewards of that of \citet{2016MNRAS.458..624C}. See also \cref{fig:vinf_vinf}.} 
    \label{fig:vinf_thiswork_crow}
\end{figure}

\subsection{Terminal velocity measurements\label{dis:terminal}}

With spectral fitting we are able to break the degeneracy between \vinf and \vwindturb. While based on the edge velocity alone one cannot disentangle \vinf and \vwindturb, the shape of the absorption component of \CIVline is affected differently by the two parameters, and we are able to distinguish between the two. We note that narrow absorption components, unresolved in our spectra, may have contributed to absorption near the terminal velocity of non-saturated line profiles \citep{1990ApJ...361..607P}. As the equivalent width of these absorption components is small, we do not expect a significant effect on our measurements. 

The terminal velocities we find from spectral fitting are systematically larger than those of \citet{2016MNRAS.458..624C}, who used the exact same dataset but obtained \vinf by inspecting the blue wing of the \CIVline P-Cygni line (\cref{fig:vinf_vinf}). 
For 29 sources they identify the maximum blueward extent of (near) saturated absorption profiles, $\varv_{\rm black}$, and assume $\varv_\infty = \varv_{\rm black}$  \citep[following][]{1990ApJ...361..607P}. For these 29 sources they also identify the velocity at which the violet absorption meets the continuum, $\varv_{\rm edge}$, and with this derive $\varv_\infty = 0.8 \varv_{\rm edge}$. With the latter relation they estimate \vinf by identifying $\varv_{\rm edge}$ for the remaining 15 sources, that have less strong wind lines. 

\Cref{fig:vinf_vinf} shows the \vinf values derived from spectral fitting versus those found from $\varv_{\rm black}$ or $\varv_{\rm edge}$, which we call $\varv_{\infty \rm, blue}$. 
We find an average difference of 283$\pm 30$ km~s$^{-1}$ between the two methods, taking into account uncertainties on the measurements, and excluding three sources for which we could not obtain a good fit to the \CIVline profile (see \cref{sec:res:terminal}).  
Moreover, a trend is visible in the difference: on average the difference between \vinf derived with the two methods increases for lower terminal velocities.  \Cref{fig:vinf_thiswork_crow} shows \CIVline profiles for three sources with different mass-loss rates and terminal velocities. The location of \vinf according to this work and that of \citet{2016MNRAS.458..624C} is indicated. According to the spectral models, the terminal velocity does not coincide with the point of strongest absorption, but lies more towards the blue. 
Some desaturation thus occurs close to the terminal velocity already in H48, the star with the strongest P-Cygni profile of these three. If the approach of \citet{2016MNRAS.458..624C} is followed for obtaining \vinf, the obtained values $\varv_{\infty \rm, blue}$ could be corrected for this effect by using the relation shown as the blue solid line in \cref{fig:vinf_vinf}: 
\begin{equation}
    \varv_\infty = \varv_{\infty \rm, blue} + 0.27 \cdot (3700 - \varv_{\infty \rm, blue}),
\end{equation}
(velocities in km~s$^{-1}$) for $1000 {\ \rm km}~{\rm s}^{-1} > \varv_{\infty \rm, blue} > 3500 {\ \rm km}~{\rm s}^{-1}$. One could use this equation for getting a more accurate first order estimate of \vinf from reading off $\varv_{\rm blue}$ from spectra with relatively poor resolution.

Upon comparing the edge velocities of \citet{2016MNRAS.458..624C}, obtained by reading off the velocity at the wavelength where the blue edge of \CIVline meets the continuum, to the edge velocities we obtain by spectral modelling and assuming $\varv_{\rm edge} = \varv_{\infty} + \varv_{\rm windturb}$, we find that they are generally in good agreement. For our sample we find the following average: 
\begin{align*}
\varv_{\rm windturb} &= 0.14 \pm 0.06 \ \varv_\infty \hspace{1cm} \mathrm{or, equivalently:}\\
\varv_\infty &= 0.88 \pm 0.04 \ \varv_{\rm edge} \numberthis \label{eq:windturb}
\end{align*}
Had \citet{2016MNRAS.458..624C} assumed this value instead of $\varv_\infty = 0.8 \varv_{\rm edge}$, our respective values for \vinf would lie closer together for the weaker wind sources (triangles in \cref{fig:vinf_vinf}), although part of the discrepancy would still remain. We stress that \cref{eq:windturb} is based on fits of the sources for which we carried out 12-free parameter fits, that is, the sources with stronger winds. Data of higher \snr and resolution is required to disentangle \vinf and \vwindturb for the sources with weaker winds.

\section{Summary \& Outlook\label{sec:conclusion_outlook}}

We have simultaneously analysed optical and UV spectroscopy of a population of 56 stars in the core of the R136 star cluster, nine members of which have masses $M\gtrsim100$~\Msun. 
For the first time we investigate the wind structure parameters of a large range of spectral types while fitting the interclump density, the wind turbulence and the effects of optically thick clumps such as velocity-porosity. By taking into account these effects we improve the accuracy of mass-loss determinations of the most massive stars. Moreover, the derived mass-loss rates are no longer affected by the well known mass-loss/clumping dichotomy, but are actual values. 
Our main findings are the following: \begin{itemize}
\item The HRD-positions of the sources suggest a cluster age of 1-2.5~Myr, in line with the findings of \citet{Bestenlehner2020}. The ages of the highly nitrogen enriched WNh stars are in line with the age of the rest of the population. 
\item Our conservative estimate for the initial mass of R136a1, the most massive star in our sample, is 250~\Msun. The initial masses of R136a2 and R136a3 well exceed 150~\Msun. The spectroscopic masses of these sources, which we measure here for the first time, further support this conclusion. 
\item We compare the theoretical predictions of \citet{2001A&A...369..574V}, \citet{2018A&A...612A..20K}, \citet{2021A&A...648A..36B} and \citet{2021MNRAS.504.2051V} to the observed mass-loss rates and terminal velocities and find that none of the predictions satisfactorily reproduces both quantities. The largest discrepancies for the terminal velocities are found for stars with $\log L/{\rm L}_\odot \lesssim 5.3$ for the \citet{2021A&A...648A..36B} predictions, and stars with $\log L/{\rm L}_\odot \gtrsim 5.3$ for the \citet{2018A&A...612A..20K} predictions.
\item Overall, the mass-loss recipe of \citet{2018A&A...612A..20K} best matches the observed mass-loss rates of the stars in our sample. The predictions of \citet{2021A&A...648A..36B} match almost as good, performing better in the low-luminosity regime ($\log L/{\rm L}_\odot \lesssim 5.3$), but worse for the higher luminosities. The prescriptions of \citet{2001A&A...369..574V} and \citet{2021MNRAS.504.2051V} overpredict the mass-loss rates for all luminosity regimes. 
\item The stellar winds of the stars in our sample are highly clumped, with an average clumping factor of \fcl~$= 29\pm15$. 
\item  We find tentative trends in the wind structure parameters as a function of mass-loss, where the stars with the highest mass-loss rates seem to have smoother, albeit still clumpy, winds (but see below). 
\item We provide a prescription for the mass-loss rates of the most massive stars as a function of the electron scattering Eddington parameter, following the work of \citet{2020MNRAS.493.3938B}. For this, we have used our best fit values of the CAK force multiplier parameter $\alpha_{\rm eff}= \alpha - \delta = 0.46\pm0.04$ and transition mass-loss rate $\log \dot{M}_{\rm e,trans} = -5.19\pm0.10$. 
\item The point with the strongest absorption in a  P-Cygni profile does typically not correspond with the terminal velocity, which lies more bluewards. 
We provide an equation that quantifies this effect. 
\end{itemize}
This is the first investigation of trends in wind structure parameters of massive stars. 
While the measurements of \citet{Hawcroft21} are not in contradiction with our results, neither can they confirm the trends we observe, given the limited span of mass-loss rates in their sample. 
Further investigation of these trends is thus necessary. \citet{Hawcroftinprep2} are undertaking such a study, analysing optical and UV spectroscopy of a sample of about 30 LMC and SMC stars covering most O-type subclasses (O3-O9). Furthermore, our study can be considered a pilot for a larger investigation making use of the high-quality HST UV spectra of the ULLYSES project\footnote{Hubble UV Legacy Library of
Young Stars as Essential Standards, a Director’s Discretionary program in progress ( \url{https://ullyses.stsci.edu/index.html})} \citep[][]{2020RNAAS...4..205R}. The ULLYSES sample, when complete, will consist of $\sim$250 massive stars (mostly in the Magellanic clouds), including $\sim$150 O-stars, covering all O-star subtypes and luminosity classes. Complemented by the optical XshootU program\footnote{\url{https://massivestars.org/xshootu/}}, ULLYSES will provide an excellent opportunity for further study of the structure of massive star winds. 

\begin{acknowledgements}

We like to thank the anonymous referee, whose constructive criticism has helped us to improve the presentation of this work. This publication is part of the project `Massive stars in low-metallicity environments: the progenitors of massive black holes' with project number OND1362707 of the research TOP-programme, which is (partly) financed by the Dutch Research Council (NWO). Observations were taken with the NASA/ESA \textit{HST}, obtained from the data archive at the Space Telescope Institute. Computations were carried out on the Dutch national e-infrastructure with the support of SURF Cooperative. SAB would like to thank the late Lykle Voort for his help and patience. FAD and JOS acknowledge support from the Odysseus program of the Belgian Research Foundation Flanders (FWO) under grant G0H9218N. TS acknowledges support from the European Union’s Horizon 2020 under the Marie Sk\l{}odowska-Curie grant agreement No 101024605. This work has received funding from the European Research Council (ERC) under the European Union’s Horizon 2020 research and innovation programme (Grant agreement No.\ 945806). This work is supported by the Deutsche Forschungsgemeinschaft (DFG, German Research Foundation) under Germany’s Excellence Strategy EXC 2181/1-390900948 (the Heidelberg STRUCTURES Excellence Cluster). GG acknowledges support from Deutsche Luft- und Raumfahrt (DLR) grant No. 50\,OR\,2009. 

\end{acknowledgements}

\bibliography{references}

\clearpage 
\newpage


\begin{appendices}
\crefalias{section}{appendix}
\renewcommand{\thetable}{\Alph{section}.\arabic{table}}
\renewcommand{\thefigure}{\Alph{section}.\arabic{figure}}
\setcounter{figure}{0}
\setcounter{table}{0}

\section{Data quality \label{sec:app:dataquality}}

\subsection{\snr\label{sec:app:dataquality_snr}}

\begin{figure}
    \centering
    \includegraphics[width=0.49\textwidth]{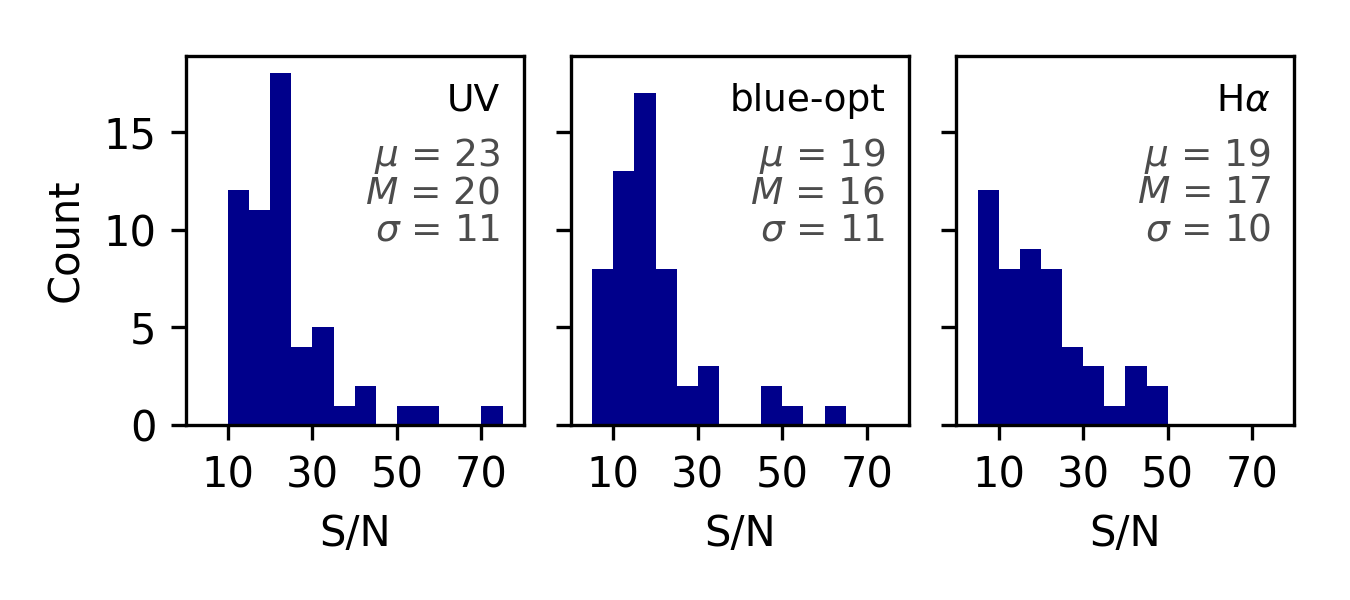}
    \caption{\snr distribution for each wavelength range: UV (left), blue-optical (middle) and \halpha (right). Mean $\mu$, median $M$ and standard devition $\sigma$ are given for each range.}
    \label{fig:app:snr_dist}
\end{figure}

Distributions of \snr for all sources in our sample are shown in \cref{fig:app:snr_dist}. We see that generally the UV spectra have the highest \snr, but note that typically \cuvline, which is very close to the blue grating edge and is also slightly affected by Ly-$\alpha$ absorption, has a lower \snr than the values shown in the figure. Also, in the most red part of the UV grating, \heiiuvline and \nivuvline (if used at all) are typically a bit more noisy than the other lines. 

\subsection{Wavelength correction\label{sec:app:dataquality_rv}}

\begin{figure}
    \centering
    \includegraphics[width=0.50\textwidth]{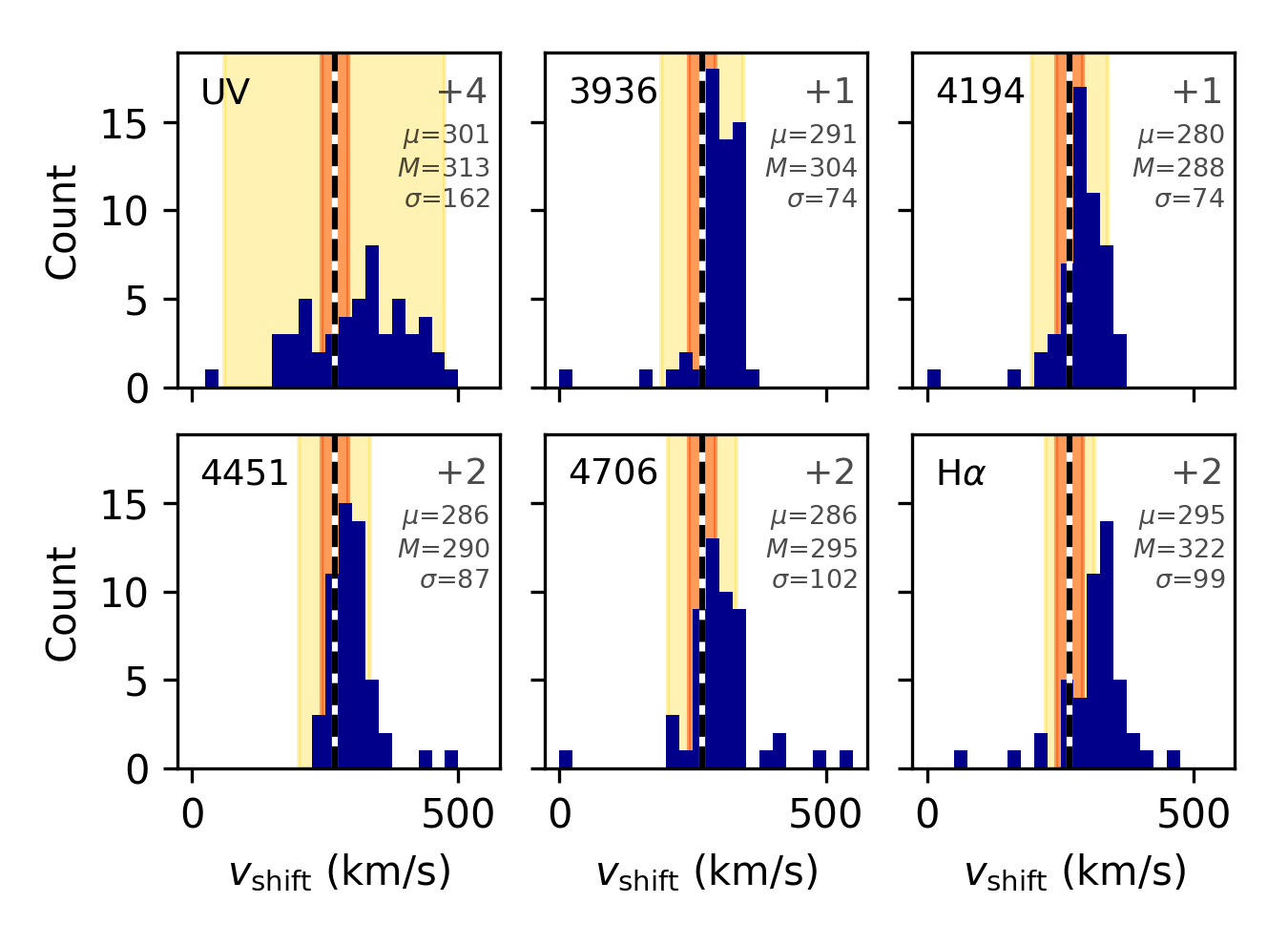}
    \caption{Distribution of measured wavelength corrections $\varv_{\rm shift}$ in km~s$^{-1}$, for each grating (dark blue). We indicate the mean R136 velocity of 267.7~km~s$^{-1}$ as measured by \citet[][black dashed lines]{2012A&A...546A..73H}, together with the non-binary-corrected velocity dispersion of $\sigma = \pm$25 km~s$^{-1}$ (orange-red shaded), as well as a velocity range corresponding $\pm$2 pixels around the R136 mean (yellow shaded). The numbers in the upper right corner of each plot denote the amount of stars that fall out of the plotted velocity range. Below that, we show the inferred mean $\mu$, median $M$ and standard deviation $\sigma$ of each grating (all in km~s$^{-1}$).}
    \label{fig:app:radial_velocity_distR136}
\end{figure}

The radial velocity shifts inferred for all stars, for each grating, are shown in \cref{fig:app:radial_velocity_distR136}. If our radial velocity measurements were accurate, we would expect that they have a mean velocity of $267.7 \pm 25$~km~s$^{-1}$, as measured by \citet{2012A&A...546A..73H}, who analysed high-quality spectra of stars around the R136 core. Due to the calibration related wavelength offset however, the dispersion we find is expected to be higher. Indeed, the standard deviation on the radial velocities is on the order of $\pm2$~pixels for most gratings. The exception is \halpha, where the typical offset is 3 to 4 pixels. 

\begin{figure*}
    \centering
    \includegraphics[width=\textwidth]{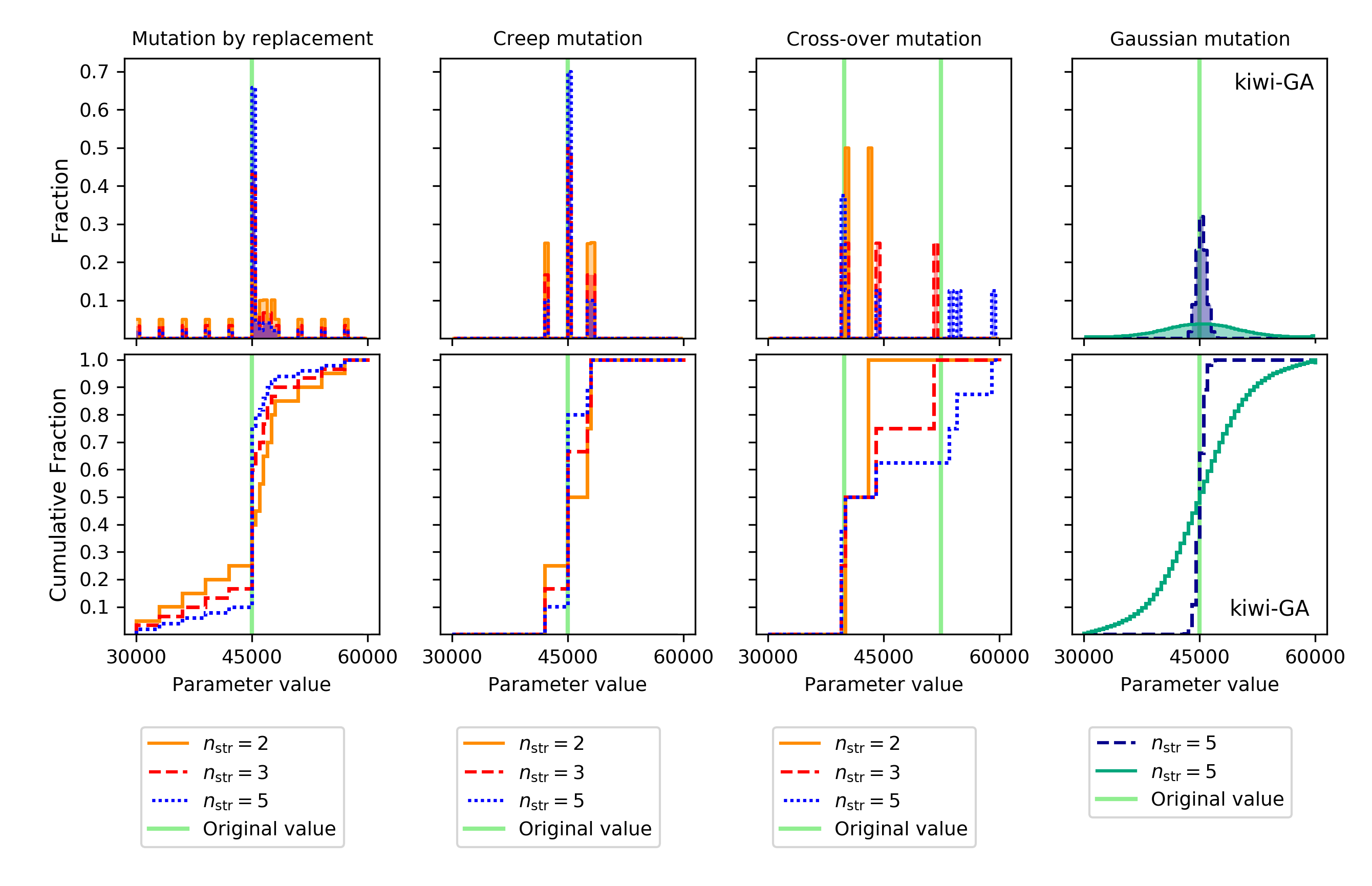}
    \caption{Distribution (top panels) and cumulative distribution (bottom panels) of mutation sizes of different schemes. The panels in the first three columns on the left show the behaviour of different types of mutation that result from modifying strings that represent the parameters in three different ways: by replacing one string digit by a random other digit (mutation by replacement), by increasing or decreasing the value of one string digit with 1 (creep mutation), and by changing one or more digits of a string by slicing it during the recombination process (cross-over mutation). The green vertical line indicates the original value of the parameter (in the third panels from the left this are two lines: one for each parent), in orange, red and blue the distribution of parameter values after the mutation has taken place. The different colours refer to the amount of digits that decode one parameter. The rightmost panels show the behaviour when the mutation size is described by a Gaussian distribution, as is the case in \pyGA.}
    \label{fig:app:mutation_details1}
\end{figure*}

\begin{figure*}
    \centering
    \includegraphics[width=1.0\textwidth]{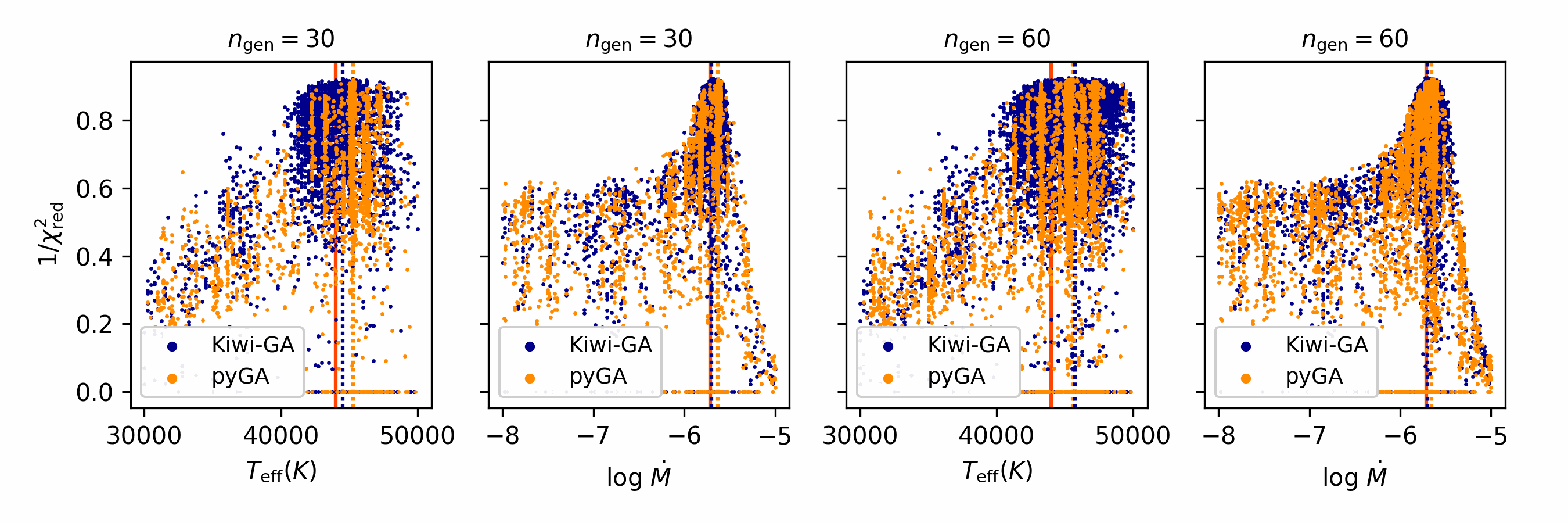}
    \caption{Comparison of the fitness distributions of \teff and log~\mdot of two genetic algorithm runs. The dark blue points show the models of a \pyGA run, the yellow points the models of a {\sc pyGA} run. Setup of the runs was identical: both runs had the same data (\fw model with simulated \snr of 30), parameter space, number of individuals, and number of generations. The left two plots show the run states after 30 generations, the right two plots after 60 generations. The ``true values'', that is, the value of the model used for simulating the data, is indicated with a red line. Dark blue and yellow dotted lines indicate the best fit models.}
    \label{fig:app:chi2compareGA}
\end{figure*}

\section{Technical details of \PyGA \label{sec:app:pyga}}

We will now discuss how \pyGA differs from the algorithm of \citet{2005A&A...441..711M} and \citet{2021A&A...651A..96A}, who follow the approach of the former.  
First of all, \pyGA applies operations of recombination (mixing of parameters of two models) and mutation (addition of random variations to a subset of the parameters) directly on the model parameters. \citet{2005A&A...441..711M} store the parameter values in the form of one string per model, where each character of the string can have a value from 0...9 and different parts of the string indirectly represent the values of all parameters. On these strings the mutation and recombination operations are applied.  
The fact that this concept is abandoned in \pyGA has consequences for the way that mutation and recombination are implemented. 
With this in mind, we list here the most significant changes of our algorithm compared to that of \citet{2005A&A...441..711M}:

\begin{itemize}
    \item Recombination is carried out per parameter. Given two sets of parameters (parents), the new parameters (offspring) are chosen by, for each parameter, picking the value of one of the parents. The genome can thus be split in more locations than only one or two, but not in the middle of a parameter. This implies that mutation due to recombination can never occur. 
    \item No parents are cloned, that is, recombination always occurs\footnote{Though not recommended, the user can choose to include cloned models by adjusting the clone fraction to $0 \leq f_{\rm clone} \leq 1 $.}. All new models are thus the product of both recombination and mutation, and none of mutation alone. This increases the diversity of the population. 
    \item Two types of mutation are introduced, each having a different rate of occurrence. In each case the size of the mutation, or the amount with which the parameter value changes, follows a Gaussian distribution. Small mutations, where the value of the mutated parameter lies close to the original value, occur frequently: after recombination, each parameter has a large chance to undergo a small mutation. Larger mutations occur with a lower frequency. The exact frequencies can be set by the user, in this work we assumed a rate of 0.5 for the small mutations, and a maximum rate of 1/$n_{\rm free}$ for the larger mutations. The latter means that on average one parameter per model undergoes a large mutation. This manner of mutating parameters is very different from the scheme used in \citet[][see \cref{sec:app:pyga:mutation} for details]{2005A&A...441..711M}.
    \item Reinsertion of new individuals into the parent population is done according to an elitist-fitness scheme, where the parent population can be larger than the amount of individuals in one generation. This means that after each generation, the fitness of both parent and offspring models is assessed and the top models of each sub-population will form the new parent population. An elitist-fitness scheme leads to a fitter parent population, while diversity is ensured by always including a certain fraction of offspring models\footnote{The exact value can be set by the user, we chose a fraction 0.75.}.
\end{itemize}
Together, these adaptions result in a faster parameter space exploration, that is, less generations need to be computed. Especially the exploration around the best fit, done in order to assess the uncertainty regions, benefits from the algorithm adaptions. 

Apart from these major points there are other additions, such as the fact that H-{\sc i}, He-{\sc i} and He-{\sc ii} ionising fluxes and uncertainties thereof can be given as an output, the possibility to make the first generation larger compared to the rest of the generations\footnote{This aids the initial exploration of the parameter space, with only a small increase of computation time.}, the option to treat X-ray parameters as free parameters, and the option to estimate an appropriate volume filling fraction $f_{\rm X}$ for each model such that \loglx (see \cref{sec:app:xrayassumptions}). Furthermore, \pyGA can be used in combination with \fw version 11, which can treat radiative transfer of the full spectrum in the co-moving frame \citep{2020A&A...642A.172P}.

\subsection{Mutation in \PyGA \label{sec:app:pyga:mutation}}

\Cref{fig:app:mutation_details1} shows several mutation schemes implemented in the algorithms of \citet{2005A&A...441..711M} and \citet{2021A&A...651A..96A}, next to the mutation scheme used in \pyGA. The scheme used in \pyGA results in a mutation distribution that covers the parameter space more regularly. In practice, this gives the user more control over the ratio of small to large mutations. On the one hand, parameters close to fit models (created by small mutations) are expected to be more successful (because the original model was already fit), but on the other hand parameters very different from the fit models can cause larger improvement in fitness. This is the reason that we chose for small mutations with a high probability, and large mutations with a low probability. 

Optimising hyper parameters of a genetic algorithm, such as mutation rate, is not trivial. An extensive study costs a lot of resources, so we limited ourselves to educated guesses and a few test runs. Nonetheless, the choices we have made seem to work well in practice, when the parameter space is well behaved. In the end, the goal is to map for each parameter the envelope of the fitness distribution (i.e., as a function of each free parameter, find the lowest \chisq), but without computing every model in the parameter space. In the grid-fitting approach each model in the parameter space is computed, and one can be certain that the envelope of the fitness distribution is completely mapped. With a genetic algorithm, the mapping is only approximate; one might miss certain models in the parameter space. 
In \cref{fig:app:chi2compareGA} we compare fitness distributions of \pyGA to those of {\sc pyGA}, using runs with identical setup. The \pyGA distributions are smoother (have less gaps), 
meaning that that set of models provide a closer representation of the true fitness envelope.
We see that both algorithms find very similar best fit solutions, that are in both cases close to the ``true values'', that is, the ones that the original model have (mind that deviations from this could be caused by the simulated data: a \snr of 30 was imposed on the synthetic spectra). 
In practice, this thus means that the algorithms work the same and give similar outputs, but that with \pyGA one needs to compute less models, that is, needs a lower amount of computation time to trace the envelope of the fitness distribution. We suspect that the reason for this lies mostly in the mutation scheme, but also other changes in the algorithm could have contributed to the improved performance. 
We note that \pyGA has the option to use the mutation operators used in the algorithms of \citet[][demonstrated in \cref{fig:app:mutation_details1}]{1995ApJS..101..309C,2005A&A...441..711M}. 
We note furthermore that, regardless of the genetic algorithm or mutation operators used, the amount of generations must be increased with higher quality data. 

\section{Comparison of methods\label{app:comparisons}}

\begin{figure*}
    \centering
    \includegraphics[width=1.0\textwidth]{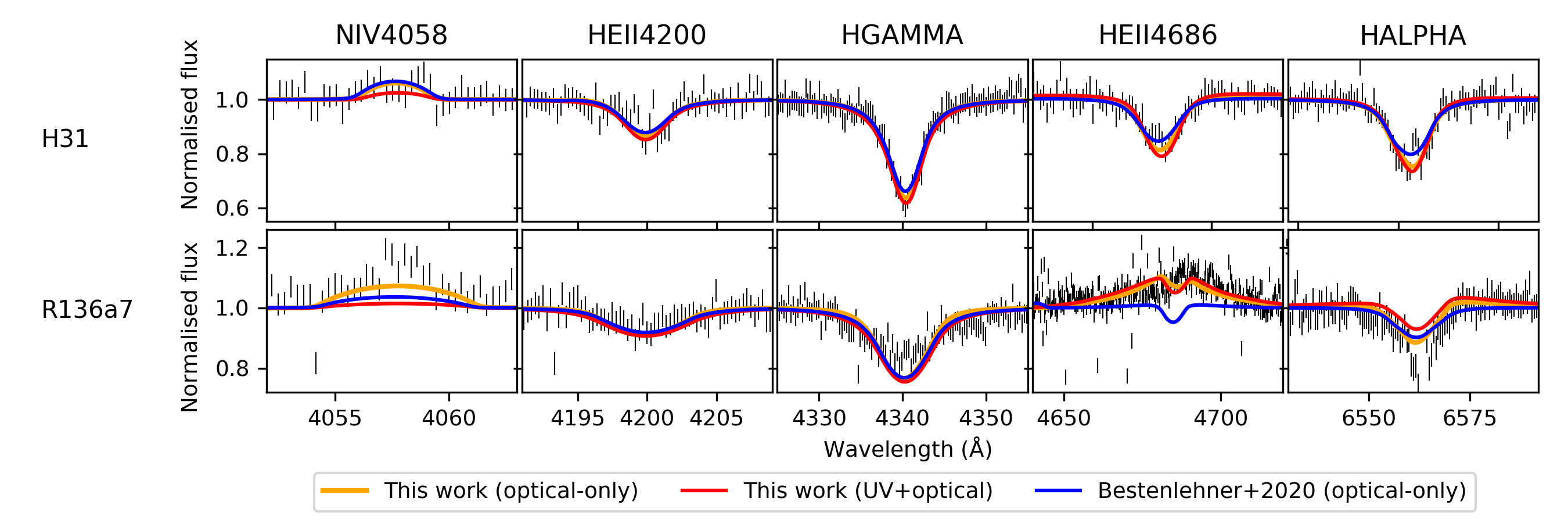}
    \caption{Comparison of optical lines of fits obtained with different methods. The top row shows lines of H31, for which the the parameters of \citet{Bestenlehner2020} and this work agree well: this level of agreement is representative for most stars in the sample. The bottom row shows lines of R136a7, where there is some discrepancy between the parameters of \citet{Bestenlehner2020} and our results. The difference is especially clear in \heiiline, and seems related to differences in \teff and \mdot. }  
    \label{fig:app:compareH31_R136a7}
\end{figure*}

In this section we compare the results of different methods. We stress that the error bars have to be taken into account when comparing the different methods; with a few exceptions, the agreement is good. To illustrate typical differences in the best fit models of the different methods, we show in \Cref{fig:app:compareH31_R136a7} a selection of optical lines for two stars: one star where there is some discrepancy between parameters, another star for which the agreement between parameters is good; the latter being representative for the sample. 

We derive the parameters \teff, \logg, \mdot, \logll, and in some cases also $\beta$ and nitrogen abundance, in two different manners: the optical-only and optical~+~UV runs. Additionally, we obtain a third measure for \teff by fitting \cmfgen models to the iron pseudo-continuum in the UV. Lastly, \citet{Bestenlehner2020} have already carried out an optical-only analysis of the same spectra, but using a different method. For the O-stars, \citet{Bestenlehner2020} use the IACOB-GBAT tool \citet{2011JPhCS.328a2021S} instead of \fw and \pyGA, while for the WNh stars, they use \cmfgen. For a few stars, values where obtained both with \fw and \cmfgen, in which case we adopt, in this section, the result of their \fw analysis. Furthermore, in our analysis nitrogen abundance was a free parameter that could range from 6.9 to 10.0, while \citet{Bestenlehner2020} use only three grid values, namely 7.1, 8.2 and 8.5. In this section we compare the outcomes of the different analyses.

\subsection{Temperatures\label{app:teff_discussion}}

\begin{figure}
    \centering
    \includegraphics[width=0.43\textwidth]{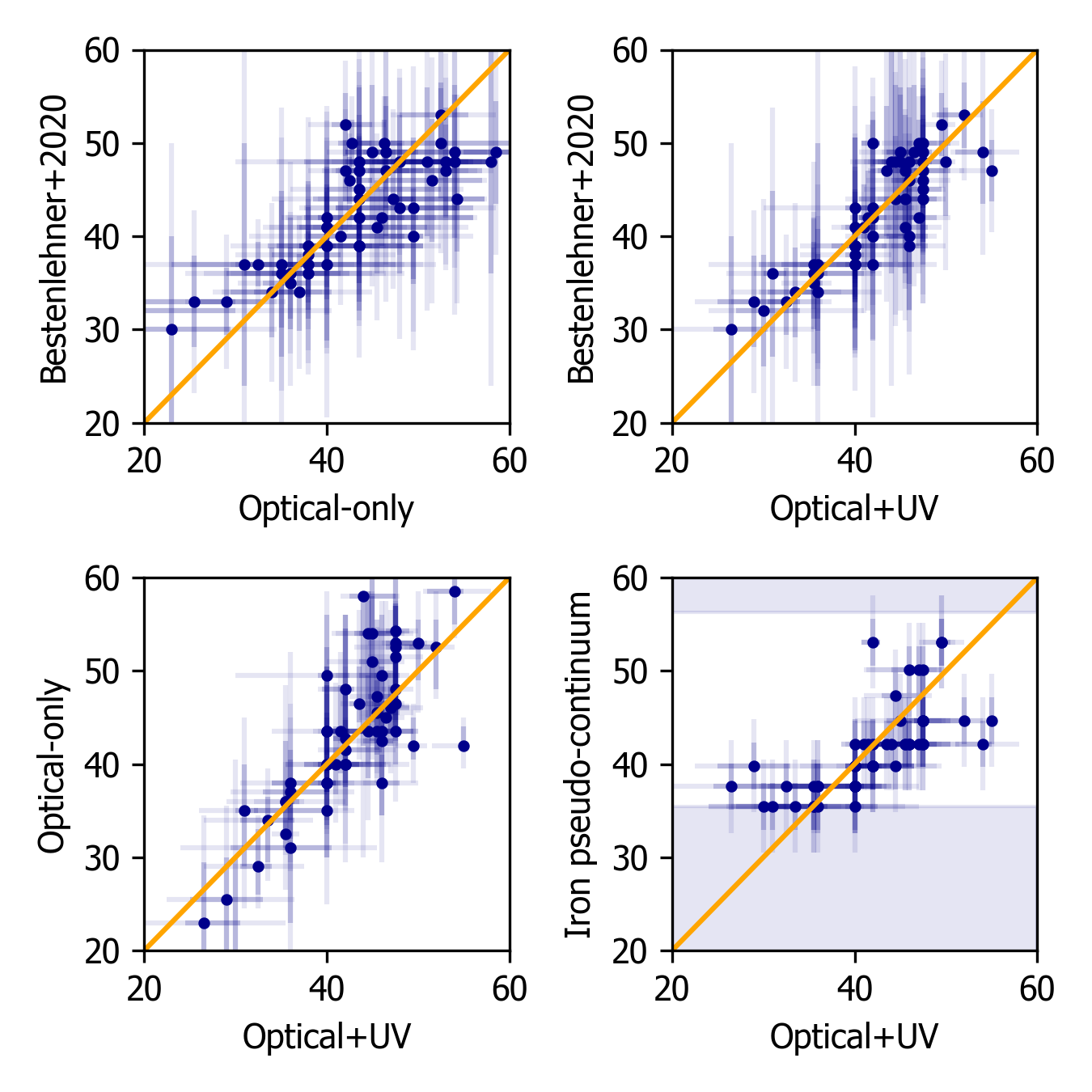}
    \caption{Comparison of different temperature analyses in units of 1000~K. Top: the optical-only analysis of \citet{Bestenlehner2020} against our optical-only (left) and our optical~+~UV analysis (right). Bottom: our optical-only (left) and our iron pseudo-continuum analysis (right) against our optical~+~UV analysis. The shaded are in the bottom right plot indicates the region that is not covered by the \cmfgen grid that we used \citep{2014A&A...570A..38B}. 
    }
    \label{fig:app:teff:compare}
\end{figure}

\begin{figure}
    \centering
    \includegraphics[width=0.48\textwidth]{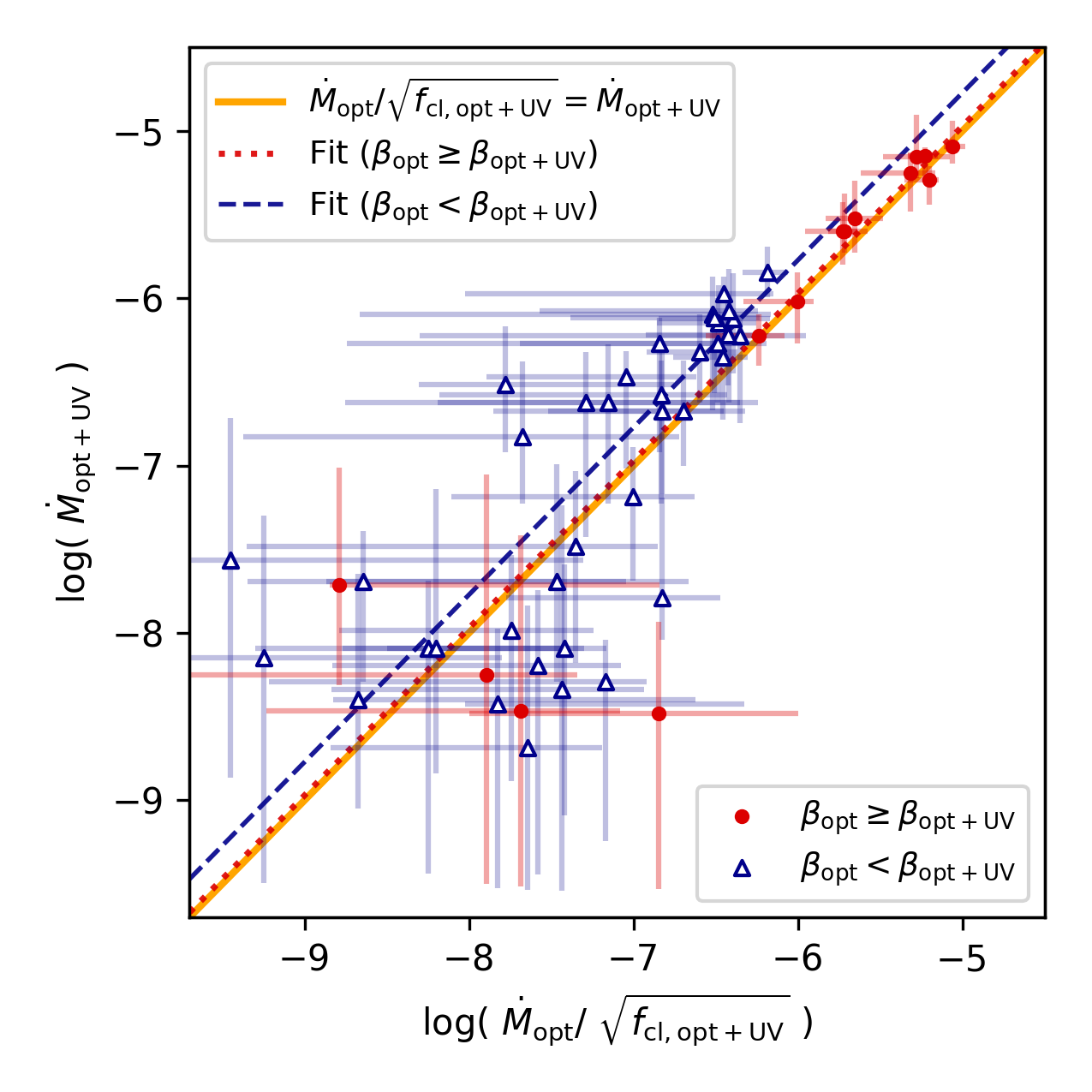}
    \caption{Comparison of the mass-loss rates derived from the optical~+~UV data, versus those of the optical-only data (fitted assuming a smooth wind) corrected for clumping using the individual \fcl values obtained from our optical~+~UV fits. The difference between the two seems to be related to different values of $\beta$ (see text). 
      \label{fig:app:fclumping_correct}  }
    
\end{figure}

\begin{figure}
    \centering
    \includegraphics[width=0.43\textwidth]{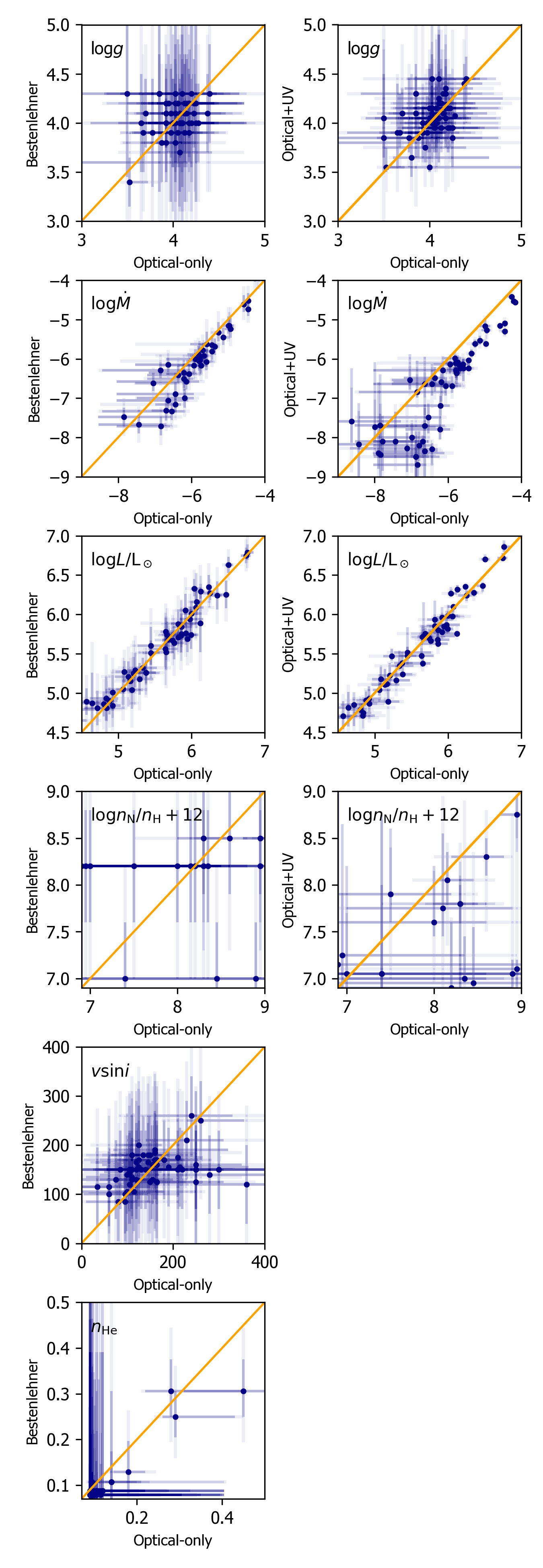}
    \caption{Comparison of different analyses. Left column: optical-only analysis with \pyGA (this work) against the optical-only analysis with IACOB-GBAT \citep{Bestenlehner2020}, right column: optical-only analysis with \pyGA vs. optical~+~UV analysis with \pyGA (both this work).}
    \label{fig:app:optical:compare}
\end{figure}

\Cref{fig:app:teff:compare} shows a comparison of the three \teff measurements from this paper to each other and to those of the optical-only analysis of \citet{Bestenlehner2020}. 
For the hotter stars (\teff~$\gtrsim 45000$~K) we find systematically higher temperatures from our optical-only analysis than does \citet[][top left panel]{Bestenlehner2020}, though within errors also there the agreement between the methods is good. Inspecting the top right and lower left panels, we see that these higher temperatures we find from our optical-only analysis are not found from the optical~+~UV fit. Assuming that the optical~+~UV analysis is the most reliable method because it takes into account the most spectral information, the temperatures of \citet{Bestenlehner2020} seem to be more reliable than our optical-only values, although the difference is small. It may be the case that the nitrogen lines do not affect strongly the goodness of fit that \pyGA uses for fitting, while this should be the case. Namely, for measuring the temperature for stars hotter than $\sim 45000$~K one relies mostly on (weak) nitrogen lines, as He-{\sc i}/He-{\sc ii} ratio can no longer be used since all He-{\sc i} is gone at those temperatures. Upon inspecting our fits and especially the fits to the nitrogen lines, we do not see cases where the nitrogen lines have an especially bad fit. The reason for the slight, albeit significant overestimation of our optical-only temperatures is thus not understood. 

The bottom right panel of \cref{fig:app:teff:compare} contains a comparison of our optical~+~UV temperatures and the temperatures we get from fitting the iron lines with the \cmfgen grid. Clearly, the iron (Fe) lines show a strong temperature dependence that is consistent with the temperatures of the H-He-C-N-O lines. At the lower end the Fe lines indicate a systematically higher temperature.  This can be explained by the fact that the lowest temperature of the \cmfgen grid is 35481~K. At the higher temperature end the points are scattered. This could be due to the fact that in the wavelength range from $1600-1700$~\AA, where most of the Fe~{\sc iv} lines lie, the \snr of our spectra is poor. Lastly, we note that both Fe lines as well as H-He-C-N-O lines are sensitive to micro-turbulence, but that a fixed value for micro-turbulence was assumed in both the \cmfgen grid and in the optical~+~UV runs (10 km~s$^{-1}$ in both cases). 

\subsection{Mass-loss rates \& $\beta$}

We see a systematic offset offset between the mass-loss rates obtained by our optical-only analysis, and that of  \citet{Bestenlehner2020}. This can be explained by the different assumptions and or fit-values for the wind acceleration parameter $\beta$. For most sources, in the optical-only analysis we assume $\beta = 0.9$, while \citet{Bestenlehner2020} leaves this parameter free, and finds for most sources a value of 1.0 (16 sources) or 1.2 (28 sources)\footnote{Values obtained through personal communication with the author.}. When basing the mass-loss only on optical lines without strong emission features, the parameter $\beta$ is degenerate with mass-loss rate, where a lower mass-loss rate or a lower $\beta$ are having the same effect on the wind line. This explains the observed discrepancy in measured mass-loss rates: we find or assume lower values for $\beta$, resulting in higher values for mass-loss rate, compared to \citet{Bestenlehner2020}. For deducing additional uncertainties on \mdot related to lack of knowledge of the value of $\beta$, one should vary $\beta$ and assess for each value the corresponding \mdot. \citet{2004A&A...413..693M} and \citet{2004A&A...415..349R} carry out such an analysis and find typical uncertainties of $0.1-0.3$ dex. 

The discrepancy between the optical-only and optical~+~UV measurements of the mass-loss rate is related predominantly to clumping. We investigate whether the observed differences is in line with expectations by using the individual \fcl values obtained from our optical~+~UV fits to do a clumping correction for the mass-loss rates from our optical-only fits, where we assumed a smooth wind, that is, \fcl~$=1$. In other words, we check whether
\begin{equation}\label{eq:clumpcheck}
    \dot{M}_{\rm opt+UV} = \dot{M}_{\rm opt}/\sqrt{f_{\rm cl, opt+UV}}
\end{equation}
is satisfied\footnote{From this analysis we exclude the three WNh stars, for which we assumed $f_{\rm cl} = 10$ in the optical-only runs.}. In theory, the density of the $\rho^2$-sensitive optical recombination lines from which the optical-only mass-loss rates are derived should scale with $\sqrt{f_{\rm cl}}$ and the above should hold, assuming the adopted clumping factor is spatially constant.  Upon applying the clumping correction to $\dot{M}_{\rm opt}$ we find that on average the shifted values are too low compared to $\dot{M}_{\rm opt+UV}$. Closer inspection reveals that this discrepancy is related to  different values of $\beta$ that are assumed or derived for the optical-only and the optical~+~UV runs. Again, the degeneracy of $\beta$ and \mdot plays a role: For sources where $\beta_{\rm opt} < \beta_{\rm opt+UV}$ the correction with $\sqrt{f_{\rm cl, opt+UV}}$ is too large, while for sources where $\beta_{\rm opt} \geq \beta_{\rm opt+UV}$ \cref{eq:clumpcheck} is satisfied (see \cref{fig:app:fclumping_correct}). The latter group includes the sources where \halpha is in emission and $\beta$ could be measured from the optical-only run. Lastly, we note that after taking into account the different values of $\beta$ there remains a group where the correction with $\sqrt{f_{\rm cl, opt+UV}}$ is not large enough. An explanation for this could be that the mass-loss rates found from the optical-only analysis are higher than the true mass-loss rates. For these sources we probably derive an upper limit rather than the actual mass-loss rate from the optical-only data. 

\subsection{Other parameters}

\Cref{fig:app:optical:compare} shows a comparison of the optical-only analysis of this paper to the optical-UV analysis, and the analysis of \citet{Bestenlehner2020}. The most striking differences can be seen in the nitrogen abundance. 
The match of the \citet{Bestenlehner2020} analysis with our optical-only analysis is not particularly good, but note that \citet{Bestenlehner2020} stress that the nitrogen abundance value that they provide is an indication rather than a precise measurement\footnote{Note that they measure only the nitrogen abundance for stars with $T\gtrsim 45000$~K, so we include only these points in the comparison plot.} and that the error bars on our optical-only analysis are very large. Furthermore, upon closer inspection of individual sources we see that agreement is good for all six sources that have significant overabundance (R136a1, R136a2, R136a3, R136a4, R136a5, R136b, H36, H48). This is also the case for these sources if one compares the optical-only versus optical~+~UV analysis, though there is a slight systematic towards higher abundances from the optical-only analysis. 

For \logg, \logll, \vsini, \yhe there is generally good agreement between the different runs. The exception are a few \vsini values of our analysis that seem to be very high, however this concerns a few very low \snr stars for which \vsini is essentially unconstrained. Another point that shows a clear mismatch is the one with the highest He abundance in the bottom plot: this is a3, where we from the optical only analysis found a temperature of 42750~K where with the UV data we assessed that 50000~K is likely closer to the true value, given the N{\sc v} and O{\sc v} lines in the UV. With a higher temperature, we find a lower He abundance. 

\bigskip 

The above shows that, even if the same code is used for the analysis, the outcome (at least the best fit parameters) can strongly depend on the analysis method, at least when the data quality is not superb.

\section{Stellar masses and ages \label{sec:app:bonnsai}}

\begin{figure}
    \centering
    \includegraphics[width=0.46\textwidth]{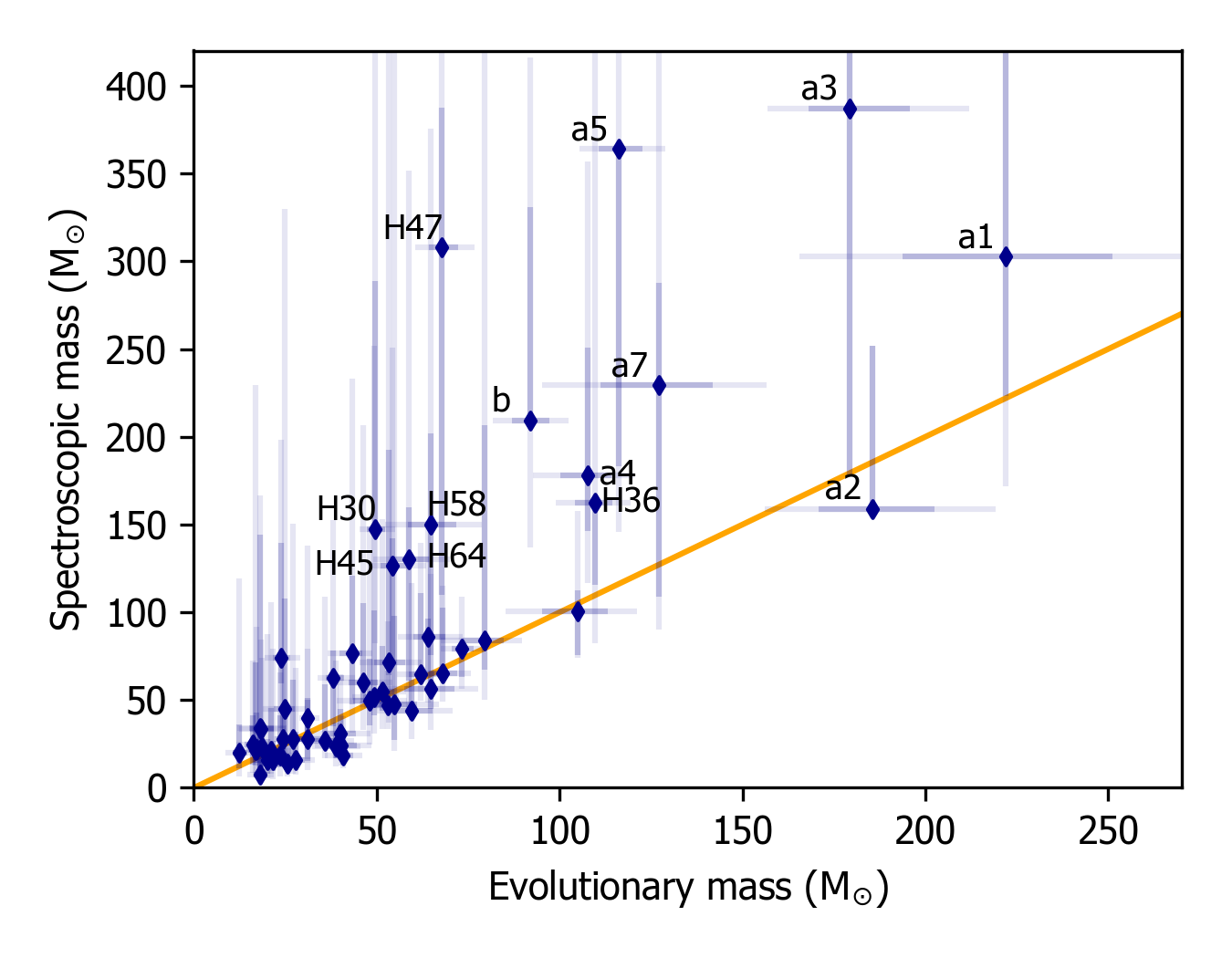}
    \caption{For more massive stars spectroscopic masses are systematically larger than evolutionary masses.}
    \label{fig:app:spec_vs_evol}
\end{figure}

\begin{figure}
    \centering
    \includegraphics[width=0.46\textwidth]{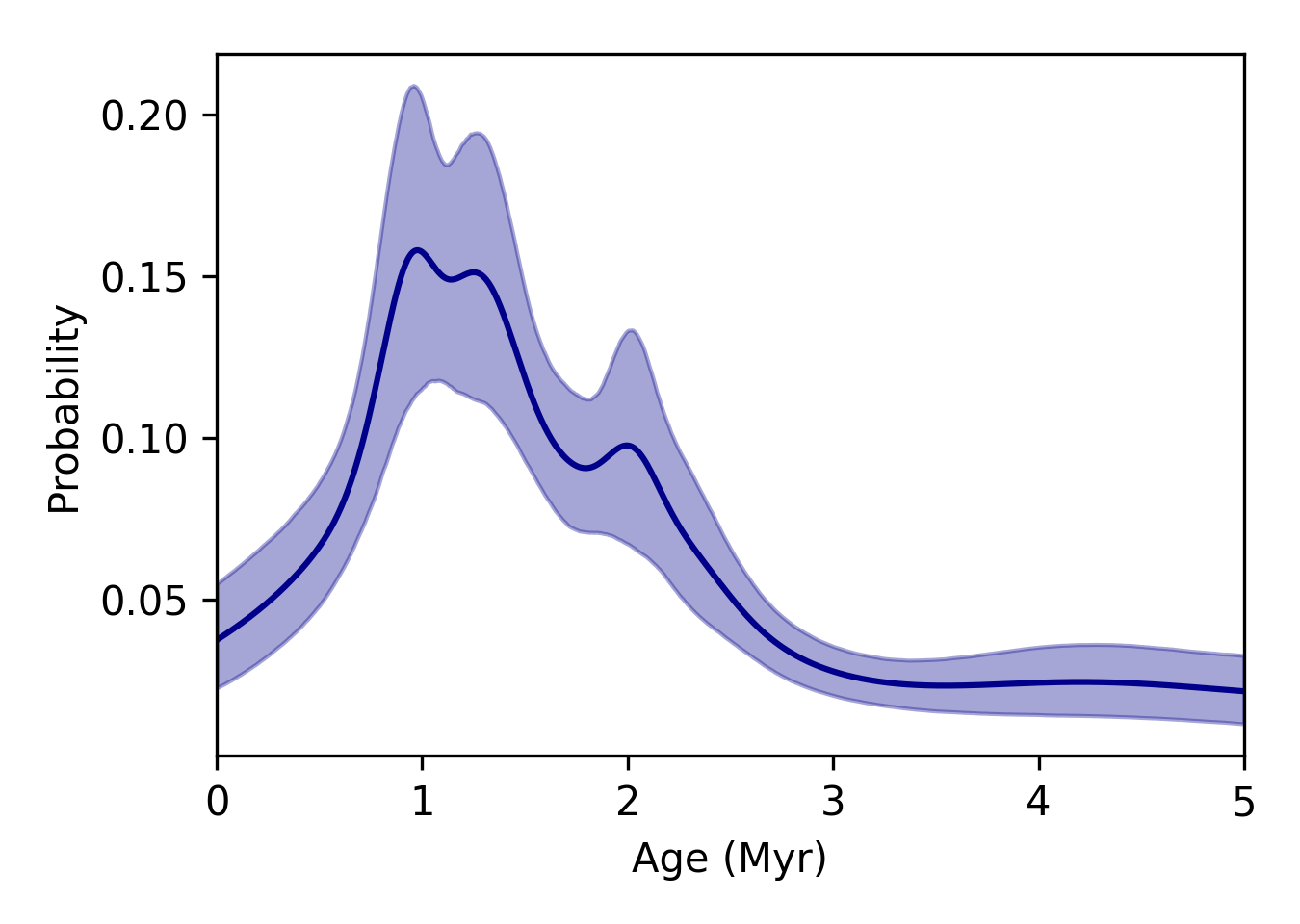}
    \caption{Age distribution of stars in the core of R136, as found using {\sc Bonnsai}. The dark blue solid line and the shaded area around it are the observed distribution and bootstrapped $2\sigma$ uncertainties. }
    \label{fig:app:age_dist}
\end{figure}

\begin{figure}
    \centering
    \includegraphics[width=0.46\textwidth]{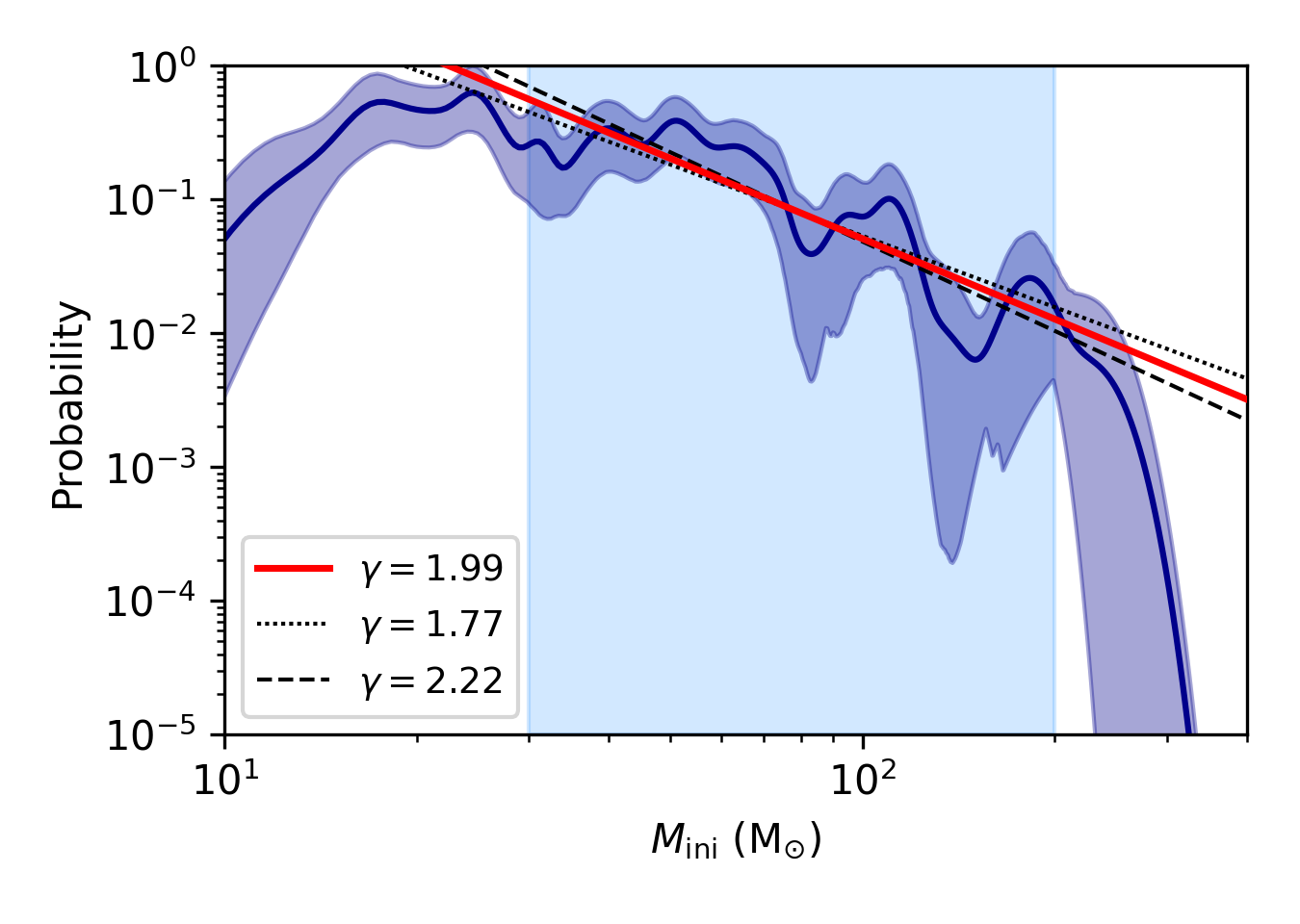}
    \caption{Distribution of initial masses of stars in the core of R136, as found using {\sc Bonnsai}. The dark blue solid line and the shaded area around it are the observed distribution and bootstrapped $2\sigma$ uncertainties. Red solid line is the best power law fit over the region 30-200${\rm M}_\odot$ (light blue background), black dashed and dotted line represent $2\sigma$ uncertainty on the slope.}
    \label{fig:app:mass_dist}
\end{figure}

\begin{figure}
    \centering
    \includegraphics[width=0.46\textwidth]{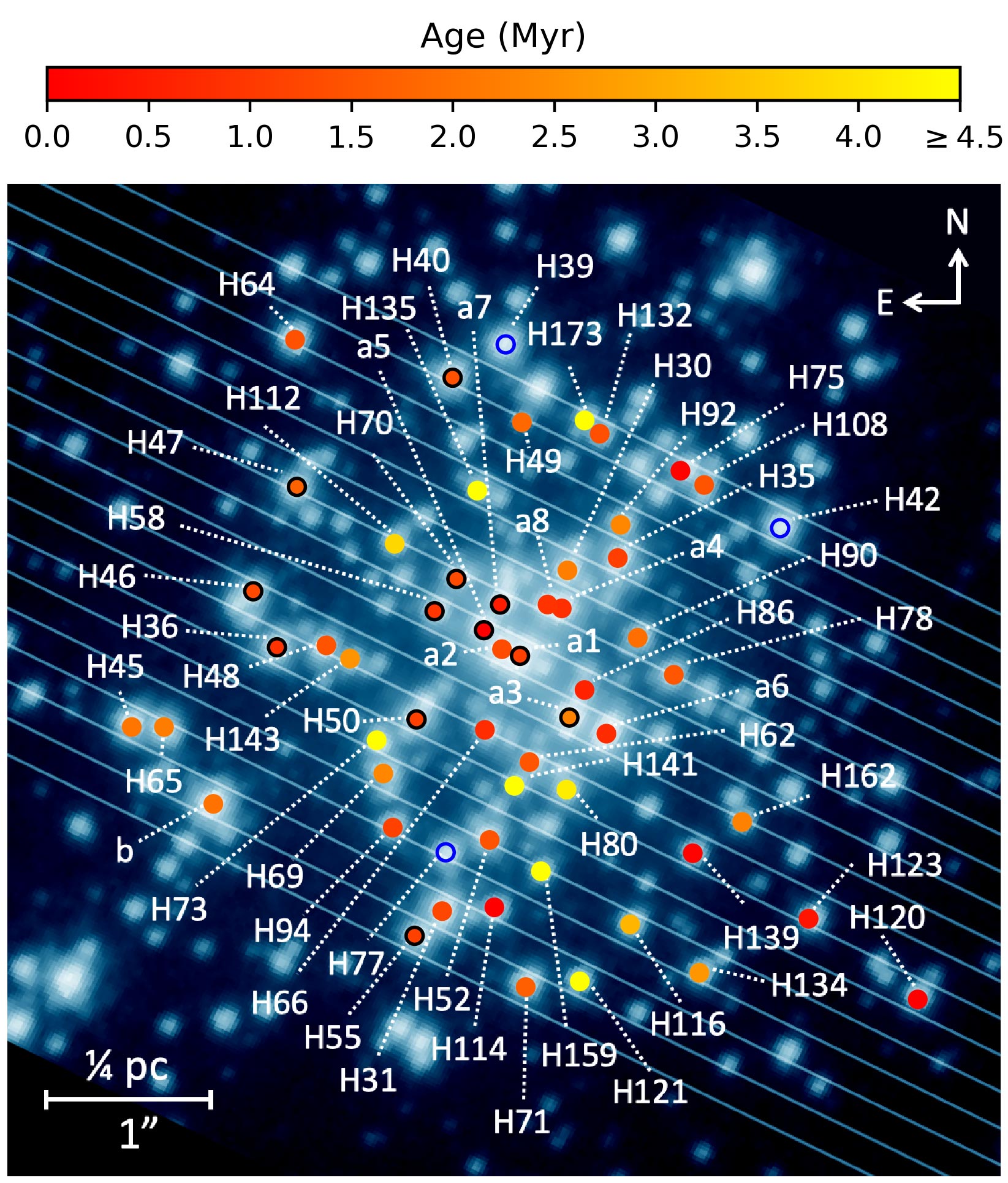}
    \caption{As \cref{fig:positions}, but colour coding after stellar age instead of mass.}
    \label{fig:app:positions:age}
\end{figure}

The power law slope of the initial mass function that we derive, $\gamma = 1.99\pm0.11$ in the range $30-200~{\rm M}_{\odot}$, is consistent with the value of 1.90$\pm^{0.37}_{0.26}$ derived for massive stars in the wider 30 Doradus region \citep{2018Sci...359...69S}. The sample of \citet{2018Sci...359...69S} excludes the core of R136 and the most massive star in their sample is 203$\pm^{40}_{44}$~\Msun. The highest initial mass that we derive, 273$\pm^{25}_{36}$~${\rm M}_{\odot}$ for R136a1, is consistent with the stochastic sampling analysis of \citet{2018Sci...359...69S}, that excludes maximum initial masses of over 500~\Msun in 30 Doradus with 90\% certainty. We note that for our {\sc Bonnsai} runs we used the \citet{1955ApJ...121..161S} initial mass-function, that, with a slope of 2.35, prefers lower masses. This can impact the slope that we derive, albeit in a conservative way, that is, without this prior we would have found an even shallower slope. Regarding the ages; while we derive a cluster age of $1-2.5$~Myr, there is a tail of older stars, suggesting that the stars in R136 are not coeval; this could be either the result of a projected view onto this dense central region in 30 Doradus, or is related to the formation of R136 \citep[see also][]{2016MNRAS.458..624C,2018A&A...618A..73S,2020MNRAS.493.3938B}. 

A comparison of the evolutionary and spectroscopic masses (see also table \cref{tab:app:more_uv_restuls}) can be found in \cref{fig:app:spec_vs_evol}. The masses agree well within errors, but for $M \gtrsim 70 {\rm M}_\odot$ the spectroscopic masses are, with a few exceptions, systematically higher than the evolutionary masses. \citet{Bestenlehner2020} find the same (their Figure 3), although in some cases (e.g., R136a5, R136b) the current analysis gives larger values. Furthermore, note that in addition we derive spectroscopic masses for the WNh stars R136a1-a3. For an in depth discussion of the mass discrepancy, see \citet{Bestenlehner2020}. 

The age distribution we derive for R136 based on the {\sc Bonnsai} values is shown in \cref{fig:app:age_dist}. We find a strong peak at 1.1-1.3~Myr, consistent with the results of \citet{Bestenlehner2020}. A second peak peak may be suggestive of a second burst of star formation around 2.0~Myr, though given the uncertainties the evidence for this is not strong. 

\Cref{fig:app:mass_dist} shows the initial mass distribution we derive from the {\sc Bonnsai} values. We fit a power law in the form $\xi(M) \propto M^{-\gamma}$ to the initial mass distribution in the mass range 30-200${\rm M}_{\odot}$ and find a slope of $\gamma = 1.99\pm0.11$. This is consistent with the results of \citet{Bestenlehner2020} of the R136 core ($\gamma = 2.0 \pm 0.3$), and the findings of \citet[][$\gamma = 1.90\pm^{0.37}_{0.26}$]{2018Sci...359...69S} for the rest of the Tarantula Nebula,
but steeper than the standard Salpeter initial mass function.

Lastly, we compare the age of our sources with the spatial distribution (\cref{fig:app:positions:age}). While many young sources are found close to the centre, several of them are also found in the outskirts. Moreover, the old sources are scattered throughout the cluster. The fact that we do not observe them in the very centre might be an observational bias: the very bright WNh stars could dominate the older, lower mass stars in that region. We thus conclude that we do not see a strong correlation between the spatial position of the stars and their age. This is in line with what \citet{Bestenlehner2020} conclude (their Figure 11).

\section{Notes on individual sources\label{app:notespersource}}

\subsection{R136a1\label{sec:app:a1_analysis}}
Possibly contaminated by H17, which lies north of R136a1 with an angular separation of 75 mas and a flux ratio of 0.112 in the K-band \citep{2017A&A...602A..56K}. Due to the low flux ratio and the fact that H17 lies mostly outside the slit we assume the spectral analysis of R136a1 is not affected by H17.

For this star \ovline is poorly reproduced -- something that is seen also for the other WNh stars of the sample (R136a2, R136a3). In particular, the line is not broad enough compared to the data, implying that there is not enough O~{\sc v} in the outer wind. We explored whether this could be related to our (canonical) X-ray assumptions by probing the X-ray parameter space, and ruled this out as a cause. Another possibility is that something might go wrong with the \fw treatment particularly for very dense winds, either due to indirect line overlap effects in the outer wind, or something else. Lastly, it could be that the wind structure that is assumed (i.e., the clumping stratification) is not representative for what actually happens in the wind, and it could be that this discrepancy especially shows, or, is largest for, the WNh stars. The exact cause of the issue is currently unidentified. We note that \ovline is sensitive to clumping, however we do not believe the poor fit of \ovline results in spurious clumping factors for the WNh stars. Namely, for this particular line a higher clumping factor leads to a weaker profile which is the opposite behaviour of some other of the clumping factor sensitive lines (e.g. \CIVline). In the WNh stars we find rather high clumping factors, where a better fit of \ovline (deeper profile) would be obtained with a lower clumping factor. It thus seems that other clumping sensitive lines dominate the value that we obtain. 

\subsection{R136a2}

\ovline was poorly reproduced; see explanation in \cref{sec:app:a1_analysis}. 

\subsection{R136a3 \label{sec:app:a3_analysis}}

\begin{figure*}
    \centering
    \includegraphics[width=0.98\textwidth]{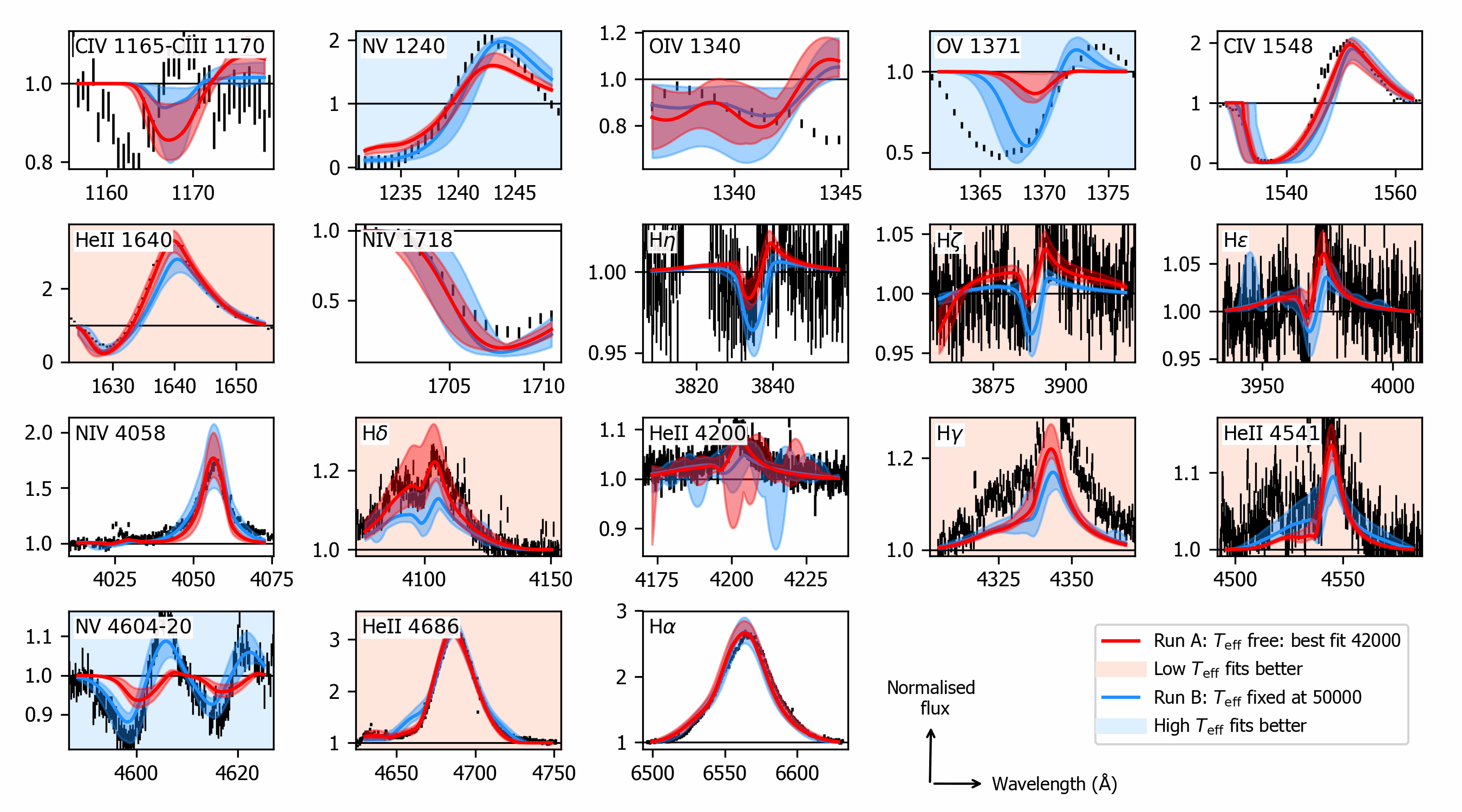}
    \caption{Two \pyGA runs for R136a3. In blue the original run, having a setup identical to that of all other stars (resulting in \teff$=42$~kK). In red the run where we fixed \teff to 50~kK, all other aspects of the setup held the same. For some lines the higher temperature results in a better fit than the original run (light red background), while for other lines the original run resulted in better fits (light blue background). }
    \label{fig:a3_hot_cold}
\end{figure*}

\begin{figure}
    \centering
    \includegraphics[width=0.320\textwidth]{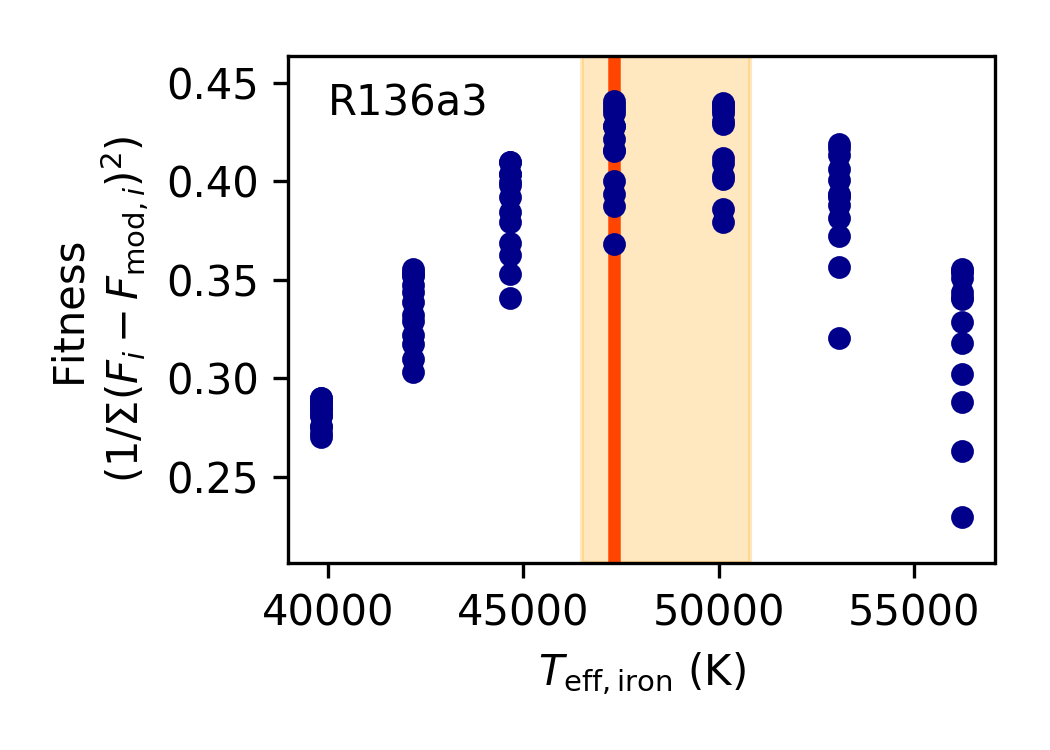}
    \caption{Fitness distribution of the iron pseudo-continuum fit of R136a3. The best fitting temperature (47315~K) is indicated. A temperature of 50000~K results in about the same fitness. The fitness is comparable to the inverse of a \chisq: it is the inverse of the sum of the square of the residuals. The residuals are the difference between the normalised observed flux $F_{i}$ and model flux $F_{{\rm mod},i}$, for all flux points of the iron pseudo-continuum.}
    \label{fig:app:a3_iron}
\end{figure}

The best fit of the optical~+~UV run of R136a3 results in a \teff of 42~kK, as opposed to 50~kK found by \citet{Bestenlehner2020} and 53~kK found by \citet{2010MNRAS.408..731C}. Upon inspecting the fit we see that while in general the fit is matching the data well, this does not hold for the higher ionisation lines \cuvivline, \ovline, and \nvopt, suggesting the derived \teff of 42~kK is too low. We carry out another fit where we fix \teff to 50~kK and note that the fits to \ovline and \nvopt improve considerably, but the fit to many other lines gets worse (\cref{fig:a3_hot_cold}). The mass-loss rate and helium abundance we find from the two runs differ significantly, we find $\log \ \dot{M} = -4.2\pm^{0.07}_{0.04}$ and $\log \ \dot{M} = -4.5\pm^{0.04}_{0.07}$, and $x_{\rm He} = 0.49\pm^{0.01}_{0.11}$ and $x_{\rm He} = 0.37\pm^{0.10}_{0.10}$ for the low and high temperature, respectively.  
Because there is no way to reproduce the strong \ovline and \nvopt lines with a lower temperature, and the fact that when assuming a low effective temperature R136a3 lies far away from the other two WNh stars in the HRD, we deem the higher temperature more likely. Further support for that comes from the fit to the iron lines in the UV presented in \cref{fig:app:a3_iron}, resulting in a temperature of around $47000-50000$~K. For these reasons, we adopt a temperature of \teff~$ = 50$kK for R136a3. The other stellar parameters we obtain from the optical~+~UV run with \teff fixed to that value (blue model in \cref{fig:a3_hot_cold}). For the \vsini we stick to the value of the optical-only run (best fit: $T_{\rm eff} = 42750$~K) as we do 
for the other stars. 

We note that \citet{2017A&A...602A..56K} resolve R136a3 into two sources, however the flux ratio is 0.124 in the J-band (0.044 in the K-band), so in practice one source will dominate our optical and UV spectroscopy. We do not see evidence for unresolved binarity of R136a3 where the source would consist of a hot and a cooler component, at least not directly: no \hei lines are visible. 

Furthermore, we note that, while we adapt the higher \teff for R136a3 the rest of our analysis for reasons explained in this section, we have to keep in mind that the low values we obtain from our original fit may point to the limits of the applicability of this \Fw version, that was not designed for situations where, e.g., \hgamma is in strong emission. 

Lastly, we note that, also in the higher \teff fit, \ovline was poorly reproduced; see explanation in \cref{sec:app:a1_analysis}.

\subsection{R136a6}

Consists of two stars with an approximately equal flux contribution (see also \cref{sec:stis_data_r136}). We carry out the spectral analysis but exclude it from the analyses of the sample as a whole regarding mass-loss and clumping properties.

\subsection{R136a8}

No optical spectra available and thus not included in the analysis of \citet{Bestenlehner2020}. The optical~+~UV run of this source was thus in fact carried out on UV data only. The setup was the same as for the other sources, but instead of fixing the \vsini and helium abundance on values derived of the optical, we fixed them at 150 km~s$^{-1}$ and $x_{\rm He} = 0.10$, respectively; these are typical values given the optical-only fits of the other sample stars. 

\subsection{H36 \label{app:sec:h36}}

From the optical-only run we find \teff$=42000$~K for this star. This value is low compared to what we find from the optical~+~UV run, 49500~K, and inconsistent with the spectral type of the star (O2 If*). The optical-only fit is good (i.e., profiles of models covering the uncertainty ranges coincide with the scatter of the data points), with the exception of \nvopt, which is too weak in our models. Upon comparison of the optical lines of the the optical-only and optical~+~UV fits we see that the the higher temperature of the  optical~+~UV run does not improve the fit of most optical lines, not even for \nvopt, which has a similar best fit profile in both our runs. Most notable is the worsening of the Balmer line fits, which generally are too deep for the optical~+~UV fit, but match the data well for the optical-only fit. In this temperature regime and with this data quality, the \hei and \heii are of not much help for constraining the temperature, between our optical-only and optical~+~UV runs these lines are, within the noise level, unchanged. \citet{Bestenlehner2020} find, based on optical data only, a high temperature for this star (52000~K). Their \nvopt seems to have a better fit than we get, this seems to be at the cost of the fitness of \heiiline, and, as for our optical~+~UV run, of the Balmer lines, which for their best model are too deep compared to what is observed. However, the higher ionisation ions, both the \niv lines in the optical and UV as well as the strong \ovline, do strongly support a high temperature, which we therefore deem more likely for this source.

\subsection{H129}

The position of H129 in the HRD is on the left of main sequence. We do not use the parameters derived for this star for the further analysis, but exclude it from the analyses of the sample as a whole regarding mass-loss and clumping properties (see also \cref{res:teffs}). For the optical-only run we estimated \vinf based on \logll from \citet{Bestenlehner2020}, but the extrapolated velocity was only 100~km~s$^{-1}$ we assumed a velocity of 500~km~s$^{-1}$ instead. 

\section{X-ray \label{app:sec:X-rays}}

\subsection{X-rays implementation in \fw}

Wind-embedded shocks and associated X-ray emission are implemented into \fw by assuming a very small fraction of the stellar wind to be very hot and emit X-rays\footnote{This is the standard approach, that is employed in several model atmosphere codes for hot stars (for references, see \citealt{2016A&A...590A..88C}). In later versions of \fw, an alternative implementation, accounting for the different effective emissivities from radiative and adiabatic shock cooling zones, is also implemented (see \citealt{2020A&A...642A.172P}).} \citep{2016A&A...590A..88C}. 
The shocks are then described by five parameters: the volume filling fraction of the X-ray emission $f_{\rm X}$, the maximum jump velocity of the shocks $u_{\infty}$ and the exponent $\gamma_X$ that relates the outflow velocity to the jump velocity ($u_{\infty}$ and $\gamma_X$ together describing the shock temperature, typically being of the order $10^6$~K), and two parameters related to the onset radius of the X-ray emission\footnote{See Eq. (8) of  \citet{2016A&A...590A..88C}.}: $R_{\rm min}^{\rm input}$ and a factor $m_{\rm X}$. The total X-ray luminosity is thus not a free parameter, but follows as output from the model. For further details on the implementation of wind-embedded shocks in \fw see \citet{2016A&A...590A..88C}.

\subsection{Assumptions regarding X-rays \label{sec:app:xrayassumptions}}

For both optical-only and optical~+~UV fits, we include wind-embedded shocks and resulting X-rays. The corresponding parameters are not fitted, but instead fixed at certain values for each star. 
For all stars, we assume $\gamma_X = 0.75$, $m_{\rm X} = 30$, and $R_{\rm min}^{\rm input} = 1.45$.  The value of $\gamma_X$ sets the gradient of the shock strength relative to the wind velocity \citep{1994A&A...283..525P}; a value of $\gamma_X < 1$ means that the relative increase in shock jump velocity as a function of radius is higher than that of the wind velocity. 
Our assumed value $\gamma_X = 0.75$ lies in between the higher value assumed by \citet{2009MNRAS.394.2065K,2016A&A...590A..88C} and the lower values adopted in models of \citet{1994A&A...283..525P}. 
For $m_{\rm X}$ we use the best fit value from \citet{2001A&A...375..161P}, and in our choice of $R_{\rm min}^{\rm input}$ we follow,  for example, \citet{1994A&A...283..525P}. 

The X-ray parameters having the most profound influence on the ionisation fractions are the maximum jump velocity of the shocks $u_{\infty}$ and the volume filling fraction of the X-ray emission 
$f_{\rm X}$ \citep{2016A&A...590A..88C}, because they directly relate to the temperature distribution and total X-ray flux. We tailor these parameters per star; for the first, we assume $u_{\infty} = 0.3 \varv_{\infty}$ (after \citealt{2016A&A...590A..88C}, who follow \citealt{2009MNRAS.394.2065K}). Here we take \vinf values from \citet{2016MNRAS.458..624C}, or use the estimated values as discussed in \cref{sec:setupoptical}. For two stars with \vinf$\approx 3500$ (H36 and H46) we assume $u_{\infty} = 0.2 \varv_{\infty}$, based on LDI simulations of winds of luminous stars with high terminal velocities (F.~Driessen, priv.~comm.). The typical maximum jump velocities we assume translate to maximum shock temperatures ranging from $0.6 - 28.0 \cdot 10^6$~K. 

Then, given these assumptions, we choose $f_{\rm X}$ such that the total X-ray luminosity in the ROSAT band (0.1-2.5~keV) equals $L_{\rm X}/L=10^{-7}$, the canonical ratio for O-stars \citep[e.g.,][for Galactic O-stars; \citealt{Crowtherinprep} confirm that this is a reasonable assumption for our sample stars, finding an average of $\log L_{\rm X}/L=-6.6\pm0.3$ for nine X-ray sources associated with R136a.]{1989ApJ...341..427C,1997A&A...322..167B,2006MNRAS.372..661S,2011ApJS..194....7N,2015ApJS..221....1R}. However, $L_{\rm X}$ is not a direct input parameter of \fw. Instead, we estimate $f_{\rm X}$ as to give $L_{\rm X}$ close to $L_{\rm X}/L=10^{-7}$. This is done for each model individually during the \pyGA run. Based on \mdot and \vinf of the model, $f_{\rm X}$ is computed to satisfy:

\begin{equation}
\log (f_{\rm X}) = -5.45 - 1.05 \log(\dot{M}_6 / v_{\infty}) \label{eq:xrays}, 
\end{equation}
where $\dot{M}_6 = \dot{M} / 10^{-6}$, with $\dot{M}$ in ${\rm M}_\odot$~yr$^{-1}$. This equation is derived from the observational values of \citet[][their Figure~6]{1996rftu.proc....9K}. Note that we extrapolate the relation that they find to weaker winds (lower $\dot{M}_6 / v_{\infty}$). The input and output X-ray parameters for each source can be found in \Cref{tab:app:xrayin_out}. 

\subsection{Validity of the X-ray assumptions}

\begin{figure}
    \centering
    \includegraphics[width=0.45\textwidth]{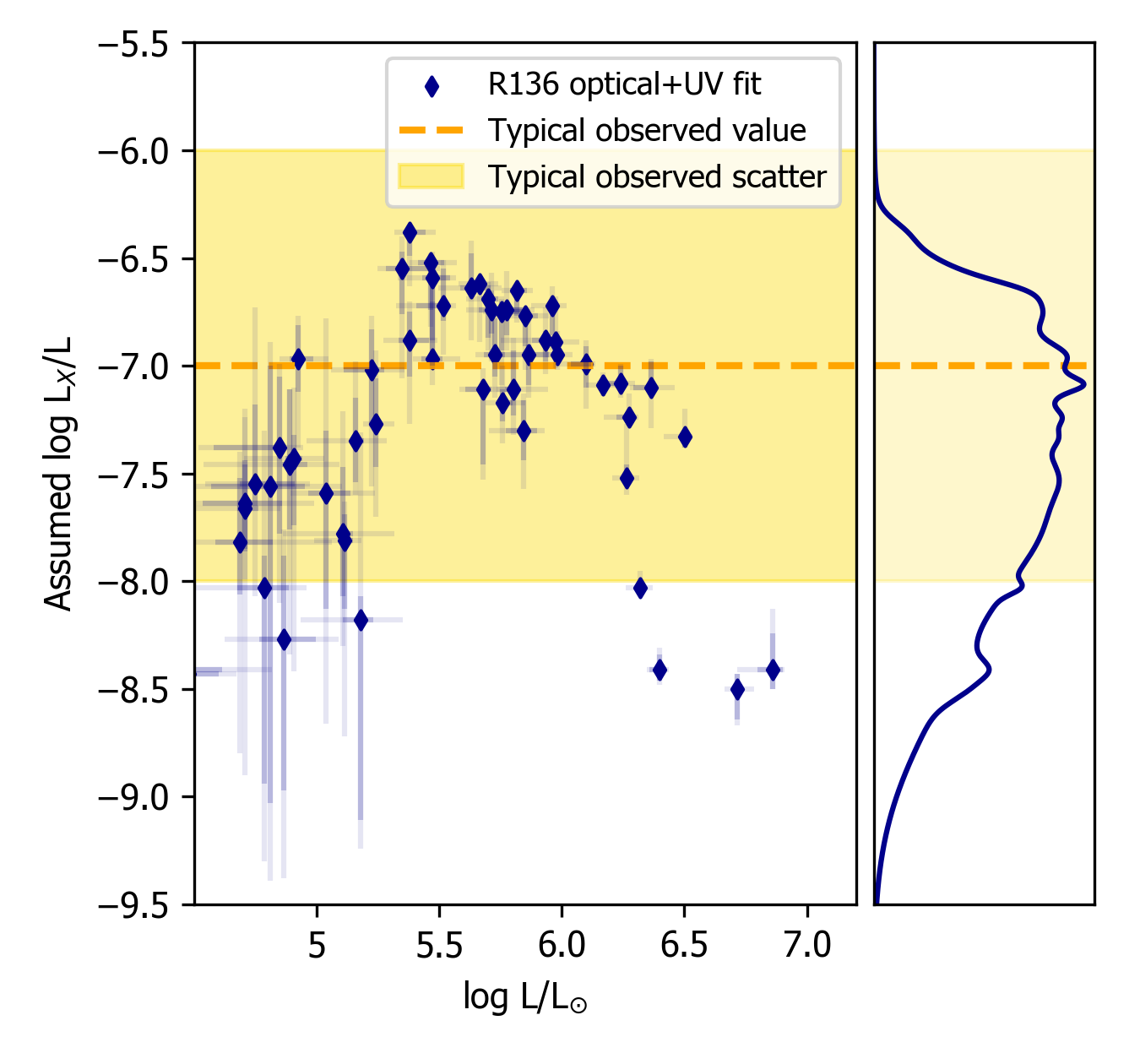}
    \caption{\emph{Left:} $L_{\rm X} / {L} $ against ${L}$ of the our best fitting models (dark blue diamonds with $2\sigma$ error bars). \emph{Right:} Distribution of the $L_{\rm X} / {L} $ values from our best fitting models. We indicate the typical observed value of $L_{\rm X} / {L} $ (orange dashed line) and the typical scatter around this value (yellow band). We stress that the X-ray luminosity $L_{\rm X} / {L} $ is not a free parameter in our fitting and the  $L_{\rm X} / {L} $ values shown here do \emph{not} represent the actual X-ray flux of the given stars: this plot serves solely as a check on our assumptions (see text). } 
    \label{fig:app:xrayoutput}
\end{figure}

We check the validity of the assumptions for our X-ray setup by comparing the output $\log L_{\rm X} / {L}  $ of each optical~+~UV run to the canonical value, $\log L_{\rm X} / {L}  = -7$. 
\Cref{fig:app:xrayoutput} shows that indeed, generally the output values are close to canonical. In the six cases where it is not, we underestimated the assumed X-ray flux. Exact values of X-ray input and output can be found in \Cref{tab:app:xrayin_out}. 

\begin{table}[p]
    \centering
    \caption{X-ray input ($u_\infty$) and output ($\log f_{\rm X}$ and $\log L_{\rm X} / {L} $) values for the optical~+~UV runs. Error bars are $1\sigma$. }
    \label{tab:app:xrayin_out}

    \begin{tabular}{p{2cm} l r@{$\pm$}l r@{$\pm$}l }
\hline \hline
Source & $u_\infty$ (km/s) & \multicolumn{2}{c}{$\log f_X$} & \multicolumn{2}{c}{$\log L_X / {L}$} \\ \hline 
R136a1   & 780    & -3.34  & $^{0.02}_{0.15}$ & -8.41  & $^{0.17}_{0.09}$ \\
R136a2   & 728    & -3.53  & $^{0.05}_{0.12}$ & -8.50  & $^{0.07}_{0.14}$ \\
R136a3   & 720    & -3.42  & $^{0.04}_{0.11}$ & -8.59  & $^{0.13}_{0.15}$ \\
R136a4   & 743    & -1.91  & $^{0.07}_{0.03}$ & -7.24  & $^{0.03}_{0.01}$ \\
R136a5   & 846    & -2.71  & $^{0.07}_{0.07}$ & -7.47  & $^{0.05}_{0.08}$ \\
R136a6   & 795    & -2.16  & $^{0.06}_{0.07}$ & -7.08  & $^{0.08}_{0.01}$ \\
R136a7   & 813    & -2.29  & $^{0.07}_{0.16}$ & -7.10  & $^{0.12}_{0.02}$ \\
R136a8   & 743    & -1.96  & $^{0.05}_{0.07}$ & -7.09  & $^{0.02}_{0.02}$ \\
R136b    & 846    & -2.91  & $^{0.05}_{0.00}$ & -8.03  & $^{0.03}_{0.02}$ \\
H30      & 747    & -1.72  & $^{0.27}_{0.03}$ & -7.17  & $^{0.06}_{0.09}$ \\
H31      & 844    & -1.61  & $^{0.01}_{0.00}$ & -6.95  & $^{0.00}_{0.01}$ \\
H35      & 831    & -1.79  & $^{0.19}_{0.02}$ & -6.65  & $^{0.01}_{0.12}$ \\
H36      & 700    & -2.41  & $^{0.08}_{0.02}$ & -7.52  & $^{0.06}_{0.03}$ \\
H40      & 825    & -1.61  & $^{0.15}_{0.11}$ & -6.88  & $^{0.04}_{0.10}$ \\
H45      & 786    & -1.47  & $^{0.31}_{0.00}$ & -7.11  & $^{0.18}_{0.12}$ \\
H46      & 688    & -2.58  & $^{0.03}_{0.14}$ & -6.99  & $^{0.08}_{0.11}$ \\
H47      & 914    & -2.50  & $^{0.08}_{0.07}$ & -6.89  & $^{0.04}_{0.04}$ \\
H48      & 914    & -2.16  & $^{0.09}_{0.03}$ & -6.72  & $^{0.00}_{0.14}$ \\
H49      & 894    & -1.49  & $^{0.31}_{0.10}$ & -6.75  & $^{0.05}_{0.18}$ \\
H50      & 786    & -1.66  & $^{0.23}_{0.00}$ & -6.77  & $^{0.00}_{0.14}$ \\
H52      & 846    & -1.55  & $^{0.17}_{0.00}$ & -6.69  & $^{0.02}_{0.18}$ \\
H55      & 864    & -1.46  & $^{0.18}_{0.06}$ & -6.74  & $^{0.01}_{0.12}$ \\
H58      & 894    & -1.22  & $^{0.20}_{0.11}$ & -6.95  & $^{0.05}_{0.14}$ \\
H62      & 831    & -1.74  & $^{0.12}_{0.09}$ & -6.64  & $^{0.16}_{0.04}$ \\
H64      & 531    & -1.20  & $^{0.23}_{0.22}$ & -7.30  & $^{0.14}_{0.14}$ \\
H65      & 762    & -1.74  & $^{0.17}_{0.10}$ & -6.95  & $^{0.17}_{0.10}$ \\
H66      & 777    & -1.58  & $^{0.02}_{0.07}$ & -6.62  & $^{0.02}_{0.00}$ \\
H68      & 573    & -1.17  & $^{0.28}_{0.10}$ & -7.11  & $^{0.04}_{0.35}$ \\
H69      & 774    & -0.90  & $^{0.10}_{0.37}$ & -6.97  & $^{0.27}_{0.03}$ \\
H70      & 801    & -1.50  & $^{0.15}_{0.10}$ & -6.74  & $^{0.09}_{0.11}$ \\
H71      & 743    & -1.23  & $^{0.47}_{0.01}$ & -6.59  & $^{0.05}_{0.29}$ \\
H75      & 765    & -1.32  & $^{0.33}_{0.04}$ & -6.52  & $^{0.00}_{0.22}$ \\
H78      & 713    & -1.11  & $^{0.16}_{0.29}$ & -6.72  & $^{0.17}_{0.07}$ \\
H80      & 497    &  0.51   & $^{0.35}_{0.12}$ & -7.81  & $^{0.12}_{0.32}$ \\
H86      & 743    & -1.44  & $^{0.18}_{0.00}$ & -6.38  & $^{0.00}_{0.11}$ \\
H90      & 743    & -1.15  & $^{0.34}_{0.16}$ & -6.55  & $^{0.08}_{0.21}$ \\
H92      & 624    & -0.06  & $^{0.25}_{0.29}$ & -7.27  & $^{0.21}_{0.20}$ \\
H94      & 747    & -0.61  & $^{0.21}_{0.21}$ & -6.88  & $^{0.13}_{0.17}$ \\
H143     & 444    & -0.11  & $^{0.22}_{0.38}$ & -7.35  & $^{0.20}_{0.19}$ \\
H73      & 603    & 0.96   & $^{0.99}_{0.09}$ & -8.18  & $^{0.11}_{0.93}$ \\
H108     & 312    & 0.04   & $^{0.35}_{0.43}$ & -7.46  & $^{0.35}_{0.30}$ \\
H112     & 565    & 0.34   & $^{0.38}_{0.37}$ & -7.78  & $^{0.31}_{0.29}$ \\
H114     & 531    & -0.39  & $^{0.40}_{0.27}$ & -7.02  & $^{0.19}_{0.25}$ \\
H116     & 288    & 0.83   & $^{0.74}_{0.50}$ & -8.27  & $^{0.39}_{0.70}$ \\
H120     & 313    & 0.01   & $^{0.49}_{0.29}$ & -7.38  & $^{0.33}_{0.40}$ \\
H121     & 344    & 0.94   & $^{0.88}_{0.31}$ & -8.03  & $^{0.15}_{0.91}$ \\
H123     & 485    & -0.20   & $^{0.25}_{0.16}$ & -6.97  & $^{0.16}_{0.15}$ \\
H129     & 150    & 0.02   & $^{0.88}_{0.28}$ & -8.43  & $^{0.21}_{0.92}$ \\
H132     & 464    & 0.29   & $^{0.70}_{0.35}$ & -7.59  & $^{0.29}_{0.54}$ \\
H134     & 351    & 0.22   & $^{0.25}_{0.34}$ & -7.55  & $^{0.37}_{0.10}$ \\
H135     & 363    & -0.30   & $^{0.58}_{0.11}$ & -7.64  & $^{0.20}_{0.22}$ \\
H139     & 369    & 0.13   & $^{0.40}_{0.14}$ & -7.43  & $^{0.11}_{0.31}$ \\
H141     & 300    & 0.05   & $^{0.51}_{0.74}$ & -7.82  & $^{0.30}_{0.24}$ \\
H159     & 388    & 0.18   & $^{0.33}_{0.29}$ & -7.66  & $^{0.20}_{0.19}$ \\
H162     & 350    & -0.72  & $^{2.72}_{0.73}$ & -7.56  & $^{0.56}_{1.47}$ \\
H173     & 212    & 0.61   & $^{0.75}_{0.88}$ & -8.41  & $^{0.33}_{0.56}$ \\
\hline
    \end{tabular}

\end{table}

\section{Trends in wind structure\label{sec:app:windstructure}} 

\begin{figure*}
    \centering
    \includegraphics[width=0.95\textwidth]{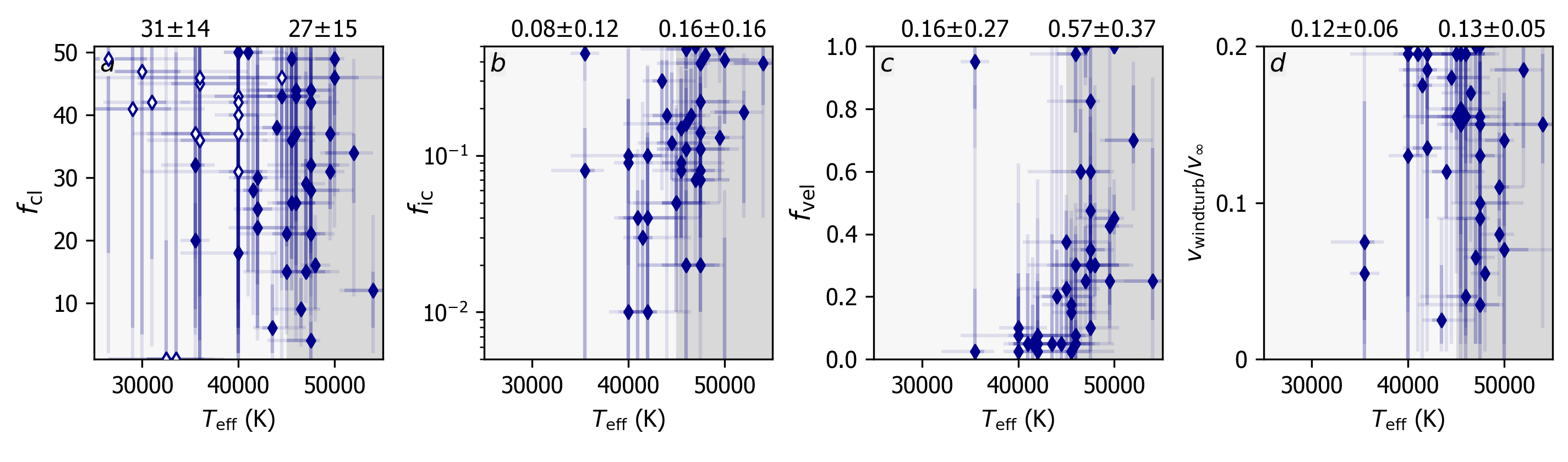}
    \caption{As \cref{fig:windstructure1} but as a function of \teff. The cutoff value for the `low' and `high' regime is at $T_{\rm eff} = 45000$~K.}
    \label{fig:app:windstructure_teff}
\end{figure*}

\begin{figure*}
    \centering
    \includegraphics[width=0.95\textwidth]{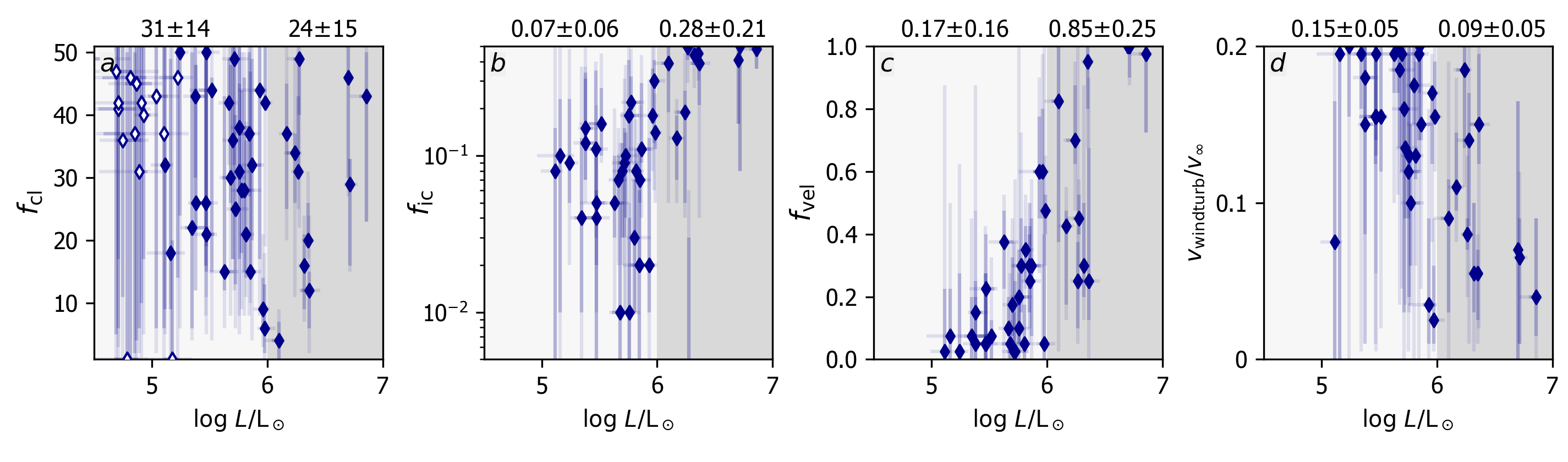}
    \caption{As \cref{fig:windstructure1} but as a function of \logll. The cutoff value for the `low' and `high' regime is at $\log \ L/{\rm L}_\odot = 6$.}
    \label{fig:app:windstructure_luminosity}
\end{figure*}

\begin{figure*}
    \centering
    \includegraphics[width=0.95\textwidth]{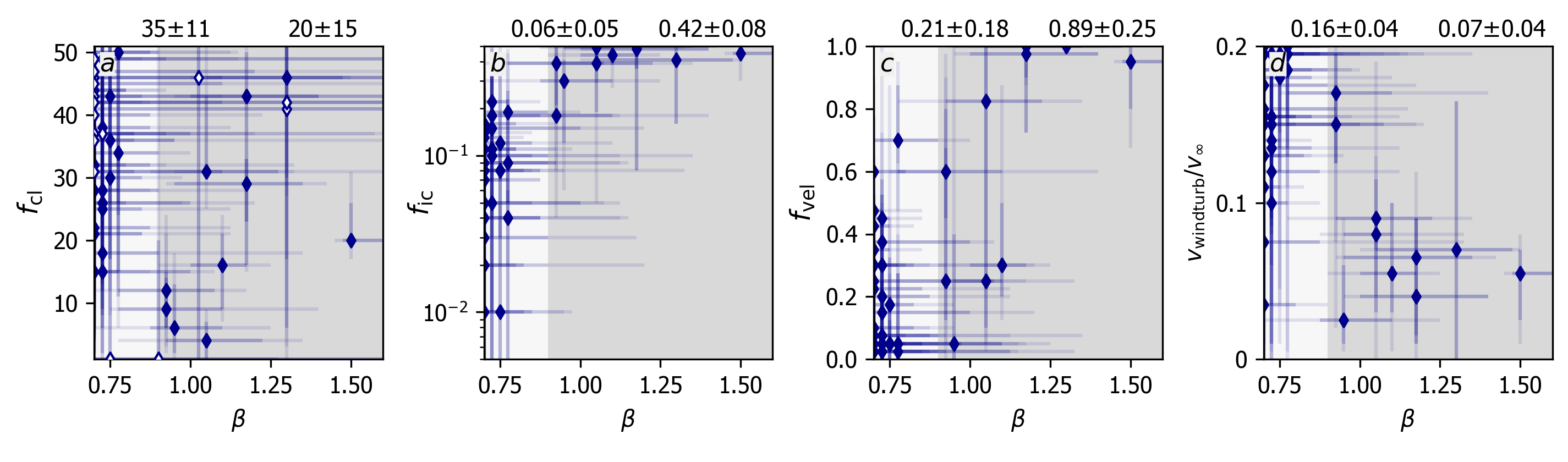}
    \caption{As \cref{fig:windstructure1} but as a function of $\beta$. The cutoff value for the `low' and `high' regime is at $\beta = 0.9$.}
    \label{fig:app:windstructure_beta}
\end{figure*}

\begin{figure*}
    \centering
    \includegraphics[width=0.95\textwidth]{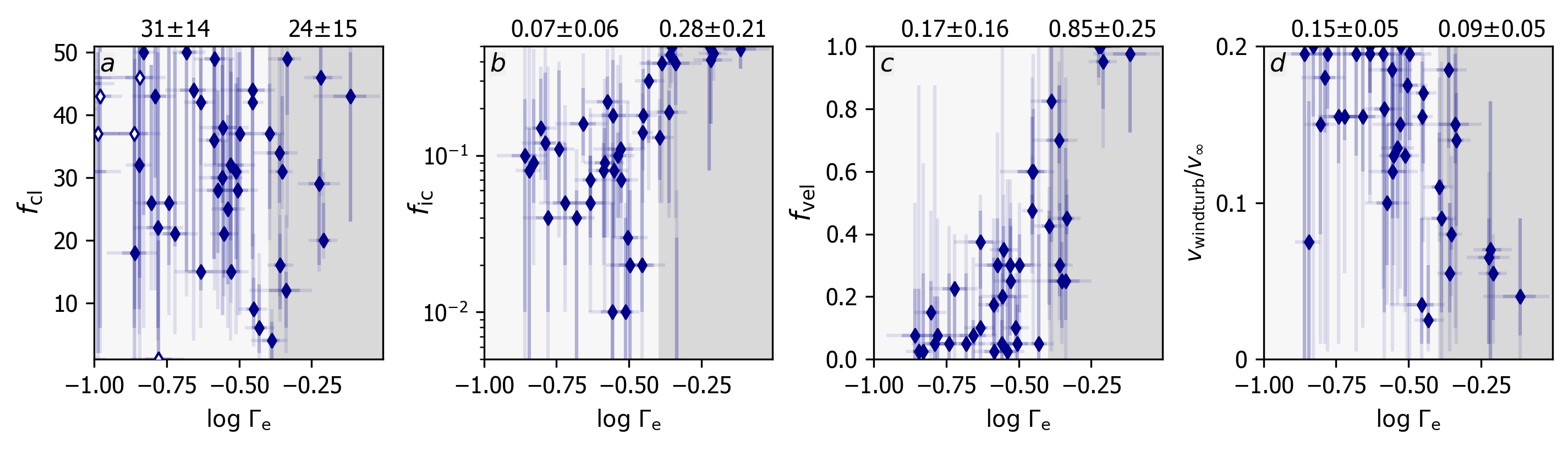}
    \caption{As \cref{fig:windstructure1} but as a function of $\log \Gamma_{\rm Edd,e}$. The cutoff value for the `low' and `high' regime is at $\log \Gamma_{\rm Edd,e} = -0.40$ corresponding to $\Gamma_{\rm Edd,e} = 0.40$. Note that $\Gamma_{\rm Edd,e}$ is proportional to $T_{\rm eff}^4 / g$. }
    \label{fig:app:windstructure_gamma}
\end{figure*}

\Cref{fig:app:windstructure_teff} to \ref{fig:app:windstructure_gamma} show the trends in the wind structure parameters as a function of effective temperature, luminosity, $beta$, and the Eddington parameter for electron scattering. As these parameters are strongly correlated with each other, in all these plots similar trends are visible (see \cref{sec:windstructuretrends}). 

\section{Additional tables \label{sec:app:fitvalues}}

\Cref{tab:app:bestfit:optical} contains the values of the optical-only runs. \Cref{tab:app:more_uv_restuls} contains stellar masses, ages and ionising fluxes based on the parameters derived from optical~+~UV runs. Furthermore, it contains the temperatures derived based on the iron continuum. Note that the uncertainties we quote for $M_{\rm evol}$, $M_{\rm ini}$ and age are only statistical uncertainties that result from the {\sc Bonnsai} tool. Systematic uncertainties, that is, those resulting from the chosen input-physics of the evolutionary model, for example, the assumed mixing scheme, or mass-loss rate prescription, are not included. We discuss the systematic uncertainties on evolutionary masses for the WNh stars in \cref{dis:WNh_stars}. 
\Cref{tab:diaglines} lists for each star the diagnostic lines that were used in the analysis. 

\newpage 
\clearpage

\begin{table*}
\small 
\caption{Additional parameters and $1 \sigma$ uncertainties: spectral types, initial, evolutionary, spectroscopic and spectroscopic masses, ages, H-{\sc i}, He-{\sc i} and He-{\sc ii} ionising radiation $Q_0$, $Q_1$, $Q_2$, the H-{\sc i} ionising luminosity, the Eddington factors for electron scattering $\Gamma_{\rm Edd,e}$, 
and the effective temperature derived from the iron continuum and used for the normalisation of the UV spectra $T_{\rm iron}$. This table is complimentary to \Cref{tab:bestfit:12free,tab:bestfit:6free} and is based on the optical~+~UV runs.
\label{tab:app:more_uv_restuls}}

\begin{tabular}{l  l r@{$\pm$}l r@{$\pm$}l r@{$\pm$}l r@{$\pm$}l l l l l r@{$\pm$}l l}
\hline \hline
Source  & Spectral Type$^{a)}$ & \multicolumn{2}{c}{M$_{\rm ini}$} & \multicolumn{2}{c}{M$_{\rm evol}$} & \multicolumn{2}{c}{M$_{\rm spec}$} & \multicolumn{2}{c}{Age} & log $Q_0$ & log $Q_1$ & log $Q_2$ & $L_{\rm 13.6}$~$^{b)}$ & \multicolumn{2}{c}{$\Gamma_{\rm Edd,e}$} & T$_{\rm iron}$ \\
  & [-] & \multicolumn{2}{c}{[M$_\odot$]} & \multicolumn{2}{c}{[M$_\odot$]} & \multicolumn{2}{c}{[M$_\odot$]} & \multicolumn{2}{c}{[Myr]} & [s$^{-1}$] & [s$^{-1}$] & [s$^{-1}$] & [-] & \multicolumn{2}{c}{[-]} & [kK] \\ \hline
R136a1 & WN5h & 273& $^{25}_{36}$ & 222& $^{29}_{28}$ & 303 & $^{123}_{79}$ & 1.14& $^{0.17}_{0.14}$ & 50.71 & 50.13 & 45.24 & 6.66 & 0.77& $^{0.10}_{0.10}$ & 50 \\
R136a2 & WN5h & 221& $^{16}_{12}$ & 186& $^{17}_{15}$ & 159 & $^{93}_{5}$ & 1.34& $^{0.13}_{0.18}$ & 50.59 & 50.05 & 45.19 & 6.55 & 0.60& $^{0.05}_{0.05}$ & 50 \\
R136a3$^{c)}$ & WN5h & 213& $^{12}_{11}$ & 179& $^{16}_{11}$ & 387 & $^{2949}_{208}$ & 1.28& $^{0.17}_{0.21}$ & 50.56 & 50.04 & 44.97 & 6.52 & 0.61& $^{0.06}_{0.04}$ & 47 \\
R136a4$^{e)}$ & O3V ((f*))(n) & 113& $^{7}_{7}$ & 108& $^{6}_{7}$ & 178 & $^{73}_{32}$ & 0.84& $^{0.19}_{0.08}$ & 50.10 & 49.55 & 45.65 & 6.05 & 0.46& $^{0.03}_{0.03}$ & 47$^{d)}$ \\
R136a5$^{e)}$ & O2 I(n)f* & 123& $^{6}_{6}$ & 116& $^{6}_{5}$ & 364 & $^{1}_{181}$ & 0.98& $^{0.11}_{0.09}$ & 50.13 & 49.56 & 43.71 & 6.08 & 0.44& $^{0.02}_{0.02}$ & 44 \\
R136a6$^{f)}$ & O2 I(n)f*p & 109& $^{8}_{10}$ & 105& $^{8}_{10}$ & 100 & $^{12}_{25}$ & 0.76& $^{0.28}_{0.28}$ & 50.10 & 49.59 & 45.98 & 6.06 & 0.44& $^{0.03}_{0.04}$ & 44 \\
R136a7$^{e)}$ & O3 III(f*) & 131& $^{14}_{14}$ & 127& $^{15}_{16}$ & 230 & $^{58}_{121}$ & 0.50 & $^{0.37}_{0.46}$ & 50.21 & 49.73 & 46.16 & 6.18 & 0.46& $^{0.05}_{0.06}$ & 42 \\
R136a8 & - & 100& $^{6}_{5}$ & 96& $^{6}_{6}$ & 179& $^{140}_{66}$ & 0.90 & $^{0.21}_{0.24}$ & 49.98 & 49.43 & 45.26 & 5.93 & 0.40 & $^{0.02}_{0.02}$ & 53 \\
R136b & O4 If & 107& $^{6}_{5}$ & 92& $^{5}_{5}$ & 209 & $^{121}_{1}$ & 2.02& $^{0.10}_{0.14}$ & 50.00 & 49.04 & 40.86 & 5.88 & 0.62& $^{0.04}_{0.03}$ & 35 \\
H30 & O6.5 Vz & 52& $^{3}_{2}$ & 50& $^{3}_{2}$ & 147 & $^{141}_{2}$ & 2.22& $^{0.21}_{0.24}$ & 49.38 & 48.57 & 42.14 & 5.28 & 0.31& $^{0.02}_{0.01}$ & 39 \\
H31 & O2 V((f*)) & 76& $^{3}_{3}$ & 73& $^{3}_{3}$ & 79 & $^{1}_{16}$ & 1.26& $^{0.13}_{0.19}$ & 49.79 & 49.20 & 45.27 & 5.74 & 0.35& $^{0.02}_{0.01}$ & 44 \\
H35 & O3 V & 63& $^{4}_{3}$ & 62& $^{4}_{3}$ & 65 & $^{46}_{1}$ & 1.10 & $^{0.26}_{0.48}$ & 49.62 & 49.03 & 45.19 & 5.56 & 0.28& $^{0.02}_{0.01}$ & 44 \\
H36 & O2 If* & 115& $^{5}_{5}$ & 110& $^{5}_{5}$ & 162 & $^{2}_{47}$ & 0.90 & $^{0.12}_{0.11}$ & 50.10 & 49.55 & 43.94 & 6.06 & 0.44& $^{0.02}_{0.02}$ & 53 \\
H40 & O3 V & 67& $^{7}_{7}$ & 65& $^{6}_{7}$ & 56 & $^{7}_{11}$ & 1.44& $^{0.44}_{0.51}$ & 49.76 & 49.15 & 45.00 & 5.70 & 0.35& $^{0.04}_{0.04}$ & 50 \\
H45$^{g)}$ & O4: Vz & 57& $^{5}_{3}$ & 54& $^{5}_{3}$ & 127 & $^{15}_{55}$ & 2.06& $^{0.26}_{0.41}$ & 49.48 & 48.75 & 43.31 & 5.39 & 0.31& $^{0.03}_{0.02}$ & 42 \\
H46$^{e)}$ & O2-3 III(f*) & 84& $^{5}_{6}$ & 80& $^{5}_{6}$ & 84 & $^{122}_{17}$ & 1.30 & $^{0.25}_{0.14}$ & 49.94 & 49.36 & 45.51 & 5.88 & 0.41& $^{0.03}_{0.03}$ & 47$^{d)}$ \\
H47 & O2 V((f*)) & 72& $^{5}_{4}$ & 68& $^{4}_{4}$ & 308 & $^{79}_{198}$ & 1.74& $^{0.17}_{0.24}$ & 49.70 & 49.05 & 43.15 & 5.63 & 0.37& $^{0.02}_{0.02}$ & 42 \\
H48 & O2-3 III(f*) & 71& $^{4}_{4}$ & 68& $^{4}_{4}$ & 65 & $^{37}_{4}$ & 1.46& $^{0.22}_{0.21}$ & 49.77 & 49.16 & 44.05 & 5.71 & 0.35& $^{0.02}_{0.02}$ & 50$^{d)}$ \\
H49 & O3 & 57& $^{6}_{4}$ & 55& $^{6}_{5}$ & 48 & $^{51}_{20}$ & 1.84& $^{0.33}_{0.84}$ & 49.53 & 48.86 & 44.68 & 5.46 & 0.28& $^{0.03}_{0.02}$ & 42 \\
H50$^{e)}$ & O3-4 V((f*)) & 66& $^{5}_{4}$ & 64& $^{5}_{4}$ & 86 & $^{10}_{20}$ & 1.14& $^{0.32}_{0.52}$ & 49.64 & 49.04 & 44.62 & 5.58 & 0.30& $^{0.02}_{0.02}$ & 42 \\
H52 & O3-4 Vz & 53& $^{3}_{3}$ & 52& $^{3}_{3}$ & 55 & $^{26}_{11}$ & 1.46& $^{0.37}_{0.50}$ & 49.48 & 48.84 & 43.58 & 5.41 & 0.26& $^{0.02}_{0.01}$ & 42 \\
H55 & O2 V((f*))z & 61& $^{6}_{4}$ & 60& $^{6}_{5}$ & 44 & $^{18}_{5}$ & 1.16& $^{0.37}_{0.69}$ & 49.59 & 48.99 & 45.06 & 5.53 & 0.27& $^{0.03}_{0.02}$ & 44 \\
H58$^{e)}$ & O2-3 V: & 66& $^{7}_{6}$ & 65& $^{7}_{6}$ & 150 & $^{52}_{74}$ & 0.96& $^{0.47}_{0.65}$ & 49.64 & 49.06 & 44.79 & 5.58 & 0.30& $^{0.03}_{0.03}$ & 42 \\
H62$^{g)}$ & O2-3 V & 49& $^{5}_{4}$ & 48& $^{5}_{4}$ & 50 & $^{24}_{14}$ & 1.50& $^{0.52}_{0.88}$ & 49.40 & 48.76 & 42.95 & 5.33 & 0.23& $^{0.03}_{0.02}$ & 50$^{d)}$ \\
H64 & O4-5 V: & 61& $^{5}_{5}$ & 59& $^{6}_{5}$ & 130 & $^{30}_{69}$ & 1.44& $^{0.47}_{0.57}$ & 49.60 & 48.99 & 44.55 & 5.54 & 0.32& $^{0.03}_{0.03}$ & 42 \\
H65 & O4 VC16 & 51& $^{4}_{3}$ & 49& $^{3}_{3}$ & 52 & $^{50}_{11}$ & 2.18& $^{0.28}_{0.37}$ & 49.46 & 48.74 & 42.54 & 5.38 & 0.29& $^{0.02}_{0.02}$ & 39 \\
H66 & O2 V-III(f*) & 54& $^{3}_{3}$ & 53& $^{3}_{3}$ & 47 & $^{19}_{0}$ & 0.78& $^{0.44}_{0.51}$ & 49.46 & 48.87 & 45.02 & 5.41 & 0.23& $^{0.01}_{0.01}$ & 42 \\
H68 & O4-5 Vz & 48& $^{4}_{4}$ & 46& $^{4}_{4}$ & 60 & $^{45}_{11}$ & 2.16& $^{0.42}_{0.50}$ & 49.40 & 48.67 & 43.37 & 5.31 & 0.28& $^{0.02}_{0.02}$ & 39 \\
H69 & O4-5 Vz & 39& $^{3}_{2}$ & 38& $^{3}_{2}$ & 62 & $^{16}_{27}$ & 2.36& $^{0.45}_{0.71}$ & 49.13 & 48.38 & 41.96 & 5.04 & 0.21& $^{0.02}_{0.01}$ & 42 \\
H70$^{g)}$ & O5 Vz & 54& $^{4}_{4}$ & 53& $^{4}_{4}$ & 72 & $^{121}_{8}$ & 1.32& $^{0.46}_{0.66}$ & 49.48 & 48.85 & 44.29 & 5.41 & 0.26& $^{0.02}_{0.02}$ & 47$^{d)}$ \\
H71 & O2-3 V((f*)) & 41& $^{4}_{4}$ & 40& $^{4}_{4}$ & 24 & $^{9}_{7}$ & 1.68& $^{0.70}_{0.94}$ & 49.26 & 48.60 & 43.85 & 5.19 & 0.19& $^{0.02}_{0.02}$ & 44 \\
H75$^{g)}$ & O6 V & 43& $^{3}_{3}$ & 43& $^{3}_{3}$ & 77 & $^{45}_{29}$ & 0.06& $^{1.04}_{0.07}$ & 49.21 & 48.62 & 44.12 & 5.15 & 0.18& $^{0.01}_{0.01}$ & 39$^{d)}$ \\
H78 & O4: V & 41& $^{4}_{4}$ & 40& $^{4}_{4}$ & 31 & $^{14}_{8}$ & 1.52& $^{0.80}_{1.00}$ & 49.30 & 48.68 & 44.26 & 5.24 & 0.22& $^{0.02}_{0.02}$ & 42 \\
H80 & O8 V & 25& $^{1}_{1}$ & 24& $^{1}_{1}$ & 28 & $^{15}_{10}$ & 4.18& $^{0.66}_{0.51}$ & 48.56 & 47.33 & 41.51 & 4.42 & 0.14& $^{0.01}_{0.01}$ & 37 \\
H86$^{e)}$ & O5: V & 41& $^{3}_{2}$ & 41& $^{3}_{2}$ & 18 & $^{2}_{5}$ & 0.60& $^{0.53}_{0.59}$ & 49.19 & 48.55 & 43.95 & 5.12 & 0.16& $^{0.01}_{0.01}$ & 42 \\
H90 & O4: V: & 36& $^{4}_{3}$ & 36& $^{4}_{3}$ & 27 & $^{33}_{3}$ & 1.94& $^{0.63}_{1.21}$ & 49.06 & 48.34 & 42.92 & 4.98 & 0.17& $^{0.02}_{0.01}$ & 42 \\
H92 & O6 Vz & 32& $^{2}_{1}$ & 31& $^{2}_{1}$ & 40 & $^{10}_{17}$ & 2.36& $^{0.44}_{0.72}$ & 48.87 & 48.05 & 42.42 & 4.77 & 0.15& $^{0.01}_{0.01}$ & 42 \\
H94 & O4-5 Vz & 39& $^{3}_{3}$ & 39& $^{3}_{3}$ & 24 & $^{13}_{7}$ & 1.16& $^{0.69}_{0.90}$ & 49.15 & 48.48 & 43.89 & 5.08 & 0.16& $^{0.01}_{0.01}$ & 47 \\
H143 & O8-9 V-III & 28& $^{3}_{3}$ & 28& $^{3}_{3}$ & 16 & $^{16}_{4}$ & 2.60& $^{1.22}_{1.49}$ & 48.85 & 48.05 & 42.74 & 4.76 & 0.14& $^{0.01}_{0.01}$ & 37 \\
H73 & O9.7-B0 V & 24& $^{3}_{2}$ & 24& $^{3}_{2}$ & 74 & $^{65}_{15}$ & 4.46& $^{0.98}_{1.02}$ & 48.25 & 46.16 & 41.05 & 4.10 & 0.17& $^{0.02}_{0.02}$ & 37 \\
H108 & O Vn & 22& $^{3}_{3}$ & 22& $^{3}_{3}$ & 16 & $^{11}_{6}$ & 1.46& $^{2.22}_{1.47}$ & 48.53 & 47.73 & 42.37 & 4.43 & 0.10& $^{0.01}_{0.01}$ & 35 \\
H112 & O7-9 Vz & 25& $^{3}_{3}$ & 25& $^{3}_{3}$ & 45 & $^{63}_{16}$ & 3.78& $^{1.30}_{1.84}$ & 48.50 & 47.21 & 41.47 & 4.36 & 0.14& $^{0.02}_{0.02}$ & 35 \\
H114 & O5-6 V & 32& $^{3}_{3}$ & 31& $^{4}_{3}$ & 28 & $^{23}_{13}$ & 0.00& $^{2.06}_{0.01}$ & 48.96 & 48.31 & 43.62 & 4.89 & 0.14& $^{0.02}_{0.01}$ & 39 \\
H116 & O7 V & 21& $^{3}_{2}$ & 21& $^{3}_{2}$ & 21 & $^{24}_{9}$ & 3.20& $^{1.56}_{2.29}$ & 48.30 & 47.06 & 41.05 & 4.16 & 0.09& $^{0.01}_{0.01}$ & 37 \\
H120 & - & 19& $^{3}_{3}$ & 18& $^{3}_{3}$ & 34 & $^{2}_{22}$ & 0.00& $^{4.92}_{0.01}$ & 48.43 & 47.60 & 42.19 & 4.33 & 0.10& $^{0.02}_{0.02}$ & 37 \\
H121 & O9.5 V & 19& $^{2}_{2}$ & 19& $^{2}_{2}$ & 23 & $^{11}_{11}$ & 4.56& $^{2.01}_{2.61}$ & 47.98 & 46.08 & 40.98 & 3.82 & 0.09& $^{0.01}_{0.01}$ & 35 \\
H123 & O6 V & 26& $^{2}_{1}$ & 26& $^{2}_{1}$ & 14 & $^{11}_{5}$ & 0.36& $^{1.30}_{0.37}$ & 48.57 & 47.77 & 42.39 & 4.48 & 0.09& $^{0.01}_{0.00}$ & 37 \\
H129 & - & 18& $^{2}_{2}$ & 18& $^{2}_{2}$ & 7 & $^{15}_{4}$ & 0.00& $^{2.20}_{0.01}$ & 48.16 & 47.44 & 42.38 & 4.07 & 0.04& $^{0.00}_{0.00}$ & 39 \\
H132 & O7: V & 27& $^{3}_{2}$ & 27& $^{3}_{3}$ & 28 & $^{34}_{12}$ & 1.40& $^{1.02}_{1.41}$ & 48.65 & 47.82 & 42.36 & 4.55 & 0.10& $^{0.01}_{0.01}$ & 37 \\
H134 & O7 Vz & 20& $^{3}_{2}$ & 20& $^{3}_{2}$ & 16 & $^{12}_{8}$ & 2.64& $^{1.62}_{2.16}$ & 48.18 & 46.99 & 40.88 & 4.05 & 0.07& $^{0.01}_{0.01}$ & 37 \\
H135 & B & 16& $^{2}_{2}$ & 16& $^{2}_{2}$ & 25 & $^{17}_{11}$ & 7.18& $^{2.70}_{2.96}$ & 47.22 & 44.43 & 40.37 & 3.05 & 0.08& $^{0.01}_{0.01}$ & 39 \\
H139 & - & 24& $^{2}_{2}$ & 24& $^{2}_{2}$ &  19 & $^{23}_{7}$ & 0.06& $^{2.52}_{0.07}$ & 48.53 & 47.73 & 42.38 & 4.43 & 0.09& $^{0.01}_{0.01}$ & 39 \\
H141 & O5-6 V & 17& $^{3}_{2}$ & 17& $^{3}_{2}$ & 21 & $^{50}_{9}$ & 6.54& $^{2.14}_{3.73}$ & 47.44 & 44.88 & 40.38 & 3.28 & 0.08& $^{0.01}_{0.01}$ & 35 \\
H159 & - & 17& $^{1}_{1}$ & 17& $^{1}_{1}$ & 22 & $^{20}_{8}$ & 6.46& $^{1.74}_{1.96}$ & 47.58 & 45.13 & 40.56 & 3.41 & 0.08& $^{0.01}_{0.01}$ & 35 \\
H162 & - & 18& $^{4}_{3}$ & 18& $^{4}_{3}$ & 34 & $^{110}_{22}$ & 2.28& $^{3.04}_{2.29}$ & 48.19 & 46.86 & 40.50 & 4.05 & 0.10& $^{0.02}_{0.02}$ & 35 \\
H173 & O9 + V & 12& $^{2}_{2}$ & 12& $^{2}_{2}$ & 20 & $^{16}_{9}$ & 9.52& $^{4.59}_{5.34}$ & 46.44 & 43.35 & 40.05 & 2.25 & 0.06& $^{0.01}_{0.01}$ & 37 \\
\\
\multicolumn{17}{p{16.9cm}}{\tiny$^{a)}$ Spectral types from \citet[][WN5h stars]{2016MNRAS.458..624C} and \citet[][all other stars]{Saidainprep}. \tiny$^{b)}$ This quantity is expressed as $L_{\rm 13.6} = \log \ L_{E > 13.6 {\rm eV}} / {\rm L}_\odot$. \tiny$^{c)}$ Formal uncertainties assuming a \teff of 50~kK. In reality uncertainties on all parameters are larger due to the uncertain \teff. \tiny$^{d)}$ No good fit obtained to iron pseudo-continuum; assumed value closest to \citet{Bestenlehner2020} for UV normalisation. \tiny$^{e)}$ Cross-contamination of the spectrum as a result of crowding (as in \citealt{Bestenlehner2020}, with the addition of R136a8). \tiny$^{f)}$ Severe cross-contamination of the spectrum as a result of crowding: R136a6 consists of H19 and H26, see \cref{sec:stis_data_r136}. \tiny$^{g)}$ Potential spectroscopic binary (as in \citealt{Bestenlehner2020}). } \\
\hline
\end{tabular}

\end{table*}

\begin{table*}[p]
\small 
\caption{Best fit parameters and $1\sigma$ error bars for the optical-only fits for all stars except R136a8, for which optical data is not available. A smooth wind ($f_{\rm cl}= 1$) was assumed during the fit for all stars except the WNh stars R136a1, R136a2, R136a3, where we assumed $f_{\rm cl}= 10$.  \label{tab:app:bestfit:optical}}

\begin{tabular}{l  r@{$\pm$}l r@{$\pm$}l r@{$\pm$}l r@{$\pm$}l r@{$\pm$}l r@{$\pm$}l r@{$\pm$}l r@{$\pm$}l}
 \hline  \hline
 Source  & \multicolumn{2}{c}{$\log L/{\rm L}_{\odot}$} & \multicolumn{2}{c}{$T_{\rm eff}$(K)} & \multicolumn{2}{c}{$\log g$} & \multicolumn{2}{c}{$R_{\star}/{\rm R}_{\odot}$} & \multicolumn{2}{c}{$\log\dot{M}$} & \multicolumn{2}{c}{$\beta$} & \multicolumn{2}{c}{$\varv_{\rm eq} \sin i$} & \multicolumn{2}{c}{$n_{\rm He}/n_{\rm H}$} \\ \hline 
R136a1 & 6.76& $^{0.08}_{0.03}$ & 42500& $^{2500}_{500}$ & 3.80& $^{0.85}_{0.20}$ & 44.9& $^{0.8}_{1.6}$ & -4.24& $^{0.04}_{0.04}$ & 0.93& $^{0.07}_{0.12}$ & 160& $^{160}_{90}$ & 0.29& $^{0.14}_{0.03}$ \\
R136a2 & 6.75& $^{0.06}_{0.07}$ & 46250& $^{2000}_{2500}$ & 4.00& $^{1.05}_{0.45}$ & 37.1& $^{1.3}_{1.1}$ & -4.31& $^{0.06}_{0.11}$ & 0.93& $^{0.23}_{0.12}$ & 140& $^{370}_{90}$ & 0.28& $^{0.17}_{0.06}$ \\
R136a3 & 6.51& $^{0.06}_{0.05}$ & 42750& $^{2000}_{1500}$ & 3.50& $^{0.70}_{0.20}$ & 32.9& $^{0.9}_{1.1}$ & -4.24& $^{0.07}_{0.12}$ & 0.95& $^{0.28}_{0.10}$ & 300& $^{210}_{300}$ & 0.45& $^{0.06}_{0.14}$ \\
R136a4 & 6.35& $^{0.06}_{0.10}$ & 53000& $^{2750}_{4000}$ & 4.22& $^{0.17}_{0.15}$ & 18.0& $^{0.8}_{0.5}$ & -5.34& $^{0.10}_{0.08}$ & \multicolumn{2}{c}{-} & 145& $^{35}_{40}$ & 0.10& $^{0.06}_{0.01}$ \\
R136a5 & 6.13& $^{0.05}_{0.04}$ & 42000& $^{2000}_{1250}$ & 4.17& $^{0.30}_{0.45}$ & 22.1& $^{0.5}_{0.6}$ & -4.46& $^{0.05}_{0.03}$ & 0.89& $^{0.03}_{0.06}$ & 95& $^{58}_{40}$ & 0.18& $^{0.04}_{0.04}$ \\
R136a6 & 6.25& $^{0.07}_{0.11}$ & 52500& $^{3000}_{4500}$ & 4.08& $^{0.12}_{0.10}$ & 16.3& $^{0.7}_{0.5}$ & -4.95& $^{0.05}_{0.05}$ & 0.70& $^{0.04}_{0.01}$ & 165& $^{28}_{35}$ & 0.10& $^{0.03}_{0.02}$ \\
R136a7 & 6.47& $^{0.04}_{0.08}$ & 58500& $^{2000}_{3500}$ & 4.15& $^{0.28}_{0.15}$ & 17.0& $^{0.6}_{0.4}$ & -5.12& $^{0.13}_{0.09}$ & 0.88& $^{0.12}_{0.17}$ & 260& $^{95}_{42}$ & 0.14& $^{0.14}_{0.04}$ \\
R136a8 & \multicolumn{2}{c}{-} & \multicolumn{2}{c}{-} & \multicolumn{2}{c}{-} & \multicolumn{2}{c}{-} & \multicolumn{2}{c}{-} & \multicolumn{2}{c}{-} & \multicolumn{2}{c}{-} & \multicolumn{2}{c}{-} \\
R136b & 6.24& $^{0.03}_{0.02}$ & 32500& $^{750}_{250}$ & 3.52& $^{0.20}_{0.05}$ & 41.8& $^{0.8}_{0.9}$ & -4.58& $^{0.05}_{0.08}$ & 1.15& $^{0.12}_{0.06}$ & 95& $^{12}_{20}$ & 0.14& $^{0.06}_{0.01}$ \\
H30 & 5.69& $^{0.08}_{0.10}$ & 38000& $^{2500}_{2750}$ & 4.12& $^{0.20}_{0.28}$ & 16.3& $^{0.7}_{0.6}$ & -5.77& $^{0.20}_{1.08}$ & \multicolumn{2}{c}{-} & 125& $^{50}_{45}$ & 0.09& $^{0.05}_{0.01}$ \\
H31 & 6.00& $^{0.14}_{0.11}$ & 48000& $^{5500}_{4000}$ & 3.92& $^{0.17}_{0.15}$ & 14.6& $^{0.6}_{0.8}$ & -5.67& $^{0.15}_{0.19}$ & \multicolumn{2}{c}{-} & 120& $^{30}_{40}$ & 0.09& $^{0.04}_{0.01}$ \\
H35 & 5.99& $^{0.07}_{0.21}$ & 54250& $^{3000}_{8000}$ & 4.28& $^{0.12}_{0.30}$ & 11.3& $^{0.9}_{0.3}$ & -5.79& $^{0.18}_{0.79}$ & \multicolumn{2}{c}{-} & 110& $^{45}_{55}$ & 0.10& $^{0.06}_{0.01}$ \\
H36 & 6.04& $^{0.08}_{0.06}$ & 42000& $^{3000}_{2000}$ & 3.70& $^{0.25}_{0.11}$ & 20.0& $^{0.6}_{0.7}$ & -4.45& $^{0.05}_{0.10}$ & 0.70& $^{0.11}_{0.03}$ & 165& $^{50}_{52}$ & 0.10& $^{0.03}_{0.01}$ \\
H40 & 5.82& $^{0.24}_{0.12}$ & 43500& $^{9000}_{3750}$ & 3.67& $^{0.40}_{0.12}$ & 14.4& $^{0.7}_{1.3}$ & -5.57& $^{0.13}_{0.23}$ & \multicolumn{2}{c}{-} & 120& $^{45}_{65}$ & 0.10& $^{0.12}_{0.01}$ \\
H45 & 5.86& $^{0.13}_{0.13}$ & 43500& $^{5000}_{4500}$ & 4.08& $^{0.28}_{0.25}$ & 15.2& $^{0.8}_{0.8}$ & -6.12& $^{0.35}_{1.90}$ & \multicolumn{2}{c}{-} & 165& $^{60}_{60}$ & 0.09& $^{0.13}_{0.01}$ \\
H46 & 6.07& $^{0.10}_{0.07}$ & 46500& $^{3750}_{2500}$ & 3.92& $^{0.29}_{0.17}$ & 16.9& $^{0.5}_{0.7}$ & -4.98& $^{0.15}_{0.15}$ & 0.93& $^{0.15}_{0.20}$ & 190& $^{72}_{55}$ & 0.12& $^{0.10}_{0.03}$ \\
H47 & 6.06& $^{0.13}_{0.09}$ & 46500& $^{5000}_{3000}$ & 4.03& $^{0.38}_{0.17}$ & 16.7& $^{0.6}_{0.8}$ & -4.93& $^{0.10}_{0.15}$ & 0.70& $^{0.14}_{0.03}$ & 120& $^{70}_{90}$ & 0.10& $^{0.11}_{0.01}$ \\
H48 & 5.92& $^{0.16}_{0.08}$ & 45000& $^{6250}_{3000}$ & 3.92& $^{0.33}_{0.17}$ & 15.1& $^{0.5}_{0.9}$ & -5.25& $^{0.10}_{0.11}$ & \multicolumn{2}{c}{-} & 130& $^{50}_{65}$ & 0.10& $^{0.09}_{0.01}$ \\
H49 & 6.12& $^{0.02}_{0.41}$ & 58000& $^{500}_{16000}$ & 4.25& $^{0.50}_{0.51}$ & 11.5& $^{2.3}_{0.2}$ & -5.56& $^{0.25}_{0.98}$ & \multicolumn{2}{c}{-} & 215& $^{150}_{160}$ & 0.09& $^{0.20}_{0.01}$ \\
H50 & 5.82& $^{0.13}_{0.11}$ & 46000& $^{5250}_{4000}$ & 4.03& $^{0.23}_{0.20}$ & 12.9& $^{0.6}_{0.7}$ & -5.92& $^{0.18}_{0.45}$ & \multicolumn{2}{c}{-} & 125& $^{50}_{52}$ & 0.09& $^{0.05}_{0.01}$ \\
H52 & 5.74& $^{0.10}_{0.14}$ & 47250& $^{3750}_{5000}$ & 4.10& $^{0.16}_{0.30}$ & 11.2& $^{0.7}_{0.4}$ & -5.65& $^{0.13}_{0.25}$ & \multicolumn{2}{c}{-} & 135& $^{58}_{50}$ & 0.09& $^{0.08}_{0.01}$ \\
H55 & 5.92& $^{0.10}_{0.22}$ & 53000& $^{4000}_{8250}$ & 4.05& $^{0.17}_{0.35}$ & 10.9& $^{1.0}_{0.4}$ & -5.77& $^{0.20}_{0.60}$ & \multicolumn{2}{c}{-} & 110& $^{60}_{55}$ & 0.09& $^{0.14}_{0.01}$ \\
H58 & 5.98& $^{0.18}_{0.14}$ & 52500& $^{8000}_{5500}$ & 4.38& $^{0.20}_{0.25}$ & 11.9& $^{0.7}_{0.8}$ & -7.03& $^{0.65}_{0.50}$ & \multicolumn{2}{c}{-} & 105& $^{55}_{80}$ & 0.11& $^{0.08}_{0.03}$ \\
H62 & 5.87& $^{0.13}_{0.24}$ & 54000& $^{6000}_{9500}$ & 4.00& $^{0.21}_{0.33}$ & 9.9& $^{1.0}_{0.5}$ & -5.42& $^{0.05}_{0.18}$ & 0.70& $^{0.17}_{0.03}$ & 160& $^{60}_{85}$ & 0.10& $^{0.15}_{0.02}$ \\
H64 & 5.94& $^{0.03}_{0.20}$ & 49500& $^{1000}_{7500}$ & 4.17& $^{0.28}_{0.35}$ & 12.9& $^{1.1}_{0.2}$ & -6.04& $^{0.23}_{0.63}$ & \multicolumn{2}{c}{-} & 150& $^{95}_{90}$ & 0.09& $^{0.08}_{0.01}$ \\
H65 & 5.67& $^{0.43}_{0.08}$ & 40000& $^{16000}_{2500}$ & 3.77& $^{0.60}_{0.25}$ & 14.3& $^{0.5}_{2.2}$ & -5.72& $^{0.18}_{0.58}$ & \multicolumn{2}{c}{-} & 155& $^{75}_{110}$ & 0.09& $^{0.11}_{0.01}$ \\
H66 & 5.77& $^{0.13}_{0.14}$ & 51500& $^{5500}_{5500}$ & 4.15& $^{0.19}_{0.28}$ & 9.7& $^{0.6}_{0.5}$ & -5.42& $^{0.08}_{0.16}$ & 0.70& $^{0.12}_{0.03}$ & 35& $^{85}_{35}$ & 0.09& $^{0.06}_{0.01}$ \\
H68 & 5.86& $^{0.16}_{0.19}$ & 48000& $^{6500}_{7000}$ & 4.25& $^{0.35}_{0.35}$ & 12.4& $^{1.0}_{0.8}$ & -6.42& $^{0.50}_{1.60}$ & \multicolumn{2}{c}{-} & 230& $^{85}_{90}$ & 0.09& $^{0.06}_{0.01}$ \\
H69 & 5.45& $^{0.13}_{0.11}$ & 40000& $^{4375}_{3250}$ & 4.03& $^{0.25}_{0.28}$ & 11.1& $^{0.5}_{0.6}$ & -6.82& $^{0.80}_{1.60}$ & \multicolumn{2}{c}{-} & 155& $^{45}_{50}$ & 0.09& $^{0.05}_{0.01}$ \\
H70 & 5.65& $^{0.16}_{0.08}$ & 43500& $^{6000}_{2750}$ & 4.17& $^{0.30}_{0.30}$ & 11.9& $^{0.5}_{0.7}$ & -5.75& $^{0.13}_{0.16}$ & \multicolumn{2}{c}{-} & 145& $^{72}_{62}$ & 0.09& $^{0.09}_{0.01}$ \\
H71 & 5.64& $^{0.16}_{0.23}$ & 51000& $^{7000}_{8500}$ & 3.98& $^{0.20}_{0.25}$ & 8.6& $^{0.8}_{0.5}$ & -6.17& $^{0.25}_{1.30}$ & \multicolumn{2}{c}{-} & 100& $^{70}_{60}$ & 0.12& $^{0.14}_{0.04}$ \\
H75 & 5.23& $^{0.13}_{0.16}$ & 38000& $^{4250}_{4250}$ & 4.40& $^{0.23}_{0.30}$ & 9.6& $^{0.7}_{0.5}$ & -6.34& $^{0.28}_{0.85}$ & 0.80& $^{1.15}_{0.12}$ & 105& $^{78}_{55}$ & 0.09& $^{0.10}_{0.01}$ \\
H78 & 5.45& $^{0.32}_{0.18}$ & 43500& $^{12500}_{6000}$ & 4.03& $^{0.42}_{0.30}$ & 9.4& $^{0.7}_{1.1}$ & -5.87& $^{0.25}_{0.80}$ & \multicolumn{2}{c}{-} & 100& $^{85}_{100}$ & 0.09& $^{0.15}_{0.01}$ \\
H80 & 5.17& $^{0.07}_{0.11}$ & 36000& $^{2125}_{3000}$ & 3.92& $^{0.17}_{0.28}$ & 9.9& $^{0.5}_{0.3}$ & -6.42& $^{0.15}_{1.03}$ & \multicolumn{2}{c}{-} & 125& $^{48}_{35}$ & 0.09& $^{0.05}_{0.01}$ \\
H86 & 5.38& $^{0.14}_{0.11}$ & 45500& $^{5500}_{4000}$ & 3.88& $^{0.35}_{0.23}$ & 7.9& $^{0.4}_{0.5}$ & -5.75& $^{0.10}_{0.15}$ & \multicolumn{2}{c}{-} & 210& $^{55}_{98}$ & 0.09& $^{0.07}_{0.01}$ \\
H90 & 5.34& $^{0.19}_{0.17}$ & 41500& $^{7250}_{5000}$ & 4.12& $^{0.28}_{0.25}$ & 9.1& $^{0.7}_{0.7}$ & -6.62& $^{0.70}_{0.90}$ & \multicolumn{2}{c}{-} & 75& $^{78}_{75}$ & 0.09& $^{0.07}_{0.01}$ \\
H92 & 5.38& $^{0.13}_{0.13}$ & 43500& $^{5000}_{4500}$ & 4.20& $^{0.30}_{0.25}$ & 8.7& $^{0.5}_{0.5}$ & -6.62& $^{0.55}_{1.40}$ & \multicolumn{2}{c}{-} & 85& $^{68}_{60}$ & 0.10& $^{0.11}_{0.01}$ \\
H94 & 5.66& $^{0.10}_{0.23}$ & 54000& $^{4500}_{9000}$ & 4.22& $^{0.30}_{0.35}$ & 7.8& $^{0.7}_{0.3}$ & -6.19& $^{0.25}_{1.00}$ & \multicolumn{2}{c}{-} & 180& $^{75}_{90}$ & 0.10& $^{0.15}_{0.01}$ \\
H143 & 5.29& $^{0.19}_{0.20}$ & 43500& $^{7000}_{6500}$ & 3.95& $^{0.45}_{0.45}$ & 7.9& $^{0.7}_{0.6}$ & -6.20& $^{0.28}_{0.95}$ & \multicolumn{2}{c}{-} & 250& $^{55}_{110}$ & 0.10& $^{0.21}_{0.01}$ \\
H73 & 5.08& $^{0.15}_{0.12}$ & 29000& $^{4000}_{3000}$ & 3.85& $^{0.38}_{0.23}$ & 13.9& $^{0.7}_{0.8}$ & -6.85& $^{0.65}_{1.15}$ & \multicolumn{2}{c}{-} & 250& $^{65}_{80}$ & 0.10& $^{0.14}_{0.02}$ \\
H108 & 5.19& $^{0.15}_{0.23}$ & 49500& $^{6500}_{8500}$ & 4.17& $^{0.28}_{0.30}$ & 5.4& $^{0.5}_{0.3}$ & -6.67& $^{0.15}_{0.68}$ & \multicolumn{2}{c}{-} & 240& $^{90}_{80}$ & 0.09& $^{0.08}_{0.01}$ \\
H112 & 5.14& $^{0.20}_{0.19}$ & 36000& $^{6500}_{5000}$ & 4.03& $^{0.60}_{0.47}$ & 9.6& $^{0.8}_{0.8}$ & -7.89& $^{1.80}_{0.15}$ & \multicolumn{2}{c}{-} & 250& $^{130}_{100}$ & 0.09& $^{0.18}_{0.01}$ \\
H114 & 5.21& $^{0.18}_{0.15}$ & 43500& $^{7000}_{5000}$ & 4.12& $^{0.35}_{0.30}$ & 7.1& $^{0.5}_{0.5}$ & -6.52& $^{0.35}_{1.90}$ & \multicolumn{2}{c}{-} & 60& $^{80}_{55}$ & 0.09& $^{0.16}_{0.01}$ \\
H116 & 4.93& $^{0.12}_{0.17}$ & 37000& $^{4000}_{5000}$ & 4.08& $^{0.35}_{0.38}$ & 7.1& $^{0.5}_{0.4}$ & -6.82& $^{0.25}_{1.20}$ & \multicolumn{2}{c}{-} & 160& $^{95}_{80}$ & 0.09& $^{0.16}_{0.01}$ \\
H120 & 4.72& $^{0.21}_{0.21}$ & 35000& $^{6750}_{5000}$ & 4.10& $^{0.45}_{0.41}$ & 6.3& $^{0.6}_{0.5}$ & -6.96& $^{0.35}_{1.05}$ & \multicolumn{2}{c}{-} & 220& $^{110}_{80}$ & 0.09& $^{0.15}_{0.01}$ \\
H121 & 4.81& $^{0.09}_{0.17}$ & 34000& $^{2750}_{4500}$ & 4.05& $^{0.30}_{0.40}$ & 7.4& $^{0.5}_{0.3}$ & -7.82& $^{1.30}_{0.20}$ & \multicolumn{2}{c}{-} & 110& $^{70}_{70}$ & 0.10& $^{0.15}_{0.01}$ \\
H123 & 4.92& $^{0.17}_{0.12}$ & 40000& $^{6250}_{3500}$ & 3.98& $^{0.39}_{0.30}$ & 6.1& $^{0.3}_{0.4}$ & -7.84& $^{1.30}_{0.70}$ & \multicolumn{2}{c}{-} & 105& $^{65}_{75}$ & 0.10& $^{0.15}_{0.02}$ \\
H129 & 4.45& $^{0.21}_{0.19}$ & 40000& $^{7500}_{5250}$ & 4.10& $^{0.70}_{0.45}$ & 3.5& $^{0.3}_{0.3}$ & -7.77& $^{0.65}_{0.25}$ & \multicolumn{2}{c}{-} & 110& $^{120}_{100}$ & 0.09& $^{0.16}_{0.01}$ \\
H132 & 5.07& $^{0.17}_{0.11}$ & 40000& $^{6250}_{3250}$ & 4.10& $^{0.40}_{0.35}$ & 7.2& $^{0.3}_{0.5}$ & -6.62& $^{0.35}_{1.35}$ & \multicolumn{2}{c}{-} & 60& $^{90}_{60}$ & 0.12& $^{0.14}_{0.03}$ \\
H134 & 4.84& $^{0.13}_{0.16}$ & 38000& $^{4250}_{4250}$ & 4.08& $^{0.35}_{0.30}$ & 6.2& $^{0.4}_{0.3}$ & -7.42& $^{0.75}_{1.10}$ & \multicolumn{2}{c}{-} & 115& $^{68}_{58}$ & 0.09& $^{0.15}_{0.01}$ \\
H135 & 4.57& $^{0.20}_{0.28}$ & 25500& $^{5000}_{5500}$ & 3.65& $^{0.50}_{0.55}$ & 10.0& $^{1.2}_{0.8}$ & -7.09& $^{0.40}_{1.55}$ & \multicolumn{2}{c}{-} & 150& $^{80}_{70}$ & 0.12& $^{0.17}_{0.04}$ \\
H139 & 4.88& $^{0.14}_{0.15}$ & 38000& $^{4750}_{4000}$ & 4.00& $^{0.40}_{0.40}$ & 6.4& $^{0.4}_{0.4}$ & -6.77& $^{0.35}_{1.25}$ & \multicolumn{2}{c}{-} & 80& $^{75}_{70}$ & 0.10& $^{0.15}_{0.01}$ \\
H141 & 3.80& $^{0.88}_{0.12}$ & 14500& $^{15500}_{1500}$ & 2.50& $^{1.65}_{0.05}$ & 12.7& $^{0.8}_{4.2}$ & -8.62& $^{1.35}_{0.40}$ & \multicolumn{2}{c}{-} & 360& $^{210}_{140}$ & 0.09& $^{0.25}_{0.01}$ \\
H159 & 4.84& $^{0.16}_{0.22}$ & 35000& $^{5000}_{6000}$ & 4.20& $^{0.40}_{0.45}$ & 7.2& $^{0.7}_{0.5}$ & -7.98& $^{1.65}_{0.05}$ & \multicolumn{2}{c}{-} & 210& $^{115}_{100}$ & 0.09& $^{0.16}_{0.01}$ \\
H162 & 4.64& $^{0.32}_{0.33}$ & 31000& $^{10500}_{8000}$ & 3.85& $^{0.47}_{0.60}$ & 7.3& $^{1.1}_{0.8}$ & -6.85& $^{0.35}_{1.05}$ & \multicolumn{2}{c}{-} & 250& $^{180}_{110}$ & 0.09& $^{0.18}_{0.01}$ \\
H173 & 4.27& $^{0.30}_{0.21}$ & 23000& $^{6500}_{4000}$ & 3.50& $^{0.65}_{0.50}$ & 8.7& $^{0.9}_{1.0}$ & -8.40& $^{1.05}_{1.10}$ & \multicolumn{2}{c}{-} & 280& $^{130}_{120}$ & 0.09& $^{0.16}_{0.01}$ \\ \hline
    \end{tabular}
    \label{tab:all_params_opt}

\end{table*}

\setlength{\tabcolsep}{2.1pt}
\begin{sidewaystable*}
\caption{Overview of diagnostic lines used for the analysis of each source. If a line was included in the fitting, the cell is marked with `x' (STIS data) or `o' (GHRS archival data). Note that if lines are very close together or blended, they may not have a separate panel in the plots showing the best fits (\cref{fig:outGA_R136a1_UV} to \ref{fig:outGA_H173_UV}). \label{tab:diaglines}}             
\label{tab:app:linelist}      
\centering

\begin{tabular}{l p{0.0cm} c  c  c  c  c  c  c  c  c  c  c  c  c  c  c  c  c  c  c  c  c  c  c  c  c   p{0.8cm} l p{0.0cm} c  c  c  c  c  c  c  c  c  c  c  c  c  c  c  c  c  c  c  c  c  c  c  c  c   }
 Source & & \rotatebox{90}{C~\sc{iv}~$\lambda$1170}  & \rotatebox{90}{C~\sc{iii}~$\lambda$1175}  & \rotatebox{90}{N~\sc{v}~$\lambda$1240}  & \rotatebox{90}{O~\sc{iv}~$\lambda$1338-42-43}  & \rotatebox{90}{O~\sc{v}~$\lambda$1371}  & \rotatebox{90}{Si~\sc{iv}~$\lambda$1393}  & \rotatebox{90}{C~\sc{iv}~$\lambda$1548}  & \rotatebox{90}{He~\sc{ii}~$\lambda$1640}  & \rotatebox{90}{N~\sc{iv}~$\lambda$1718}  & \rotatebox{90}{H$\eta$}  & \rotatebox{90}{H$\zeta$}  & \rotatebox{90}{H$\epsilon$}  & \rotatebox{90}{He~\sc{i-ii}~$\lambda$4026}  & \rotatebox{90}{N~\sc{iv}~$\lambda$4058}  & \rotatebox{90}{H$\delta$}  & \rotatebox{90}{He~\sc{ii}~$\lambda$4200}  & \rotatebox{90}{H$\gamma$}  & \rotatebox{90}{He~\sc{i}~$\lambda$4387}  & \rotatebox{90}{He~\sc{i}~$\lambda$4471}  & \rotatebox{90}{He~\sc{ii}~$\lambda$4541}  & \rotatebox{90}{N~\sc{iii}~$\lambda$4634-40-41}  & \rotatebox{90}{N~\sc{v}~$\lambda$4603-19}  & \rotatebox{90}{He~\sc{ii}~$\lambda$4686}  & \rotatebox{90}{He~\sc{i}~$\lambda$4713}  & \rotatebox{90}{H$\alpha$} & & Source &  & \rotatebox{90}{C~\sc{iv}~$\lambda$1170}  & \rotatebox{90}{C~\sc{iii}~$\lambda$1175}  & \rotatebox{90}{N~\sc{v}~$\lambda$1240}  & \rotatebox{90}{O~\sc{iv}~$\lambda$1338-42-43}  & \rotatebox{90}{O~\sc{v}~$\lambda$1371}  & \rotatebox{90}{Si~\sc{iv}~$\lambda$1393}  & \rotatebox{90}{C~\sc{iv}~$\lambda$1548}  & \rotatebox{90}{He~\sc{ii}~$\lambda$1640}  & \rotatebox{90}{N~\sc{iv}~$\lambda$1718}  & \rotatebox{90}{H$\eta$}  & \rotatebox{90}{H$\zeta$}  & \rotatebox{90}{H$\epsilon$}  & \rotatebox{90}{He~\sc{i-ii}~$\lambda$4026}  & \rotatebox{90}{N~\sc{iv}~$\lambda$4058}  & \rotatebox{90}{H$\delta$}  & \rotatebox{90}{He~\sc{ii}~$\lambda$4200}  & \rotatebox{90}{H$\gamma$}  & \rotatebox{90}{He~\sc{i}~$\lambda$4387}  & \rotatebox{90}{He~\sc{i}~$\lambda$4471}  & \rotatebox{90}{He~\sc{ii}~$\lambda$4541}  & \rotatebox{90}{N~\sc{iii}~$\lambda$4634-40-41}  & \rotatebox{90}{N~\sc{v}~$\lambda$4603-19}  & \rotatebox{90}{He~\sc{ii}~$\lambda$4686}  & \rotatebox{90}{He~\sc{i}~$\lambda$4713}  & \rotatebox{90}{H$\alpha$}  \\ \cline{1-27} \cline{29-55}  \noalign{\vskip\doublerulesep \vskip-\arrayrulewidth} \cline{1-27} \cline{29-55}
R136a1 & & x & x & x & x & x &  & x & x & x & x & x & x &  & x & x & x & x &  &  & x & x & x & x & x & x &  & H69 & & x & x & x & x & x &  & x & x &  & x & x & x & x &  & x & x & x & x & x & x &  &  & x & x & x \\ 
R136a2 & & x & x & x & x & x &  & x & x & x & x & x & x & x & x & x & x & x &  &  & x & x & x & x & x & x & & H70 & & x & x & x & x & x &  & x & x & o & x & x & x & x & x & x & x & x & x & x & x &  &  & x &  & x \\ 
R136a3 & & x & x & x & x & x &  & x & x & x & x & x & x & x & x & x & x & x &  &  & x & x & x & x & x & x & & H71 & & x & x & x & x & x &  & x & x & x & x & x & x & x &  & x & x & x & x & x & x &  &  & x & x & x \\ 
R136a4 & & x & x & x & x & x &  & x & x & x & x & x & x & x & x & x & x & x & x &  & x & x & x & x & x & x & & H75 & & x & x & x & x & x &  & x & x &  & x & x & x & x &  & x & x & x & x & x & x &  &  & x & x & x \\ 
R136a5 & &   & x & x & x & x &  & x & x & o & x & x & x & x & x & x & x & x & x & x & x & x & x & x & x & x & & H78 & & x & x & x & x & x &  & x & x &  & x & x & x & x & x & x & x & x & x & x & x & x &  & x & x &  \\ 
R136a6 & & x & x & x & x & x &  & x & x & x & x & x & x & x & x & x & x & x & x & x & x & x & x & x & x & x & & H80 & & x & x & x & x &  &  & x & x &  &  & x & x & x &  & x & x & x & x & x & x &  &  & x & x & x \\ 
R136a7 & & x & x & x & x & x &  & x & x & o & x & x & x & x & x & x & x & x & x & x & x & x & x & x & x & x & & H86 & & x & x & x & x & x &  & x & x & x & x & x & x & x & x & x & x & x & x & x & x & x & x & x &  & x \\ 
R136a8 & & x & x & x & x & x &  & x & x &  &  &  &  &  &  &  &  &  &  &  &  &  &  &  &  & x & & H90 & & x & x & x & x & x &  & x & x &  & x & x & x & x &  & x & x &  &  &  &  &  &  &  &  &  \\ 
R136b & & x & x & x & x &  & x & x & x & o & x & x & x & x & x & x & x & x & x & x & x & x &  & x & x & x & & H92 & & x & x & x & x &  &  & x & x &  & x & x & x & x &  & x & x & x & x & x & x &  &  & x &  &  \\ 
H30 & & x & x & x & x &  &  & x & x &  & x & x & x & x &  & x & x & x & x & x & x &  &  & x & x &  & & H94 & & x & x & x & x & x &  & x & x &  & x & x & x & x &  & x & x & x & x & x & x &  &  & x & x & x \\ 
H31 & & x & x & x & x & x &  & x & x &  & x & x & x & x & x & x & x & x & x & x & x &  &  & x &  & x & & H143 & & x & x & x & x &  &  & x & x & x & x & x & x & x &  & x & x & x & x & x & x &  &  & x & x &  \\ 
H35 & & x & x & x & x & x &  & x & x &  & x & x & x & x & x & x & x & x & x & x & x & x &  & x & x & x & & H73 & & x & x & x & x &  &  & x & x &  & x & x & x & x & x & x & x & x & x & x & x &  &  & x &  & x \\ 
H36 & & x & x & x & x & x &  & x & x & o & x & x & x & x & x & x & x & x & x & x & x & x & x & x & x & x & & H108 & & x & x & x & x &  &  & x & x &  & x & x & x & x & x & x & x & x &  &  & x &  &  & x &  & x \\ 
H40 & & x & x & x & x & x &  & x & x & x & x & x & x & x & x & x & x & x & x &  & x &  &  & x &  & x & & H112 & & x & x & x & x &  &  & x & x &  &  & x & x & x &  & x & x & x & x & x & x &  &  & x & x &  \\ 
H45 & & x & x & x & x & x &  & x & x & x & x & x & x & x &  & x & x & x & x & x & x &  &  & x & x & x & & H114 & & x & x & x & x & x &  & x & x &  & x & x & x & x & x & x & x & x & x & x & x &  &  & x &  & x \\ 
H46 & & x & x & x & x & x &  & x & x & o &  &  &  &  & x & x & x & x & x &  & x & x & x & x & x & x & & H116 & & x & x & x & x &  &  & x & x &  & x & x & x & x &  & x & x & x & x & x & x &  &  & x & x & x \\ 
H47 & & x & x & x & x & x &  & x & x & o & x & x & x & x & x & x & x & x &  & x & x &  & x & x & x & x & & H120 & & x & x & x & x &  &  & x & x &  & x & x & x & x &  & x & x & x & x & x & x &  &  & x & x & x \\ 
H48 & & x & x & x & x & x &  & x & x & o & x & x & x & x & x & x & x & x & x & x & x & x & x & x & x & x & & H121 & & x & x & x & x &  &  & x & x &  & x & x & x & x &  & x & x & x & x & x & x &  &  & x &  &  \\ 
H49 & &  &  & x & x & x &  & x & x &  & x & x & x & x &  & x & x & x &  &  &  &  &  & x & x & x & & H123 & & x & x & x & x &  &  & x & x &  & x & x & x & x &  & x & x & x & x & x & x &  &  & x & x & x \\ 
H50 & & x & x & x & x & x &  & x & x & o & x & x & x & x & x & x & x & x & x & x & x &  &  & x & x & x & & H129 & & x & x & x & x & x &  & x & x &  & x & x & x & x &  &  & x & x & x & x & x &  &  & x &  &  \\ 
H52 & & x & x & x & x & x &  & x & x &  & x & x & x & x & x & x & x & x & x & x & x & x & x & x &  & x & & H132 & & x & x & x & x &  &  & x & x &  & x & x & x & x &  & x & x & x & x & x & x &  &  & x &  & x \\ 
H55 & & x & x & x & x & x &  & x & x & o & x & x & x & x & x & x & x & x & x & x & x & x &  & x & x & x & & H134 & & x & x & x & x &  &  & x & x &  & x & x & x & x & x & x & x & x & x & x & x & x &  & x & x & x \\ 
H58 & & x & x & x & x & x &  & x & x & o & x & x & x & x &  & x & x & x & x & x & x &  & x & x &  &  & & H135 & & x & x & x & x &  &  & x & x &  & x & x & x & x &  & x & x & x & x & x & x &  &  &  &  & x \\ 
H62 & & x & x & x & x & x &  & x & x &  & x & x & x & x & x & x & x & x & x & x & x & x &  & x &  & x & & H139 & & x & x & x & x &  &  & x & x &  & x & x & x & x &  & x & x & x & x & x & x &  &  & x &  & x \\ 
H64 & & x & x & x & x & x &  & x & x &  & x & x & x & x &  & x & x & x & x & x & x &  &  & x &  & x & & H141 & &  & x & x & x &  &  & x & x &  &  & x & x & x &  &  &  & x & x & x & x &  &  & x &  & x \\ 
H65 & & x & x & x & x & x &  & x & x &  & x & x & x & x &  & x & x & x & x & x & x &  &  & x &  & x & & H159 & & x & x & x & x &  &  & x & x &  & x & x & x & x &  &  &  & x & x & x & x &  &  & x & x &  \\ 
H66 & & x & x & x & x & x &  & x & x &  & x & x & x & x & x & x & x & x & x & x & x &  & x & x &  & x & & H162 & & x & x & x & x &  &  & x & x &  & x & x & x & x &  & x & x & x & x & x & x &  &  & x & x & x \\ 
H68 & & x & x & x & x & x &  & x & x & x & x & x & x & x &  & x & x & x & x & x & x &  &  & x &  &  & & H173 & & x & x & x & x &  &  & x & x &  & x & x & x & x &  & x & x & x & x & x &  &  &  & x &  & x \\ 
\cline{1-27} \cline{29-55}
\end{tabular}

\end{sidewaystable*}

\newpage
\clearpage

\section{\pyGA output summaries \label{sec:app:GA_summaries}}

\begin{figure*}
    \centering
    \includegraphics[width=0.95\textwidth]{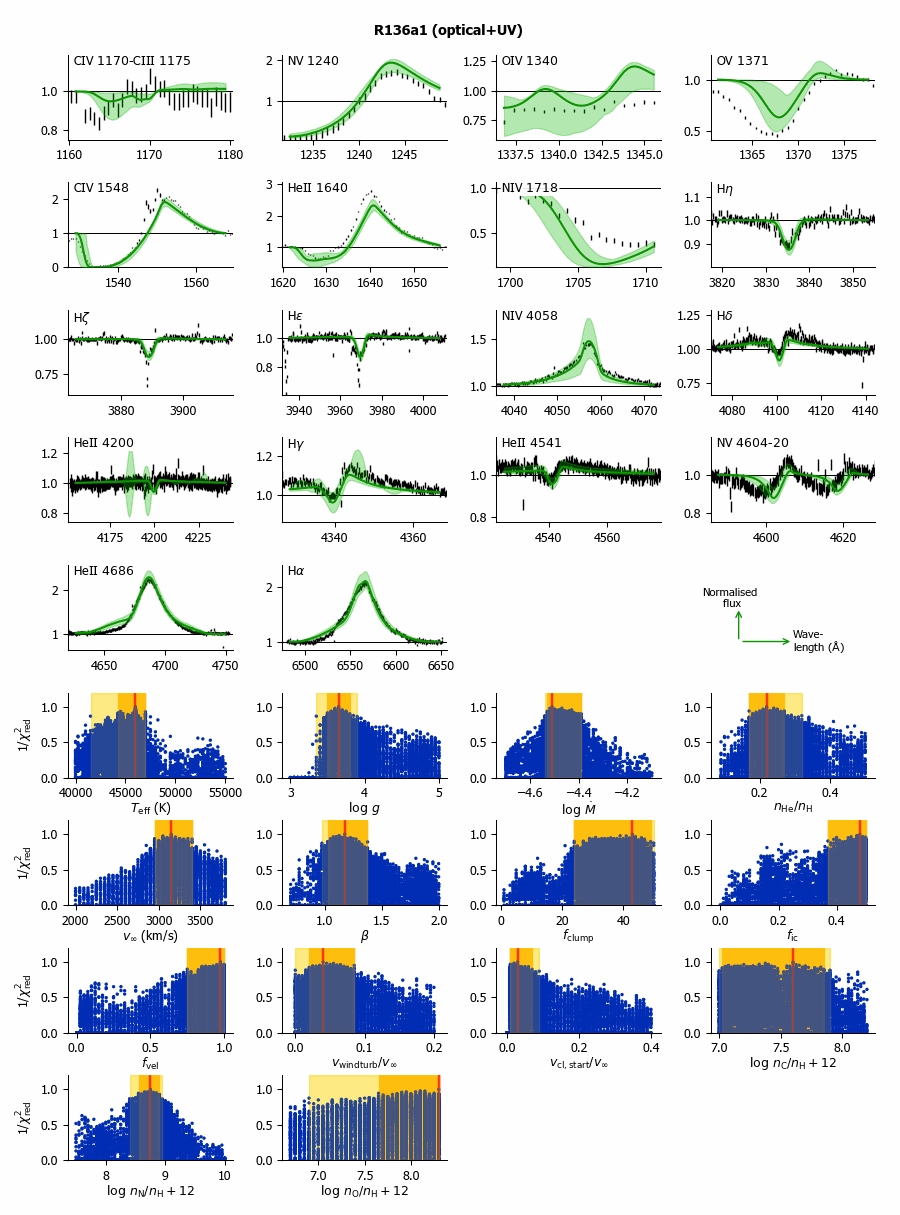}
    \caption{\PyGA output summary for the optical+UV run of R136a1 (as \Cref{fig:fitspec_example_H35}).}
    \label{fig:outGA_R136a1_UV}
\end{figure*}

\begin{figure*}
    \centering
    \includegraphics[width=0.95\textwidth]{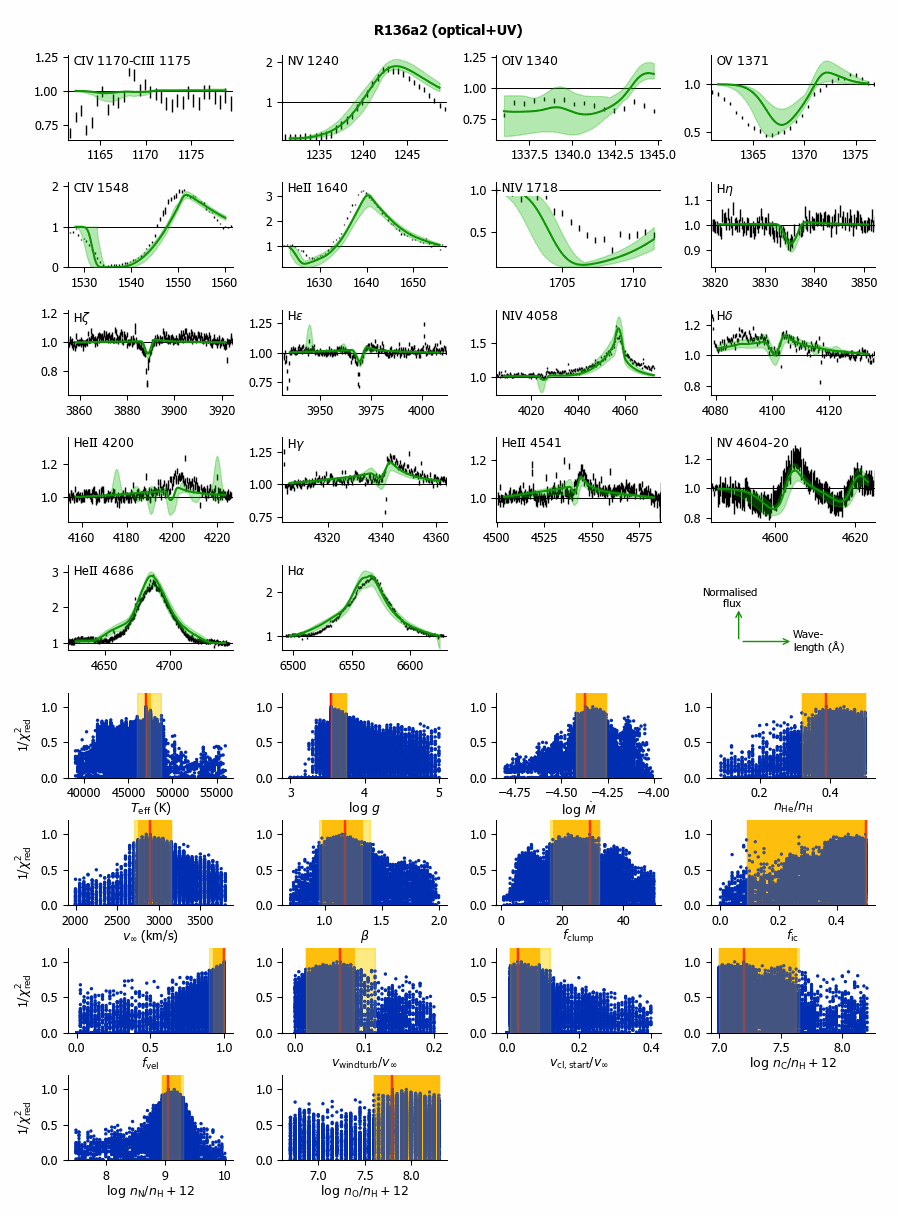}
    \caption{\PyGA output summary for the optical+UV run of R136a2 (as \Cref{fig:fitspec_example_H35}).}
    \label{fig:outGA_R136a2_UV}
\end{figure*}

\begin{figure*}
    \centering
    \includegraphics[width=0.95\textwidth]{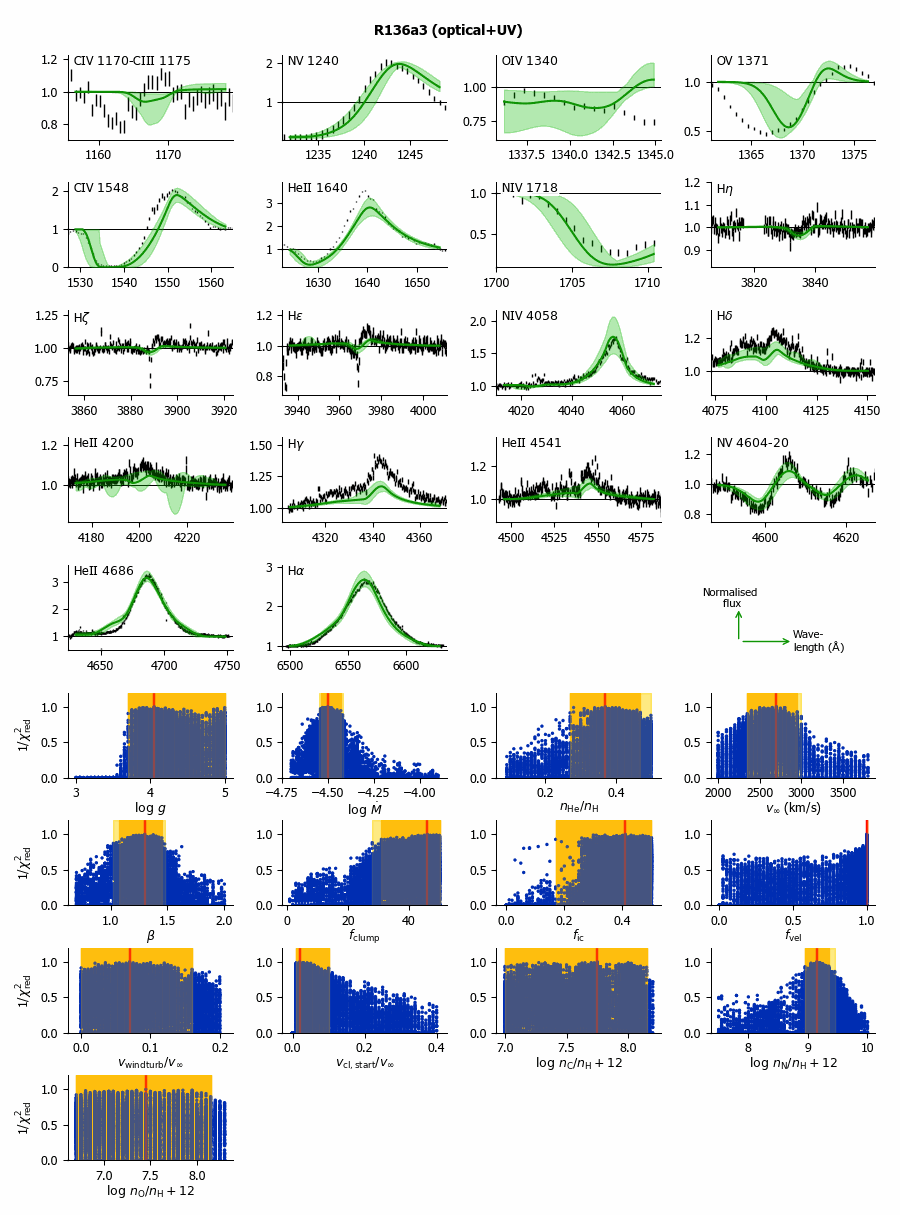}
    \caption{\PyGA output summary for the optical+UV run of R136a3 (as \Cref{fig:fitspec_example_H35}). For this model, the \teff was fixed to 50000~K. See \Cref{sec:app:a3_analysis} for more details and a comparison of the 50000~K and the 42000~K models.}
    \label{fig:outGA_R136a3_UV}
\end{figure*}

\begin{figure*}
    \centering
    \includegraphics[width=0.95\textwidth]{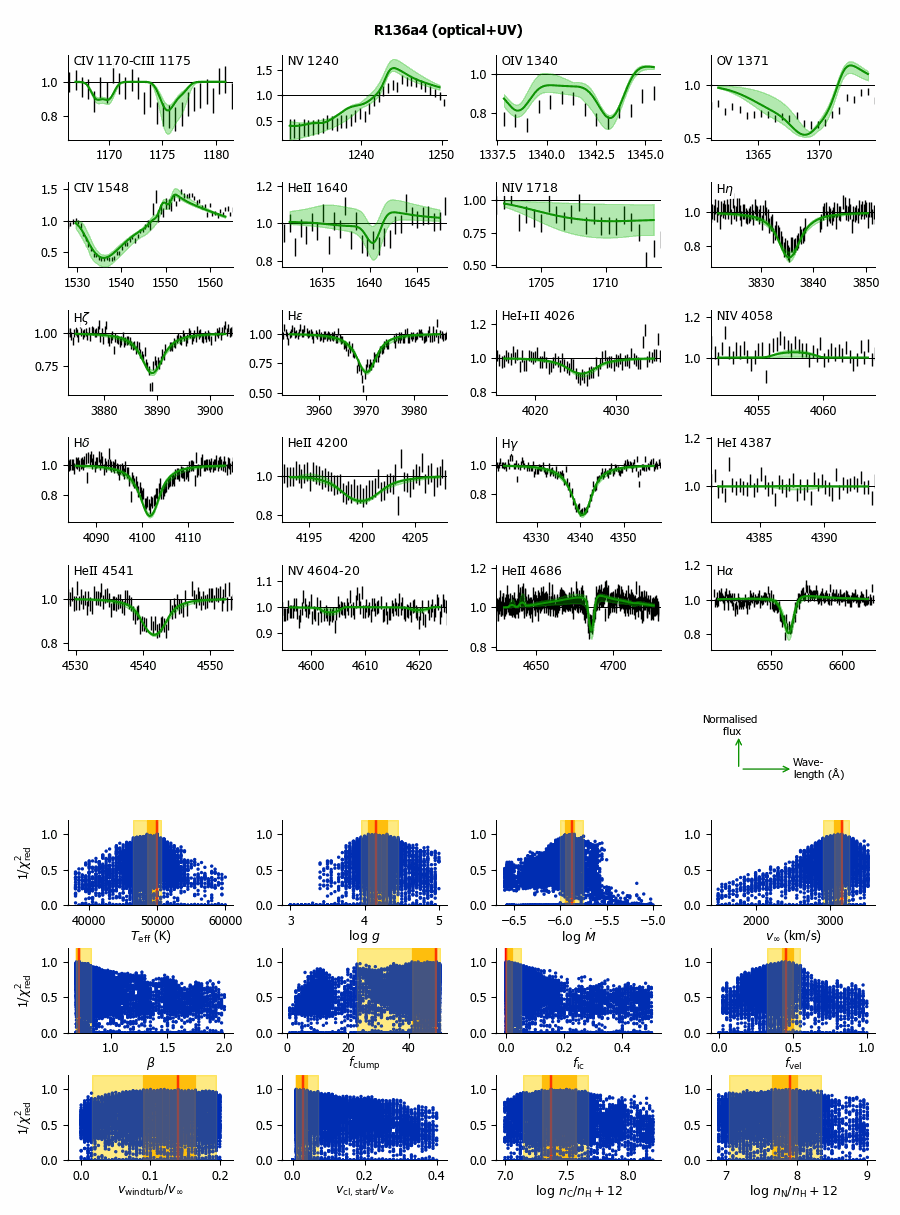}
    \caption{\PyGA output summary for the optical+UV run of R136a4 (as \Cref{fig:fitspec_example_H35}).}
    \label{fig:outGA_R136a4_UV}
\end{figure*}

\begin{figure*}
    \centering
    \includegraphics[width=0.95\textwidth]{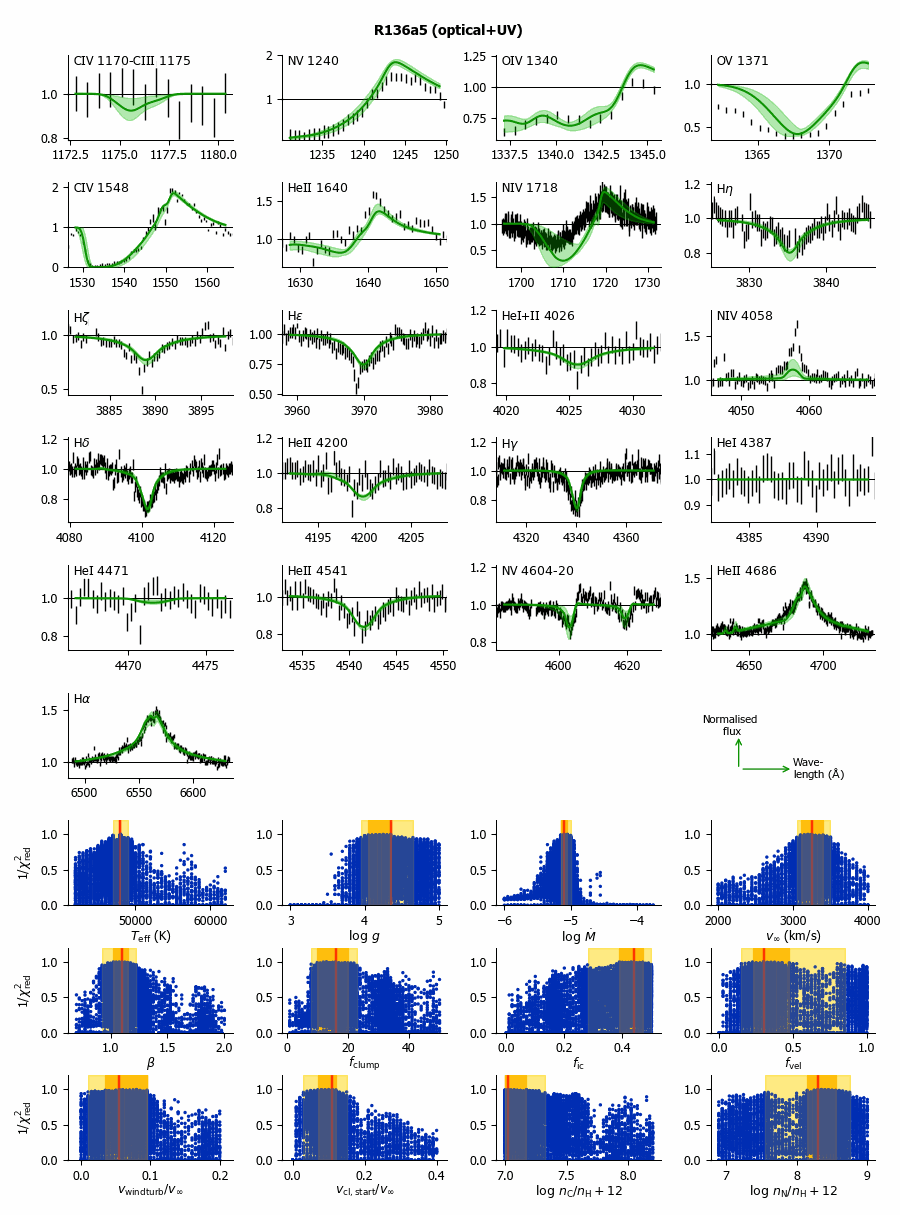}
    \caption{\PyGA output summary for the optical+UV run of R136a5 (as \Cref{fig:fitspec_example_H35}).}
    \label{fig:outGA_R136a5_UV}
\end{figure*}

\begin{figure*}
    \centering
    \includegraphics[width=0.95\textwidth]{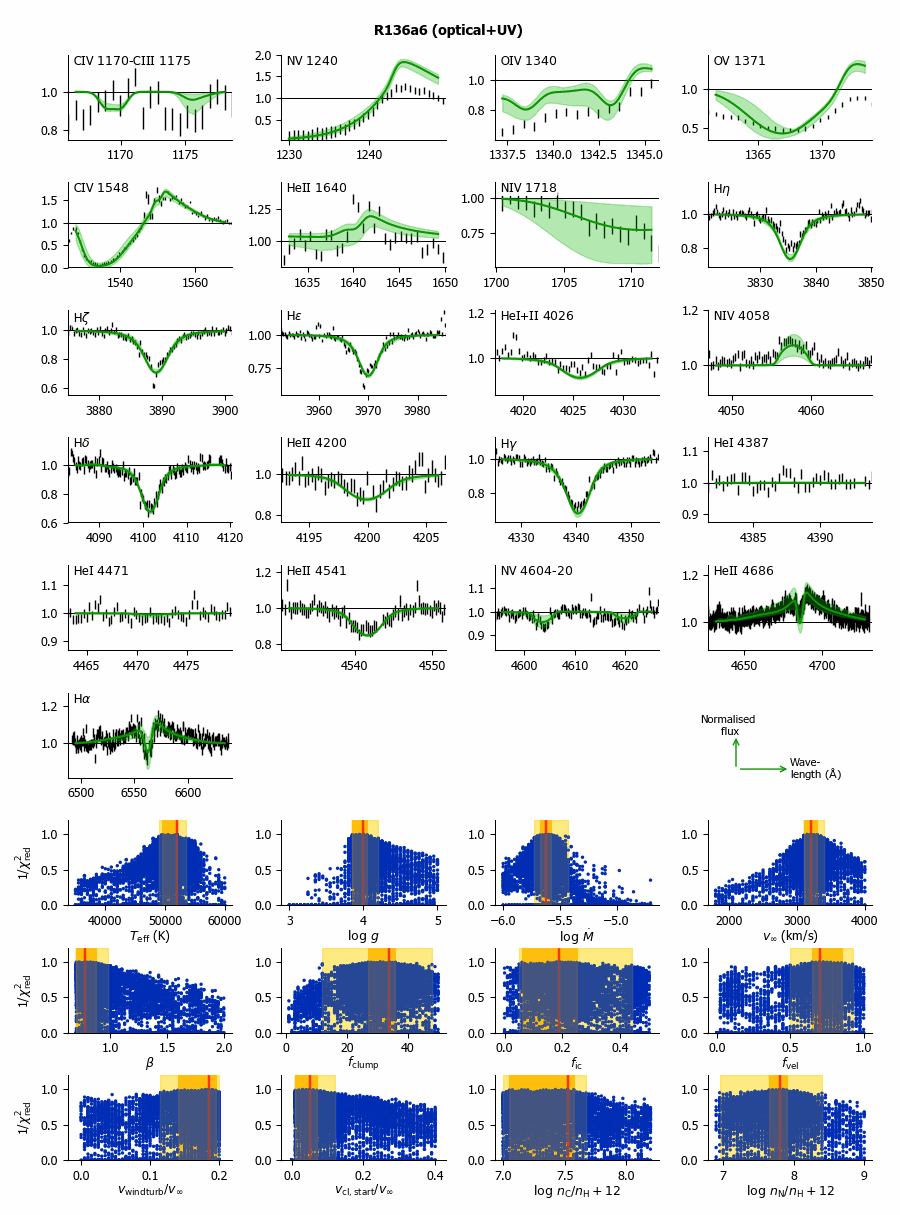}
    \caption{\PyGA output summary for the optical+UV run of R136a6 (as \Cref{fig:fitspec_example_H35}).}
    \label{fig:outGA_R136a6_UV}
\end{figure*}

\begin{figure*}
    \centering
    \includegraphics[width=0.95\textwidth]{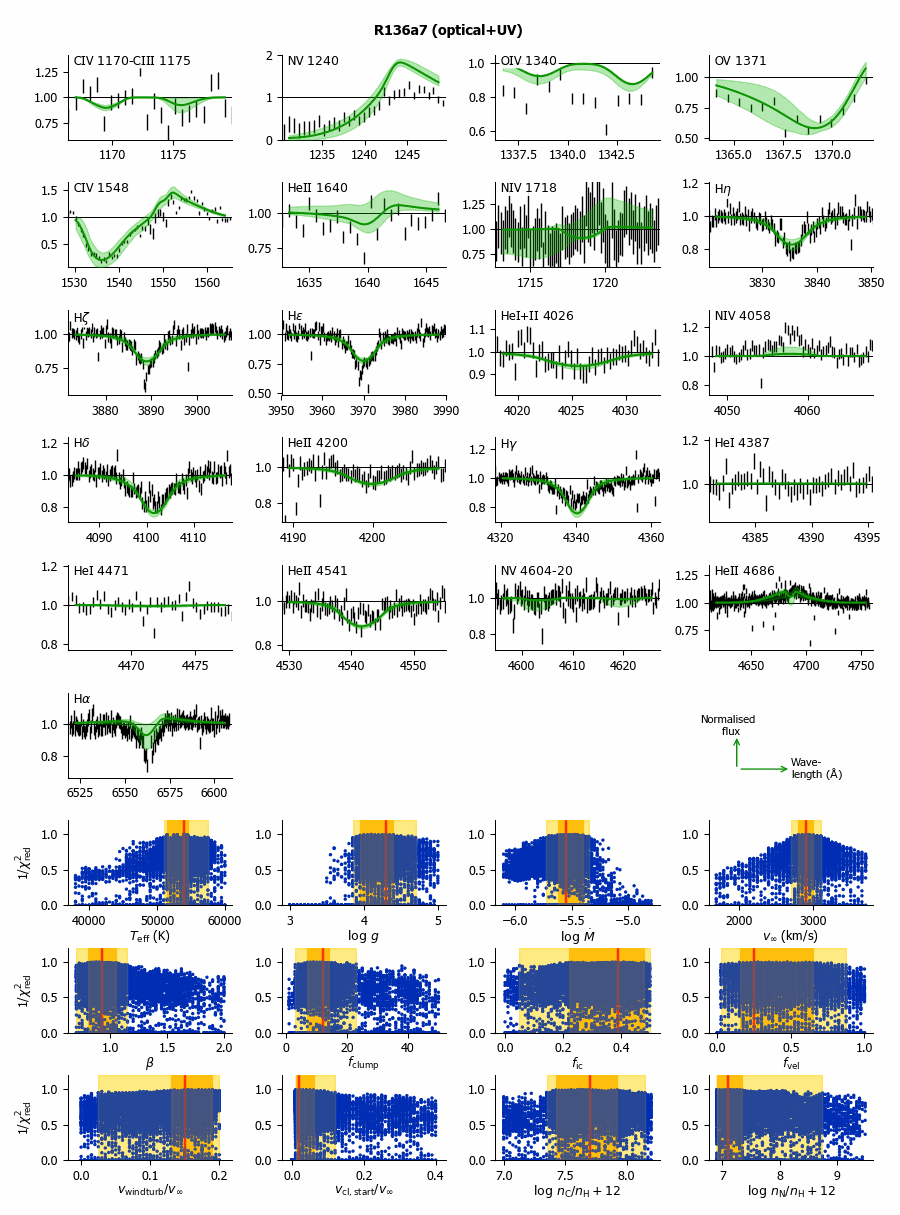}
    \caption{\PyGA output summary for the optical+UV run of R136a7 (as \Cref{fig:fitspec_example_H35}).}
    \label{fig:outGA_R136a7_UV}
\end{figure*}

\begin{figure*}
    \centering
    \includegraphics[width=0.95\textwidth]{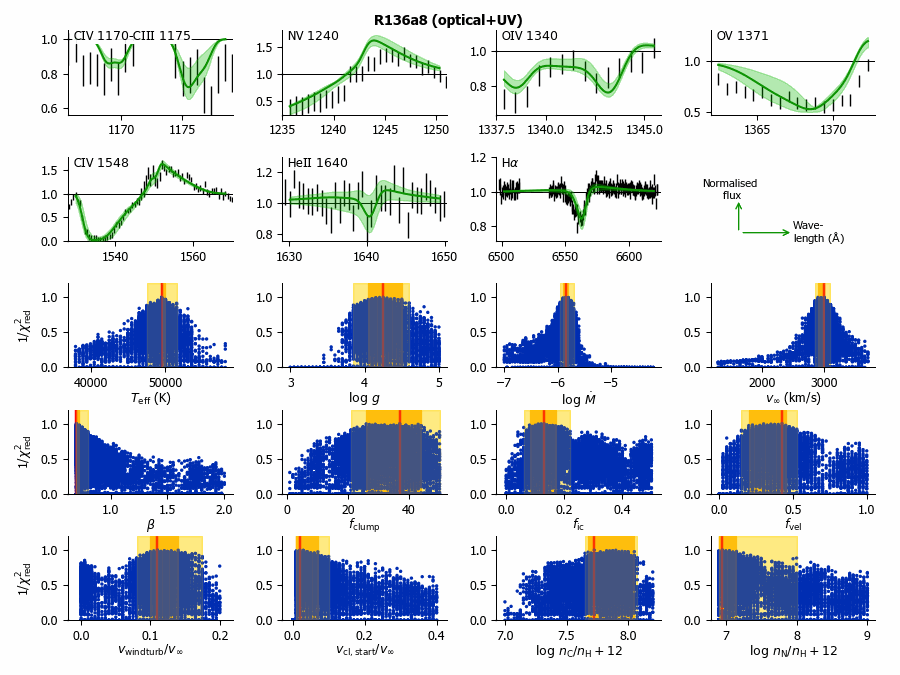}
    \caption{\PyGA output summary for the optical+UV run of R136a8 (as \Cref{fig:fitspec_example_H35}).}
    \label{fig:outGA_R136a8_UV}
\end{figure*}

\begin{figure*}
    \centering
    \includegraphics[width=0.95\textwidth]{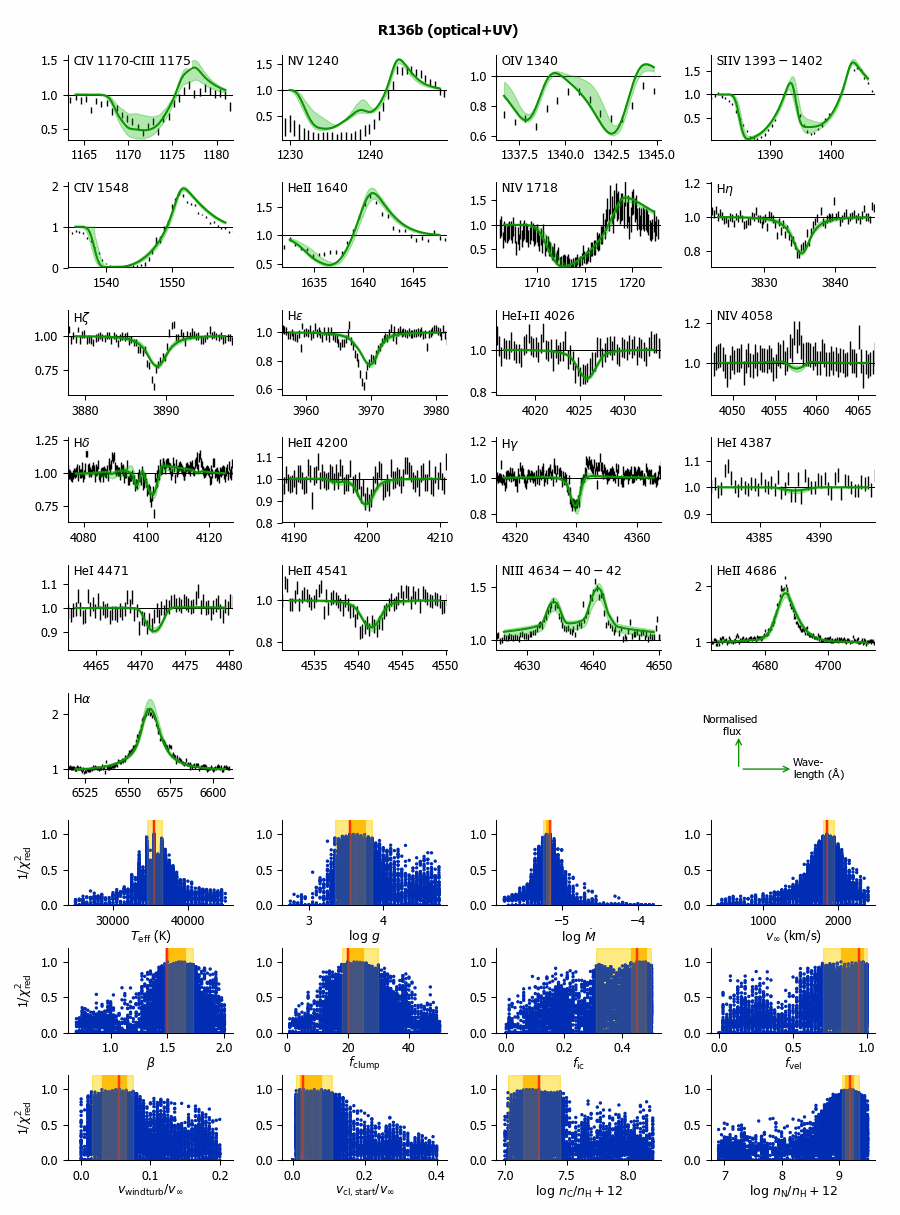}
    \caption{\PyGA output summary for the optical+UV run of R136b (as \Cref{fig:fitspec_example_H35}).}
    \label{fig:outGA_R136b_UV}
\end{figure*}

\begin{figure*}
    \centering
    \includegraphics[width=0.95\textwidth]{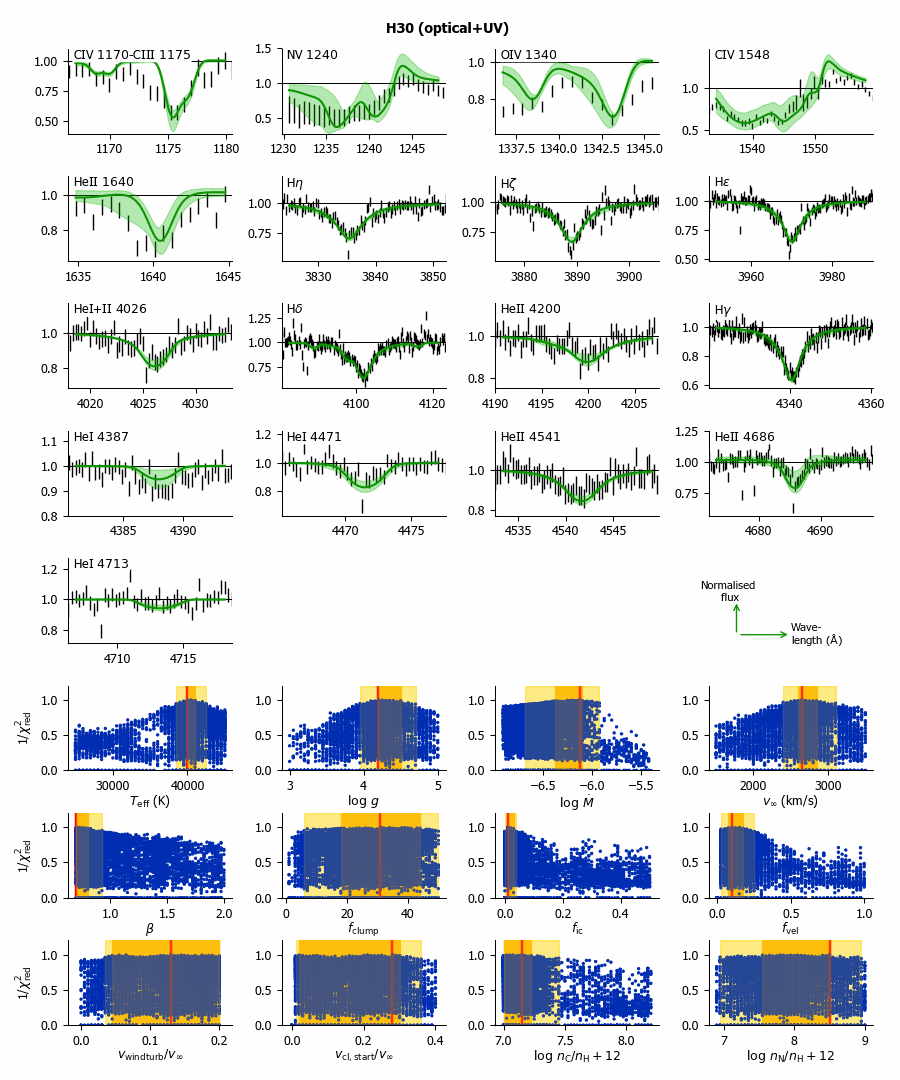}
    \caption{\PyGA output summary for the optical+UV run of H30 (as \Cref{fig:fitspec_example_H35}).}
    \label{fig:outGA_H30_UV}
\end{figure*}

\begin{figure*}
    \centering
    \includegraphics[width=0.95\textwidth]{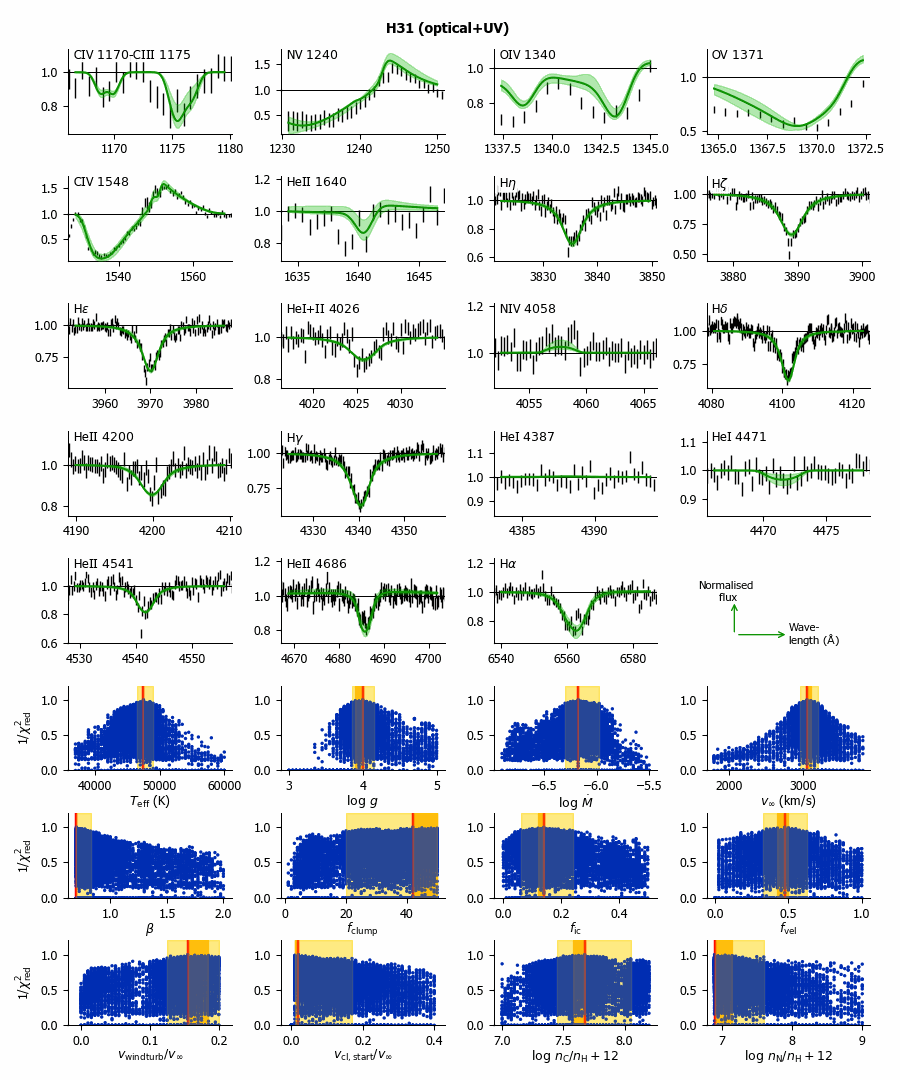}
    \caption{\PyGA output summary for the optical+UV run of H31 (as \Cref{fig:fitspec_example_H35}).}
    \label{fig:outGA_H31_UV}
\end{figure*}

\begin{figure*}
    \centering
    \includegraphics[width=0.95\textwidth]{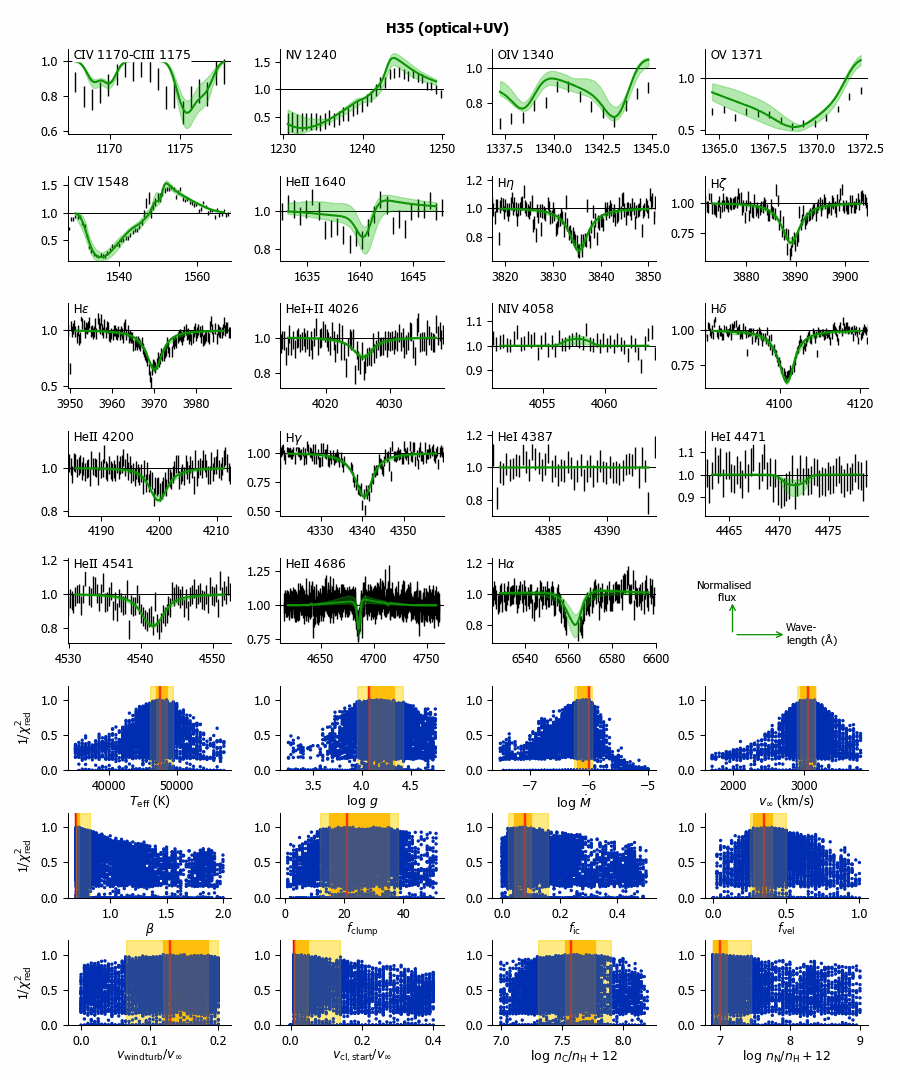}
    \caption{\PyGA output summary for the optical+UV run of H35 (as \Cref{fig:fitspec_example_H35}).}
    \label{fig:outGA_H35_UV}
\end{figure*}

\begin{figure*}
    \centering
    \includegraphics[width=0.95\textwidth]{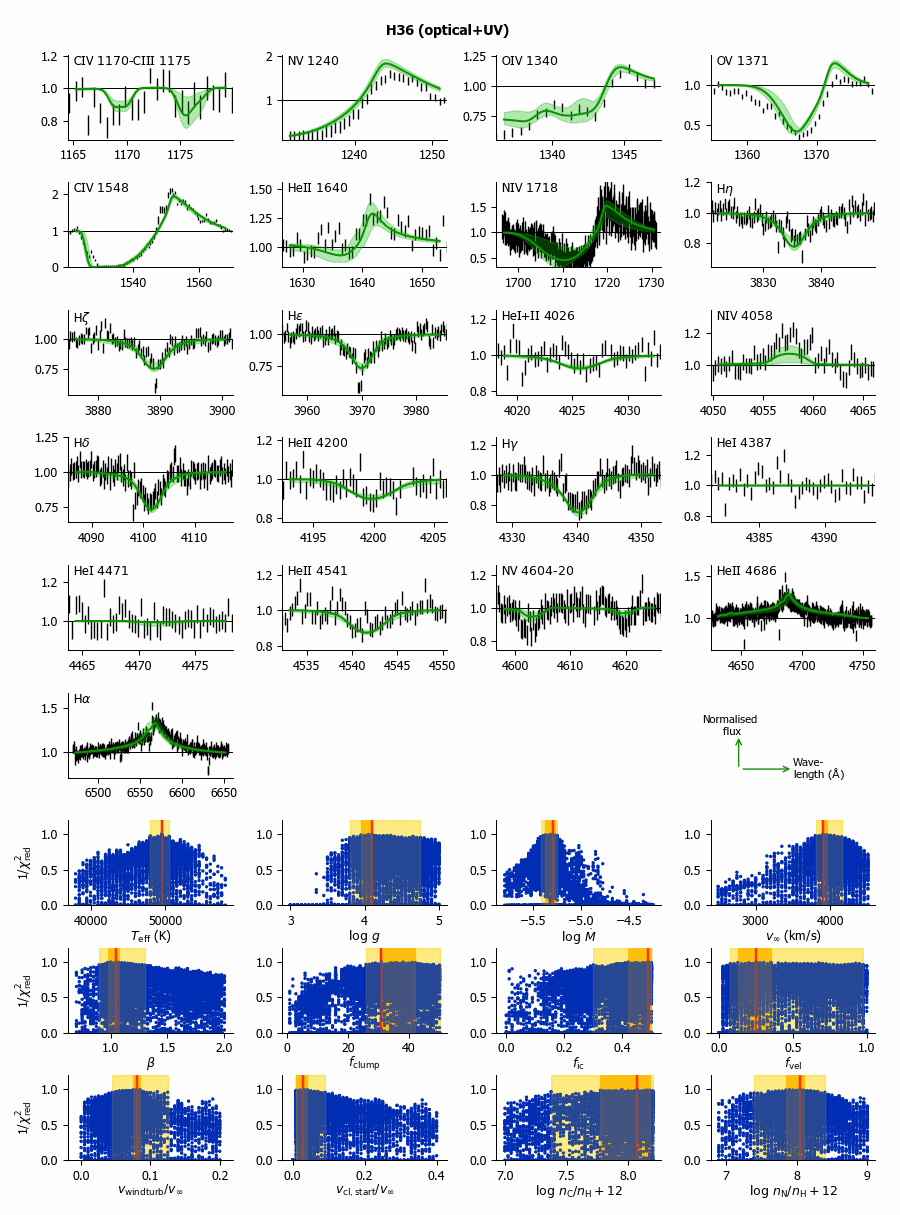}
    \caption{\PyGA output summary for the optical+UV run of H36 (as \Cref{fig:fitspec_example_H35}).}
    \label{fig:outGA_H36_UV}
\end{figure*}

\begin{figure*}
    \centering
    \includegraphics[width=0.95\textwidth]{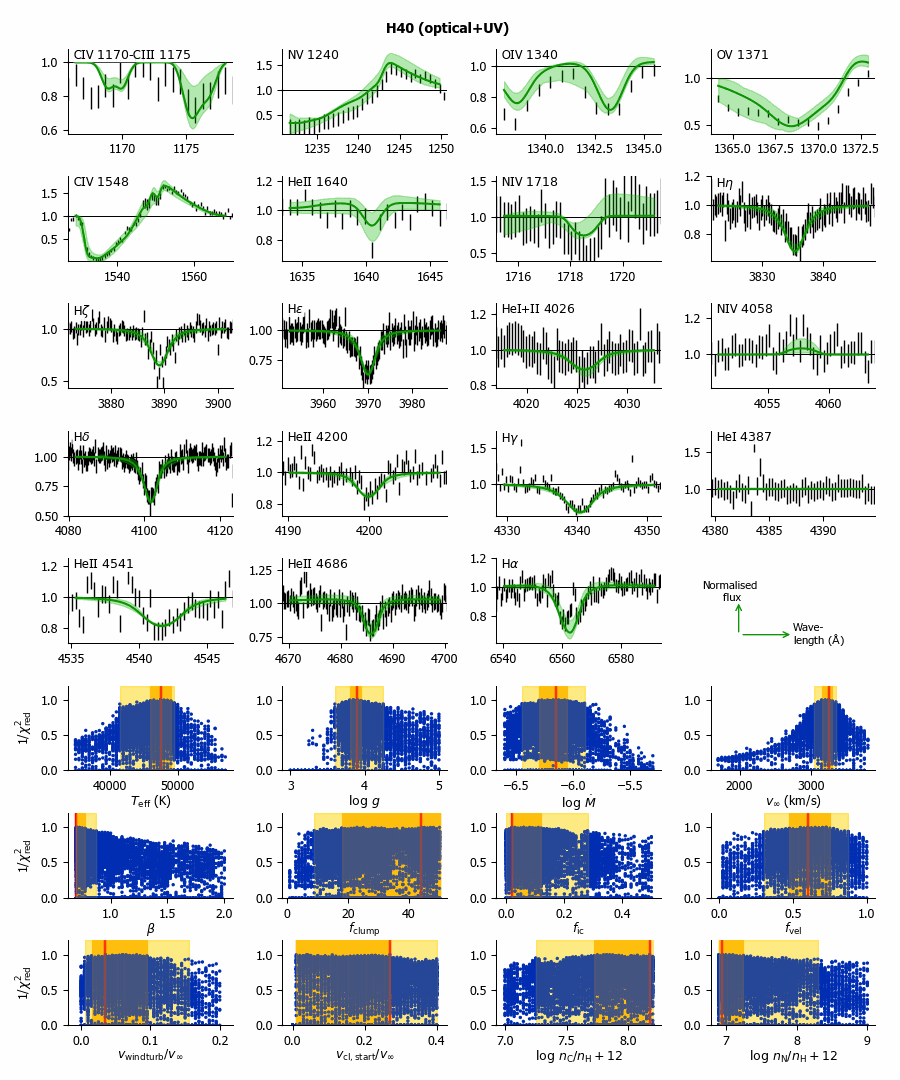}
    \caption{\PyGA output summary for the optical+UV run of H40 (as \Cref{fig:fitspec_example_H35}).}
    \label{fig:outGA_H40_UV}
\end{figure*}

\begin{figure*}
    \centering
    \includegraphics[width=0.95\textwidth]{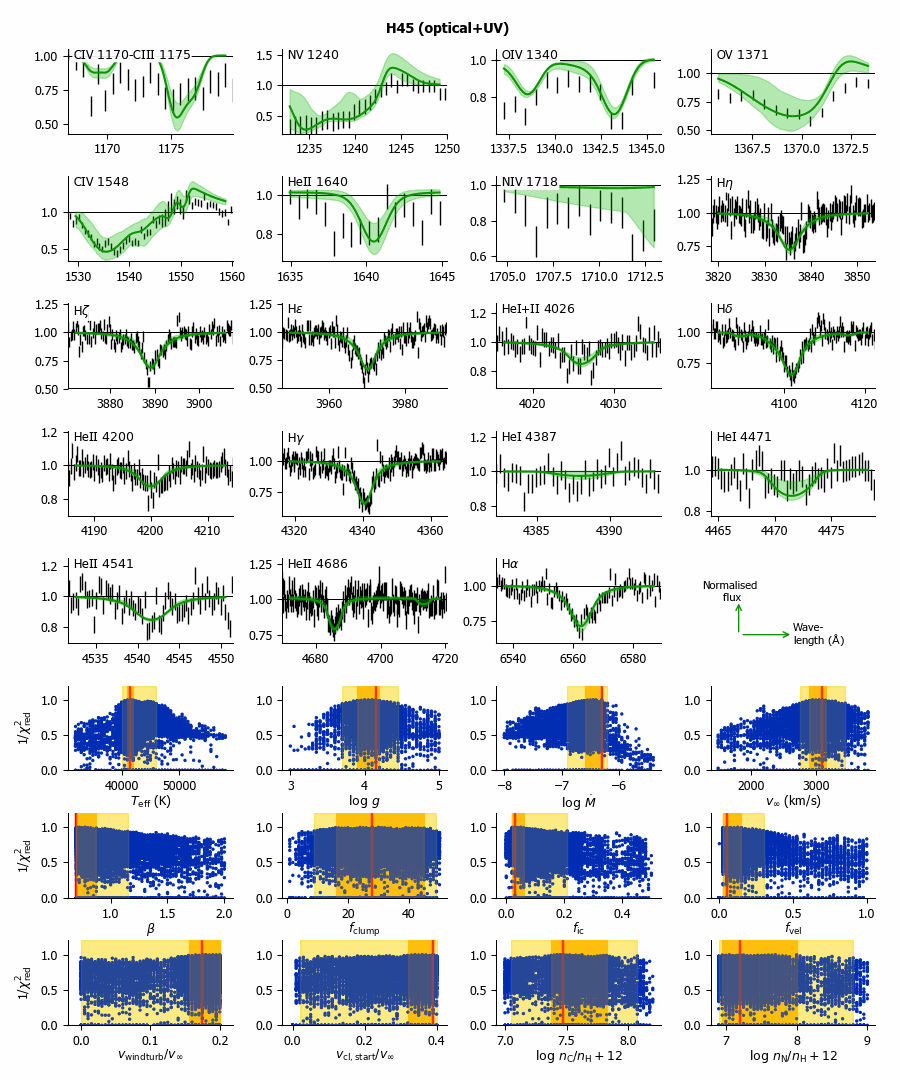}
    \caption{\PyGA output summary for the optical+UV run of H45 (as \Cref{fig:fitspec_example_H35}).}
    \label{fig:outGA_H45_UV}
\end{figure*}

\begin{figure*}
    \centering
    \includegraphics[width=0.95\textwidth]{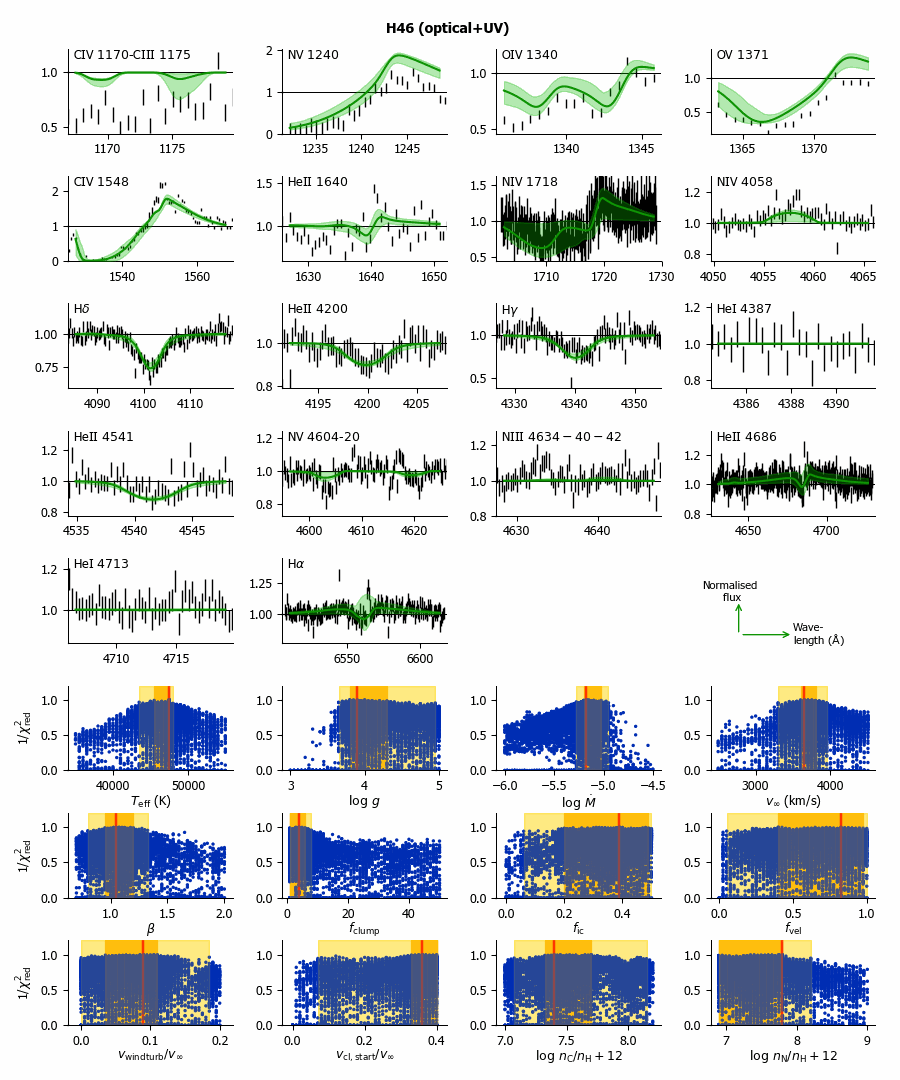}
    \caption{\PyGA output summary for the optical+UV run of H46 (as \Cref{fig:fitspec_example_H35}).}
    \label{fig:outGA_H46_UV}
\end{figure*}

\begin{figure*}
    \centering
    \includegraphics[width=0.95\textwidth]{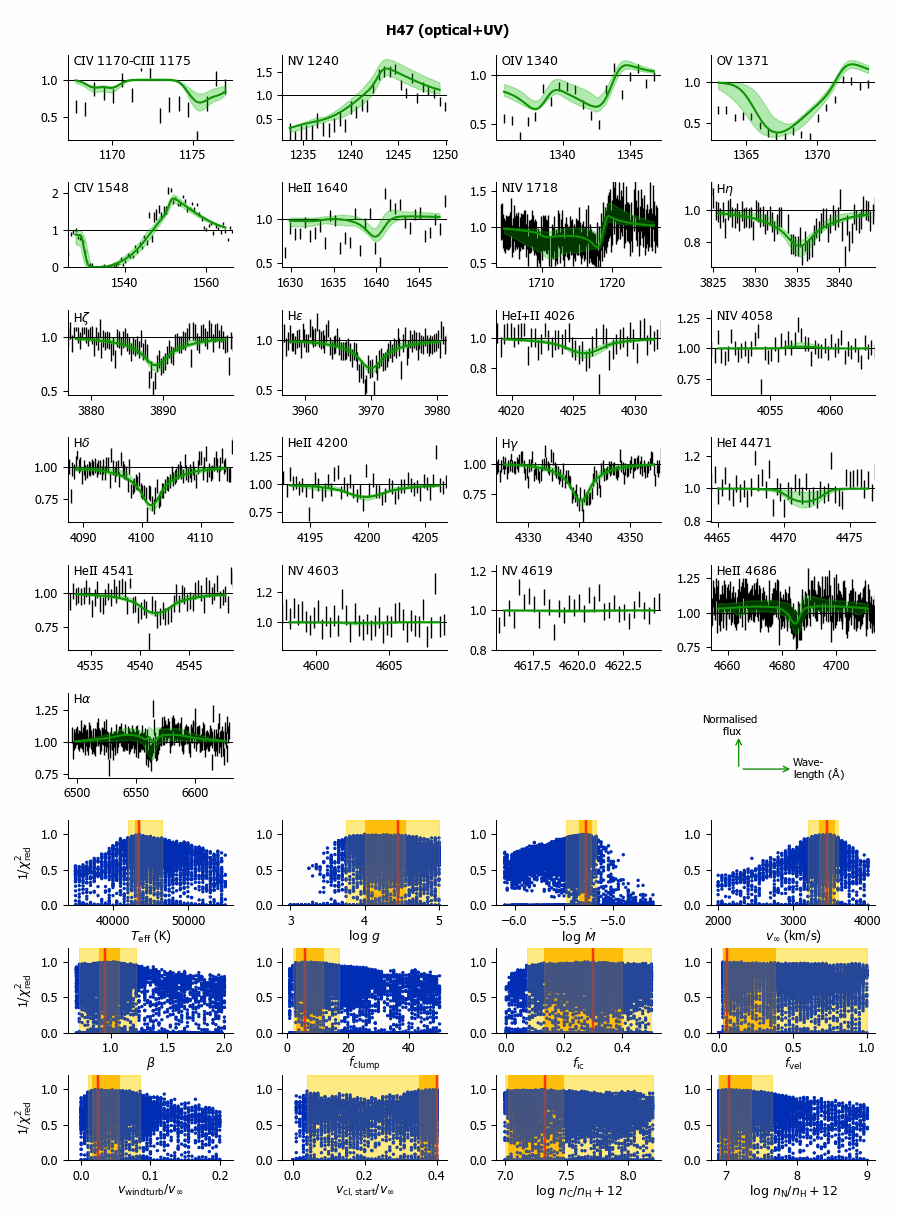}
    \caption{\PyGA output summary for the optical+UV run of H47 (as \Cref{fig:fitspec_example_H35}).}
    \label{fig:outGA_H47_UV}
\end{figure*}

\begin{figure*}
    \centering
    \includegraphics[width=0.95\textwidth]{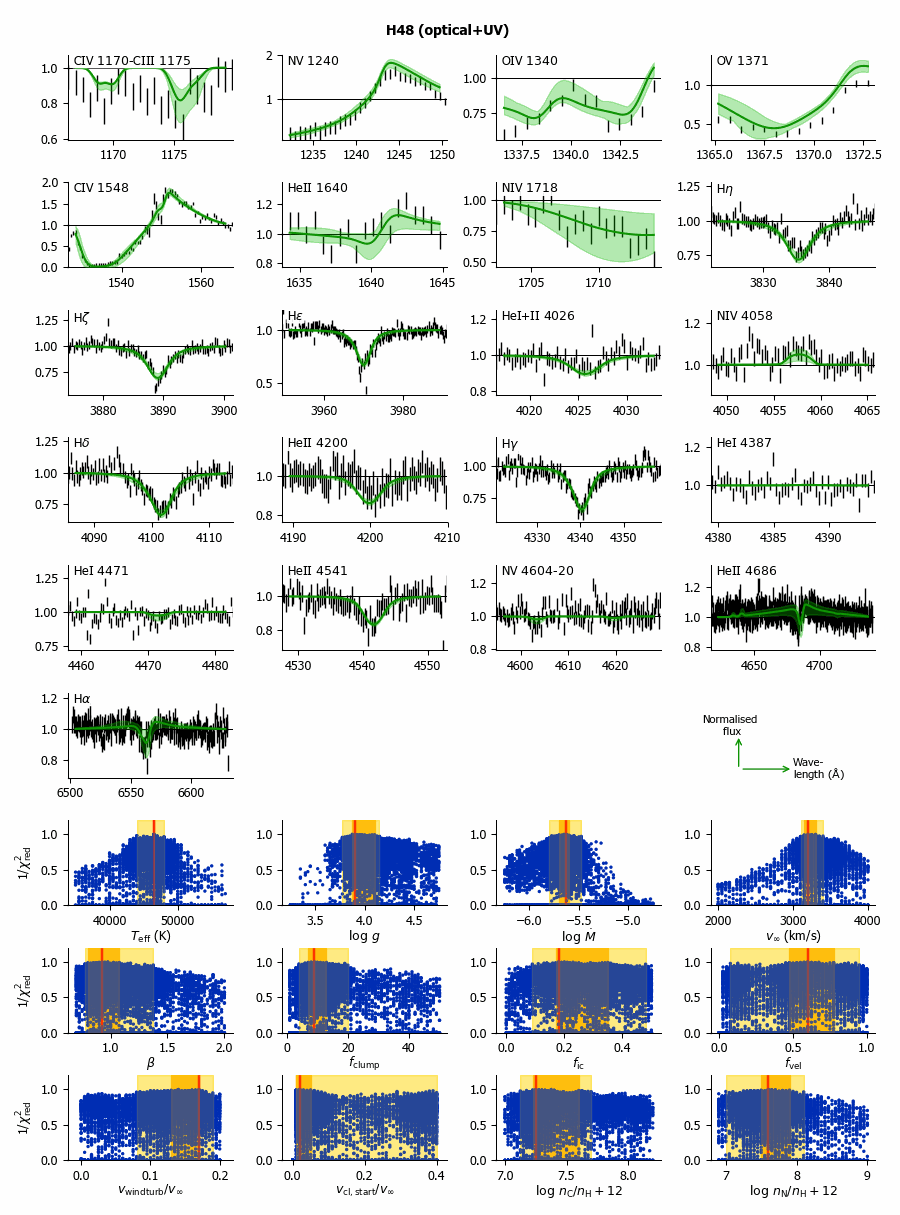}
    \caption{\PyGA output summary for the optical+UV run of H48 (as \Cref{fig:fitspec_example_H35}).}
    \label{fig:outGA_H48_UV}
\end{figure*}

\begin{figure*}
    \centering
    \includegraphics[width=0.95\textwidth]{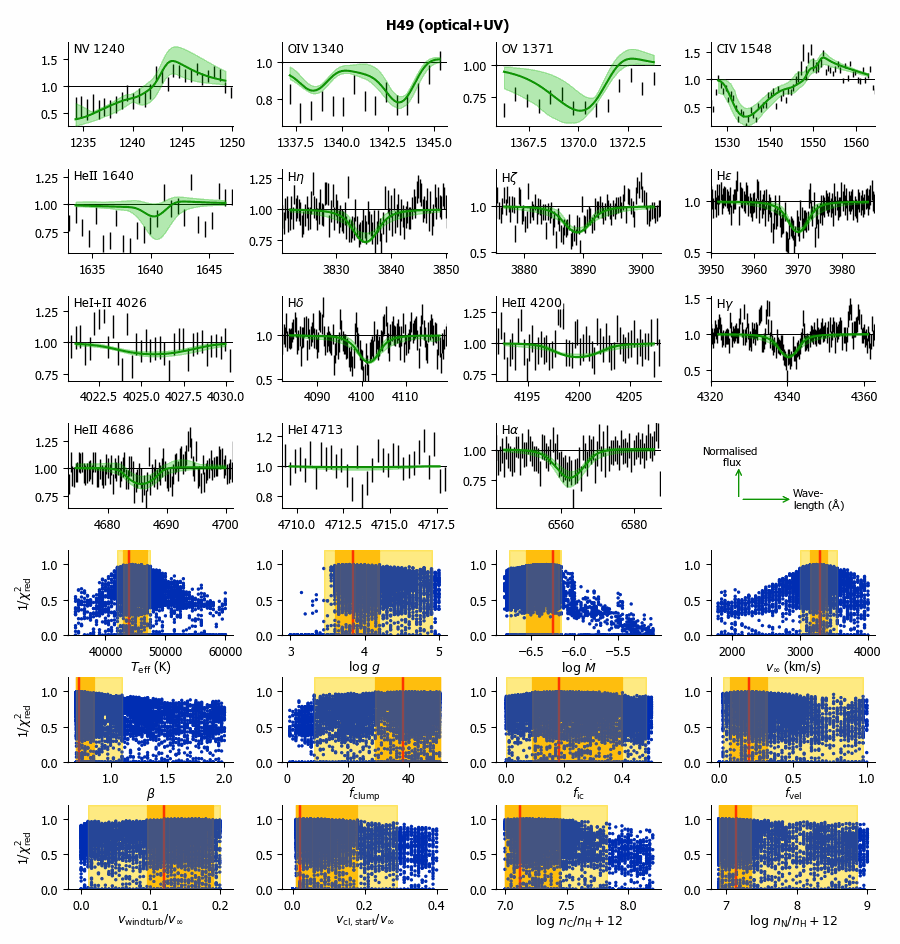}
    \caption{\PyGA output summary for the optical+UV run of H49 (as \Cref{fig:fitspec_example_H35}).}
    \label{fig:outGA_H49_UV}
\end{figure*}

\begin{figure*}
    \centering
    \includegraphics[width=0.95\textwidth]{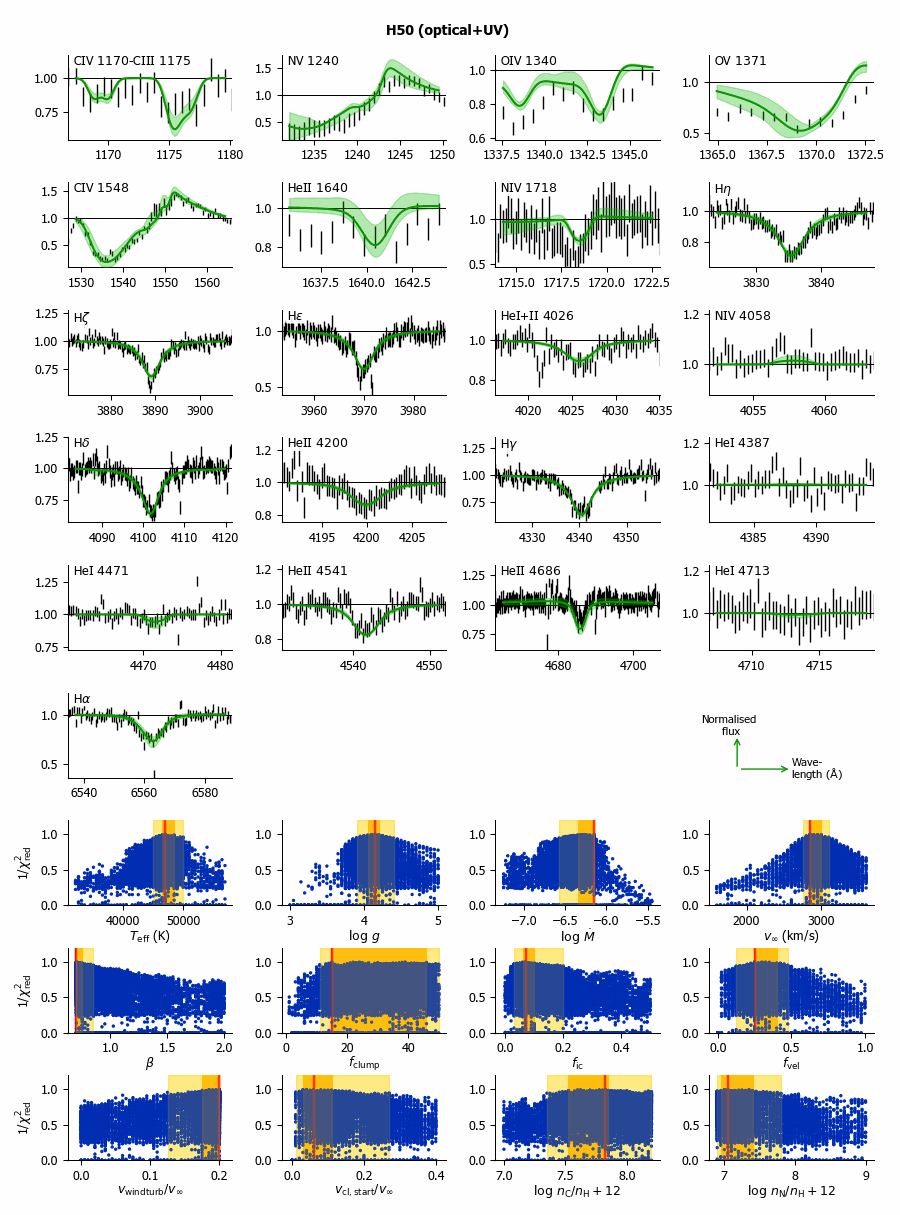}
    \caption{\PyGA output summary for the optical+UV run of H50 (as \Cref{fig:fitspec_example_H35}).}
    \label{fig:outGA_H50_UV}
\end{figure*}

\begin{figure*}
    \centering
    \includegraphics[width=0.95\textwidth]{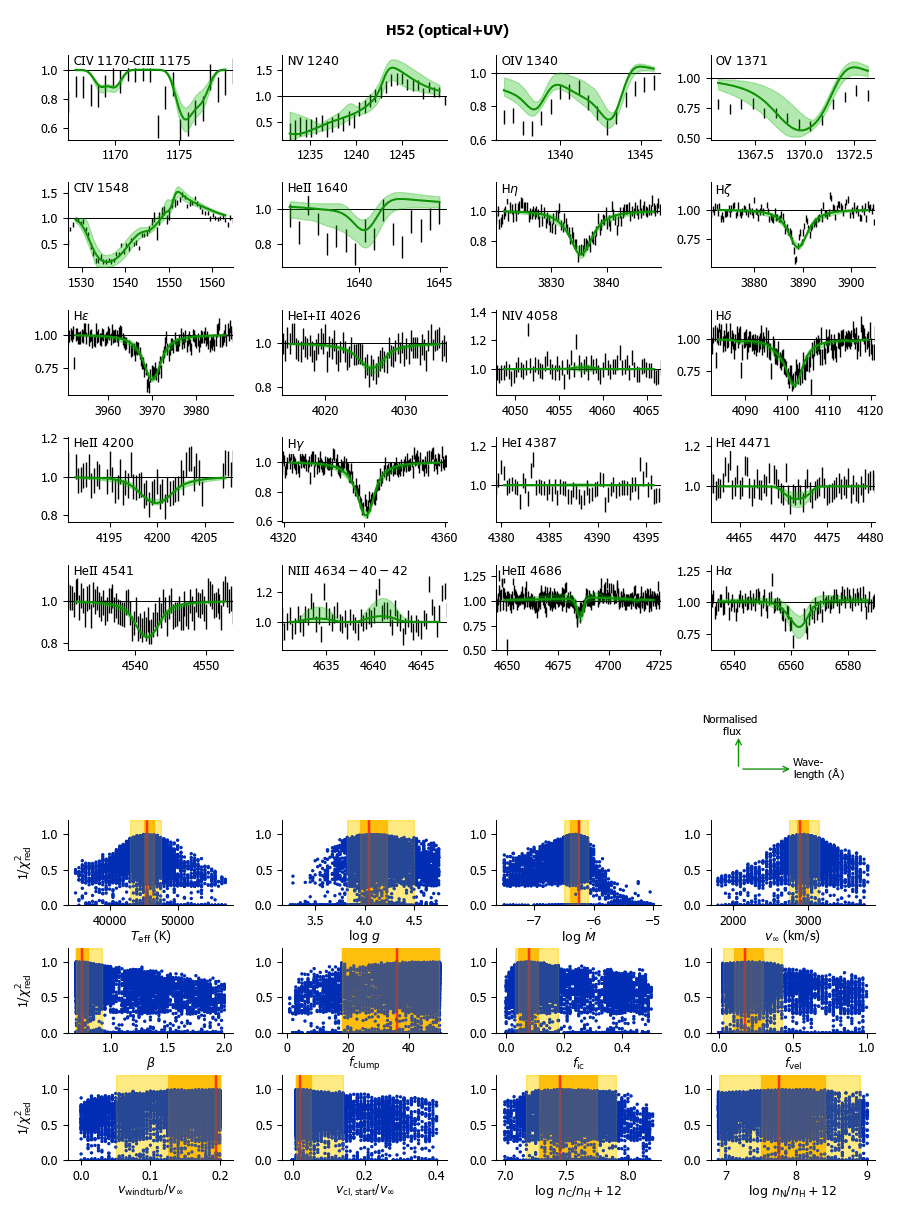}
    \caption{\PyGA output summary for the optical+UV run of H52 (as \Cref{fig:fitspec_example_H35}).}
    \label{fig:outGA_H52_UV}
\end{figure*}

\begin{figure*}
    \centering
    \includegraphics[width=0.95\textwidth]{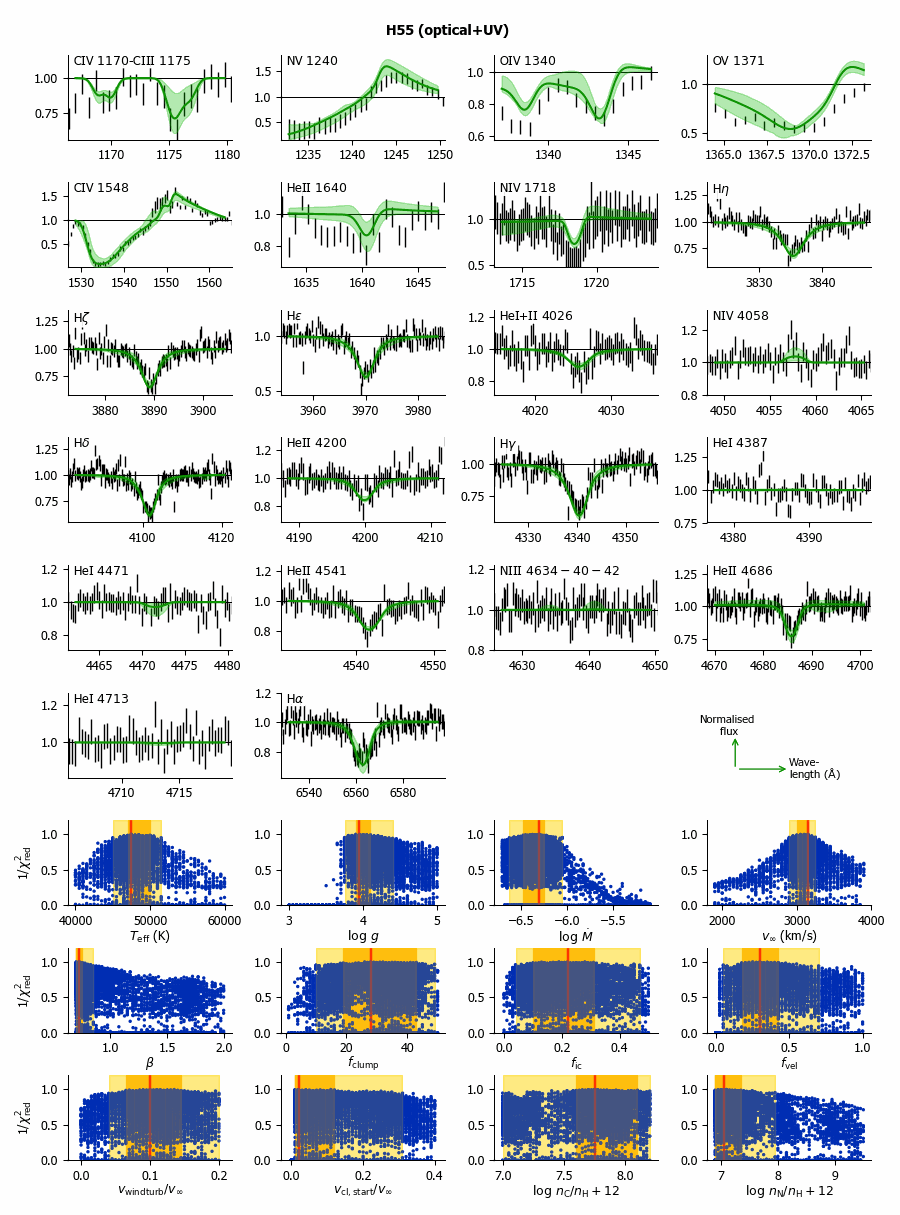}
    \caption{\PyGA output summary for the optical+UV run of H55 (as \Cref{fig:fitspec_example_H35}).}
    \label{fig:outGA_H55_UV}
\end{figure*}

\begin{figure*}
    \centering
    \includegraphics[width=0.95\textwidth]{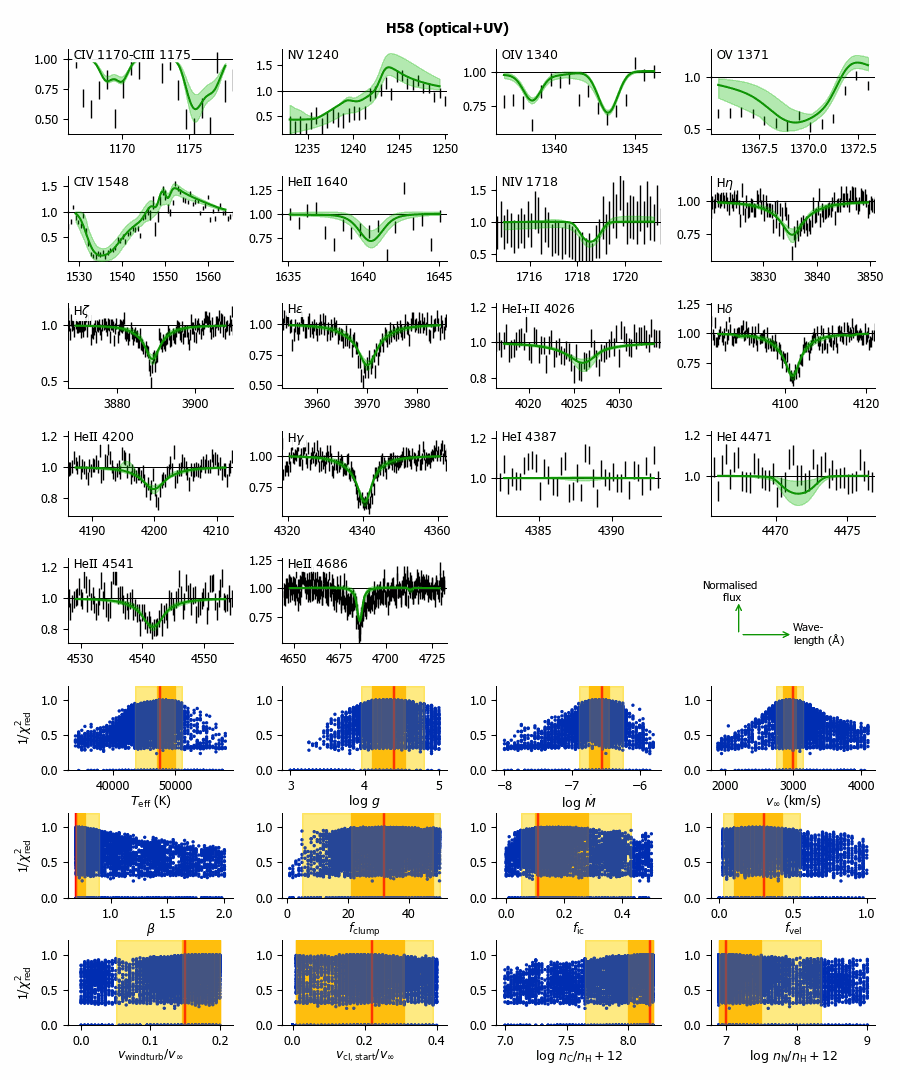}
    \caption{\PyGA output summary for the optical+UV run of H58 (as \Cref{fig:fitspec_example_H35}).}
    \label{fig:outGA_H58_UV}
\end{figure*}

\begin{figure*}
    \centering
    \includegraphics[width=0.95\textwidth]{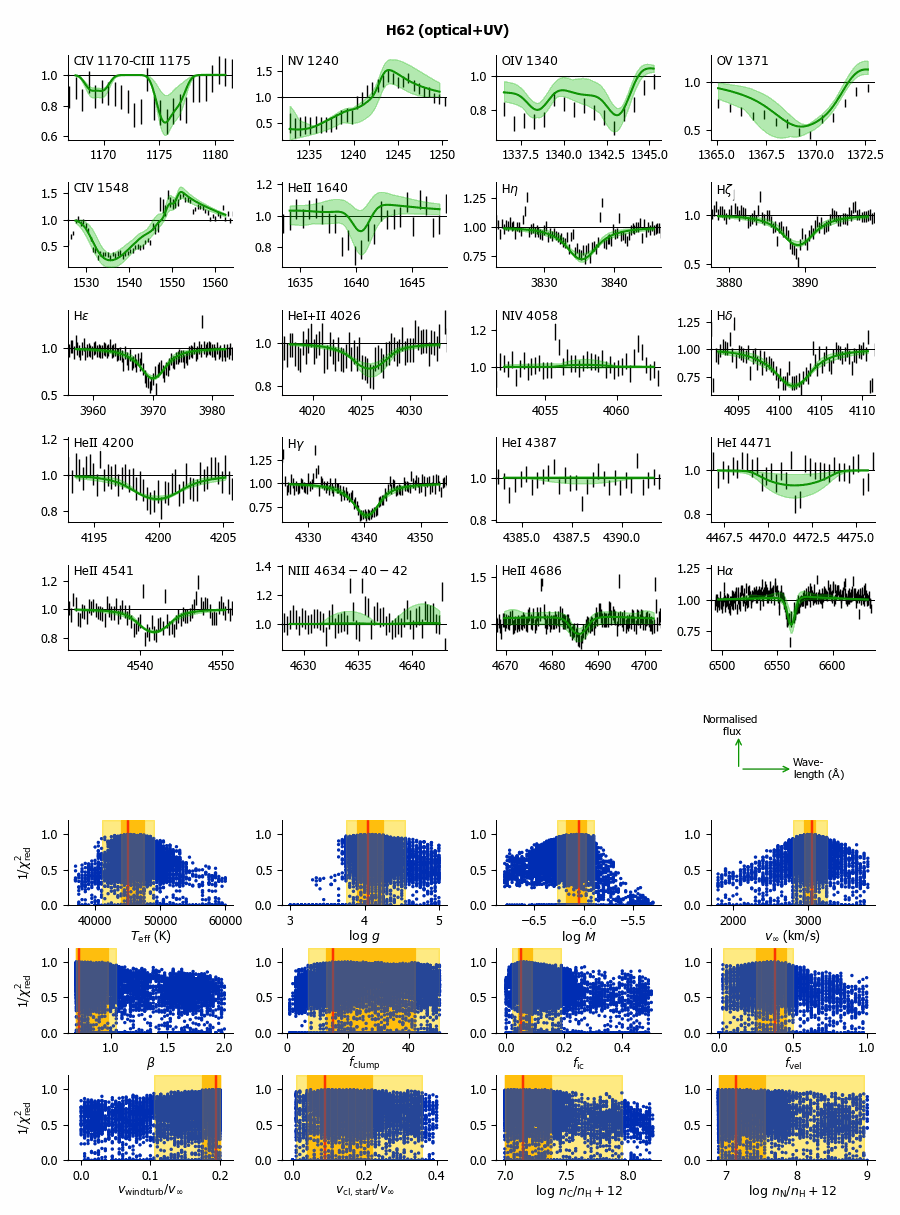}
    \caption{\PyGA output summary for the optical+UV run of H62 (as \Cref{fig:fitspec_example_H35}).}
    \label{fig:outGA_H62_UV}
\end{figure*}

\begin{figure*}
    \centering
    \includegraphics[width=0.95\textwidth]{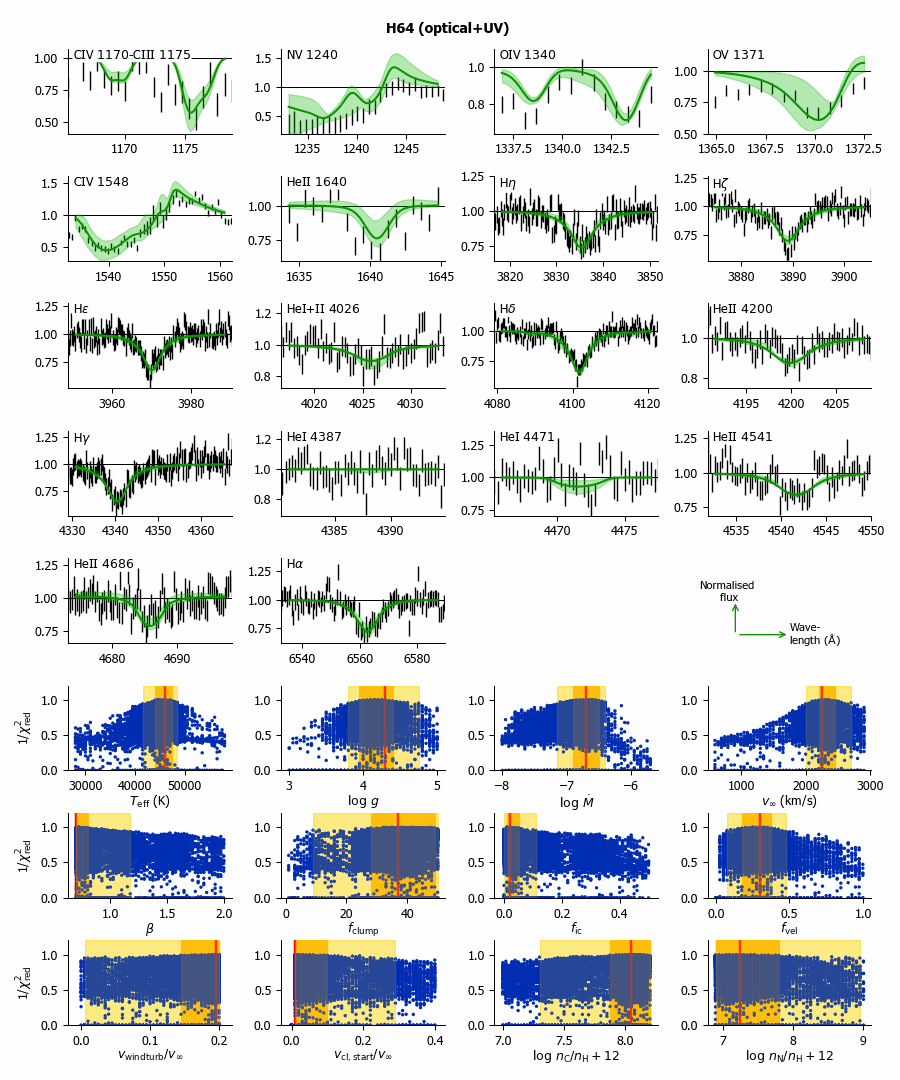}
    \caption{\PyGA output summary for the optical+UV run of H64 (as \Cref{fig:fitspec_example_H35}).}
    \label{fig:outGA_H64_UV}
\end{figure*}

\begin{figure*}
    \centering
    \includegraphics[width=0.95\textwidth]{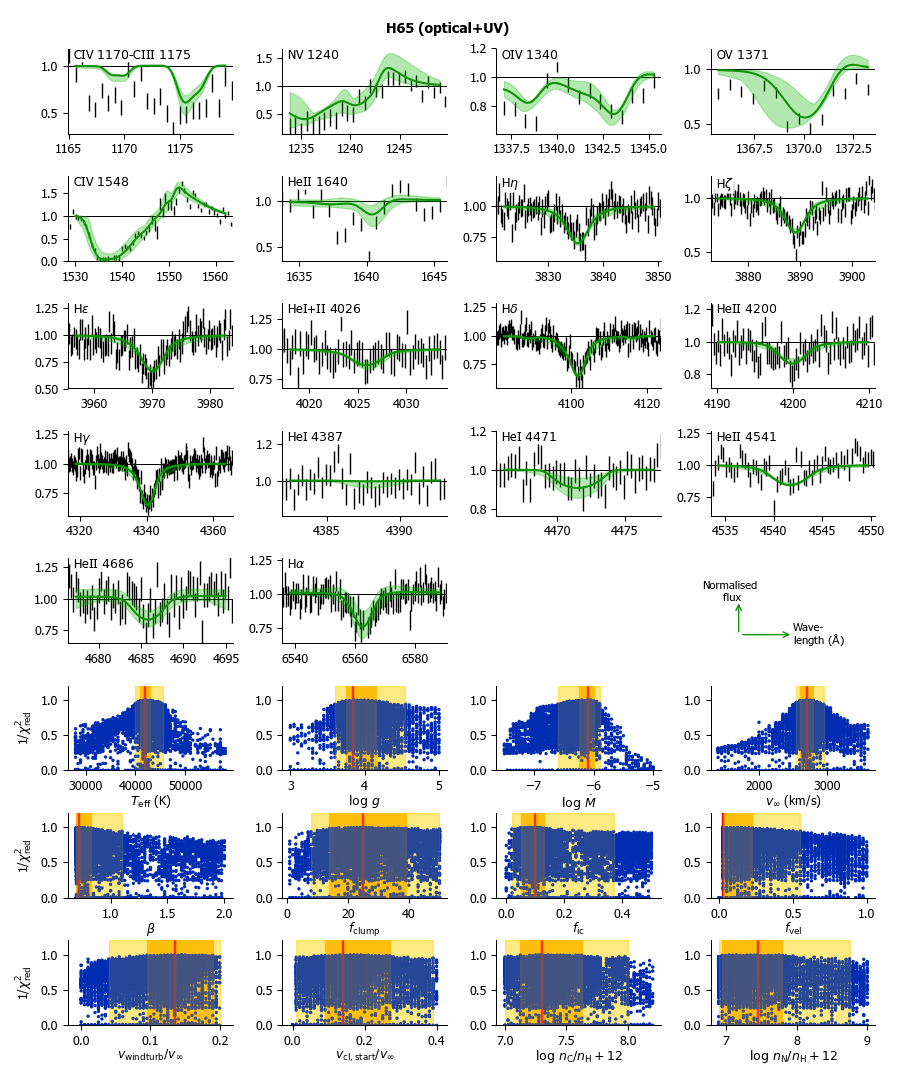}
    \caption{\PyGA output summary for the optical+UV run of H65 (as \Cref{fig:fitspec_example_H35}).}
    \label{fig:outGA_H65_UV}
\end{figure*}

\begin{figure*}
    \centering
    \includegraphics[width=0.95\textwidth]{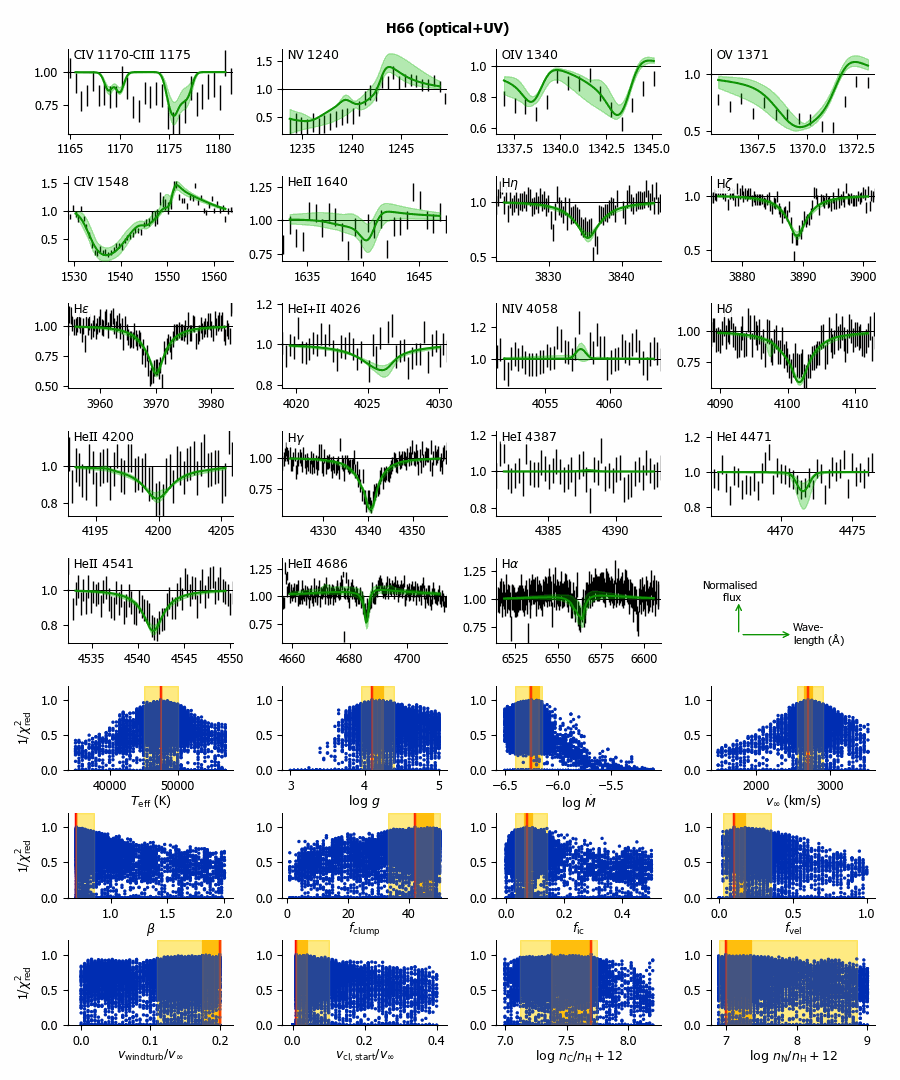}
    \caption{\PyGA output summary for the optical+UV run of H66 (as \Cref{fig:fitspec_example_H35}).}
    \label{fig:outGA_H66_UV}
\end{figure*}

\begin{figure*}
    \centering
    \includegraphics[width=0.95\textwidth]{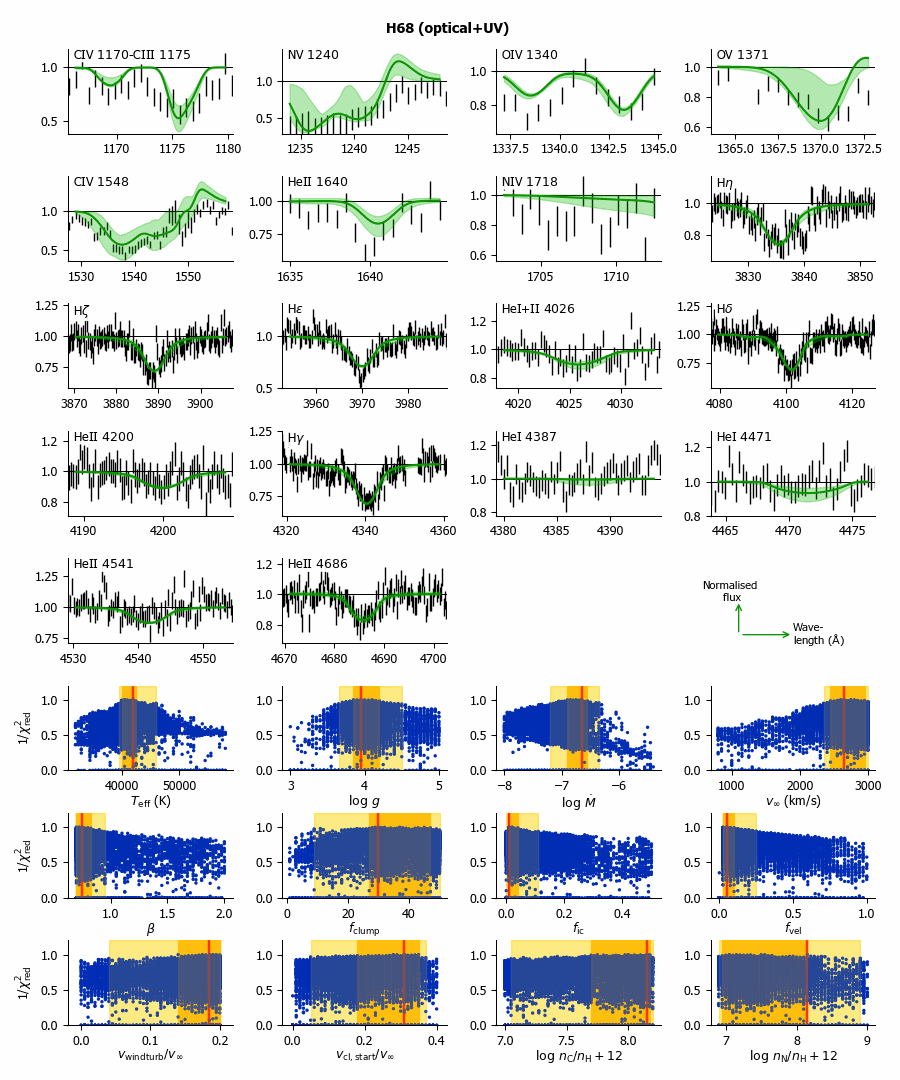}
    \caption{\PyGA output summary for the optical+UV run of H68 (as \Cref{fig:fitspec_example_H35}).}
    \label{fig:outGA_H68_UV}
\end{figure*}

\begin{figure*}
    \centering
    \includegraphics[width=0.95\textwidth]{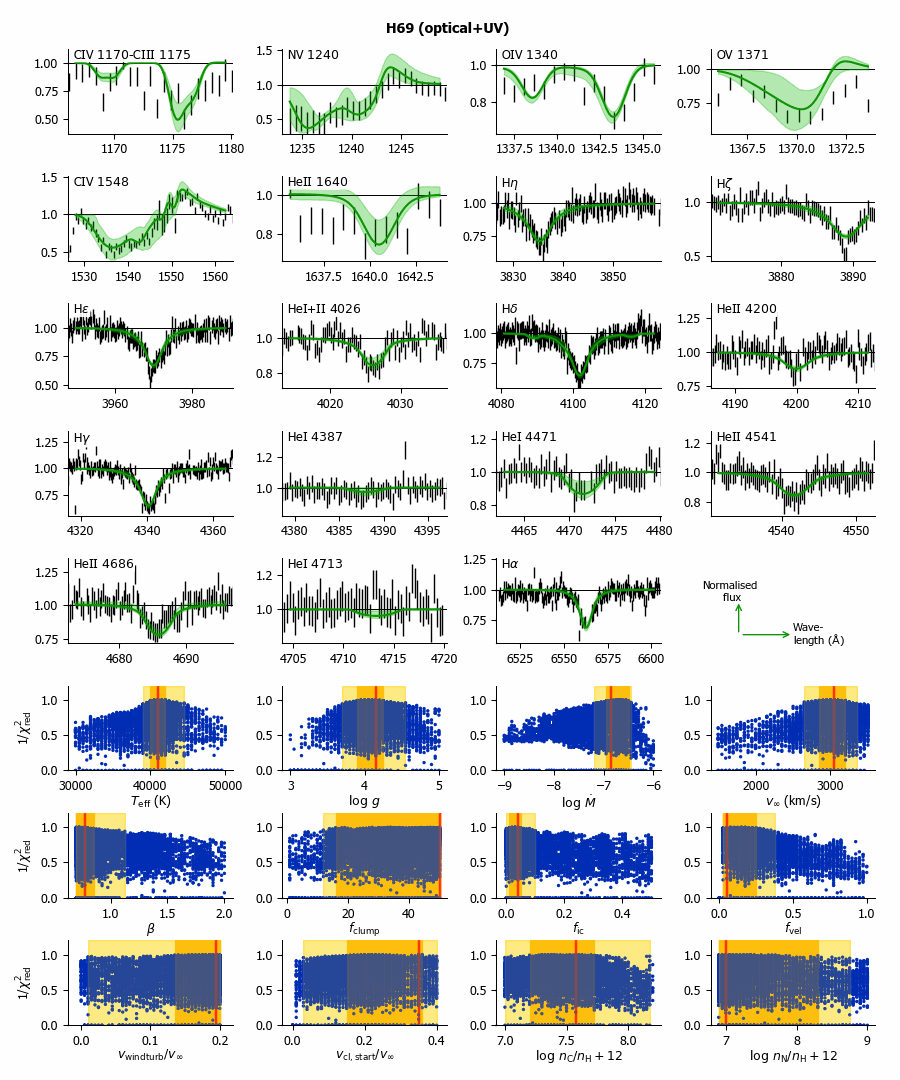}
    \caption{\PyGA output summary for the optical+UV run of H69 (as \Cref{fig:fitspec_example_H35}).}
    \label{fig:outGA_H69_UV}
\end{figure*}

\begin{figure*}
    \centering
    \includegraphics[width=0.95\textwidth]{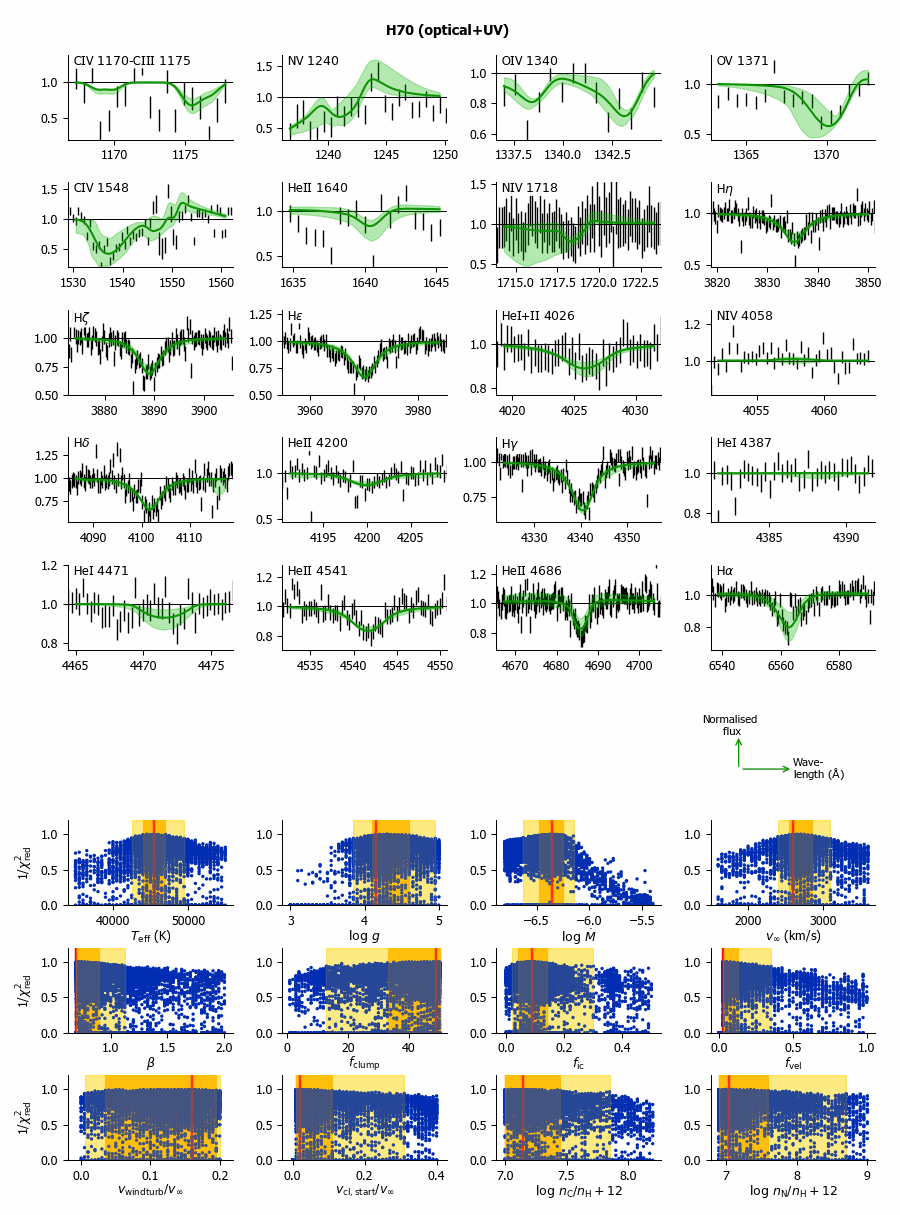}
    \caption{\PyGA output summary for the optical+UV run of H70 (as \Cref{fig:fitspec_example_H35}).}
    \label{fig:outGA_H70_UV}
\end{figure*}

\begin{figure*}
    \centering
    \includegraphics[width=0.95\textwidth]{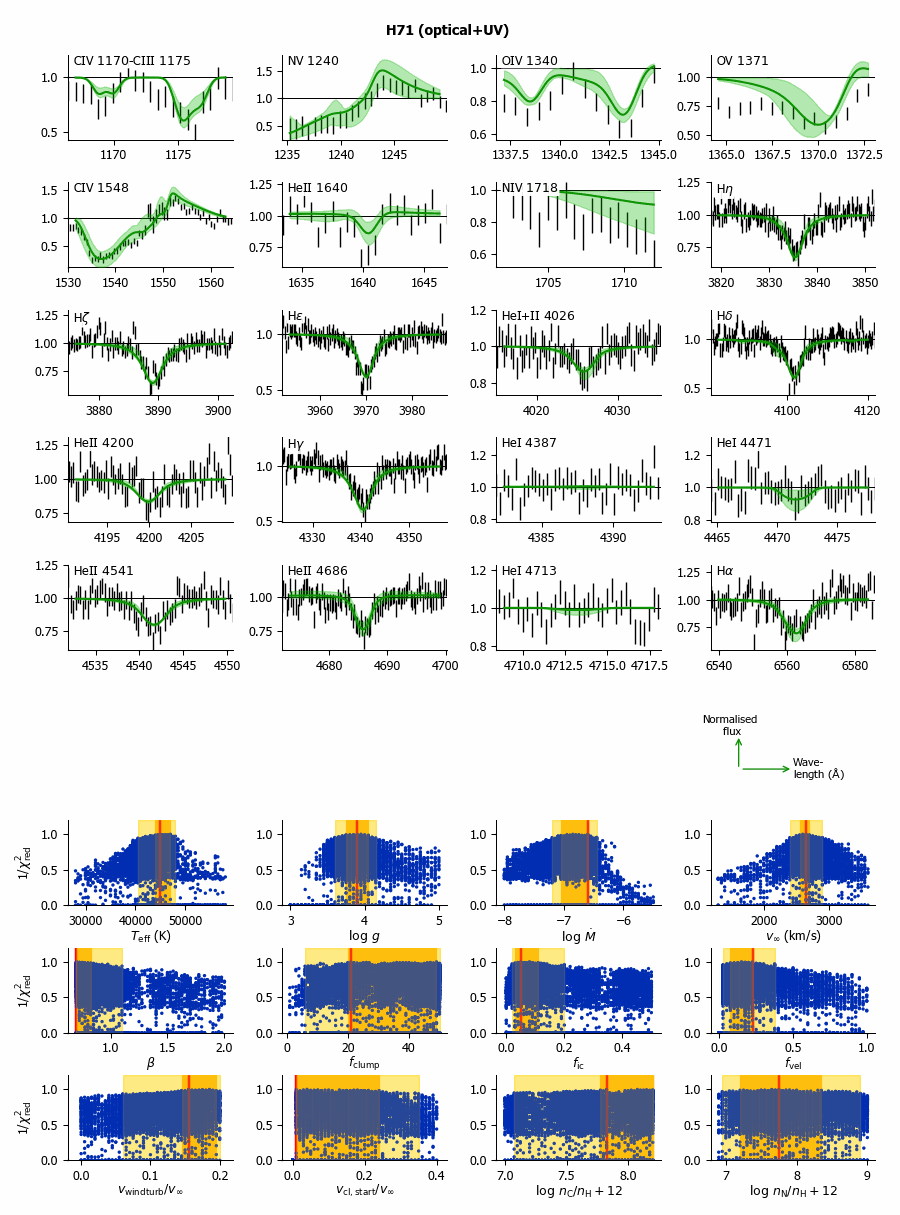}
    \caption{\PyGA output summary for the optical+UV run of H71 (as \Cref{fig:fitspec_example_H35}).}
    \label{fig:outGA_H71_UV}
\end{figure*}

\begin{figure*}
    \centering
    \includegraphics[width=0.95\textwidth]{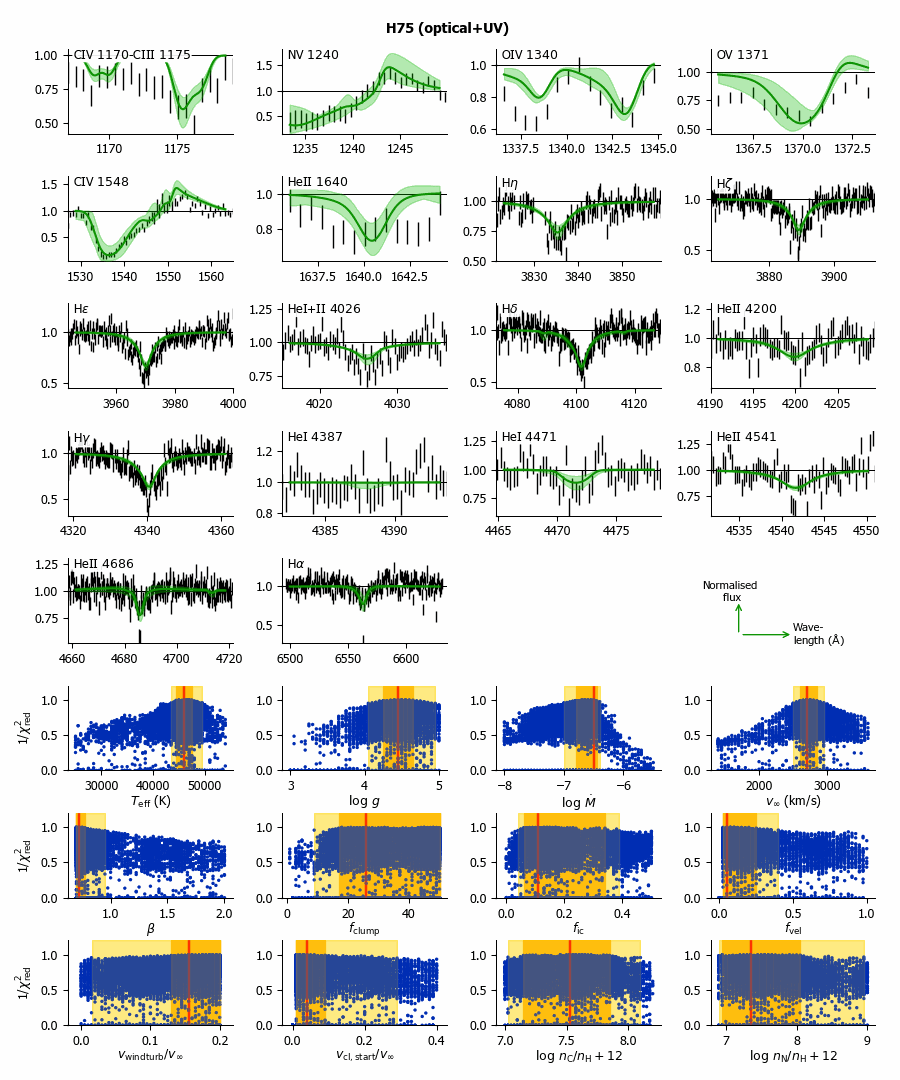}
    \caption{\PyGA output summary for the optical+UV run of H75 (as \Cref{fig:fitspec_example_H35}).}
    \label{fig:outGA_H75_UV}
\end{figure*}

\begin{figure*}
    \centering
    \includegraphics[width=0.95\textwidth]{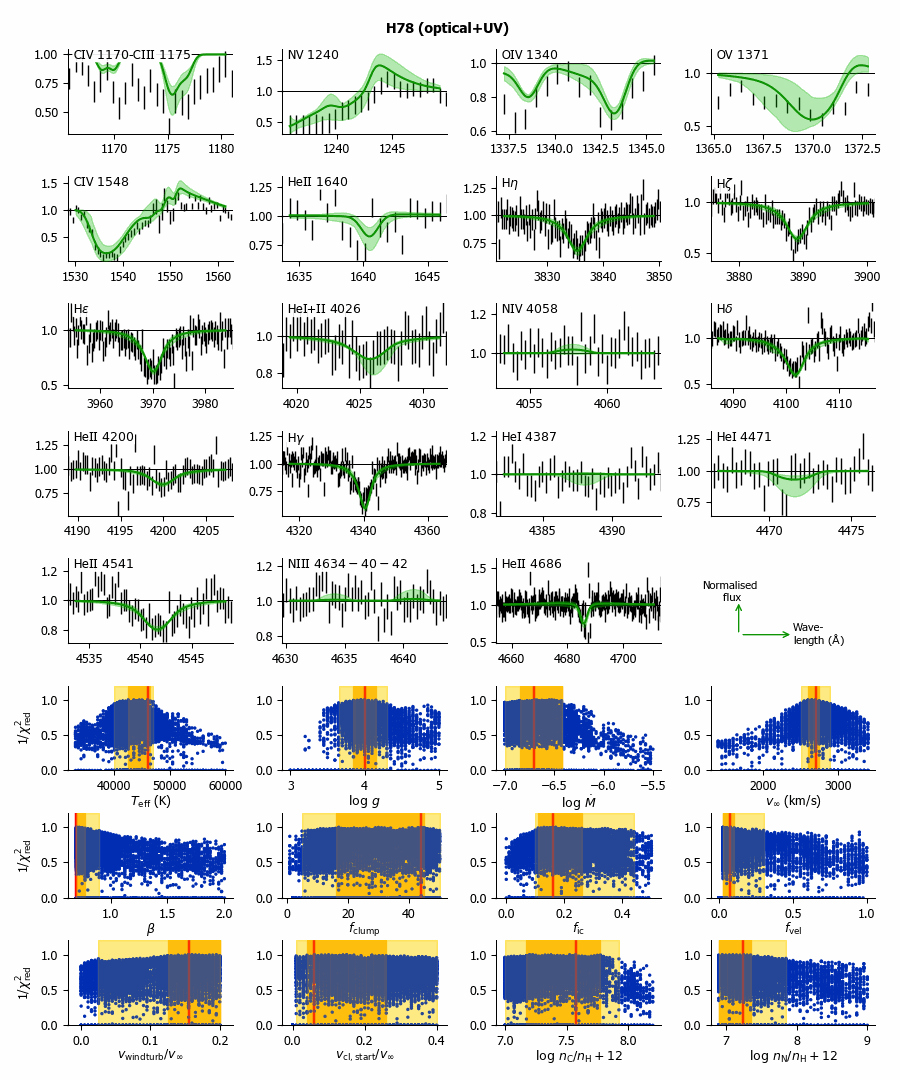}
    \caption{\PyGA output summary for the optical+UV run of H78 (as \Cref{fig:fitspec_example_H35}).}
    \label{fig:outGA_H78_UV}
\end{figure*}

\begin{figure*}
    \centering
    \includegraphics[width=0.95\textwidth]{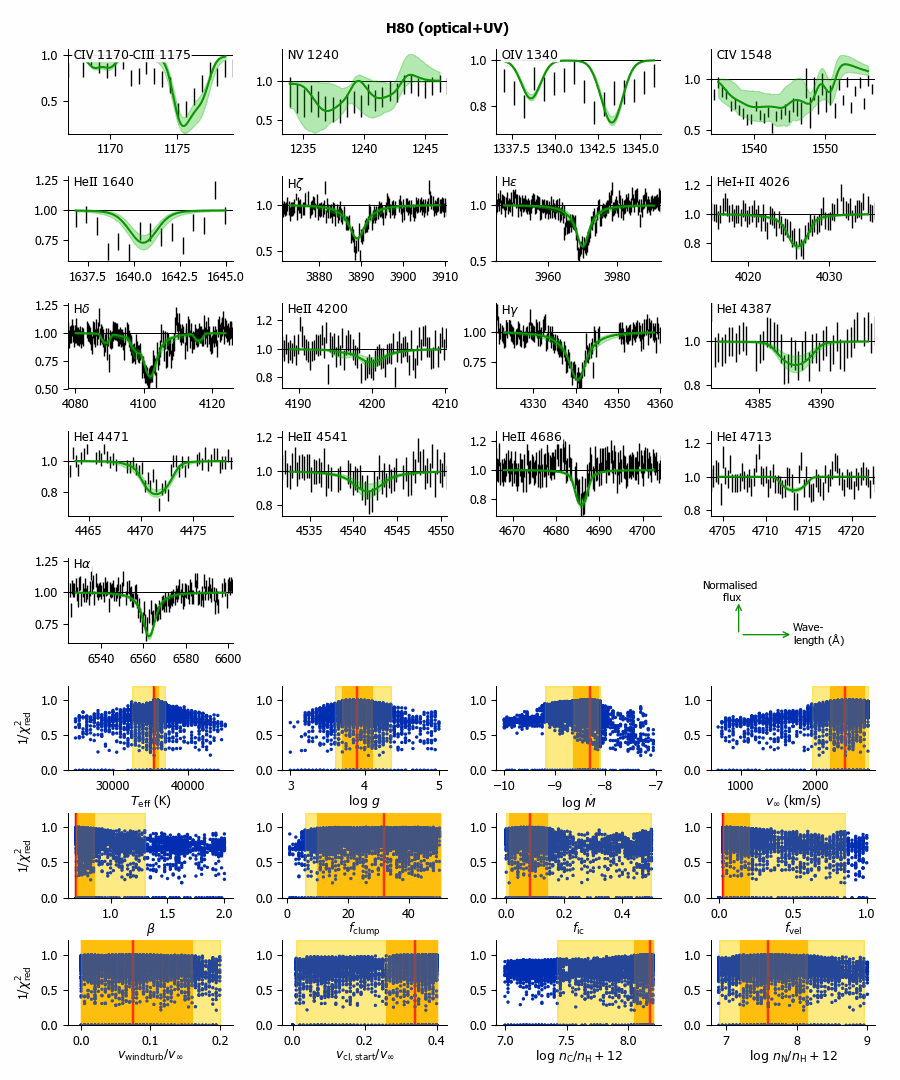}
    \caption{\PyGA output summary for the optical+UV run of H80 (as \Cref{fig:fitspec_example_H35}).}
    \label{fig:outGA_H80_UV}
\end{figure*}

\begin{figure*}
    \centering
    \includegraphics[width=0.95\textwidth]{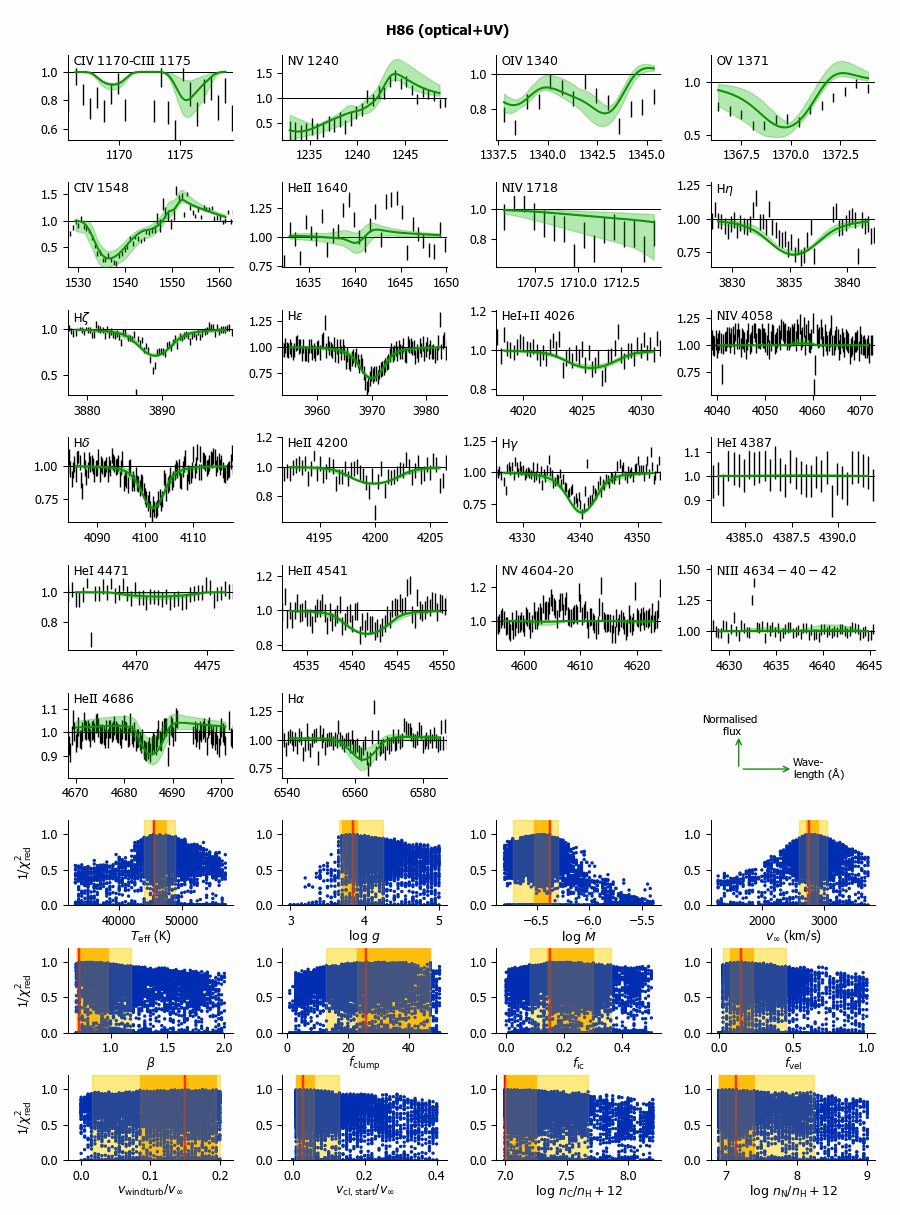}
    \caption{\PyGA output summary for the optical+UV run of H86 (as \Cref{fig:fitspec_example_H35}).}
    \label{fig:outGA_H86_UV}
\end{figure*}

\begin{figure*}
    \centering
    \includegraphics[width=0.95\textwidth]{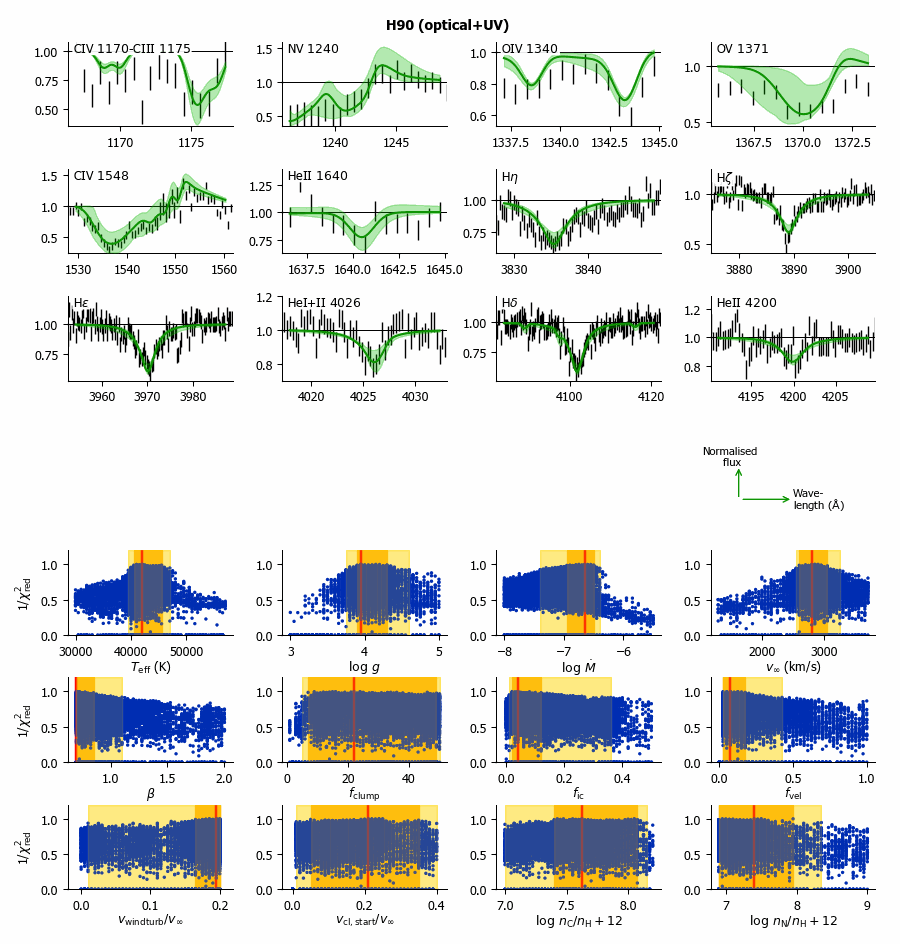}
    \caption{\PyGA output summary for the optical+UV run of H90 (as \Cref{fig:fitspec_example_H35}).}
    \label{fig:outGA_H90_UV}
\end{figure*}

\begin{figure*}
    \centering
    \includegraphics[width=0.95\textwidth]{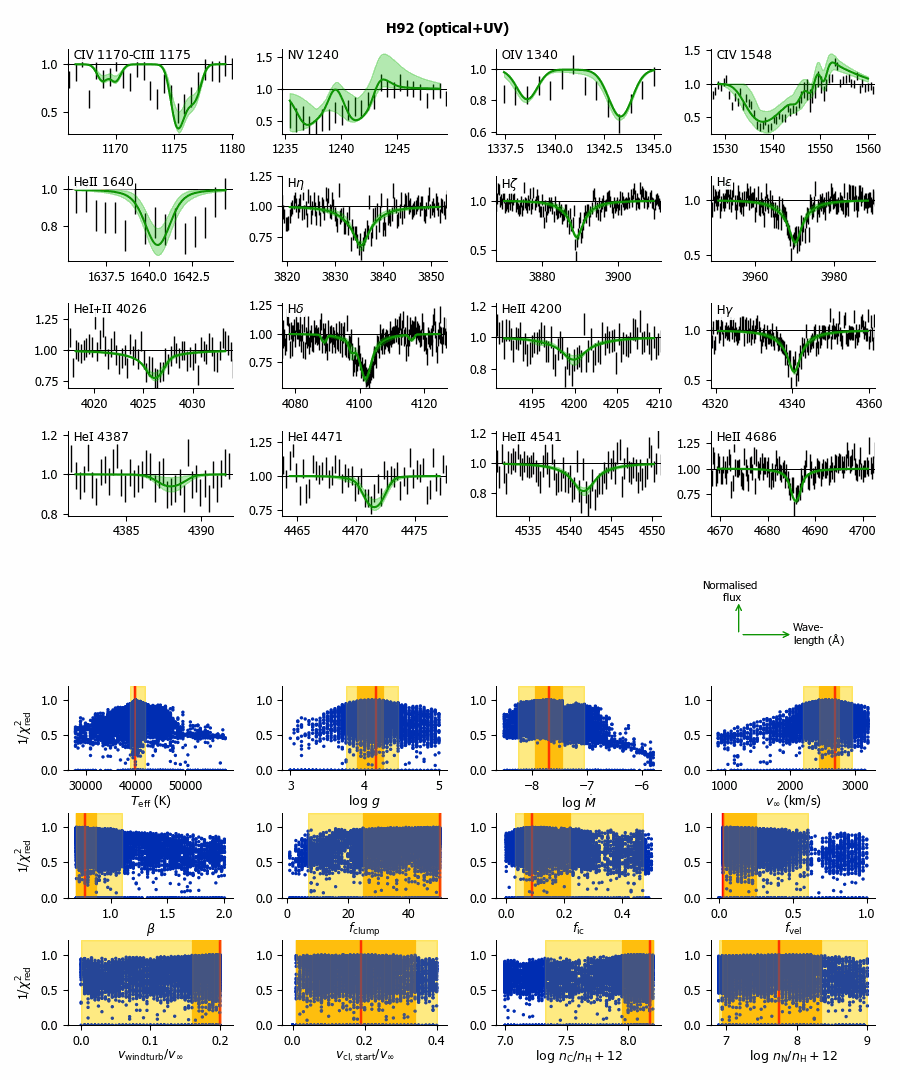}
    \caption{\PyGA output summary for the optical+UV run of H92 (as \Cref{fig:fitspec_example_H35}).}
    \label{fig:outGA_H92_UV}
\end{figure*}

\begin{figure*}
    \centering
    \includegraphics[width=0.95\textwidth]{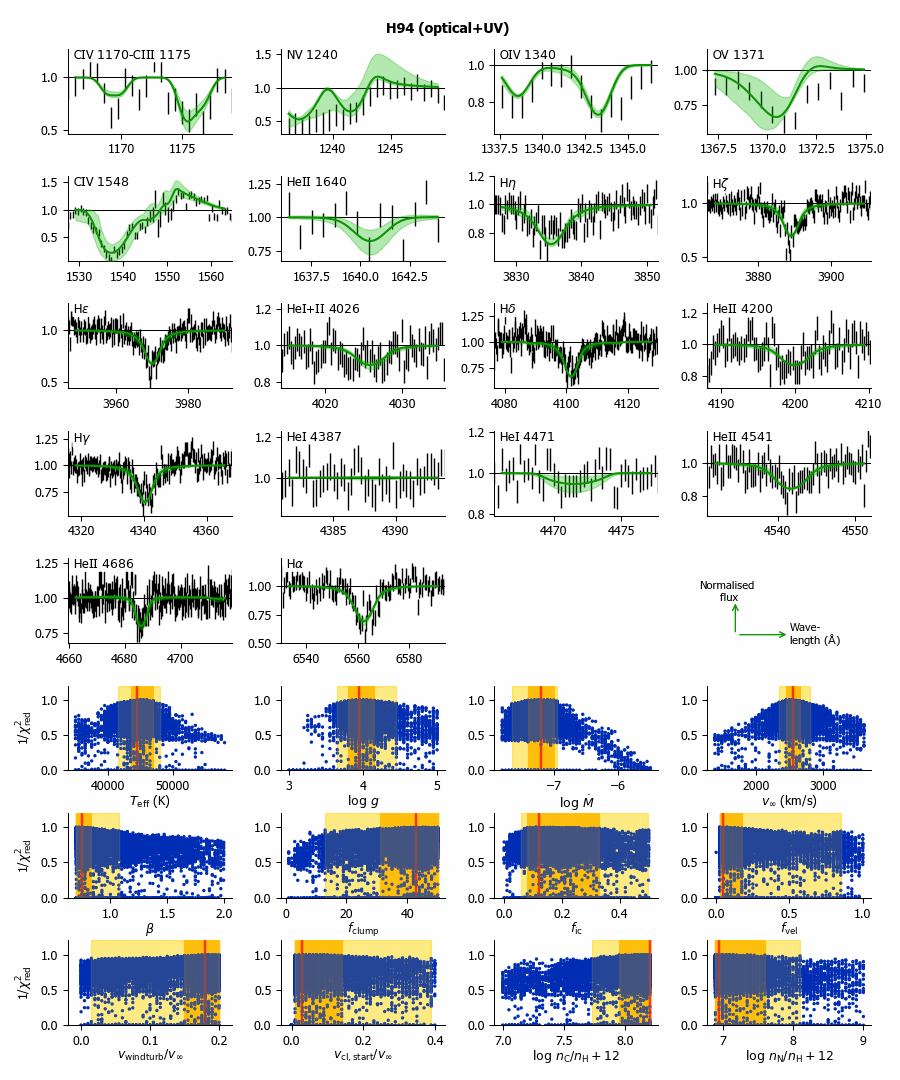}
    \caption{\PyGA output summary for the optical+UV run of H94 (as \Cref{fig:fitspec_example_H35}).}
    \label{fig:outGA_H94_UV}
\end{figure*}

\begin{figure*}
    \centering
    \includegraphics[width=0.95\textwidth]{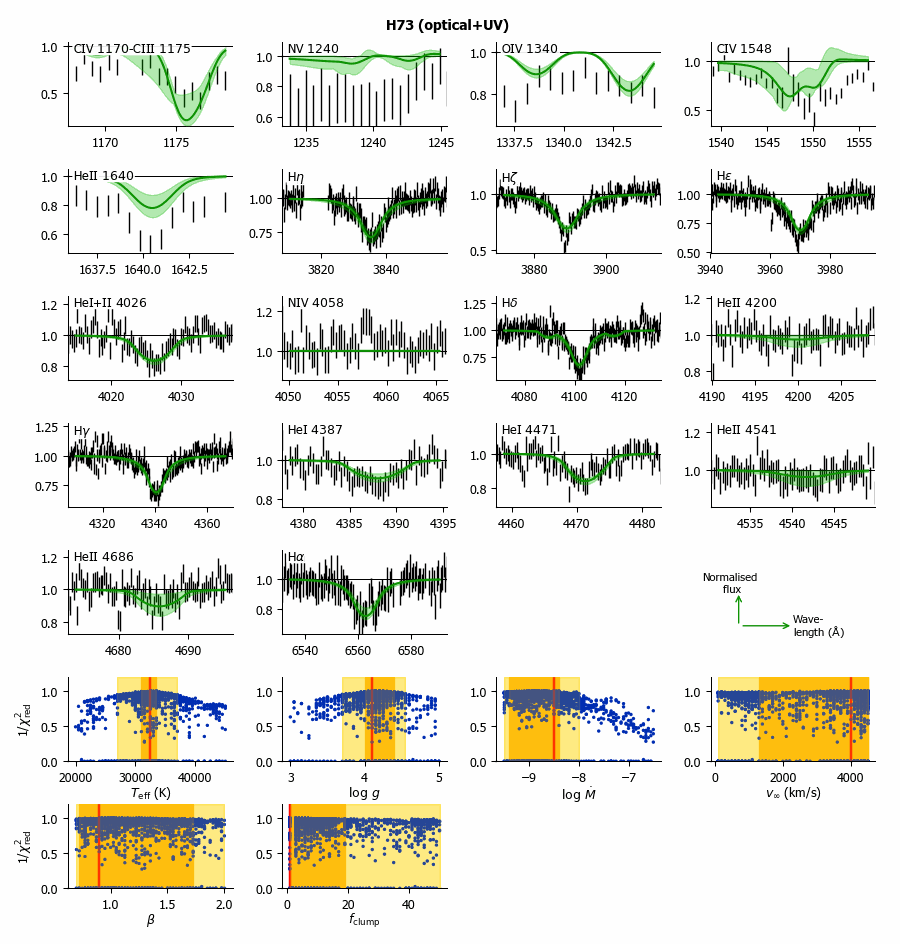}
    \caption{\PyGA output summary for the optical+UV run of H73 (as \Cref{fig:fitspec_example_H35}).}
    \label{fig:outGA_H73_UV}
\end{figure*}

\begin{figure*}
    \centering
    \includegraphics[width=0.95\textwidth]{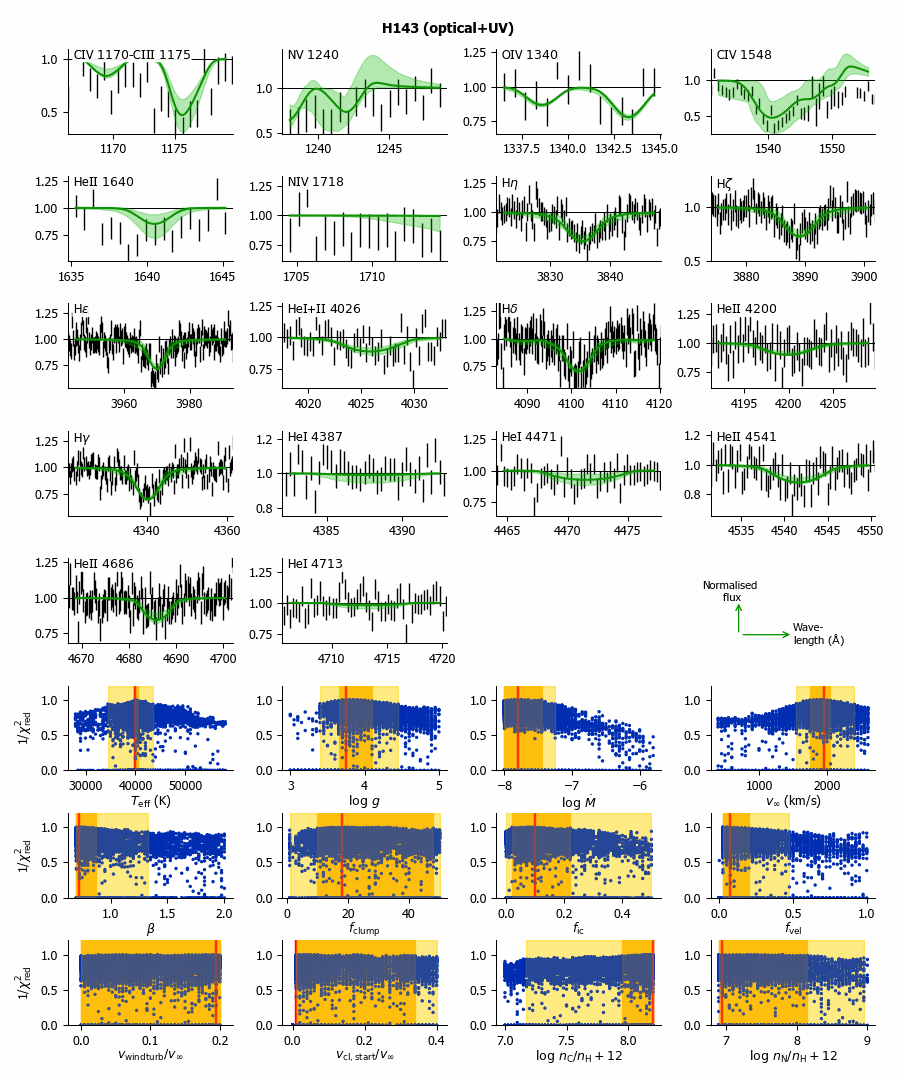}
    \caption{\PyGA output summary for the optical+UV run of H143 (as \Cref{fig:fitspec_example_H35}).}
    \label{fig:outGA_H143_UV}
\end{figure*}

\begin{figure*}
    \centering
    \includegraphics[width=0.95\textwidth]{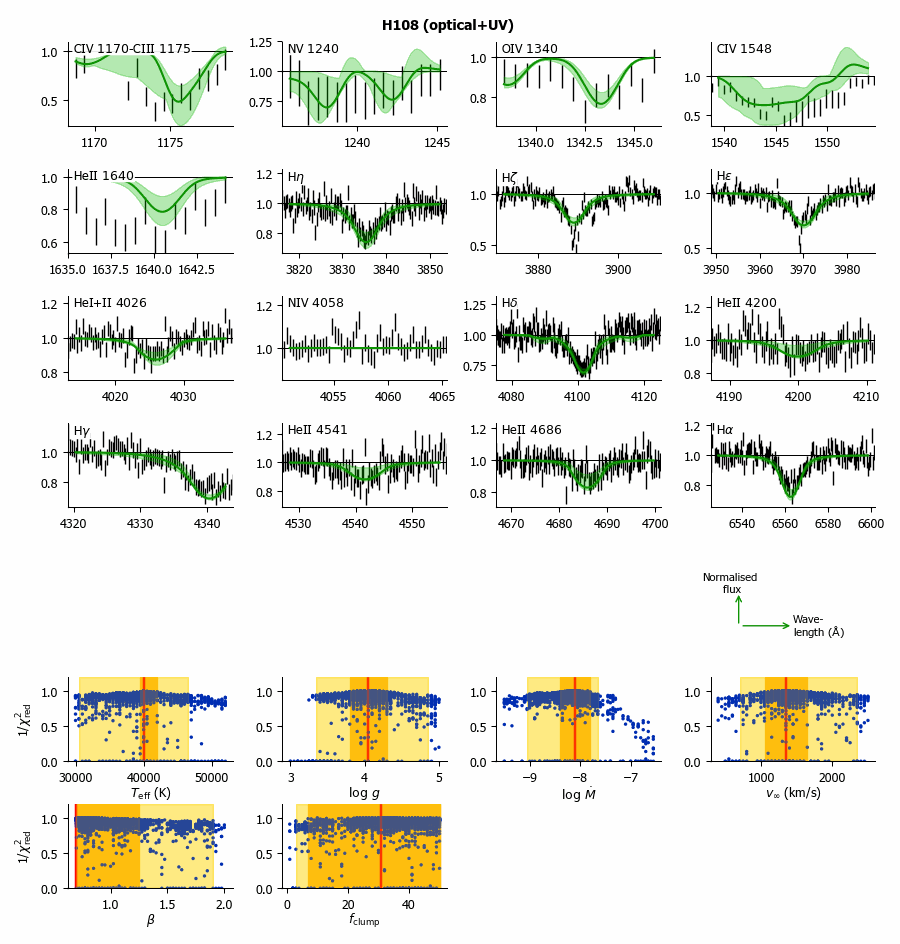}
    \caption{\PyGA output summary for the optical+UV run of H108 (as \Cref{fig:fitspec_example_H35}).}
    \label{fig:outGA_H108_UV}
\end{figure*}

\begin{figure*}
    \centering
    \includegraphics[width=0.95\textwidth]{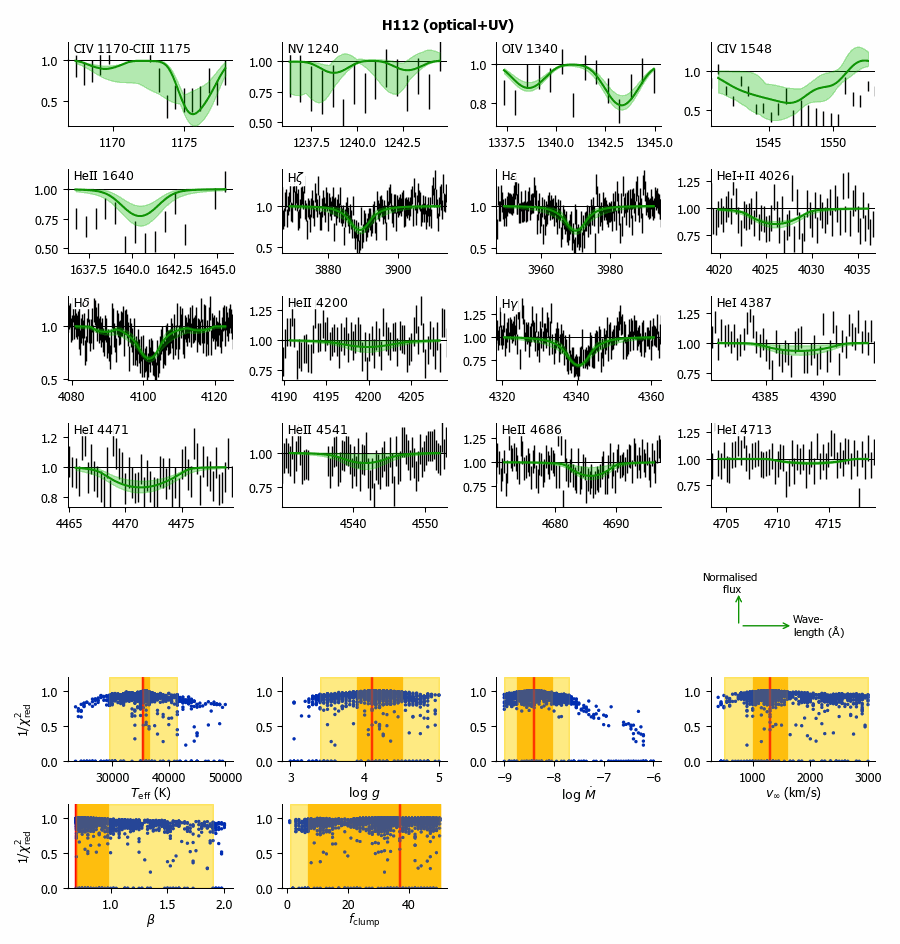}
    \caption{\PyGA output summary for the optical+UV run of H112 (as \Cref{fig:fitspec_example_H35}).}
    \label{fig:outGA_H112_UV}
\end{figure*}

\begin{figure*}
    \centering
    \includegraphics[width=0.95\textwidth]{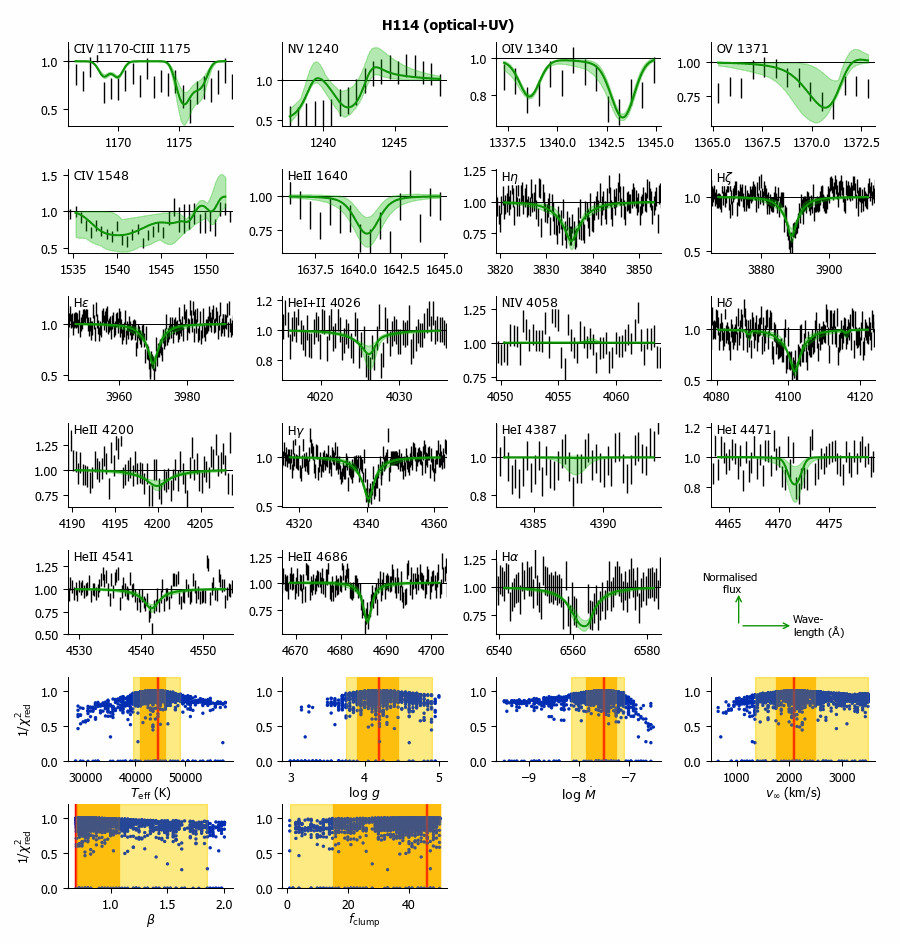}
    \caption{\PyGA output summary for the optical+UV run of H114 (as \Cref{fig:fitspec_example_H35}).}
    \label{fig:outGA_H114_UV}
\end{figure*}

\begin{figure*}
    \centering
    \includegraphics[width=0.95\textwidth]{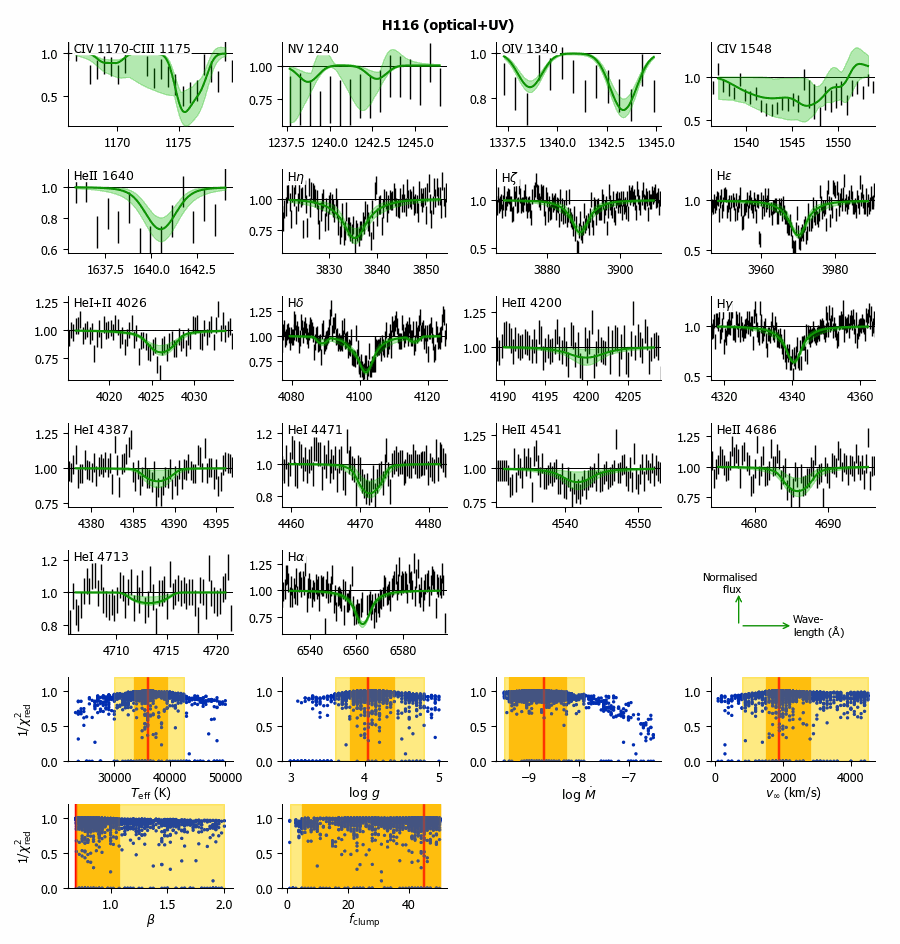}
    \caption{\PyGA output summary for the optical+UV run of H116 (as \Cref{fig:fitspec_example_H35}).}
    \label{fig:outGA_H116_UV}
\end{figure*}

\begin{figure*}
    \centering
    \includegraphics[width=0.95\textwidth]{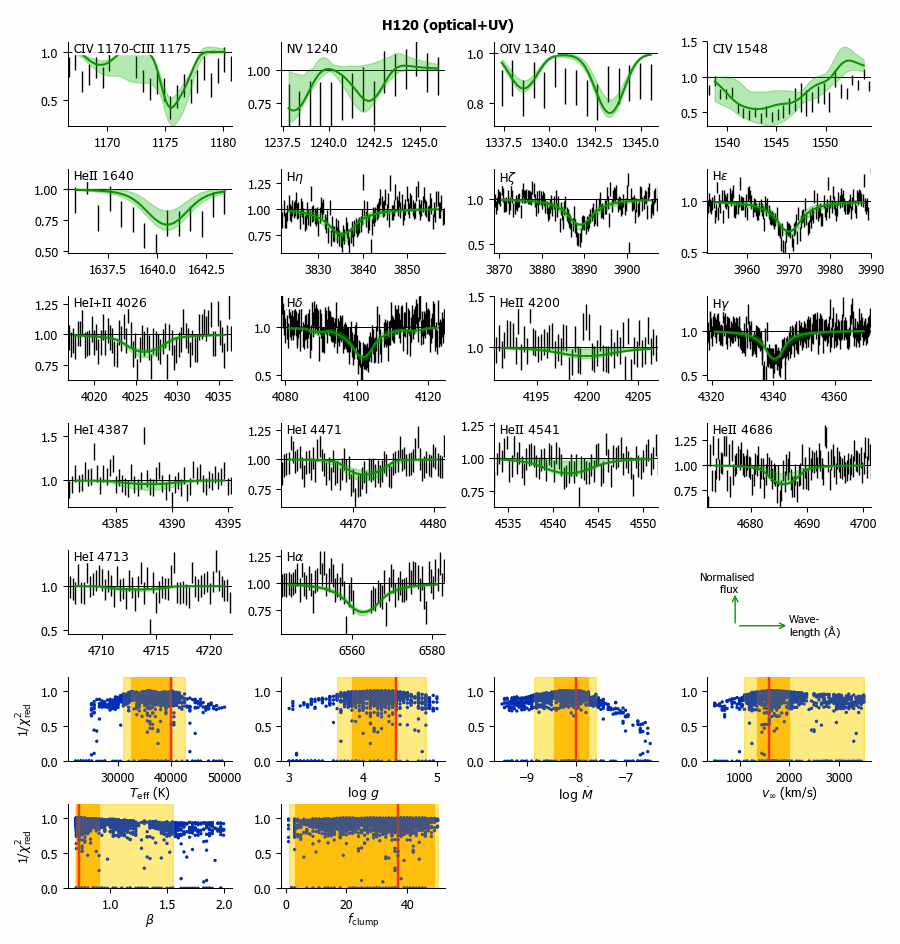}
    \caption{\PyGA output summary for the optical+UV run of H120 (as \Cref{fig:fitspec_example_H35}).}
    \label{fig:outGA_H120_UV}
\end{figure*}

\begin{figure*}
    \centering
    \includegraphics[width=0.95\textwidth]{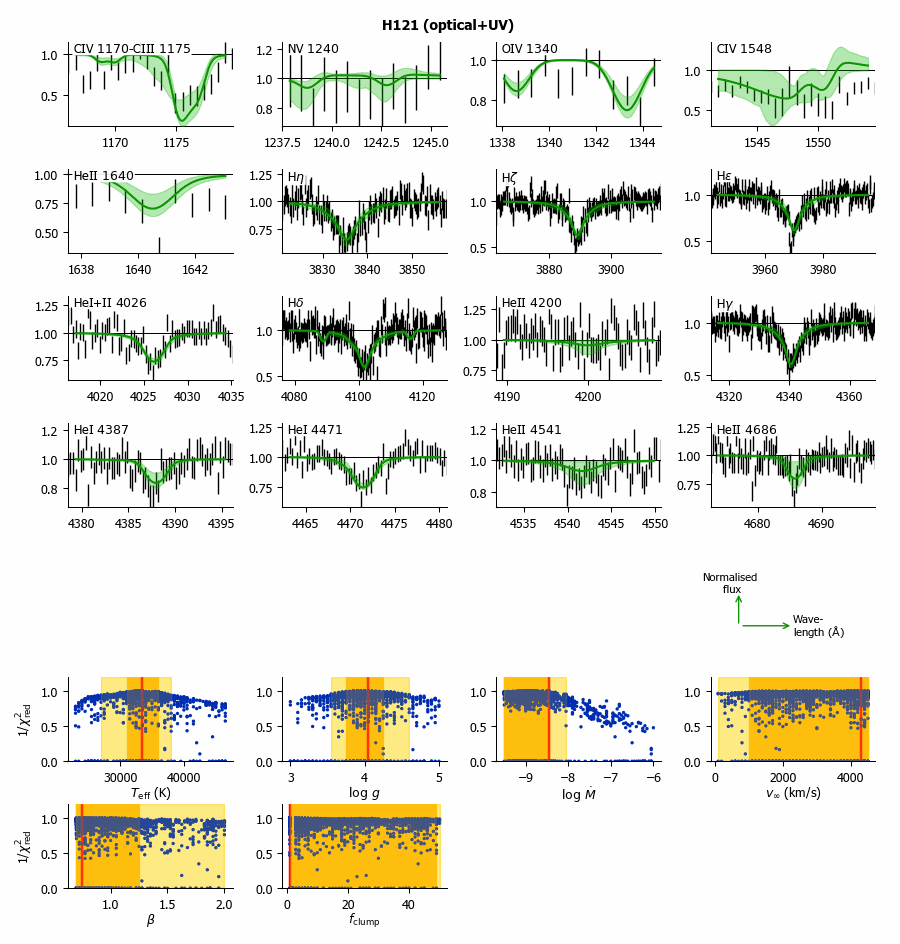}
    \caption{\PyGA output summary for the optical+UV run of H121 (as \Cref{fig:fitspec_example_H35}).}
    \label{fig:outGA_H121_UV}
\end{figure*}

\begin{figure*}
    \centering
    \includegraphics[width=0.95\textwidth]{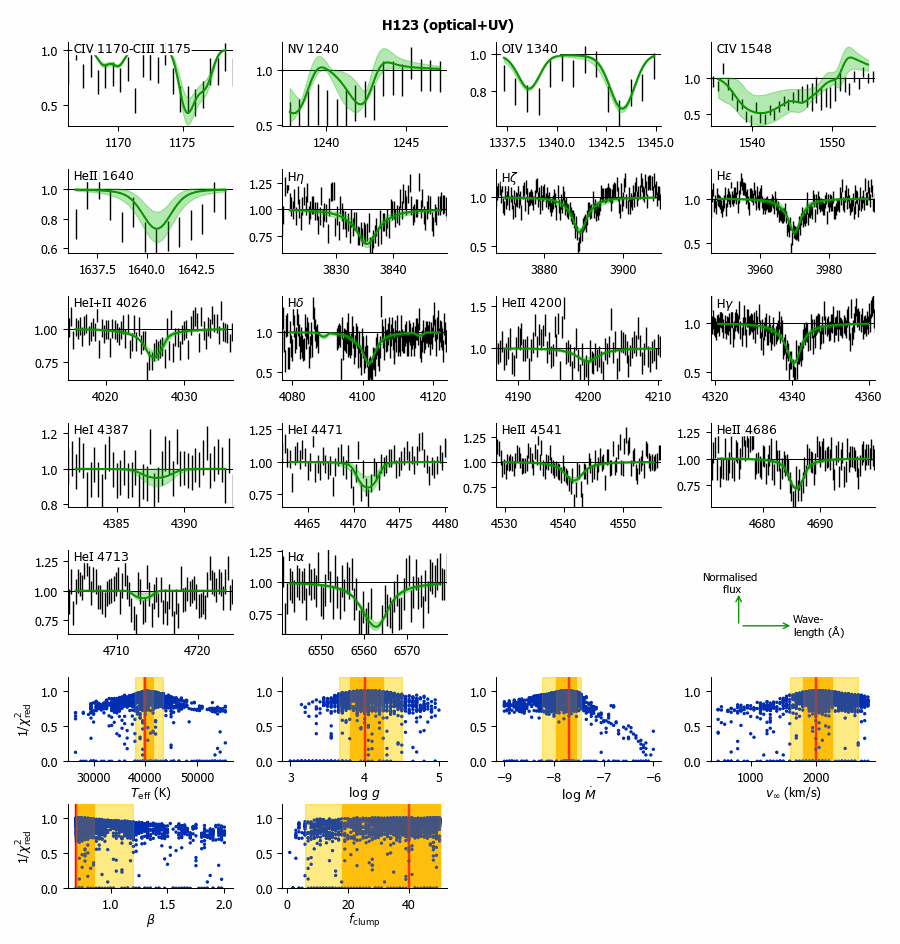}
    \caption{\PyGA output summary for the optical+UV run of H123 (as \Cref{fig:fitspec_example_H35}).}
    \label{fig:outGA_H123_UV}
\end{figure*}

\begin{figure*}
    \centering
    \includegraphics[width=0.95\textwidth]{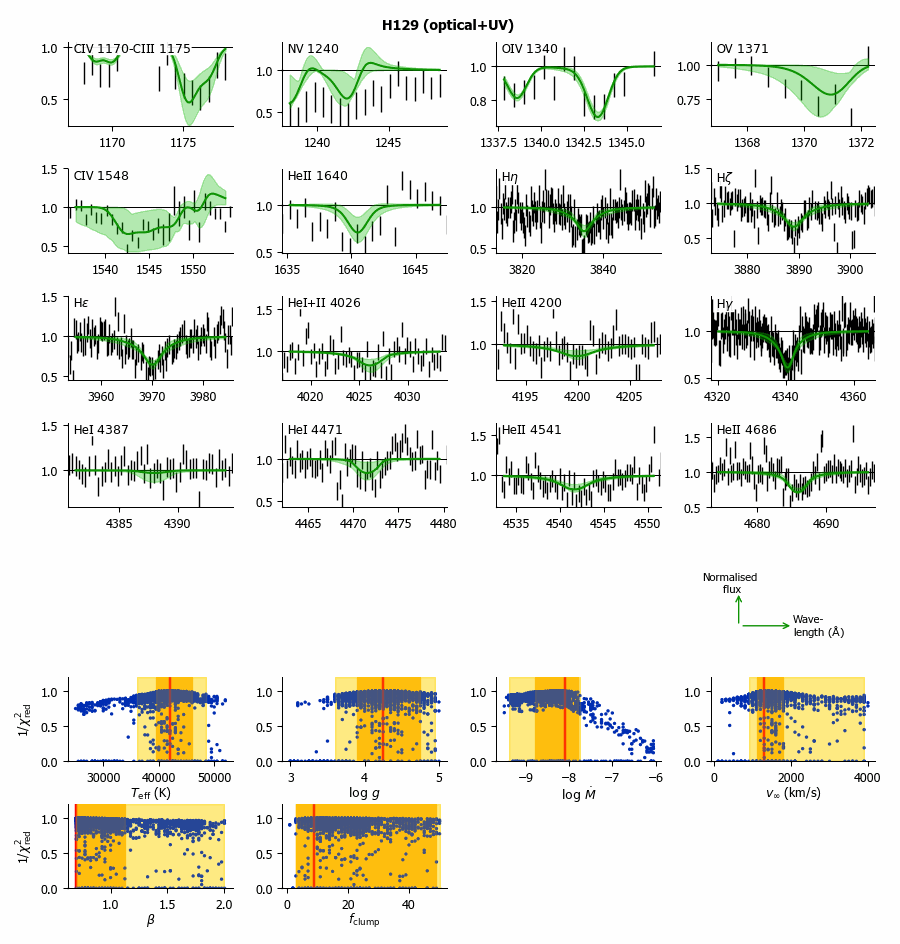}
    \caption{\PyGA output summary for the optical+UV run of H129 (as \Cref{fig:fitspec_example_H35}).}
    \label{fig:outGA_H129_UV}
\end{figure*}

\begin{figure*}
    \centering
    \includegraphics[width=0.95\textwidth]{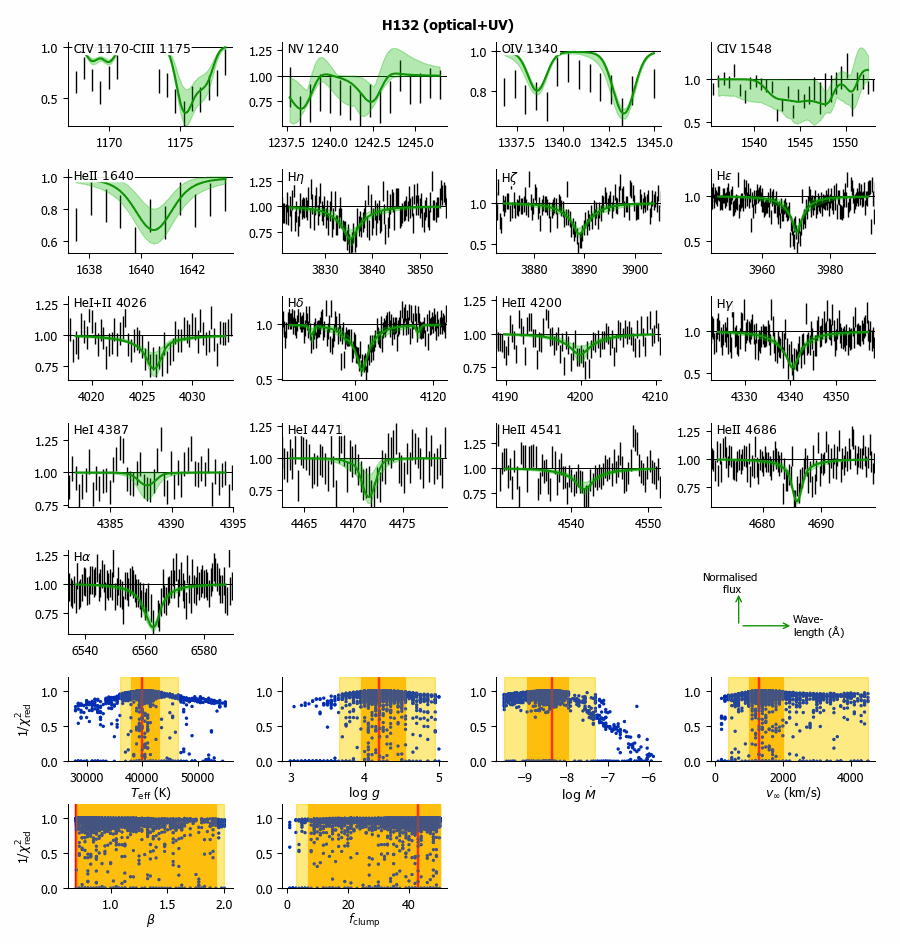}
    \caption{\PyGA output summary for the optical+UV run of H132 (as \Cref{fig:fitspec_example_H35}).}
    \label{fig:outGA_H132_UV}
\end{figure*}

\begin{figure*}
    \centering
    \includegraphics[width=0.95\textwidth]{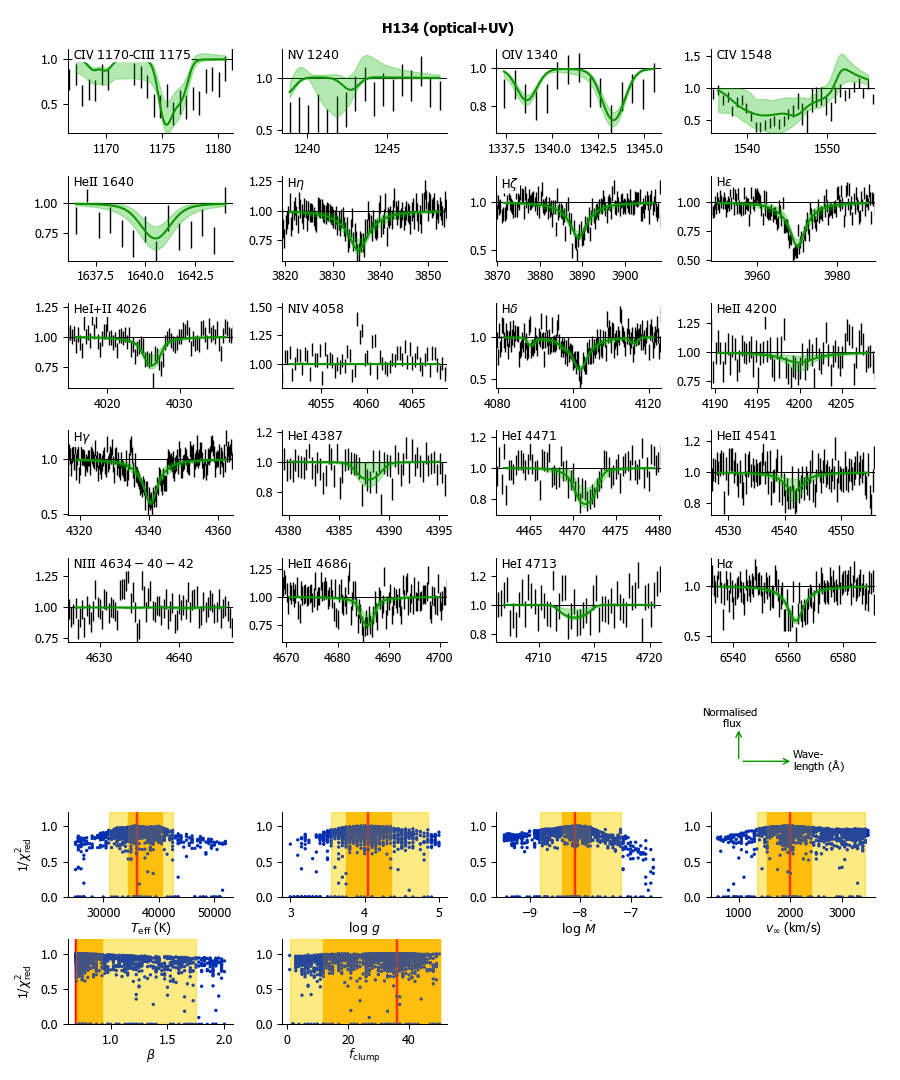}
    \caption{\PyGA output summary for the optical+UV run of H134 (as \Cref{fig:fitspec_example_H35}).}
    \label{fig:outGA_H134_UV}
\end{figure*}

\begin{figure*}
    \centering
    \includegraphics[width=0.95\textwidth]{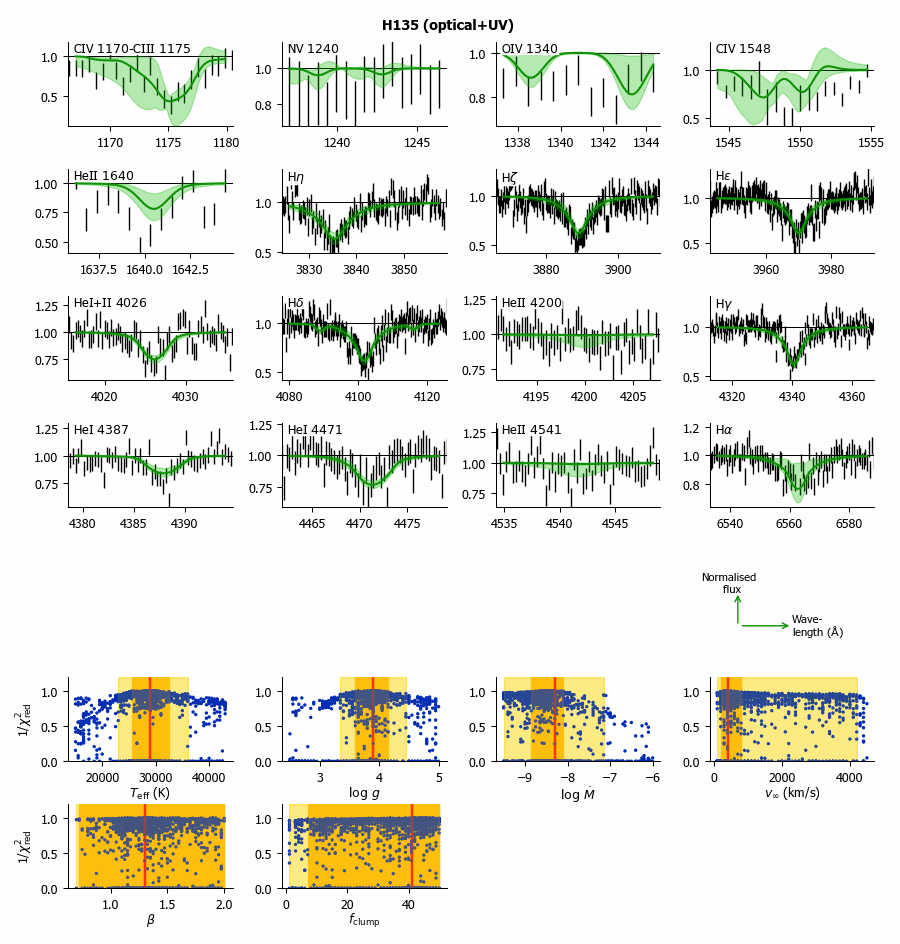}
    \caption{\PyGA output summary for the optical+UV run of H135 (as \Cref{fig:fitspec_example_H35}).}
    \label{fig:outGA_H135_UV}
\end{figure*}

Summaries of the \pyGA output of the optical~+~UV runs are presented in \cref{fig:outGA_R136a1_UV} to \ref{fig:outGA_H173_UV}. For each star we show line profiles and fitness distributions. 


\clearpage  

\clearpage  

\clearpage  

\clearpage  

\clearpage  

\clearpage  

\clearpage  

\clearpage  

\clearpage  

\clearpage  

\clearpage  

\clearpage  

\clearpage  

\clearpage  

\clearpage  

\clearpage  

\clearpage  

\clearpage  

\clearpage  

\clearpage  

\clearpage  

\clearpage  

\clearpage  

\clearpage  

\clearpage  

\clearpage  

\clearpage  

\clearpage  

\clearpage  

\clearpage  

\clearpage  

\clearpage  

\clearpage  

\clearpage  

\clearpage  

\clearpage  

\clearpage  

\clearpage  

\clearpage  

\clearpage  

\clearpage  

\clearpage  

\clearpage  

\clearpage  

\clearpage  

\clearpage  

\clearpage  

\clearpage  

\clearpage  

\clearpage  

\clearpage  

\begin{figure*}
    \centering
    \includegraphics[width=0.95\textwidth]{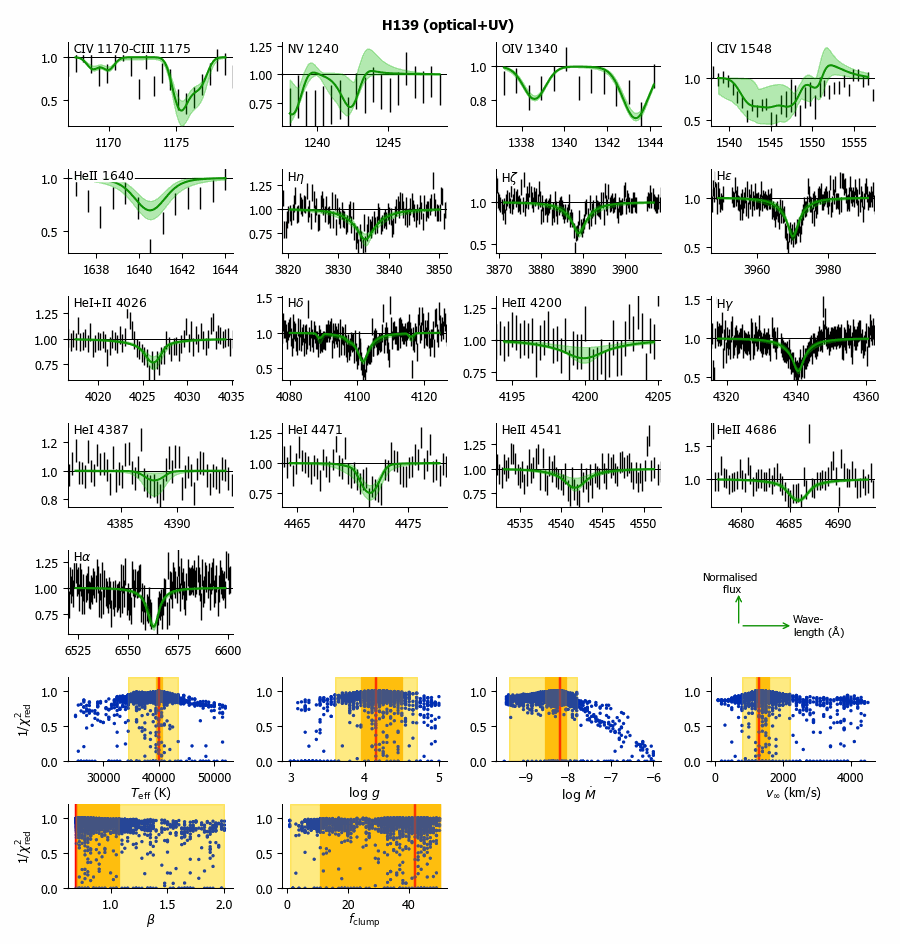}
    \caption{\PyGA output summary for the optical+UV run of H139 (as \Cref{fig:fitspec_example_H35}).}
    \label{fig:outGA_H139_UV}
\end{figure*}

\clearpage  

\begin{figure*}
    \centering
    \includegraphics[width=0.95\textwidth]{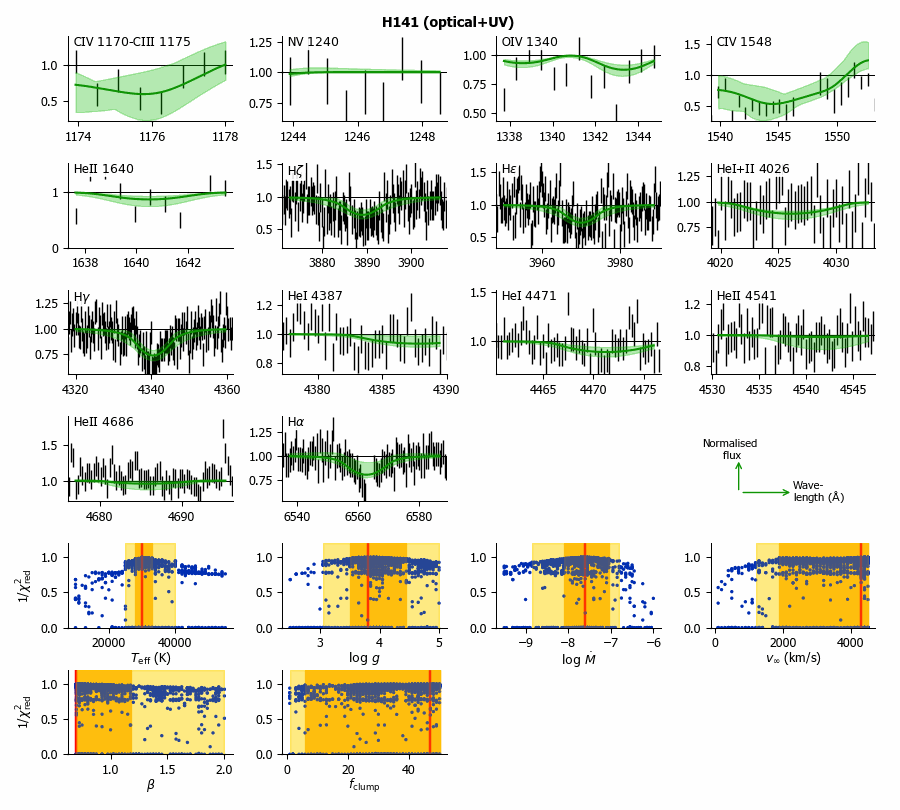}
    \caption{\PyGA output summary for the optical+UV run of H141 (as \Cref{fig:fitspec_example_H35}).}
    \label{fig:outGA_H141_UV}
\end{figure*}

\clearpage  

\begin{figure*}
    \centering
    \includegraphics[width=0.95\textwidth]{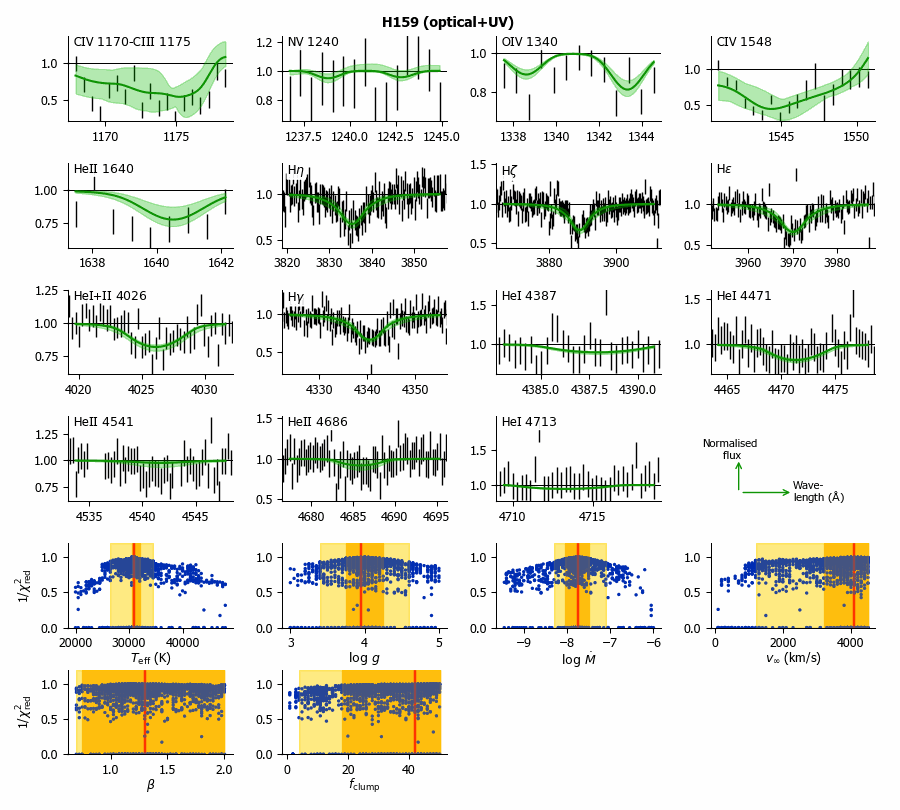}
    \caption{\PyGA output summary for the optical+UV run of H159 (as \Cref{fig:fitspec_example_H35}).}
    \label{fig:outGA_H159_UV}
\end{figure*}

\clearpage  

\begin{figure*}
    \centering
    \includegraphics[width=0.95\textwidth]{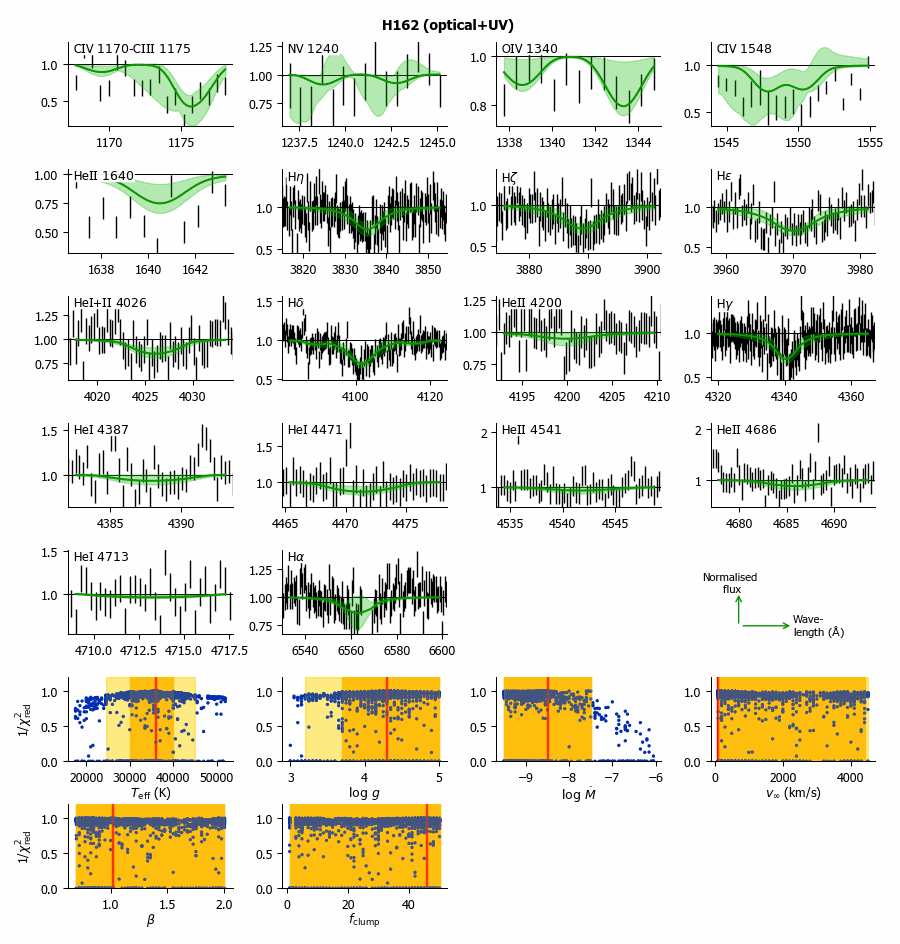}
    \caption{\PyGA output summary for the optical+UV run of H162 (as \Cref{fig:fitspec_example_H35}).}
    \label{fig:outGA_H162_UV}
\end{figure*}

\clearpage  

\begin{figure*}
    \centering
    \includegraphics[width=0.95\textwidth]{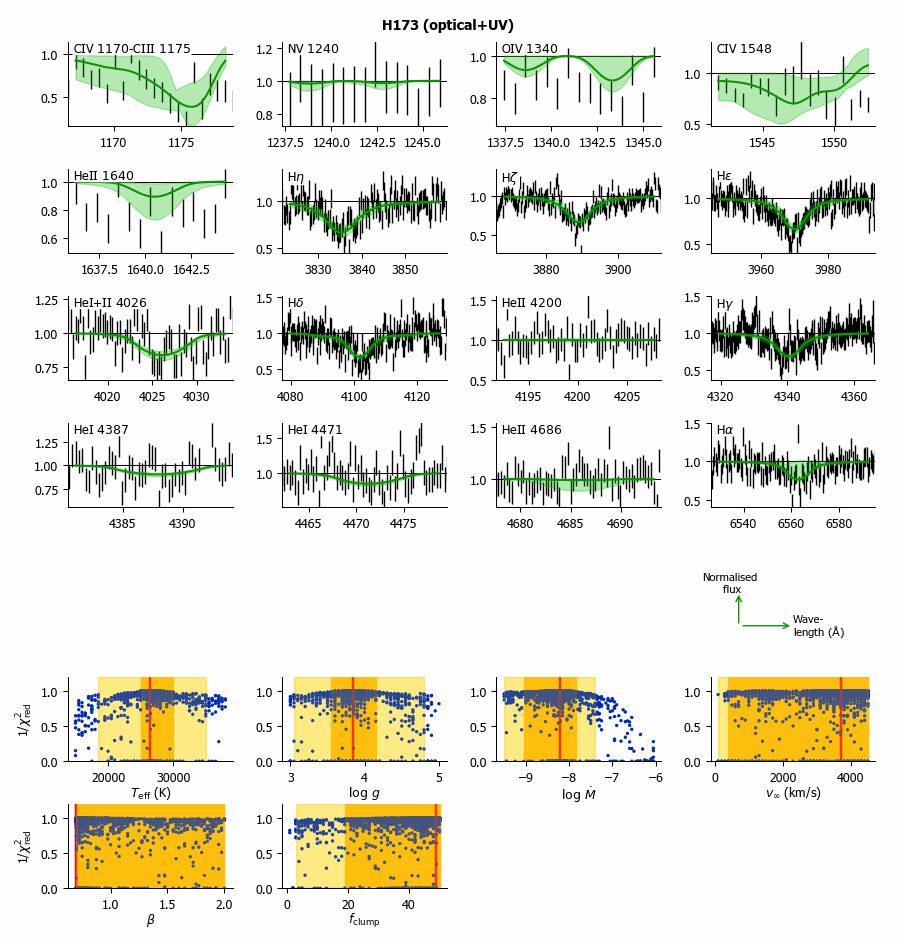}
    \caption{\PyGA output summary for the optical+UV run of H173 (as \Cref{fig:fitspec_example_H35}).}
    \label{fig:outGA_H173_UV}
\end{figure*}

\clearpage 

\end{appendices}

\end{document}